\documentclass[superscriptaddress,twocolumn,showpacs,
amssymb,amsmath,nobibnotes,aps,prd,%
nofootinbib]{revtex4-1}
\pdfoutput=1
\usepackage{graphicx,subfigure,bm,color,psfrag,hyperref}
\usepackage{amsfonts}
\usepackage{lipsum}
\usepackage{mathtools}
\usepackage{verbatim}
\usepackage[normalem]{ulem}
\usepackage[dvipsnames]{xcolor}
\hypersetup{colorlinks,linkcolor={blue},citecolor={red},urlcolor={violet}}

\begin{document}

\title{Revealing the effects of curvature on the cosmological models}

\author{Weiqiang Yang}
\email{d11102004@163.com}
\affiliation{Department of Physics, Liaoning Normal University, Dalian, 116029, P. R. China}

\author{William Giar\`e}
\email{william.giare@uniroma1.it}
\affiliation{Galileo Galilei Institute for theoretical physics, Centro Nazionale INFN di Studi Avanzati\\ Largo Enrico Fermi 2,  I-50125, Firenze, Italy}
\affiliation{INFN Sezione di Roma, P.le A. Moro 2, I-00185, Roma, Italy} 

\author{Supriya Pan}
\email{supriya.maths@presiuniv.ac.in}
\affiliation{Department of Mathematics, Presidency University, 86/1 College Street, Kolkata 700073, India}
\affiliation{Institute of Systems Science, Durban University of Technology, PO Box 1334, Durban 4000, Republic of South Africa}

\author{Eleonora Di Valentino}
\email{e.divalentino@sheffield.ac.uk}
\affiliation{School of Mathematics and Statistics, University of Sheffield, Hounsfield Road, Sheffield S3 7RH, United Kingdom}

\author{Alessandro Melchiorri}
\email{alessandro.melchiorri@roma1.infn.it}
\affiliation{Physics Department and INFN, Universit\`a di Roma ``La Sapienza'', Ple Aldo Moro 2, 00185, Rome, Italy}

\author{Joseph Silk}
\email{silk@iap.fr}
\affiliation{Institut d'Astrophysique de Paris (UMR7095: CNRS \& UPMC- Sorbonne Universities), F-75014, Paris, France}
\affiliation{Department of Physics and Astronomy, The Johns Hopkins University Homewood Campus, Baltimore, MD 21218, USA}
\affiliation{BIPAC, Department of Physics, University of Oxford, Keble Road, Oxford OX1 3RH, UK}

%--------------------------------------------------------------
\begin{abstract}
In this paper we consider the effects of adding curvature in extended cosmologies involving a free-to-vary neutrino sector and different parametrizations of Dark Energy (DE). We make use of the Planck 2018 cosmic microwave background temperature and polarization data, Baryon Acoustic Oscillations and Pantheon type Ia Supernovae data. Our main result is that a non-flat Universe cannot be discarded in light of the current astronomical data, because we find an indication for a closed Universe in most of the DE cosmologies explored in this work. On the other hand, forcing the Universe to be flat can significantly bias the constraints on the equation of state of the DE component and its dynamical nature.  
\end{abstract}

%----------------------------------------------------------------
\maketitle
%------------------------------------------------------------
\tableofcontents{}
%------------------------------------------------------------
\section{Introduction}

On large scales, our Universe is almost homogeneous and isotropic, but as far as its curvature is concerned we cannot firmly conclude that it is spatially flat~\cite{DiValentino:2020srs}. Even though the observational probes in the past years were in agreement with spatial flatness~\cite{Gaztanaga:2008de,Mortonson:2009nw,Suyu:2013kha,LHuillier:2016mtc,Chudaykin:2020ghx,Acquaviva:2021jov}, however, this result has recently been questioned by some of the experiments. The investigations with the Cosmic Microwave Background (CMB) temperature and polarization spectra from Planck 2018 team using the baseline \emph{Plik} likelihood are suggesting that our Universe could have a closed geometry at more than three standard deviations~\cite{Aghanim:2018eyx,Handley:2019tkm,DiValentino:2019qzk,DiValentino:2020hov} ($4\sigma$ when the physical curvature density is considered~\cite{Semenaite:2022unt}), and such indications are coming mainly from the temperature data affected by an otherwise inexplicable excess of lensing ($A_{\rm lens}$ problem~\cite{Calabrese:2008rt,DiValentino:2019dzu,DiValentino:2020hov}). Furthermore, also from the complementary examinations using the alternative \emph{CamSpec} likelihood~\cite{Efstathiou:2020wem,Efstathiou:2019mdh}, the closed geometry of the Universe is supported at more than $99 \%$ CL, and the indication seems to persist even considering the CMB temperature-only data of the new Planck PR4 analysis~\cite{Rosenberg:2022sdy}. Additionally, an indication for a closed universe is also present in the BAO data, using Effective Field Theories of Large Scale Structure, once the (hidden) assumptions of flatness (in the fiducial cosmology, in the reconstruction process, and in the covariance matrix) are removed from the beginning~\cite{Glanville:2022xes}. These outcomes challenged the assumptions of flatness present in the standard $\Lambda$CDM model and sparked a debate about the flatness of our Universe. In fact, if there is no solid theoretical argument, then observations in agreement with a small curvature cannot be used as a proof for a spatially flat universe~\cite{Anselmi:2022uvj}, and  the assumption of a spatially flat Universe on the spatially curved Universe may significantly affect the cosmological parameters \cite{Dossett:2012kd}.
Therefore, in this article we want to investigate, in a systematic way, some well known and classical cosmological models, leaving the curvature of the Universe as a free parameter and allow the observational data to pick up the best possibilities, removing possible biases due to the flatness assumption. 
Moreover, we are also interested in investigating what happens to the current cosmological tensions~\cite{Abdalla:2022yfr} when the curvature of the Universe is allowed into the picture (also see~\cite{Cao:2022ugh,Qi:2022sxm,Cao:2021irf,Cao:2021cix,Cao:2021zpf,Cao:2021ldv,Qi:2020rmm,Cao:2020jgu,Wang:2019yob}). In particular, we want to see what happens to the value of the Hubble constant and the $>5\sigma$ tension existing between the $\Lambda$CDM based Planck 2018 estimate~\cite{Aghanim:2018eyx} and the SH0ES (Supernovae and $H_0$ for the Equation of State of dark energy) measurement~\cite{Riess:2021jrx,Riess:2022mme}, known as the Hubble tension~\cite{DiValentino:2020zio,DiValentino:2021izs} (see also Refs.~\cite{DiValentino:2017iww,Kumar:2017dnp,Verde:2019ivm,Knox:2019rjx,Jedamzik:2020zmd,DiValentino:2017oaw,Yang:2018euj,Yang:2018uae,Pan:2019jqh,Yang:2019uog,Pan:2019gop,Poulin:2018cxd,Yang:2018qmz,Pan:2020bur,DiValentino:2019ffd,DiValentino:2019jae,Yao:2020pji,Lucca:2020zjb,Blinov:2020uvz,Anchordoqui:2021gji,Karwal:2021vpk,Freese:2021rjq,Perivolaropoulos:2021jda,Schoneberg:2021qvd,Reeves:2022aoi,Colgain:2022nlb,Naidoo:2022rda,Cruz:2022oqk,Escudero:2022rbq,Gomez-Valent:2022bku,Colgain:2022rxy,Pan:2019jqh,Yang:2019jwn,Pan:2019gop,Yang:2019uog,Yang:2020zuk,Pan:2020bur,Anchordoqui:2021gji,DiValentino:2021rjj,Yang:2022csz} aiming to alleviate the $H_0$ tension in various ways), together with the $>3\sigma$ tension in the $S_8$ parameter~\cite{DiValentino:2020vvd} defined as a combination of the amplitude of the matter power spectrum $\sigma_8$ with the matter density at present $\Omega_{m0}$ ($S_8 = \sigma_8\sqrt{\Omega_{m0}/0.3}$)  between the CMB data (within the $\Lambda$CDM assumption) and the weak lensing experiments~\cite{Heymans:2020gsg,KiDS:2020ghu,DES:2021vln,DES:2022ygi} (see Refs. \cite{FrancoAbellan:2020xnr,FrancoAbellan:2021sxk,Lucca:2021dxo,deAraujo:2021cnd,Clark:2021hlo,Anchordoqui:2021gji,BeltranJimenez:2021wbq,Heimersheim:2020aoc,DiValentino:2017oaw,Poulin:2022sgp,Amon:2022azi,Escudero:2022rbq} offering a possible route to alleviate the $S_8$ tension in various alternatives to $\Lambda$CDM).

Following this approach, in this article, we have considered a variety of cosmological models in the curved background (i.e. allowing the curvature of the Universe as a free parameter), namely, the standard $\Lambda$CDM, $w$CDM, the Chevallier-Polarski-Linder parametrization~\cite{Chevallier:2000qy,Linder:2002et}, where the dark energy has a dynamical equation of state parameter, and lastly a recently introduced emergent dark energy model, known as the phenomenologically emergent dark energy model~\cite{Li:2019yem,Pan:2019hac}~\footnote{Apart from these dark energy parametrizations, a large variants of works aiming to probe the dynamical nature of the dark energy are available in the literature, see for instance Refs.  \cite{Perenon:2022fgw,Yang:2021eud,Yang:2021flj,Menci:2020ybl,Zhao:2020ole,Zimdahl:2019pqg,Du:2018tia,Vagnozzi:2018jhn,Li:2018nlh,DiValentino:2017zyq,Pan:2017zoh,Yang:2017alx,Yang:2017amu,Zhao:2017cud,Ma:2011nc}.},  and their various extensions  by allowing a free-to-vary neutrino sector characterized by the total neutrino mass and the effective number of relativistic degrees of freedom. The models, together with their extensions in the neutrino sector, 
have been constrained consistently using the CMB data from Planck 2018, Baryon Acoustic Oscillations distance measurements, and the Pantheon sample of the Type Ia Supernovae. This work is complementary to similar approaches used in Refs.~\cite{DiValentino:2019dzu,DiValentino:2020hov,Anchordoqui:2021gji,DiValentino:2022oon}, but differs in the light of combination of parameters explored, observational datasets, and the extended analysis in the dark energy sector. In particular, for the first time, we are including in the analysis a dynamical dark energy equation of state together with a  non-zero curvature parameter that is free to vary.

The paper has been structured as follows. In \autoref{sec-2}, we introduce the framework we are exploring; in \autoref{sec-data} we describe the observational data and the method used for the statistical analysis; in \autoref{sec-results} we present our main results. Eventually, we end in \autoref{sec-sumamry} with a summary of them and our Conclusions.

\section{Dark energy in a Curved Universe}
\label{sec-2}

As argued in the ``introduction'', we start with the homogeneous and isotropic spacetime characterized by the 
Friedmann-Lema\^{i}tre-Robertson-Walker (FLRW) line element given by 
\begin{equation}
    ds^2  = -dt^2 + a^2 (t) \left[\frac{dr^2}{1-Kr^2} + r^2 (d\theta^2 + \sin^2 \theta d\phi^2)\right],
\end{equation}
where $a (t)$ is the expansion scale factor of the universe and $K$ denotes its curvature scalar. The values $K = 0, +1, -1$, respectively correspond to a spatially flat, closed and an open geometry of the universe. In the context of General Relativity, we assume that the matter distribution of the universe is minimally coupled to the gravity and none of the fluids are interacting with any component of it. Then the dynamics of the universe can be described from the Einstein's gravitational equations 

\begin{eqnarray}
H^2  = - \frac{K}{a(t)^2} + \frac{8 \pi G}{3} \; \sum \rho_i, \label{EFE1}\\
2 \dot{H} + 3 H^2  = -  \frac{K}{a(t)^2} -  8 \pi G \; \sum p_i,\label{EFE2},
\end{eqnarray}
where an overhead dot represents the time derivative; $H \equiv \dot{a} (t)/a(t)$ is the Hubble function of the FLRW universe; $\rho_{i}$ and $p_{i}$ are, respectively, the energy density and pressure term of the $i$-th component; $G$ is the Newton's gravitational constant. Now introducing the critical density, $\rho_c = 3H^2/8 \pi G$, eqn. (\ref{EFE1}) can be expressed as $1 = \Omega_K + \sum_i \Omega_i$, where $\Omega_i = \rho_i/\rho_c$ is termed as the density parameter for the $i$-th fluid  and $\Omega_K = -K/(a^2 H^2)$ is termed as the curvature density parameter.\footnote{At this point we mention that unless otherwise specified, ``$0$'' attached to any quantity will refer to its present value; for example, $\Omega_{K0}$ refers to the present value of the curvature density parameter. }  Notice that the evolution of the energy density of each component can be found by solving the balance equation, 
\begin{eqnarray}\label{cons-eqn}
\sum_{i} \dot{\rho}_i + 3 H \; \sum_{i} (p_i + \rho_i) = 0, 
\end{eqnarray}
which can be obtained from the field equations (\ref{EFE1}) and (\ref{EFE2}). Since the fluids are not interacting with each other, therefore, we have $\dot{\rho}_i + 3 H\; (p_i + \rho_i) = 0$ for each $i$.  
Here,  we consider that the total energy density of the Universe is distributed among radiation ($r$), baryons ($b$), neutrinos ($\nu$), cold dark matter (${\rm CDM}$) and a dark energy (${\rm DE}$) component. For $i= r, b, \nu, {\rm CDM}, {\rm DE}$, one can find the evolution law of the energy density by solving the conservation equation for the $i$-th fluid. 
Now, concerning the dark energy sector, using the conservation equation (\ref{cons-eqn}) for its equation-of-state $w_{\rm DE} = p_{\rm DE}/\rho_{\rm DE}$, one can find the evolution of the energy density as 

\begin{eqnarray}
\rho_{\rm DE} = \rho_{\rm DE,0} \exp \left[ 3 \; \int_{0}^{z} \frac{1+w_{\rm DE}(z^\prime)}{1+z^\prime} dz^\prime \right],
\end{eqnarray}
where $\rho_{\rm DE,0}$ is the present value of the dark energy density. Notice that for $w_{\rm DE} = -1$, the energy density becomes constant which represents a cosmological constant in the Universe. 
As the nature of dark energy is not yet clearly understood to anyone, thus, one might be interested to see how different variants of the dark energy model can affect the evolution of the expansion history along with the curvature of the Universe.  
As there are many proposals of dark energy in the literature, here, we pick up some well known choices as follows: 

\begin{itemize}
    \item The cosmological constant as the dark energy candidate characterized by its equation of state $w_{\rm DE} = -1$. 
    
    \item The dark energy with a constant equation of state other than the cosmological constant (i.e. $w_{\rm DE} \neq -1$).\footnote{Let us comment that  from now on we shall use $w$ instead of $w_{\rm DE}$ and we label the corresponding cosmological model as $w$CDM (instead of $w_{\rm DE}$CDM) as already done in the literature. }
    
    \item The case when the dark energy equation of state is dynamical and it assumes the Chevallier-Polarski-Linder parametrization~\cite{Chevallier:2000qy,Linder:2002et}:
    \begin{eqnarray}
    w_{\rm DE} = w_0 + w_a (1-a)~, 
    \end{eqnarray}
    where $w_0$ is the present value of $w_{\rm DE}$ and $w_a = dw_{\rm DE}/da$ at $a = 1$ is another free parameter in this parametrization.\footnote{Note that the corresponding cosmological model is labeled as $w_0w_a$CDM.}

    \item Finally, we consider a very exotic dark energy model, namely the phenomenologically emergent dark energy (PEDE) where DE does not have any effective presence in the past but it emerges at late times. In this model scenario, the dark energy density ($\rho_{\rm DE}$) is directly parametrized which takes the form \cite{Li:2019yem,Pan:2019hac}
    
    \begin{eqnarray}
    \frac{\rho_{\rm DE}}{\rho_{\rm c,0}} = \Omega_{\rm DE,0} \Bigl[1- \tanh (\log_{10} (1+z)) \Bigr]~,
    \end{eqnarray}
    where $\rho_{\rm c, 0}$ is the critical energy density at present time and $\Omega_{\rm DE,0}$ is the present value of the DE density parameter $\Omega_{\rm DE} \equiv \rho_{\rm DE}/\rho_{\rm crit}$. We note that even if this parametrization has a dynamical nature, this does not offer any additional degree of freedom, or free parameter. 
    
\end{itemize}

Having all the models, one can now constrain them in presence of the curvature of the Universe. In the next section we describe the observational datasets and the statistical process aimed to confront the models with the data.

\section{Observational data and statistical process}
\label{sec-data}

In this section we describe the main observational datasets that have been used to constrain the underlying cosmological models and the methodology of the statistical simulations. In what follows we first describe the observational datasets. 

\begin{itemize}

\item {\bf Cosmic Microwave Background (CMB) Observations:} We consider the measurements of the CMB observations from the final release of the Planck 2018 team. In particular, we have used the CMB temperature and polarization power spectra, that means  
 {\rm plikTTTEEE+lowl+lowE} from Planck 2018~\cite{Aghanim:2018eyx,Aghanim:2019ame}.

\item {\bf Baryon Acoustic Oscillations (BAO):} Various measurements of the BAO data have been used in this work. In particular we have used the measurements from 
6dFGS~\cite{Beutler:2011hx}, SDSS-MGS~\cite{Ross:2014qpa}, and 
BOSS DR12~\cite{Alam:2016hwk} as considered by the Planck 2018 team~\cite{Aghanim:2018eyx}. 

\item {\bf Pantheon sample from Supernovae Type Ia (SNIa):} Along with the CMB and BAO measurements, another potential cosmological probe is the SNIa, the fist observational data indicating the accelerating expansion of the Universe.  Here we took a recent collection of SNIa, known as the Pantheon sample~\cite{Scolnic:2017caz} distributed in the redshift interval $z \in [0.01, 2.3]$. 

\end{itemize}

Now we discuss the parameter space of each cosmological scenario  considered in this article.  We have considered the extensions of $\Lambda$CDM, $w$CDM, $w_0w_a$CDM and PEDE by allowing the curvature of the Universe ($\Omega_{K}$) and a free-to-vary neutrino sector. The free-to-vary neutrino sector has two free parameters:  $M_{\nu} \doteq \sum m_{\nu}$ (the total neutrino mass) and $N_{\rm eff}$ (the effective number of relativistic degrees of freedom ).  The scenarios 
$\Lambda$CDM $+$ $\Omega_K$, $\Lambda$CDM $+$ $\Omega_K$ $+$ $M_{\nu}$, $\Lambda$CDM $+$ $\Omega_K$ $+$ $N_{\rm eff}$ and $\Lambda$CDM $+$ $\Omega_K$ $+$ $M_{\nu}$ $+$ $N_{\rm eff}$,  respectively  contain  7, 8, 8, and 9 free parameters as follows:
\begin{eqnarray}
&&\mathcal{L}_1 \equiv\Bigl\{\Omega_{b}h^2, \Omega_{c}h^2, 100\theta_{MC}, \tau, n_{s}, \log[10^{10}A_{s}], \Omega_K \Bigr\}, \nonumber \\ 
&&\mathcal{L}_2 \equiv \mathcal{L}_1 \cup \{M_{\nu}\},\nonumber \\ 
&&\mathcal{L}_3 , \equiv \mathcal{L}_1 \cup \{N_{\rm eff}\},\nonumber \\ 
&&\mathcal{L}_4 \equiv \mathcal{L}_1 \cup \{M_{\nu}\} \cup \{N_{\rm eff}\}. \nonumber 
\end{eqnarray}
The scenarios 
$w$CDM $+$ $\Omega_K$, $w$CDM $+$ $\Omega_K$ $+$ $M_{\nu}$, $w$CDM $+$ $\Omega_K$ $+$ $N_{\rm eff}$ and $w$CDM $+$ $\Omega_K$ $+$ $M_{\nu}$ $+$ $N_{\rm eff}$, respectively  contain  8, 9, 9, and 10 free parameters as follows:
\begin{eqnarray}
&&\mathcal{W}_1 \equiv\Bigl\{\Omega_{b}h^2, \Omega_{c}h^2, 100\theta_{MC}, \tau, n_{s}, \log[10^{10}A_{s}], \Omega_K, w \Bigr\}, \nonumber \\ 
&&\mathcal{W}_2 \equiv \mathcal{W}_1 \cup \{M_{\nu}\},\nonumber \\ 
&&\mathcal{W}_3 , \equiv \mathcal{W}_1 \cup \{N_{\rm eff}\},\nonumber \\ 
&&\mathcal{W}_4 \equiv \mathcal{W}_1 \cup \{M_{\nu}\} \cup \{N_{\rm eff}\}. \nonumber 
\end{eqnarray}
The scenarios 
$w_0w_a$CDM $+$ $\Omega_K$, $w_0w_a$CDM $+$ $\Omega_K$ $+$ $M_{\nu}$, $w_0w_a$CDM $+$ $\Omega_K$ $+$ $N_{\rm eff}$ and $w_0w_a$CDM $+$ $\Omega_K$ $+$ $M_{\nu}$ $+$ $N_{\rm eff}$, respectively  contain  9, 10, 10, and 11 free parameters as follows:
\begin{eqnarray}
&&\mathcal{C}_1 \equiv\Bigl\{\Omega_{b}h^2, \Omega_{c}h^2, 100\theta_{MC}, \tau, n_{s}, \log[10^{10}A_{s}], \Omega_K, w_0, w_a \Bigr\}, \nonumber \\ 
&&\mathcal{C}_2 \equiv \mathcal{C}_1 \cup \{M_{\nu}\},\nonumber \\ 
&&\mathcal{C}_3 , \equiv \mathcal{C}_1 \cup \{N_{\rm eff}\},\nonumber \\ 
&&\mathcal{C}_4 \equiv \mathcal{C}_1 \cup \{M_{\nu}\} \cup \{N_{\rm eff}\}, \nonumber 
\end{eqnarray}
Finally, the scenarios, 
PEDE $+$ $\Omega_K$, PEDE $+$ $\Omega_K$ $+$ $M_{\nu}$, PEDE $+$ $\Omega_K$ $+$ $N_{\rm eff}$ and PEDE $+$ $\Omega_K$ $+$ $M_{\nu}$ $+$ $N_{\rm eff}$, respectively  contain  7, 8, 8, and 9 free parameters as follows:
\begin{eqnarray}
&&\mathcal{E}_1 \equiv\Bigl\{\Omega_{b}h^2, \Omega_{c}h^2, 100\theta_{MC}, \tau, n_{s}, \log[10^{10}A_{s}], \Omega_K \Bigr\}, \nonumber \\ 
&&\mathcal{E}_2 \equiv \mathcal{E}_1 \cup \{M_{\nu}\},\nonumber \\ 
&&\mathcal{E}_3 , \equiv \mathcal{E}_1 \cup \{N_{\rm eff}\},\nonumber \\ 
&&\mathcal{E}_4 \equiv \mathcal{E}_1 \cup \{M_{\nu}\} \cup \{N_{\rm eff}\}. \nonumber 
\end{eqnarray}

To constrain the parameter spaces of the cosmological scenarios described above, we have used the publicly available Markov Chain Monte Carlo code \texttt{CosmoMC}~\cite{Lewis:2002ah,Lewis:1999bs} (see \url{http://cosmologist.info/cosmomc/}), which supports the Planck 2018 likelihood~\cite{Aghanim:2019ame} and additionally has a convergence diagnostic following the Gelman-Rubin statistics~\cite{Gelman:1992zz}. We adopt on the parameters discussed above the flat priors listed in \autoref{tab.priors}.

\begin{table}[h]
\begin{center}
\renewcommand{\arraystretch}{1.5}
\begin{tabular}{c|cccc}
\hline
\textbf{Parameter}                    & \textbf{Prior} \\
\hline\hline
$\Omega_{\rm b} h^2$         & $[0.005\,,\,0.1]$\\
$\Omega_{\rm c} h^2$       & $[0.001\,,\,0.99]$\\
$100\,\theta_{\rm {MC}}$             & $[0.5\,,\,10]$\\
$\tau$                       & $[0.01\,,\,0.8]$\\
$\log(10^{10}A_{\rm s})$         & $[1.61\,,\,3.91]$\\
$n_s$                        & $[0.8\,,\, 1.2]$\\
$\Omega_K$       &  $[-0.3\,,\, 0.3]$\\
$w_0$                  &   $[-3\,,\,0]$\\
$w_a$                  &  $[-3\,,\,3]$\\
$M_\nu$       &  $[0\,,\, 1]$\\
$N_{\rm eff}$       &  $[2.2\,,\, 4]$\\
\hline\hline
\end{tabular}
\end{center}
\caption{List of the flat priors assumed on the independent parameters. }
\label{tab.priors}
\end{table}

\section{Observational Constraints}
\label{sec-results}

This section is entirely devoted to describing the observational constraints on various dark energy scenarios in a curved Universe. In what follows we report the constraints on each model and its various extensions in a dedicated way.   

\subsection{Non-flat $\Lambda$CDM and its extensions}

In this section we describe the various extensions of non-flat $\Lambda$CDM scenario. We note that for all extensions we have made use of the same datasets and their combinations and for each parameter either we report its 68\% and 95\% CL constraints or the upper/lower limits at 68\% and 95\% CL.  

\begingroup                                                                                                                     
\squeezetable                                                                                                                   
\begin{center}                                                                                                                  
\begin{table*}[htb]                                                                                                                  \resizebox{\textwidth}{!}{  
\begin{tabular}{cccccccc}                                                                                                           
\hline\hline                                                                                                                    
Parameters & CMB & CMB+BAO &  CMB+Pantheon & CMB+BAO+Pantheon    \\ \hline
$\Omega_c h^2$ & $    0.1181_{-    0.0015-    0.0029}^{+    0.0015+    0.0029}$ & $    0.1197_{-    0.0015-    0.0028}^{+    0.0014+    0.0029}$ & $    0.1191_{-    0.0015-    0.0029}^{+    0.0015+    0.0030}$ & $    0.1196_{-    0.0014-    0.0028}^{+    0.0014+    0.0028}$   \\

$\Omega_b h^2$ & $    0.02261_{-    0.00017-    0.00033}^{+    0.00017+    0.00033}$ & $    0.02239_{-    0.00015-    0.00031}^{+    0.00017+    0.00030}$  & $    0.02245_{-    0.00016-    0.00031}^{+    0.00016+    0.00033}$ & $    0.02240_{-    0.00015-    0.00030}^{+    0.00015+    0.00030}$   \\

$100\theta_{MC}$ & $    1.04117_{-    0.00033-    0.00067}^{+    0.00034+    0.00065}$ & $    1.04095_{-    0.00033-    0.00063}^{+    0.00032+    0.00062}$  & $ 1.04103_{-    0.00031-    0.00062}^{+    0.00032+    0.00061}$ & $    1.04098_{-    0.00031-    0.00061}^{+    0.00031+    0.00062}$     \\

$\tau$ & $    0.0487_{-    0.0076-    0.016}^{+    0.0079+    0.016}$ &   $    0.0551_{-    0.0076-    0.015}^{+    0.0076+    0.016}$   & $    0.054_{-    0.0076-    0.015}^{+    0.0076+    0.016}$ & $    0.055_{-    0.0082-    0.015}^{+    0.0074+    0.017}$   \\

$n_s$ & $    0.9706_{-    0.0046-    0.0095}^{+    0.0047+    0.0091}$ &  $    0.9660_{-    0.0045-    0.0088}^{+    0.0045+    0.00894}$   & $    0.9676_{-    0.0044-    0.0090}^{+    0.0045+    0.0093}$ & $    0.9663_{-    0.0045-    0.0087}^{+    0.0044+    0.0088}$     \\

${\rm{ln}}(10^{10} A_s)$ & $    3.028_{-    0.016-    0.035}^{+    0.016+    0.032}$ & $    3.045_{-    0.016-    0.031}^{+    0.016+    0.032}$  & $    3.042_{-    0.016-    0.032}^{+    0.016+    0.033}$ & $    3.045_{-    0.016-    0.031}^{+    0.016+    0.034}$ \\

$\Omega_{K0}$ & $   -0.043_{-    0.015-    0.034}^{+    0.018+    0.033}$ & $    0.0008_{-    0.0019-    0.0039}^{+    0.0019+    0.0040}$  & $   -0.0060_{-    0.0054-    0.012}^{+    0.0063+    0.011}$  & $    0.0008_{-    0.0019-    0.0038}^{+    0.0019+    0.0039}$    \\

$\Omega_{m0}$ & $    0.481_{-    0.067-    0.12}^{+    0.057+    0.13}$ & $    0.3096_{-    0.0066-    0.012}^{+    0.0066+    0.013}$   & $    0.336_{-    0.024-    0.041}^{+    0.020+    0.044}$ & $    0.309_{-    0.0068-    0.012}^{+    0.0062+    0.013}$  \\

$\sigma_8$ & $    0.775_{-    0.014-    0.030}^{+    0.016+    0.028}$ & $    0.8109_{-    0.0085-    0.017}^{+    0.0081+    0.016}$  & $    0.805_{-    0.0097-    0.019}^{+    0.0097+    0.019}$  & $    0.811_{-    0.0083-    0.016}^{+    0.0081+    0.016}$  \\

$H_0$ [Km/s/Mpc] & $   54.5_{-    3.9-    7.2}^{+    3.3+    7.5}$ & $   67.90_{-    0.67-    1.3}^{+    0.67+    1.3}$   & $   65.2_{-    2.2-    4.1}^{+    2.1+    4.4}$ & $   67.97_{-    0.65-    1.3}^{+    0.65+    1.3}$      \\

$S_8$ & $    0.978_{-    0.048-    0.096}^{+    0.047+    0.094}$ & $    0.824_{-    0.014-    0.024}^{+    0.012+    0.026}$   & $    0.850_{-    0.024-    0.047}^{+    0.024+    0.048}$ & $    0.823_{-    0.012-    0.024}^{+    0.012+    0.025}$   \\

$r_{\rm{drag}}$ [Mpc] & $  147.34_{-    0.31-    0.61}^{+    0.31+    0.60}$ & $  147.16_{-    0.30-    0.62}^{+    0.33+    0.58}$ & $  147.24_{-    0.31-    0.61}^{+    0.31+    0.61}$ &    $  147.177_{-    0.30-    0.59}^{+    0.30+    0.60}$  \\
\hline                                                  
%$\chi^2$  & 2762.134  &  2779.196   &  3806.448  &  3812.704  \\
\hline 

\end{tabular}                                                    }                                                               
\caption{68\% and 95\% CL constraints on various free and derived parameters of the $\Lambda$CDM $+$ $\Omega_K$ scenario for several observational datasets. }
\label{tab:LCDM}                                                                                                   
\end{table*}                                                                                                                     
\end{center}                                                                                                                    
\endgroup   

\begin{figure*}
    \centering
    \includegraphics[width=0.8\textwidth]{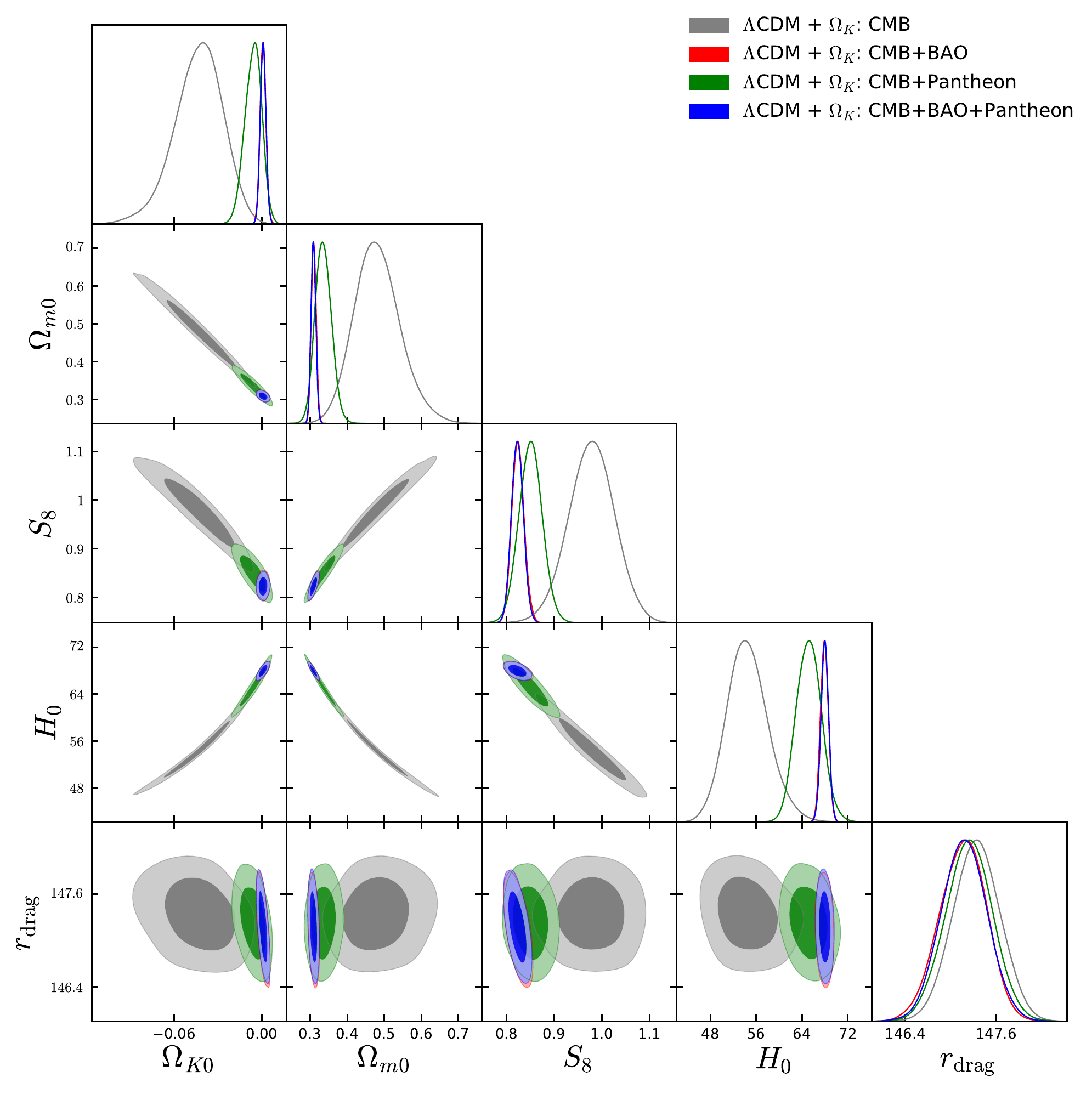}
    \caption{1-dimensional marginalized posterior distributions and the 2-dimensional joint contours for the most relevant parameters of cosmological scenario  $\Lambda$CDM $+$ $\Omega_K$ for various observational datasets.  }
    \label{fig:LCDM-Omegak}
\end{figure*}

\subsubsection{$\Lambda$CDM $+$ $\Omega_K$}

In \autoref{tab:LCDM} we show the constraints on the $\Lambda$CDM $+$ $\Omega_K$ scenario for various datasets, and in \autoref{fig:LCDM-Omegak} the 1D and 2D contour plots for a few important parameters.

We start investigating the constraints from the CMB alone case which are shown in the second column of \autoref{tab:LCDM}. 
We see that for CMB alone, a strong indication of a closed Universe is found at more than 95\% CL ($\Omega_{K0} =  -0.043_{- 0.034}^{+ 0.033}$). The Hubble constant takes a very low value ($H_0 = 54.5_{-3.9}^{+    3.3}$ km/s/Mpc at 68\% CL) increasing Hubble tension, and due to the strong correlation
between $H_0$, $\Omega_K$ and $\Omega_{m0}$, the matter density $\Omega_{m0}$ takes a higher value as expected. Also we notice that the tension in $S_8$ increases for this case. So, effectively, we see that CMB alone gives a strong indication for a closed Universe at the expense of increased tensions in both $H_0$ and $S_8$.

When BAO data are added to CMB, that means when we consider the combination CMB+BAO (see the third column of \autoref{tab:LCDM}), we see that the $\Omega_K$ becomes consistent to the spatially flat Universe and the Hubble constant goes up leading to Planck's $\Lambda$CDM value. However, this result is obtained with the combination of datasets in disagreement at more than $3\sigma$~\cite{Handley:2019tkm,DiValentino:2019qzk,Vagnozzi:2020rcz}, making the combination CMB+BAO not safe and the results are not completely reliable. 
The results for CMB+Pantheon and CMB+BAO+Pantheon remain almost similar unlike a mild lower value of the Hubble constant in the CMB+Pantheon case ($H_0 = 65.2_{-4.1}^{+2.1}$ km/s/Mpc at 68\% CL for CMB+Pantheon) but in both the cases, the spatial flatness is found to be consistent with these data. 

\begingroup                                                                                                  
\squeezetable                                                                                                                   
\begin{center}                                                                                                                  
\begin{table*}[htb]                                                                                                                  \resizebox{\textwidth}{!}{   
\begin{tabular}{ccccccccc}                                                                                                            
\hline\hline                                                                                                                    
Parameters & CMB & CMB+BAO & CMB+Pantheon &  CMB+BAO+Pantheon\\   \hline

$\Omega_c h^2$ & $    0.1183_{-    0.0015-    0.0030}^{+    0.0015+    0.0030}$ & $    0.1197_{-    0.0014-    0.0027}^{+    0.0014+    0.0027}$   & $    0.1190_{-    0.0014-    0.0028}^{+    0.0014+    0.0028}$ &  $    0.1196_{-    0.0013-    0.0028}^{+    0.0014+    0.0029}$    \\

$\Omega_b h^2$ & $    0.02254_{-    0.00018-    0.00035}^{+    0.00018+    0.00036}$ &  $    0.02238_{-    0.00015-    0.00030}^{+    0.00015+    0.00030}$   & $    0.02246_{-    0.00016-    0.00032}^{+    0.00016+    0.00032}$ & $    0.02241_{-    0.00015-    0.00029}^{+    0.00015+    0.00031}$    \\

$100\theta_{MC}$ & $    1.04101_{-    0.00033-    0.00069}^{+    0.00034+    0.00068}$ & $    1.04094_{-    0.00031-    0.00063}^{+    0.00031+    0.00061}$   & $    1.04103_{-    0.00032-    0.00065}^{+    0.00032+    0.00063}$ &  $    1.04097_{-    0.00033-    0.00063}^{+    0.00033+    0.00064}$   \\

$\tau$ & $    0.0477_{-    0.0076-    0.016}^{+    0.0077+    0.016}$ & $    0.0544_{-    0.0075-    0.015}^{+    0.0074+    0.016}$   & $    0.0544_{-    0.0077-    0.015}^{+    0.0077+    0.016}$  & $    0.055_{-    0.0081-    0.015}^{+    0.0076+    0.016}$  \\

$n_s$ & $    0.9693_{-    0.0050-    0.0099}^{+    0.0050+    0.0098}$ & $    0.9659_{-    0.0045-    0.0087}^{+    0.0045+    0.0087}$  & $    0.9678_{-    0.0048-    0.0088}^{+    0.0045+    0.0093}$  & $    0.9663_{-    0.0046-    0.0088}^{+    0.0045+    0.0091}$     \\

${\rm{ln}}(10^{10} A_s)$ & $    3.026_{-    0.016-    0.035}^{+    0.016+    0.033}$ &  $    3.044_{-    0.016-    0.031}^{+    0.016+    0.032}$  & $    3.042_{-    0.017-    0.032}^{+    0.016+    0.034}$  & $    3.045_{-    0.016-    0.031}^{+    0.015+    0.032}$  \\

$\Omega_{K0}$ & $   -0.073_{-    0.022-    0.071}^{+    0.043+    0.060}$ &  $    0.0007_{-    0.0022-    0.0041}^{+    0.0020+    0.0044}$  & $   -0.0061_{-    0.0056-    0.012}^{+    0.0064+    0.011}$   &  $    0.0007_{-    0.0024-    0.0045}^{+    0.0020+    0.0048}$    \\

$\Omega_{m0}$ & $    0.65_{-    0.21-    0.28}^{+    0.10+    0.34}$ & $    0.3102_{-    0.0069-    0.014}^{+    0.0069+    0.014}$   & $    0.334_{-    0.023-    0.042}^{+    0.022+    0.044}$ &  $    0.309_{-    0.0066-    0.013}^{+    0.0067+    0.013}$   \\

$\sigma_8$ & $    0.698_{-    0.052-    0.12}^{+    0.078+    0.11}$ &  $    0.812_{-    0.010-    0.028}^{+    0.015+    0.025}$   &  $    0.807_{-    0.011-    0.029}^{+    0.015+    0.027}$ & $    0.812_{-    0.0098-    0.029}^{+    0.015+    0.025}$    \\

$H_0$ [Km/s/Mpc] & $   48.3_{-    5.9-   11}^{+    5.7 +   11}$ & $   67.84_{-    0.67-    1.3}^{+    0.67+    1.4}$ & $   65.3_{-    2.3-    4.4}^{+    2.2+    4.4}$ & $   67.97_{-    0.66-    1.3}^{+    0.66+    1.4}$  \\

$M_\nu$ [eV] & $  <0.46\,<0.79$ & $  <0.072\,<0.17 $  & $   <0.066\,<0.16 $ &  $  <0.071\,<0.17$   \\

$S_8$ & $    1.008_{-    0.053-    0.11}^{+    0.052+    0.10}$ &  $    0.826_{-  0.014-    0.030}^{+    0.014+    0.029}$   & $    0.851_{-    0.026-    0.053}^{+    0.027+    0.050}$ &  $    0.824_{-    0.013-    0.029}^{+    0.015+    0.029}$   \\

$r_{\rm{drag}}$ [Mpc] &  $  147.25_{-    0.32-    0.64}^{+    0.32+    0.63}$  &  $  147.16_{-    0.30-    0.58}^{+    0.30+    0.57}$  & $  147.26_{-    0.30-    0.61}^{+    0.29+    0.60}$ & $  147.17_{-    0.30-    0.61}^{+    0.30+    0.61}$   \\
\hline                                                  
%$\chi^2$ & 2763.428 &  2776.266  &  3806.436  &  3814.636 \\
\hline                                                              
\end{tabular}                                                   }                                                                
\caption{68\% and 95\% CL constraints on various free and derived parameters of the $\Lambda$CDM $+$ $\Omega_K$ $+$ $M_{\nu}$ scenario for several observational datasets. }  \label{tab:LCDM-Mnu}                                                                                                   
\end{table*}                                                                                                                     
\end{center}                                                                                                                    
\endgroup                           
 
 \begin{figure*}
    \centering
    \includegraphics[width=0.8\textwidth]{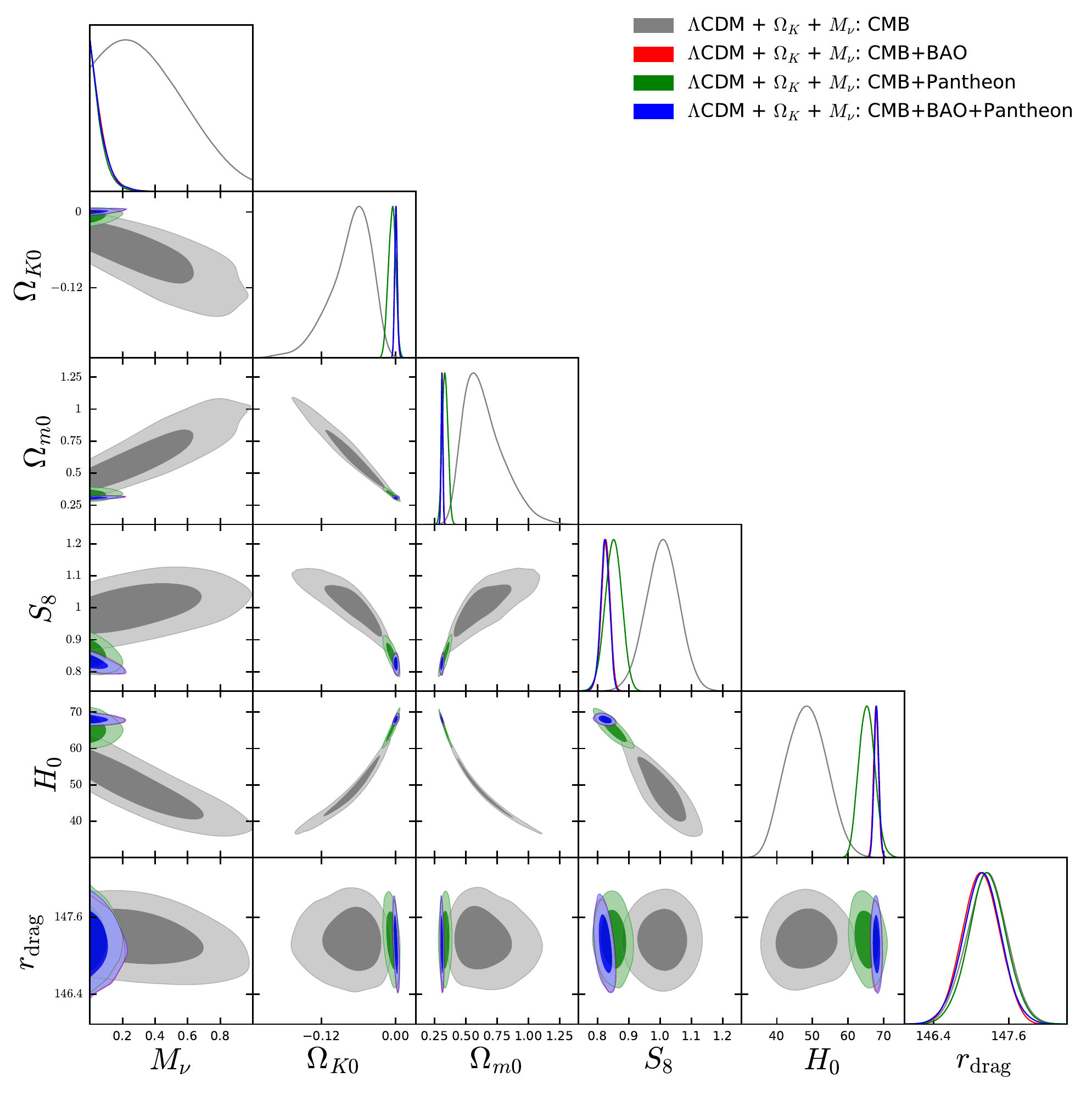}
    \caption{1-dimensional marginalized posterior distributions and the 2-dimensional joint contours for the most relevant parameters of cosmological scenario  $\Lambda$CDM $+$ $\Omega_K$ $+$ $M_{\nu}$ for various observational datasets.  }
    \label{fig:LCDM-Omegak-Mnu}
\end{figure*}

\subsubsection{$\Lambda$CDM $+$ $\Omega_K$ $+$ $M_{\nu}$}

The first non-flat $\Lambda$CDM extended scenario that we analyze involves massive neutrinos. We refer to this model as $\Lambda$CDM $+$ $\Omega_K$ $+$  $M_{\nu}$. %where the total neutrino mass $M_{\nu}\doteq \sum m_{\nu}$ is regarded as a free parameter of the cosmological model. 
In \autoref{tab:LCDM-Mnu} we provide the observational constraints on the parameters, while in \autoref{fig:LCDM-Omegak-Mnu} we show their 1D and 2D marginalized posterior distributions.

Also in this extended model, considering only the Planck satellite data on the CMB temperature anisotropies and polarization, we obtain a preference for a curved cosmological spacetime at more than 95\% CL,  with the constraint on the curvature parameter reading $\Omega_{K0}=-0.073^{+0.060}_{-0.071}$ at 95\% CL. As for the previous baseline case, this preference is strongly reduced when the Planck data are combined with the BAO datasets that are in tension with it, and their combination prefers spatial flatness within one standard deviation. As concerns the value of the expansion rate today, due to its degeneracy with the mass of the active neutrinos, in this extended parameter space the CMB measurements prefer further lower values of $H_0$ giving a less tight bound $H_0=48.3^{+5.7}_{-5.9}$ km/s/Mpc at 68\% CL, which is also in strong tension with the SH0ES independent measurement of $H_0$~\cite{Riess:2021jrx}, at the level of $\sim 4.3\sigma$. Combining the CMB measurements with the BAO and Pantheon data, the tension between the different datasets remains statistically significant. In particular the values inferred combining CMB and BAO, $H_0=67.84\pm0.67$ km/s/Mpc at 68\% CL, is similar to the result obtained in the flat $\Lambda$CDM model, while combining CMB and Pantheon we get $H_0=65.3^{+2.2}_{-2.3}$ km/s/Mpc at 68\% CL. Notice that both these values are in tension with the SH0ES collaboration at the level of $4.3\sigma$ and $3.2\sigma$, respectively. Furthermore, due to the strong anti-correlation between $H_0$ and the matter density parameter, lower values of $H_0$ prefer higher values of $\Omega_{m0}$, resulting in a severe tension also for the $S_8$ parameter. Comparing the bound of $S_8$ derived by the CMB measurements alone with respect to the constraint obtained including also the BAO and Pantheon data, we can observe a tension at the level of about $\sim3\sigma$.
Finally, concerning the neutrino masses, the CMB data alone constrain their total mass to be $M_{\nu}<0.79$ eV at 95\% CL, i.e. a factor more $3$ relaxed than the flat $\Lambda$CDM+$M_{\nu}$ scenario
However, neutrinos suppress structure formation at small scales and therefore astrophysical galaxy clustering measurements turn out to be crucial to improve the constraining power on their total mass. For this reason, the upper bounds on $M_{\nu}$ can be significantly improved to  $M_{\nu}<0.17$ eV (always at 95\% CL) including also the Baryon Acoustic Oscillation and the Pantheon datasets. This upper limit should be compared with that obtained for a flat $\Lambda$CDM+$M_{\nu}$ scenario, that is $M_{\nu}<0.13$ eV at 95\% CL for CMB+BAO. In other words, the assumption of flatness produces a much stronger upper limit on the total neutrino mass than when the curvature can vary, biasing  conclusions that are important to laboratory experiments.

\begingroup                                                                                                                     
\squeezetable                                                                                                                   
\begin{center}                                                                                                                  
\begin{table*}[htb]                                                                                                                   \resizebox{\textwidth}{!}{   
\begin{tabular}{ccccccccc}                                                                                                            
\hline\hline                                                                                                                    
Parameters & CMB & CMB+BAO & CMB+Pantheon &  CMB+BAO+Pantheon \\ \hline
$\Omega_c h^2$ & $    0.1181_{-    0.0030-    0.0060}^{+    0.0031+    0.0061}$ & $    0.1185_{-    0.0031-    0.0058}^{+    0.0030+    0.0062}$  & $    0.1187_{-    0.0030-    0.0058}^{+    0.0030+    0.0061}$  &  $  0.1185_{-    0.0032-    0.0058}^{+    0.0029+    0.0062}$    \\

$\Omega_b h^2$ & $    0.02260_{-    0.00026-    0.00049}^{+    0.00025+    0.00051}$ &  $    0.02232_{-    0.00023-    0.00046}^{+    0.00024+    0.00045}$   & $    0.02245_{-    0.00024-    0.00047}^{+    0.00024+    0.00047}$ &  $    0.02233_{-    0.00022-    0.00045}^{+    0.00022+    0.00045}$  \\

$100\theta_{MC}$ & $    1.04117_{-    0.00044-    0.00086}^{+    0.00044+    0.00088}$ & $    1.04110_{-    0.00044-    0.00086}^{+    0.00043+    0.00087}$   & $    1.04108_{-    0.00044-    0.00085}^{+    0.00044+    0.00088}$ &  $    1.04111_{-    0.00043-    0.00085}^{+    0.00042+    0.00089}$     \\

$\tau$ & $    0.0484_{-    0.0076-    0.018}^{+    0.0088+    0.017}$ &  $    0.0546_{-    0.0076-    0.015}^{+    0.0076+    0.016}$  & $    0.0545_{-    0.0083-    0.015}^{+    0.0074+    0.016}$ & $    0.055_{-    0.0076-    0.015}^{+    0.0077+    0.016}$  \\

$n_s$ & $    0.9703_{-    0.0094-    0.018}^{+    0.0093+    0.018}$ & $    0.9627_{-    0.0088-    0.018}^{+    0.0088+    0.018}$   & $    0.9669_{-    0.0091-    0.018}^{+    0.0091+    0.018}$ &  $    0.9631_{-    0.0085-    0.017}^{+    0.0090+    0.017}$ \\

${\rm{ln}}(10^{10} A_s)$ & $    3.028_{-    0.019-    0.041}^{+    0.021+    0.039}$ & $    3.041_{-    0.019-    0.036}^{+    0.019+    0.036}$  & $    3.042_{-    0.018-    0.037}^{+    0.018+    0.038}$  & $    3.042_{-    0.018-    0.036}^{+    0.019+    0.037}$   \\

$\Omega_{K0}$ & $   -0.044_{-    0.016-    0.038}^{+    0.020+    0.035}$ & $    0.0012_{-    0.0021-    0.0041}^{+    0.0021+    0.0044}$   & $   -0.0060_{-    0.0057-    0.012}^{+    0.0066+    0.011}$ & $    0.0012_{-    0.0021-    0.0041}^{+    0.0020+    0.0043}$    \\

$\Omega_{m0}$ & $    0.484_{-    0.072-    0.13}^{+    0.057+    0.14}$ & $    0.3108_{-    0.0072-    0.014}^{+    0.0073+    0.014}$   & $    0.336_{-    0.023-    0.042}^{+    0.021+    0.044}$  & $    0.310_{-    0.0071-    0.013}^{+    0.0063+    0.013}$    \\

$\sigma_8$ & $    0.774_{-    0.016-    0.034}^{+    0.018+    0.032}$ & $    0.808_{-    0.012-    0.022}^{+    0.012+    0.022}$  &  $    0.803_{-    0.012-    0.023}^{+    0.012+    0.024}$ & $    0.808_{-    0.011-    0.022}^{+    0.011+    0.022}$ \\

$H_0$ [Km/s/Mpc] & $   54.4_{-    4.0-    7.6}^{+    3.6+    7.6}$ & $   67.5_{-    1.2-    2.3}^{+    1.2+    2.4}$  & $   65.1_{-    2.4-    4.2}^{+    2.2+    4.6}$ & $   67.6_{-    1.1-    2.2}^{+    1.1+    2.2}$   \\

$N_{\rm eff}$ & $    3.04_{-    0.19-    0.39}^{+    0.20+    0.40}$ & $    2.96_{-    0.19-    0.38}^{+    0.19+    0.39}$   & $    3.03_{-    0.20-    0.38}^{+    0.20+    0.40}$ & $    2.97_{-    0.19-    0.38}^{+    0.19+    0.38}$   \\

$S_8$ & $    0.980_{-    0.049-    0.10}^{+    0.048+    0.10}$ &  $    0.822_{-    0.014-    0.026}^{+    0.013+    0.027}$  & $    0.849_{-    0.026-    0.052}^{+    0.026+    0.050}$ & $    0.821_{-    0.013-    0.025}^{+    0.013+    0.026}$   \\

$r_{\rm{drag}}$ [Mpc] &  $  147.4_{-    1.9-    3.8}^{+    1.9+    4.0}$ & $  148.0_{-    1.9-    3.8}^{+    1.9+    3.8}$  & $  147.5_{-    1.9-    3.8}^{+    1.9+    3.8}$ & $  148.0_{-    1.9-    3.7}^{+    1.9+    3.8}$   \\

\hline                                                  
%$\chi^2$ & 2763.766  &  2779.958 & 3804.290  &   3814.626  \\
\hline                                                       
\end{tabular}                                                   }                                                           
\caption{68\% and 95\% CL constraints on various free and derived parameters of the $\Lambda$CDM $+$ $\Omega_K$ $+$ $N_{\rm eff}$ scenario for several observational datasets. }
\label{tab:LCDM-Neff}    
\end{table*}  
\end{center}                                                
\endgroup   
\begin{figure*}
    \centering
    \includegraphics[width=0.8\textwidth]{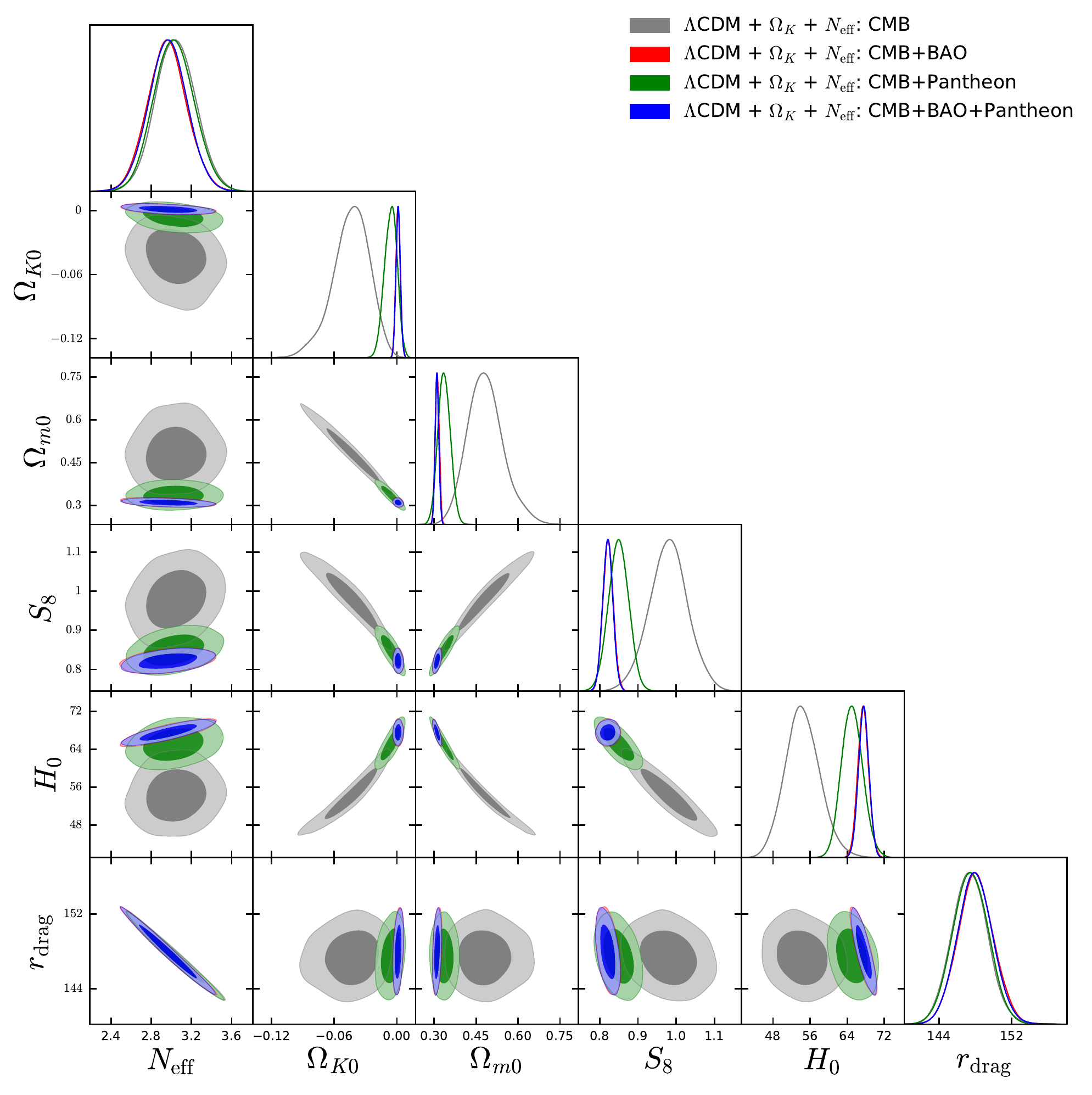}
    \caption{1-dimensional marginalized posterior distributions and the 2-dimensional joint contours for the most relevant parameters of cosmological scenario  $\Lambda$CDM $+$ $\Omega_K$ $+$ $N_{\rm eff}$ for various observational datasets. }
    \label{fig:LCDM-Omegak-Neff}
\end{figure*}

\subsubsection{$\Lambda$CDM $+$ $\Omega_K$ $+$ $N_{\rm eff}$}

The second non-flat extended cosmological model analyzed here, accounts for the possibility to have a larger effective number of relativistic degrees of freedom $N_{\rm eff}$ at recombination. We recall that $N_{\rm eff}$ is defined by the relation 
\begin{equation}	
\rho_{\rm rad } = N_{\rm eff}\frac{7}{8}\left(\frac{4}{11}\right)^{4/3}\rho_\gamma
\end{equation}
with $\rho_\gamma$ the present Cosmic Microwave Background energy density. Within the standard $\Lambda$CDM cosmological model, this parameter is fixed to $N_{\rm eff}=3.044$~ ~\cite{Mangano:2005cc,deSalas:2016ztq,Akita:2020szl,Froustey:2020mcq,Bennett:2020zkv,Archidiacono:2011gq} which consists of three massless neutrino species and an additional correction coming from the non-instantaneous neutrino decoupling. Here we regard $N_{\rm eff}$ as an additional free parameter of the $\Lambda$CDM $+$ $\Omega_K$ $+$ $N_{\rm eff}$ cosmological model. It should be noted that when testing departures from its reference value, one can probe and constrain several extended models both of cosmology and particle physics that predict extra dark radiation in the early Universe, including the cases of additional neutrino species and hot relics beyond the standard model of elementary particles~\cite{DiValentino:2011sv,DiValentino:2013qma,DiValentino:2015wba,Giare:2020vzo,Giare:2021cqr,DEramo:2022nvb,Baumann:2016wac,Gariazzo:2015rra,Archidiacono:2022ich,An:2022sva, Abdalla:2022yfr,SimonsObservatory:2019qwx}.

\autoref{tab:LCDM-Neff} summarizes the 68\% and 95\% CL cosmological constraints, while \autoref{fig:LCDM-Omegak-Neff} shows the 68\% and 95\% CL contour plots for different cosmological parameters.
As for the previous extended model, also in this case the CMB data alone suggest a non-flat background geometry with the case $\Omega_{K0}=0$ excluded at more than 95\% CL. Conversely, the inclusion of the other CMB-independent datasets eliminates such a preference for a curved spacetime, giving instead indication for flatness within 1$\sigma$, always as a result of the tension they have with Planck. As concerns the effective number of relativistic degrees of freedom, we see that the reference value $N_{\rm eff}=3.044$ is always consistent within one standard deviation for all the different data combinations analyzed. Therefore we find no significant evidence of deviations from the prediction of standard model of elementary particles and, in any case, in this extended parameter space current cosmological and astrophysical observations still constrain additional contributions to $\Delta N_{\rm eff} \doteq N_{\rm eff}-3.044\lesssim 0.4$ (at 95\% CL), analogously to the standard case of spatial flatness.\footnote{Notice that while this bound $\Delta N_{\rm eff}\lesssim 0.4$ is tight enough to severely constrain several exotic scenarios such as additional neutrino species, it is also too large for probing other extensions of the standard model involving extra relativistic species decoupled at high temperatures of the order of the top quark annihilation, whose contribution $\Delta N_{\rm eff}\sim0.027$ is much smaller than the current constraining power.} Actually, as we can see in \autoref{fig:LCDM-Omegak-Neff} there is no correlation between $N_{\rm eff}$ and $\Omega_K$. Modifying the relativistic energy density in the early Universe leads to several implications as it changes the sound horizon at recombination which is partly degenerate with the late-time geometry. In particular higher values of the effective number of relativistic species lead to smaller values of the sound horizon at recombination, resulting both in fluctuations in the amplitude of $\sigma_8$ and in a preference for higher-values of the expansion rate, possibly alleviating the Hubble tension. This is not what happens in this extended model where the same tensions between early and late time measurements of $H_0$ discussed in the previous cases can be still observed, always with a statistical significance ranging between $\sim 5\sigma$ (CMB only vs SH0ES) and $\gtrsim3.3\sigma$ (when the CMB data are combined with BAO and/or Pantheon). Similarly, the different data combinations provide values of $S_8$ that are always in tension at more than 95\% CL with each other.

\subsubsection{$\Lambda$CDM $+$ $\Omega_K$ $+$ $M_{\nu}$ $+$ $N_{\rm eff}$}
\begin{figure*}
    \centering
    \includegraphics[width=0.8\textwidth]{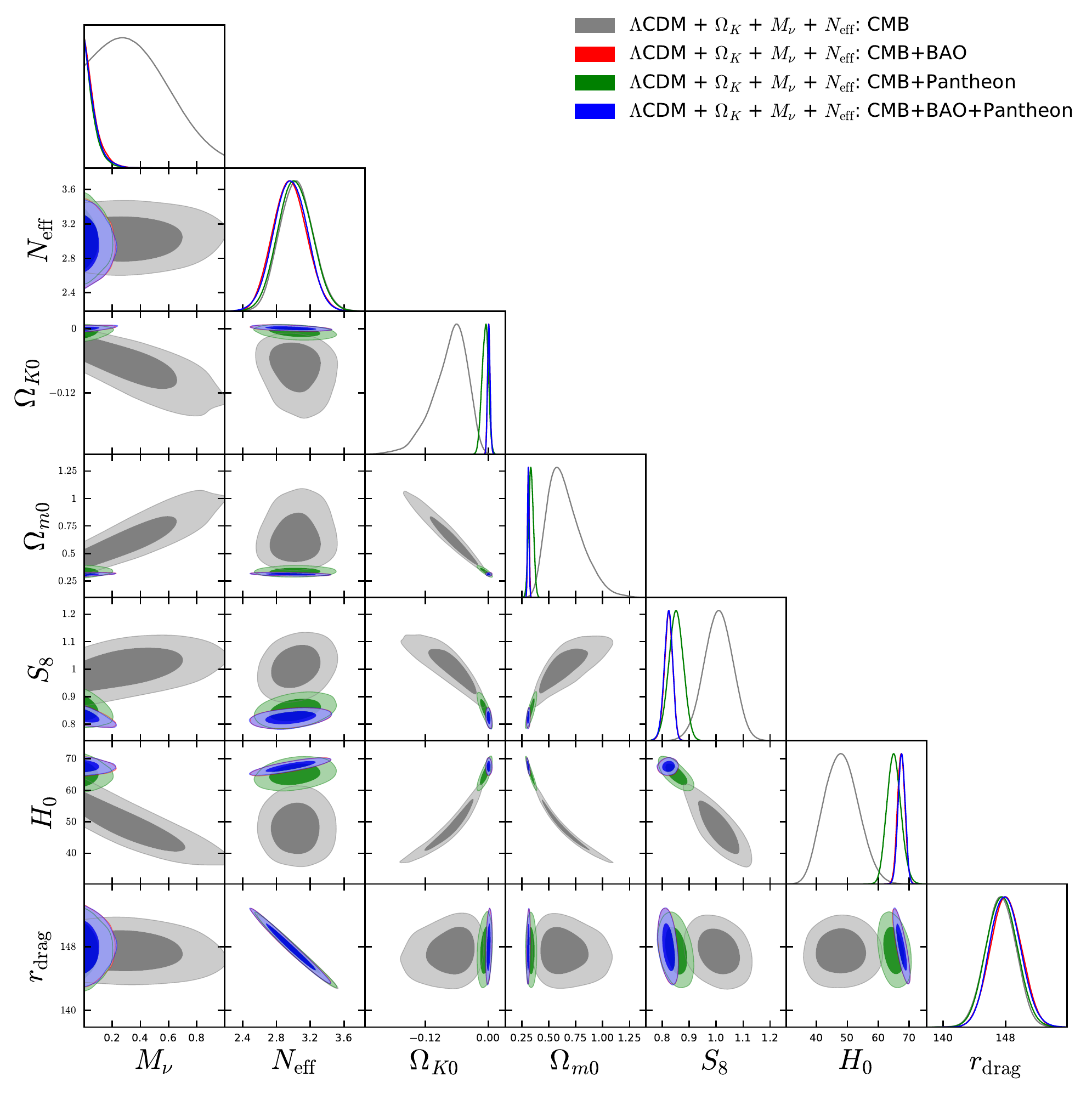}
    \caption{1-dimensional marginalized posterior distributions and the 2-dimensional joint contours for the most relevant parameters of cosmological scenario  $\Lambda$CDM $+$ $\Omega_K$ $+$ $M_{\nu}$ $+$ $N_{\rm eff}$  for various observational datasets.  }
    \label{fig:LCDM-Omegak-Mnu-Neff}
\end{figure*}

Lastly, we consider an extended non-flat cosmological model where we freely vary both the total neutrino mass and the effective number of relativistic degrees of freedom. We refer to this model as $\Lambda$CDM $+$ $\Omega_K$ $+$  $M_{\nu}$ $+$ $N_{\rm eff}$. The results obtained for this model are given in \autoref{tab:LCDM-Mnu-Neff} at 68\% and 95\% CL, while the 1D and 2D posteriors distributions of different cosmological parameters are shown in \autoref{fig:LCDM-Omegak-Mnu-Neff}.
Simultaneously varying the total neutrino mass and the effective number of relativistic degrees of freedom, any significant change is observed neither about the Planck preference for a closed Universe nor for the $H_0$-tension. In particular form the CMB data alone we get $\Omega_{K0}=-0.074^{+0.059}_{-0.070}$ at 95\% CL and $H_0=48.1^{+5.2}_{-6.0}$ km/s/Mpc at 68\% CL, in tension with local measurements at the level of $4.7\sigma$.  As for the previous models, including the BAO (Pantheon) likelihood, that disagree with Planck in this extended model, the preference for a closed Universe disappears and the constraints for the expansion rate become in tension at the level of $3.3\sigma$ with the independent result of the SH0ES collaboration. 
Also in this case the $H_0$ tension drives the values of $S_8$ to be in disagreement between 3 and 4 standard deviations with each other, depending on the different data combination considered. As concerns the bounds on the effective number of relativistic degrees of freedom, also in this case we do not find any evidence for a deviation from the value expected in the Standard Model and the constraints remain basically unchanged with respect to the previous case without neutrinos. Similarly, the upper bounds on the total neutrino mass are not affected by the inclusion of a free effective number of relativistic species and they are similar to those obtained within the $\Lambda$CDM + $\Omega_K$ + $M_{\nu}$ cosmological model. In particular exploiting the CMB data alone we can get $M_{\nu}<0.81$ eV at 95\% CL (with one sigma indication for a total neutrino mass different from zero), while including also the BAO and Pantheon data this bound can be improved to $M_{\nu}<0.18$ eV for CMB+BAO(+Pantheon) and $M_{\nu}<0.16$ eV for CMB+Pantheon, all at 95\% CL.

\begingroup                                                                                                                     
\squeezetable                                                                                                                   
\begin{center}                                                                                                                  
\begin{table*}[htb]                                                                                                                   \resizebox{\textwidth}{!}{   
\begin{tabular}{ccccccccccccc}                                                                                                            
\hline\hline                                                                                                                    
Parameters & CMB & CMB+BAO & CMB+Pantheon &  CMB+BAO+Pantheon \\ \hline 

$\Omega_c h^2$ & $    0.1182_{-    0.0032-    0.0056}^{+    0.0029+    0.0061}$ & $    0.1184_{-    0.0031-    0.0058}^{+    0.0030+    0.0061}$ & $    0.1187_{-    0.0031-    0.0059}^{+    0.0031+    0.0061}$ &  $ 0.1185_{-    0.0030-    0.0059}^{+    0.0030+    0.0059}$  \\ 

$\Omega_b h^2$ & $    0.02252_{-    0.00025-    0.00050}^{+    0.00025+    0.00050}$ & $    0.02232_{-    0.00024-    0.00046}^{+    0.00024+    0.00046}$  & $    0.02244_{-    0.00024-    0.00048}^{+    0.00024+    0.00048}$  & $    0.02233_{-    0.00023-    0.00046}^{+    0.00023+    0.00046}$ \\ 

$100\theta_{MC}$ & $    1.04102_{-    0.00046-    0.00089}^{+    0.00046+    0.00091}$ & $    1.04111_{-    0.00044-    0.00084}^{+    0.00043+    0.00087}$  & $    1.04109_{-    0.00047-    0.00087}^{+    0.00043+    0.00091}$ & $    1.04112_{-    0.00045-    0.00085}^{+    0.00044+    0.00089}$ \\ 

$\tau$ & $    0.0475_{-    0.0077-    0.017}^{+    0.0077+    0.016}$ & $    0.0544_{-    0.0083-    0.015}^{+    0.0076+    0.016}$   & $    0.0543_{-    0.0082-    0.015}^{+    0.0074+    0.016}$ & $    0.055_{-    0.0082-    0.015}^{+    0.0072+    0.016}$ \\ 

$n_s$ & $    0.9687_{-    0.0093-    0.018}^{+    0.0092 +    0.018}$ & $    0.9625_{-    0.0089-    0.018}^{+    0.0091+    0.018}$   & $    0.9667_{-    0.0091-    0.018}^{+    0.0091+    0.018}$ & $    0.9630_{-    0.0088-    0.018}^{+    0.0089+    0.017}$ \\ 

${\rm{ln}}(10^{10} A_s)$ & $    3.025_{-    0.019-    0.038}^{+    0.019+    0.037}$ & $    3.040_{-    0.019-    0.037}^{+    0.019+    0.038}$  & $    3.041_{-    0.019-    0.037}^{+    0.018+    0.038}$  & $    3.041_{-    0.018-    0.035}^{+    0.018+    0.036}$ \\ 

$\Omega_{K0}$ & $   -0.074_{-    0.023-    0.070}^{+    0.041+    0.059}$ & $    0.0011_{-    0.0025-    0.0046}^{+    0.0023+    0.0053}$   & $   -0.0061_{-    0.0057-    0.012}^{+    0.0065+    0.011}$ &  $    0.00113_{-    0.0026-    0.0048}^{+    0.0022+    0.0051}$  \\

$\Omega_{m0}$ & $    0.65_{-    0.20-    0.28}^{+    0.11+    0.33}$ & $    0.3110_{-    0.0079-    0.014}^{+    0.0072+    0.015}$   & $    0.335_{-    0.022-    0.041}^{+    0.022+    0.043}$ &  $    0.310_{-    0.0071-    0.013}^{+    0.0070+    0.014}$  \\ 

$\sigma_8$ & $    0.695_{-    0.054-    0.11}^{+    0.074+    0.11}$ & $    0.809_{-    0.013-    0.033}^{+    0.017+    0.031}$  & $    0.806_{-    0.013-    0.032}^{+    0.017+    0.030}$  & $    0.809_{-    0.012-    0.033}^{+    0.018+    0.030}$ \\

$H_0$ [Km/s/Mpc] & $   48.1_{-    6.0-   10}^{+    5.2+   11}$ & $   67.4_{-    1.2-    2.3}^{+    1.2+    2.3}$   &  $   65.1_{-    2.4-    4.2}^{+    2.1+    4.6}$ & $   67.6_{-    1.1-    2.2}^{+    1.1+    2.2}$  \\ 

$M_\nu$ [eV] & $    0.38_{-    0.36}^{+    0.12}\,<0.81$ & $  <0.073\,<0.18  $  & $  <0.067\,<0.16 $  & $  <0.072\,<0.18 $  \\ 

$N_{\rm eff}$ & $    3.04_{-    0.21-    0.38}^{+    0.19+    0.39}$ & $    2.96_{-    0.12-    0.38}^{+    0.20+    0.39}$    & $    3.02_{-    0.20-    0.38}^{+    0.20+    0.40}$ & $    2.97_{-    0.20-    0.39}^{+    0.20+    0.38}$  \\ 

$S_8$ & $    1.009_{-    0.053-    0.11}^{+    0.053+    0.10}$ & $    0.823_{-    0.015-    0.031}^{+    0.016+    0.031}$ & $    0.851_{-    0.027-    0.054}^{+    0.027+    0.053}$ & $    0.823_{-    0.015-    0.032}^{+    0.017+    0.031}$ \\ 

$r_{\rm{drag}}$ [Mpc] & $  147.4_{-    1.9-    3.8}^{+    1.9+    3.7}$ & $  148.1_{-    2.0-    3.8}^{+    2.0+    3.8}$ & $  147.5_{-    2.0-    3.8}^{+    2.0+    3.9}$ &  $  148.0_{-    2.1-    3.7}^{+    1.9+    3.9}$ \\ 

\hline                                                  
%$\chi^2$ & 2760.756  & 2777.698 & 3805.800 & 3813.204 \\
\hline 
\end{tabular}                                                    }                                                               
\caption{68\% and 95\% CL constraints on various free and derived parameters of the $\Lambda$CDM $+$ $\Omega_K$ $+$ $M_{\nu}$ $+$ $N_{\rm eff}$ scenario for several observational datasets. }\label{tab:LCDM-Mnu-Neff}                                                                                                   
\end{table*}                                                                                                                     
\end{center}                                                                                                                    
\endgroup       

\subsection{Non-flat $w$CDM and its extensions}
In this section we describe the extensions of the non-flat $w$CDM model. We mention that for all extensions of the nonflat $w$CDM model, we have made use of the same datasets and their combinations. Moreover, we also note that for each parameter we report either its 68\% and 95\% CL constraints or the upper/lower limits at 68\% and 95\% CL.  

\subsubsection{$w$CDM $+$ $\Omega_K$}

\begin{figure*}
    \centering
    \includegraphics[width=0.8\textwidth]{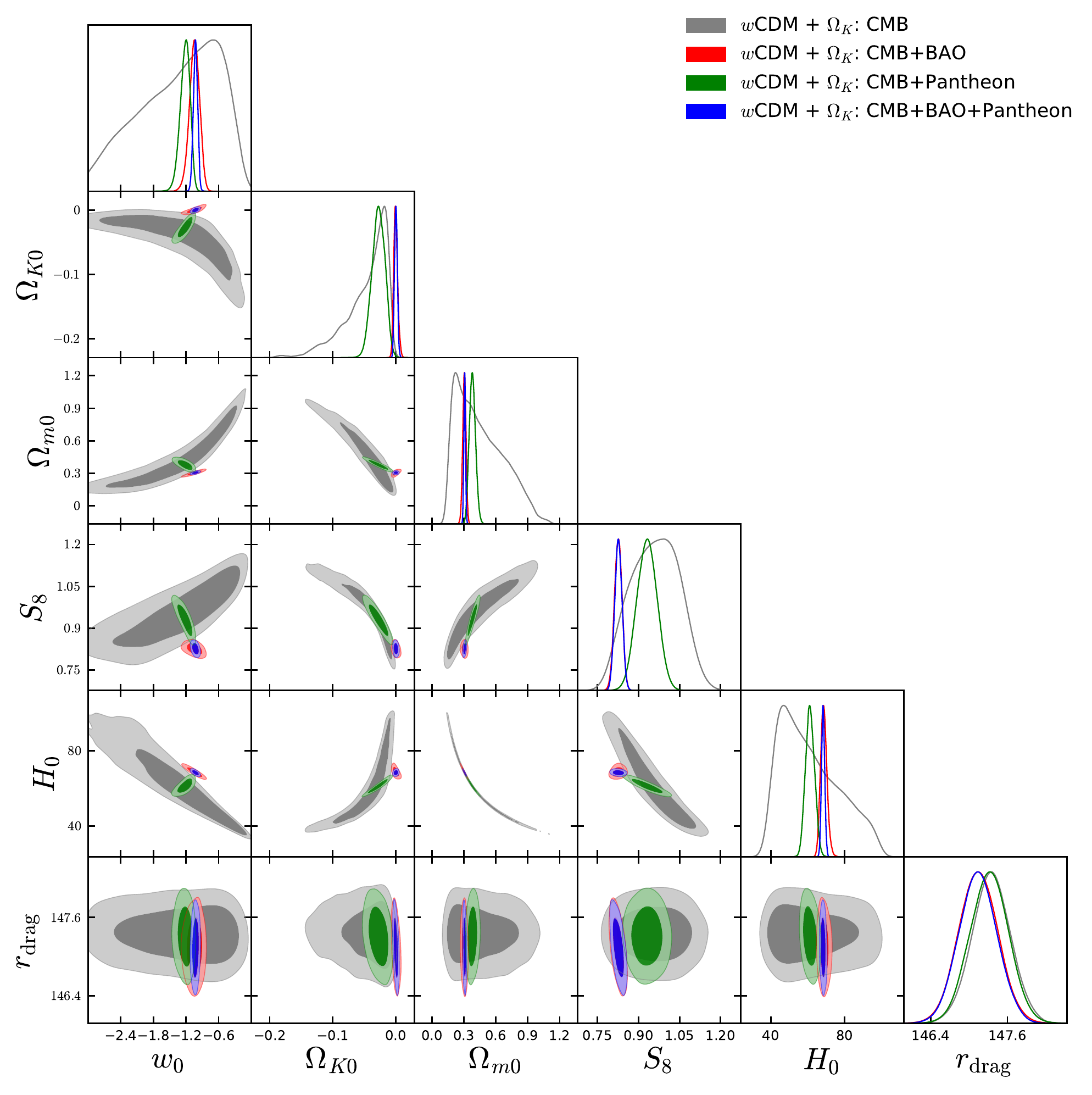}
    \caption{1-dimensional marginalized posterior distributions and the 2-dimensional joint contours for the most relevant parameters of cosmological scenario  $w$CDM $+$ $\Omega_K$  for various observational datasets.  }
    \label{fig:wCDM-Omegak}
\end{figure*}

In \autoref{tab:wCDM} we have summarized the constraints on the $w$CDM $+$ $\Omega_K$ model, and the corresponding marginalized 1D and 2D posterior distributions are shown in \autoref{fig:wCDM-Omegak}. 

We notice that for CMB alone, we get an evidence of a closed Universe at more than 68\% CL, however, within 95\% CL, the spatially flatness of the Universe is recovered. The dark energy equation of state is perfectly consistent with a cosmological constant, giving $w = -1.31^{+0.98}_{-0.49}$ at 68\% CL for CMB alone. However, even if we vary this parameter for CMB alone, unlike in phantom dark energy Universe where $w< -1$ increases the expansion rate, here this does not happen and the Hubble constant is almost unconstrained, as one can see that has a lower mean value $H_0 = 61^{+10}_{-22}$ km/s/Mpc at 68\% CL (CMB alone), but due to the large error bars, the tension with the SH0ES measurement is reduced down to $1.2\sigma$. This lowering of the mean value is due to the effects of curvature that we are witnessing in this case.   

The inclusion of BAO to CMB changes the constraints significantly, because of their mutual tension in a curved universe, and we see that this scenario becomes very close to the spatially flat $\Lambda$CDM one. However, the results for CMB+Pantheon are much more interesting. We see that for this combination we get a closed Universe at more than 95\% CL ($\Omega_K =  -0.028_{- 0.020}^{+ 0.019}$ at 95\% CL for CMB+Pantheon) together with a phantom Universe at more than 95\% CL ($w = -1.22_{-0.18}^{+0.17}$ at 95\% CL for CMB+Pantheon), i.e. an indication for a phantom closed universe, as already noticed in~\cite{DiValentino:2020hov}. Similar to the CMB alone case, here too, we get a smaller value of the Hubble constant ($H_0 =  61.2_{-2.4}^{+    2.4}$  km/s/Mpc at 68\% CL for CMB+Pantheon) and hence a bigger value of $\Omega_{m0}$ due to the positive correlation between $H_0$ and $\Omega_{m0}$. 

Finally, if we look at the combination of CMB+BAO+Pantheon (column 5 of \autoref{tab:wCDM}), because of the presence of the BAO data that are in tension with Planck, we again find that the spatial flatness is strongly suggested. At the same time $w $ is in agreement within 68\% CL with a cosmological constant $w=-1$. There is one point to note as well. Here, we see that $H_0$ is mildly increased ($H_0 = 68.30_{-    0.84}^{+    0.84}$ km/s/Mpc at 68\% CL) than its predicted value within the $\Lambda$CDM paradigm by Planck 2018~\cite{Aghanim:2018eyx}, but the tension with SH0ES is still at $3.6\sigma$.   

\begingroup                                                                                                                     
\squeezetable                                                                                                                   
\begin{center}                                                                                                                  
\begin{table*}[htb]                                                                                                                  
\resizebox{\textwidth}{!}{   
\begin{tabular}{ccccccccccccc}                                                                                                            
\hline\hline                                                                                                                    
Parameters & CMB & CMB+BAO & CMB+Pantheon &  CMB+BAO+Pantheon \\ \hline 

$\Omega_c h^2$ & $    0.1180_{-    0.0014-    0.0028}^{+    0.0014+    0.0029}$ & $    0.1197_{-    0.0014-    0.0028}^{+    0.0014+    0.0028}$  & $    0.1183_{-    0.0014-    0.0028}^{+    0.0014+    0.0028}$  & $    0.1197_{-    0.0014-    0.0027}^{+    0.0014+    0.0027}$  \\ 

$\Omega_b h^2$ & $    0.02261_{-    0.00017-    0.00032}^{+    0.00017 +    0.00034}$ & $    0.02240_{-    0.00015-    0.00031}^{+    0.00015 +    0.00030}$ & $    0.02257_{-    0.00017-    0.00032}^{+    0.00016+    0.00033}$ & $    0.02239_{-    0.00015-    0.00030}^{+    0.00015+    0.00031}$ \\ 

$100\theta_{MC}$ & $    1.04118_{-    0.00032-    0.00060}^{+    0.00031+    0.00062}$ & $    1.04096_{-    0.00032-    0.00063}^{+    0.00032+    0.00063}$   & $    1.0411_{-    0.00033-    0.00062}^{+    0.00032+    0.00063}$ & $    1.04096_{-    0.00032-    0.00061}^{+    0.00031+    0.00062}$ \\ 

$\tau$ & $    0.0485_{-    0.0076-    0.018}^{+    0.0076+    0.016}$ & $    0.0545_{-    0.0074-    0.015}^{+    0.0074+    0.015}$ & $    0.0496_{-    0.0077-    0.017}^{+    0.0079+    0.016}$  & $    0.055_{-    0.0078-    0.015}^{+    0.0073+    0.016}$\\ 

$n_s$ & $    0.9708_{-    0.0046-    0.0090}^{+    0.0045+    0.0092}$ & $    0.9660_{-    0.0045-    0.0089}^{+    0.0045+    0.0086}$  & $    0.9700_{-    0.0046-    0.0091}^{+    0.0046+    0.0091}$ & $    0.9661_{-    0.0044-    0.0088}^{+    0.0045+    0.0088}$ \\ 

${\rm{ln}}(10^{10} A_s)$ & $    3.028_{-    0.016-    0.035}^{+    0.016+    0.037}$ & $    3.044_{-    0.015-    0.031}^{+    0.015+    0.031}$  & $    3.031_{-    0.016-    0.035}^{+    0.016+    0.033}$ & $    3.045_{-    0.016-    0.030}^{+    0.015+    0.032}$ \\ 

$w$ & $   -1.31_{-    0.49-    1.4}^{+    0.98+    1.1}$ &  $   -1.046_{-    0.089-    0.19}^{+    0.099+    0.19}$  & $   -1.22_{-    0.08-    0.18}^{+    0.10+    0.17}$ & $   -1.026_{-    0.038-    0.078}^{+    0.039+    0.074}$ \\ 

$\Omega_{K0}$ & $   -0.046_{-    0.015-    0.073}^{+    0.041+    0.050}$ & $   -0.0001_{-    0.0035-    0.0061}^{+    0.0027+    0.0067}$  & $   -0.028_{-    0.009-    0.020}^{+    0.012+    0.019}$ &  $    0.0001_{-    0.0022-    0.0045}^{+    0.0022+    0.0042}$  \\ 

$\Omega_{m0}$ & $    0.46_{-    0.31-    0.36}^{+    0.14+    0.44}$ & $    0.303_{-    0.015-    0.031}^{+    0.015+    0.030}$    & $    0.380_{-    0.032-    0.054}^{+    0.027+    0.059}$ &  $    0.306_{-    0.0078-    0.015}^{+    0.0076+    0.016}$ \\ 

$\sigma_8$ & $    0.84_{-    0.21-    0.24}^{+    0.11+    0.29}$ &  $    0.822_{-    0.025-    0.049}^{+    0.025+    0.051}$  & $    0.828_{-    0.012-    0.025}^{+    0.012+    0.024}$ & $    0.818_{-    0.013-    0.025}^{+    0.013+    0.027}$  \\ 

$H_0$ [Km/s/Mpc] & $   61_{-   22-   26}^{+   10+   33}$ &  $   68.7_{-    1.9-    3.4}^{+    1.6+    3.5}$  & $   61.2_{-    2.4-    4.6}^{+    2.4+    4.8}$ & $   68.30_{-    0.84-    1.6}^{+    0.84+    1.8}$ \\ 

$S_8$ & $    0.96_{-    0.09-    0.16}^{+    0.10+    0.16}$ & $    0.826_{-    0.014-    0.028}^{+    0.014+    0.026}$  & $    0.931_{-    0.037-    0.071}^{+    0.037+    0.072}$ & $    0.827_{-    0.013-    0.025}^{+    0.013+    0.026}$ \\

$r_{\rm{drag}}$ [Mpc] & $  147.35_{-    0.30-    0.60}^{+    0.30+    0.60}$  & $  147.14_{-    0.31-    0.60}^{+    0.30+    0.60}$    & $  147.32_{-    0.30-    0.59}^{+    0.30+    0.59}$ & $  147.14_{-    0.30-    0.58}^{+    0.30+    0.60}$ \\ 

\hline                                                  
%$\chi^2$ &    2761.984   &  2779.156  &  3797.780  &   3814.422 \\
\hline                                                             
\end{tabular}                                                   }                                                                
\caption{68\% and 95\% CL constraints on various free and derived parameters of the $w$CDM $+$ $\Omega_K$ scenario using several observational datasets.  }
\label{tab:wCDM}                                                                                                   
\end{table*}                                                                                                                     
\end{center}                                                                                                                    
\endgroup   

\subsubsection{$w$CDM $+$ $\Omega_K$ $+$ $M_{\nu}$}

\begin{figure*}
    \centering
    \includegraphics[width=0.8\textwidth]{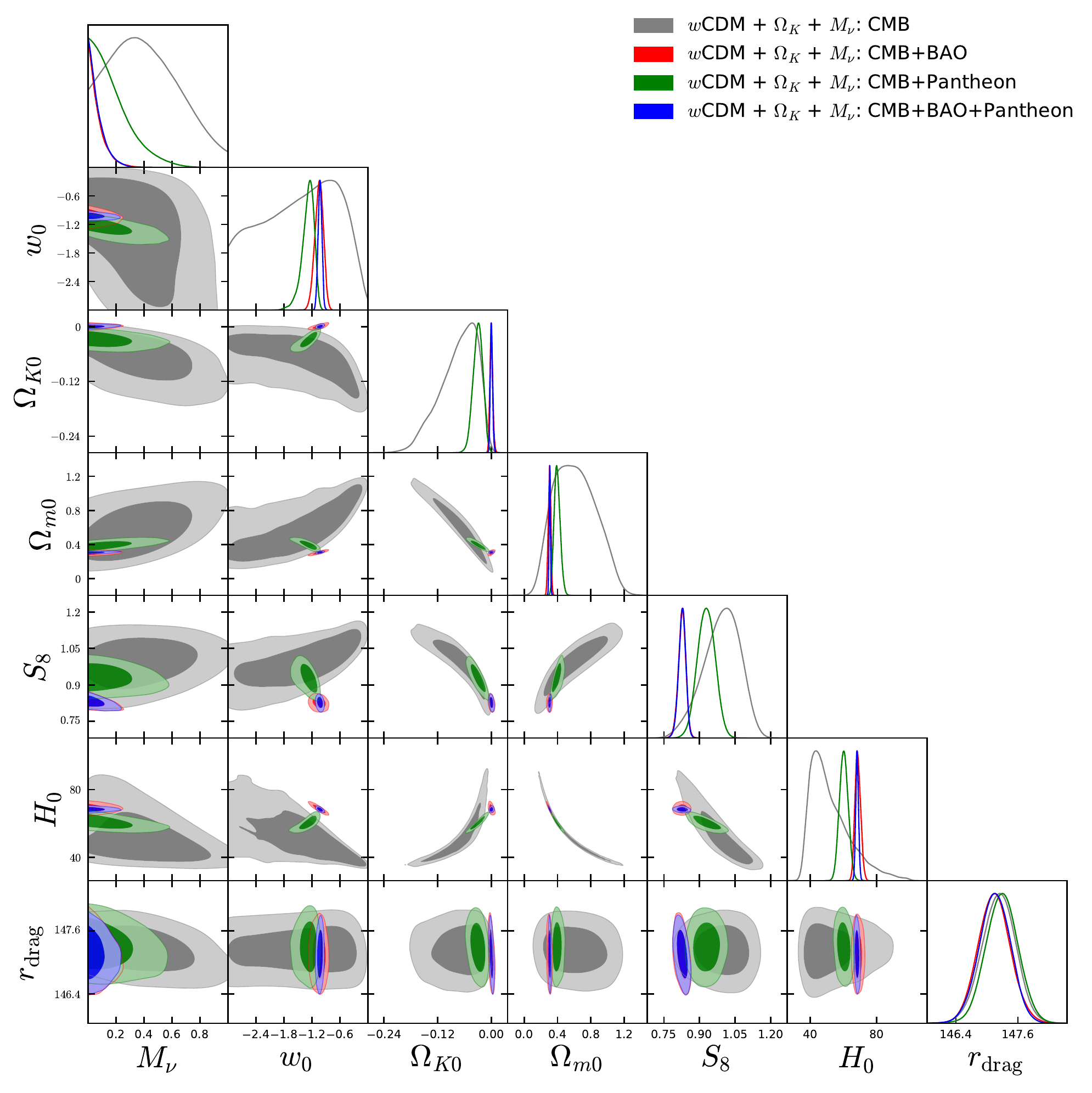}
    \caption{1-dimensional marginalized posterior distributions and the 2-dimensional joint contours for the most relevant parameters of cosmological scenario  $w$CDM  $+$ $\Omega_K$ $+$ $M_{\nu}$  for various observational datasets.  }
    \label{fig:wCDM-Omegak-Mnu}
\end{figure*}

Following the same path as the previous section, we first consider an expansion which includes the total neutrino mass as an additional free parameter of the sample. We refer to this model as $w$CDM $+$ $\Omega_K$ $+$ $M_{\nu}$ and we summarize the 68\% and 95\% CL constraints in \autoref{tab:wCDM-Mnu}, showing the correlations among the different parameters in \autoref{fig:wCDM-Omegak-Mnu}.

In this case, considering only the CMB data, a spatially flat background geometry is disfavored at slightly more than 95\% CL with the constraint on the curvature parameter reading $\Omega_{K0}=-0.073^{+0.070}_{-0.086}$ at 95\% CL. Interestingly, a strong preference for a curved spacetime is found combining CMB and Pantheon (from which we get $\Omega_{K0}=-0.030^{+0.020}_{-0.022}$ at 95\% CL). 
Conversely, including the BAO measurements, flatness becomes always consistent within one standard deviation for all the other different data combinations, but these results should be taken with caution because of the data inconsistency in a curved universe. As concerns the dark energy equation of state, using only the CMB data we find that both phantom ($w<-1$) and quintessential behaviors ($w>-1$) are allowed within the 68\% CL results. Combining CMB with BAO (+Pantheon) no significant differences are observed, as well. Conversely, it should be noted that the results for CMB+Pantheon ($w=-1.28^{+0.25}_{-0.28}$ at 95\% CL) suggest a preference for a phantom dark energy at the level of more than 95\% CL. As for the baseline $w\rm{CDM} + \Omega_K$ model, also in this extended parameter space, allowing phantom models of dark energy does not alleviate the tensions, and from CMB+Pantheon thre is indication for a phantom closed universe, as in~\cite{DiValentino:2020hov}. Indeed the value of the Hubble parameter inferred by the CMB data, $H_0=52^{+6}_{-16}$ km/s/Mpc at 68\% CL, and the independent measurement provided by the SH0ES collaboration are in tension at the level of $3.4\sigma$. This preference for lower values of $H_0$ is mostly driven by the degeneracy with the spatial curvature. Indeed, when the BAO data are combined with the CMB, no evidence is found for curvature and the constraints on $H_0$ look very different and become similar to the usual results obtained within the flat $\Lambda$CDM model. Conversely, for CMB+Pantheon, due to the preference for a curved spacetime of this dataset, we still obtain a lower value $H_0=60.2\pm2.6$ km/s/Mpc at 68\% CL, in strong tension with local measurements at about $4.6 \sigma$. Finally, it is worth pointing out that in this model the bounds on the total neutrino mass obtained using the Planck measurements ($M_{\nu}<0.83$ eV at 95\% CL) remain almost unchanged with respected to the previous scenarios, while the bounds for CMB+Pantheon are significantly relaxed to $M_{\nu}<0.46$ eV at 95\% CL. Anyway, the most significant impact on the bounds on relic neutrinos is obtained by combining CMB and BAO from which we obtain $M_{\nu}<0.19$ eV at 95\% CL. Considering CMB+BAO+Pantheon all together this bound is robust and almost unchanged.

\begingroup                                                                                                                     
\squeezetable                                                                                                                   
\begin{center}                                                                                                                  
\begin{table*}[htb]                                                                                                                   
\resizebox{\textwidth}{!}{   
\begin{tabular}{ccccccccccccc}                                                                                                            
\hline\hline                                                                                                                    
Parameters & CMB & CMB+BAO & CMB+Pantheon &  CMB+BAO+Pantheon  \\ \hline 
$\Omega_c h^2$ & $    0.1183_{-    0.0015-    0.0030}^{+    0.0015+    0.0031}$ &  $    0.1197_{-    0.0015-    0.0029}^{+    0.0015+    0.0027}$  & $    0.1184_{-    0.0015-    0.0029}^{+    0.0015+    0.0028}$ & $    0.1197_{-    0.0015-    0.0027}^{+    0.0014+    0.0029}$  \\ 

$\Omega_b h^2$ & $    0.02253_{-    0.00018-    0.00035}^{+    0.00018+    0.00036}$ &  $    0.02240_{-    0.00016-    0.00030}^{+    0.00016+    0.00031}$ & $    0.02255_{-    0.00017-    0.00034}^{+    0.00017+    0.00035}$ & $    0.02240_{-    0.00016-    0.00030}^{+    0.00016+    0.00031}$  \\ 

$100\theta_{MC}$ & $    1.04099_{-    0.00034-    0.00068}^{+    0.00034+    0.00068}$ & $    1.04095_{-    0.00034-    0.00065}^{+    0.00033+    0.00065}$  & $    1.04109_{-    0.00034-    0.00068}^{+    0.00035+    0.00064}$ & $    1.04096_{-    0.00031-    0.00064}^{+    0.00032+    0.00060}$  \\ 

$\tau$ & $    0.0472_{-    0.0071-    0.016}^{+    0.0084+    0.016}$ & $    0.0545_{-    0.0082-    0.015}^{+    0.0074+    0.016}$  &  $    0.0503_{-    0.0074-    0.017}^{+    0.0077+    0.016}$ & $    0.055_{-    0.0077-    0.014}^{+    0.0070+    0.015}$ \\ 

$n_s$ & $    0.9689_{-    0.0049-    0.0096}^{+    0.0049 +    0.0099}$ & $    0.9658_{-    0.0046-    0.0088}^{+    0.0045 +    0.0086}$  & $    0.9695_{-    0.0051-    0.0090}^{+    0.0046+    0.0095}$ & $    0.9661_{-    0.0044-    0.0089}^{+    0.0045+    0.0086}$ \\ 

${\rm{ln}}(10^{10} A_s)$ & $    3.025_{-    0.016-    0.034}^{+    0.018 +    0.034}$ & $    3.044_{-    0.017-    0.030}^{+    0.015+    0.033}$  & $    3.032_{-    0.016-    0.035}^{+    0.016+    0.033}$ & $    3.045_{-    0.016-    0.031}^{+    0.015+    0.032}$  \\ 

$w$ & $   -1.5_{-    0.6-    1.5}^{+    1.1+    1.2}$ & $   -1.046_{-    0.085-    0.19}^{+    0.099+    0.17}$  & $   -1.28_{-    0.09-    0.28}^{+    0.16+    0.25}$ & $   -1.025_{-    0.040-    0.080}^{+    0.040+    0.077}$ \\ 

$\Omega_{K0}$ & $   -0.073_{-    0.027-    0.086}^{+    0.055+    0.070}$ & $   -0.0001_{-    0.0035-    0.0062}^{+    0.0028+    0.0067}$ & $   -0.030_{-    0.010-    0.022}^{+    0.012+    0.020}$ &  $    0.0001_{-    0.0027-    0.0049}^{+    0.0023+    0.0051}$ \\ 

$\Omega_{m0}$ & $    0.61_{-    0.31-    0.44}^{+    0.22+    0.47}$ & $    0.304_{-    0.015-    0.030}^{+    0.015+    0.029}$ & $    0.396_{-    0.041-    0.069}^{+    0.032+    0.073}$ & $    0.306_{-    0.0077-    0.015}^{+    0.0076+    0.015}$ \\ 

$\sigma_8$ & $    0.74_{-    0.16-    0.20}^{+    0.07+    0.26}$ & $    0.822_{-    0.025-    0.050}^{+    0.025+    0.050}$ & $    0.809_{-    0.018-    0.060}^{+    0.033+    0.049}$ & $    0.818_{-    0.014-    0.034}^{+    0.018+    0.032}$ \\ 

$H_0$ [Km/s/Mpc] & $   52_{-   16-   20}^{+    6+   27}$ &  $   68.7_{-    1.9-    3.3}^{+    1.5+    3.5}$  &  $   60.2_{-    2.6-    4.9}^{+    2.6+    5.2}$  & $   68.27_{-    0.88-    1.6}^{+    0.81+    1.7}$  \\ 

$M_\nu$ [eV] & $    0.41_{-    0.32}^{+    0.19}\,<0.83$ &  $  <0.078\,<0.19  $ & $<0.212\,<0.46 $ & $ <0.078\,<0.18$ \\ 

$S_8$ & $    0.992_{-    0.066-    0.16}^{+    0.092+    0.15}$ &  $    0.826_{-    0.015-    0.033}^{+    0.017+    0.030}$ & $    0.928_{-    0.037-    0.072}^{+    0.037+    0.071}$ & $    0.827_{-    0.014-    0.031}^{+    0.016+    0.030}$ \\ 

$r_{\rm{drag}}$ [Mpc] & $  147.23_{-    0.32-    0.66}^{+    0.32+    0.63}$ & $  147.14_{-    0.31-    0.60}^{+    0.30+    0.63}$  & $  147.29_{-    0.31-    0.60}^{+    0.31+    0.60}$ &  $  147.15_{-    0.30-    0.61}^{+    0.30+    0.58}$\\ 

\hline                                                  
%$\chi^2$ & 2763.056   & 2778.404   &  3797.556  & 3812.69    \\
\hline                                                         
\end{tabular}                                                     }                                                              
\caption{68\% and 95\% CL constraints on various free and derived parameters of the $w$CDM  $+$ $\Omega_K$ $+$ $M_{\nu}$ scenario using several observational datasets.  }
\label{tab:wCDM-Mnu}                                                                                                   
\end{table*}                                                                                                                     
\end{center}                                                                                                                    
\endgroup         

\subsubsection{$w$CDM $+$ $\Omega_K$ $+$ $N_{\rm eff}$}

\begin{figure*}
    \centering
    \includegraphics[width=0.8\textwidth]{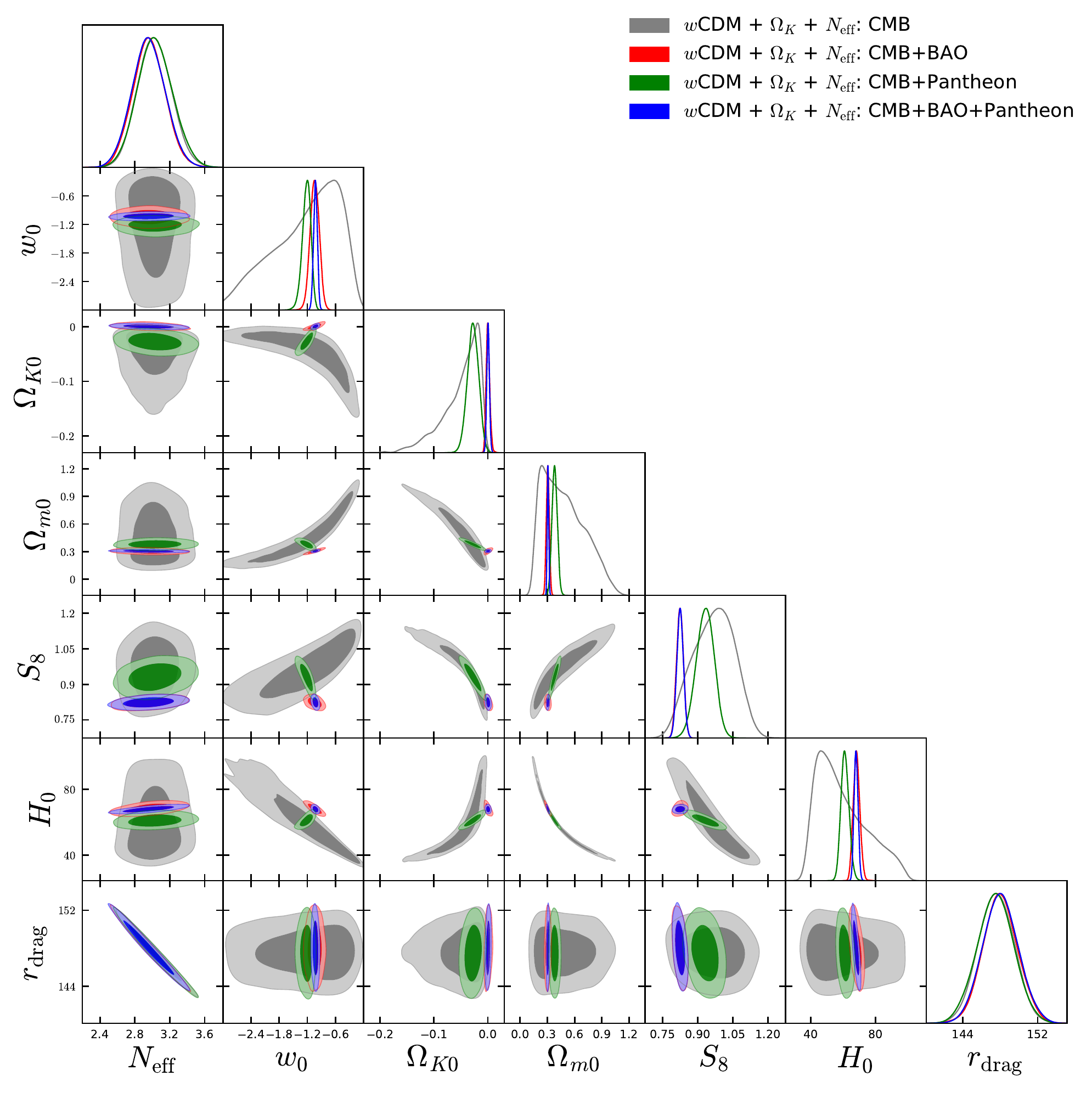}
    \caption{1-dimensional marginalized posterior distributions and the 2-dimensional joint contours for the most relevant parameters of cosmological scenario  $w$CDM  $+$ $\Omega_K$ $+$ $N_{\rm eff}$  for various observational datasets.  }
    \label{fig:wCDM-Omegak-Neff}
\end{figure*}

Now we analyze the $w$CDM $+$ $\Omega_K$ $+$ $N_{\rm eff}$ model, varying the effective number of relativistic degrees of freedom as additional parameter instead of the total neutrino mass.  The results for this model are given in \autoref{tab:wCDM-Neff}, and the triangular plots are shown in \autoref{fig:wCDM-Omegak-Neff}.

Notice that replacing the total neutrino mass with the effective number of neutrino species the results for the curvature parameter are changed and the Planck preference for a closed Universe is reduced. In this case the Planck limit on curvature  $\Omega_{K0}=-0.049^{+0.053}_{-0.076}$ at 95\% CL is consistent with flatness at the level of $1.1\sigma$. Interestingly, CMB+Pantheon combined together prefer instead a closed Universe and we obtain $\Omega_{K0}=-0.028^{+0.020}_{-0.021}$ at 95\% CL. Such a preference disappears once we introduce the BAO measurements, as noticed before. As concerns the $H_0$-tension, we see that, due to the correlation between the expansion rate and $N_{\rm eff}$ the CMB bounds are less tight, but they show in any case a preference for smaller values $H_0=59^{+9}_{-21}$ km/s/Mpc at 68\% CL. Due to the large uncertainty, the tension with the SH0ES estimation 
below $2\sigma$.
Combining CMB+BAO instead we get $H_0=68.2^{+1.9}_{-2.2}$ km/s/Mpc at 68\% CL, close to the value obtained within the flat $\Lambda$CDM model but reducing the tension with SH0ES at the level of $\sim 2.2\sigma$, due to the larger error-bars. On the other hand, considering CMB+Pantheon, the preference for a curved Universe lowers the value of $H_0$ to $H_0=61.1^{+2.3}_{-2.6}$ km/s/Mpc (at 68\% CL) while the uncertainty remains almost unchanged. As a result, the tension reaches the level of $4.8\sigma$. Notice also that these tensions strongly affect the bounds on the matter-density parameter $\Omega_{m0}$ and consequently the value of $S_{8}$ since both of them are pushed toward higher (lower) values when lower (higher) $H_0$ are preferred by the different data combinations.  Concerning the equation of state of Dark Energy, the CMB measurements give a very relaxed bound on $w$ that is completely consistent with a cosmological constant. This result remains true also combining the CMB and BAO data, while, as in the previous model, the combination CMB+Pantheon strongly suggest phantom dark energy, excluding a cosmological constant at more than 95\% CL. Finally, it should be noted that the bounds on the effective number of relativistic degrees of freedom remain basically unchanged and, exactly as in the case of (flat and curved) models with a cosmological constant,  additional contribution are constrained to $\Delta N_{\rm eff}\lesssim 0.4$ for all the different datasets.

\begingroup                                                                                                                     
\squeezetable                                                                                                                   
\begin{center}                                                                                                                  
\begin{table*}[htb]                                                                                                                  
\resizebox{\textwidth}{!}{   
\begin{tabular}{ccccccccccccc}                                                                                                            
\hline\hline                                                                                                                    
Parameters & CMB & CMB+BAO & CMB+Pantheon &  CMB+BAO+Pantheon \\ \hline 

$\Omega_c h^2$ & $    0.1179_{-    0.0030-    0.0057}^{+    0.0030+    0.0060}$ & $    0.1185_{-    0.0029-    0.0058}^{+    0.0029+    0.0058}$  & $    0.1181_{-    0.0031-    0.0058}^{+    0.0030+    0.0062}$  & $    0.1185_{-    0.0030-    0.0058}^{+    0.0030+    0.0059}$  \\ 

$\Omega_b h^2$ & $    0.02258_{-    0.00025-    0.00047}^{+    0.00024+    0.00048}$ & $    0.02232_{-    0.00023-    0.00042}^{+    0.00023+    0.00045}$ & $    0.02256_{-    0.00025-    0.00049}^{+    0.00025+    0.00051}$ & $    0.02231_{-    0.00022-    0.00045}^{+    0.00022+    0.00045}$  \\ 

$100\theta_{MC}$ & $    1.04120_{-    0.00044-    0.00085}^{+    0.00044+    0.00088}$ & $    1.04109_{-    0.00043-    0.00083}^{+    0.00043+    0.00090}$   & $    1.04117_{-    0.00048-    0.00086}^{+    0.00044+    0.00090}$  & $   1.04110_{-    0.00043-    0.00084}^{+    0.00043+    0.00086}$  \\ 

$\tau$ & $    0.0481_{-    0.0075-    0.017}^{+    0.0084+    0.017}$ & $    0.0539_{-    0.0080-    0.015}^{+    0.0076+    0.016}$  & $    0.0495_{-    0.0077-    0.017}^{+    0.0078+    0.016}$ & $    0.054_{-    0.0082-    0.015}^{+    0.0075+    0.016}$ \\

$n_s$ & $    0.9697_{-    0.0088-    0.018}^{+    0.0090+    0.018}$ & $    0.9626_{-    0.0083-    0.016}^{+    0.0085+    0.017}$ & $    0.9693_{-    0.0094-    0.018}^{+    0.0093+    0.018}$ & $    0.9624_{-    0.0086-    0.017}^{+    0.0088+    0.017}$ \\ 

${\rm{ln}}(10^{10} A_s)$ & $    3.026_{-    0.018-    0.039}^{+    0.020+    0.038}$ & $    3.039_{-    0.018-    0.035}^{+    0.018+    0.035}$ & $    3.030_{-    0.018-    0.038}^{+    0.019+    0.037}$ & $    3.041_{-    0.018-    0.036}^{+    0.018+    0.036}$ \\

$w$ & $   -1.22_{-    0.42-    1.4}^{+    0.95+    1.1}$ &  $   -1.044_{-    0.094-    0.19}^{+    0.097+    0.18}$ & $   -1.22_{-    0.08-    0.19}^{+    0.10+    0.18}$ & $   -1.028_{-    0.039-    0.080}^{+    0.040+    0.078}$  \\ 

$\Omega_{K0}$ & $   -0.049_{-    0.016-    0.076}^{+    0.044+    0.053}$ &  $    0.0003_{-    0.0038-    0.0066}^{+    0.0029+    0.0071}$ & $   -0.028_{-    0.010-    0.021}^{+    0.011+    0.020}$  & $    0.0005_{-    0.0024-    0.0047}^{+    0.0024+    0.0049}$  \\ 

$\Omega_{m0}$ & $    0.48_{-    0.32-    0.37}^{+    0.15+    0.44}$ & $    0.305_{-    0.016-    0.031}^{+    0.016+    0.031}$ & $    0.380_{-    0.029-    0.055}^{+    0.029+    0.057}$  & $    0.307_{-    0.0079-    0.016}^{+    0.0080+    0.016}$ \\

$\sigma_8$ & $    0.82_{-    0.20-    0.23}^{+    0.091+    0.29}$ & $    0.818_{-    0.027-    0.052}^{+    0.028+    0.051}$ & $    0.828_{-    0.014-    0.027}^{+    0.014+    0.027}$ & $    0.815_{-    0.015-    0.028}^{+    0.015+    0.030}$ \\ 

$H_0$ [Km/s/Mpc] & $   59_{-   21-   25}^{+    9+   33}$ & $   68.2_{-    2.2-    3.8}^{+    1.9+    4.1}$  & $   61.1_{-    2.6-    4.8}^{+    2.3+    5.0}$ & $   67.9_{-    1.2-    2.4}^{+    1.2+    2.4}$  \\

$N_{\rm eff}$ & $    3.03_{-    0.19-    0.36}^{+    0.19+    0.39}$ & $    2.96_{-    0.19-    0.36}^{+    0.18+    0.37}$ & $    3.03_{-    0.21-    0.38}^{+    0.19+    0.40}$  & $    2.96_{-    0.19-    0.37}^{+    0.19+    0.38}$ \\ 

$S_8$ & $    0.97_{-    0.08-    0.17}^{+    0.10+    0.15}$ &  $    0.824_{-    0.015-    0.028}^{+    0.014+    0.028}$  & $    0.931_{-    0.037-    0.076}^{+    0.040+    0.072}$ & $    0.824_{-    0.014-    0.027}^{+    0.014+    0.028}$ \\ 

$r_{\rm{drag}}$ [Mpc] & $  147.6_{-    1.9-    3.7}^{+    1.9+    3.6}$ &  $  148.0_{-    1.8-    3.5}^{+    1.9+    3.6}$   & $  147.5_{-    1.9-    3.9}^{+    1.9+    3.8}$   &  $  148.0_{-    1.9-    3.7}^{+    1.9+    3.8}$ \\ 

\hline                                                  
%$\chi^2$ &  2763.500 &   2779.970 &  3799.808  &  3813.752 \\
\hline

\end{tabular}                                                     }                                                              
\caption{68\% and 95\% CL constraints on various free and derived parameters of the $w$CDM $+$ $\Omega_K$ $+$ $N_{\rm eff}$ scenario using several observational datasets.  }
\label{tab:wCDM-Neff}                                                                                                   
\end{table*}                                                                                                                     
\end{center}                                                                                                                    
\endgroup   

\subsubsection{$w$CDM $+$ $\Omega_K$ $+$ $M_{\nu}$ $+$ $N_{\rm eff}$}

\begin{figure*}
    \centering
    \includegraphics[width=0.8\textwidth]{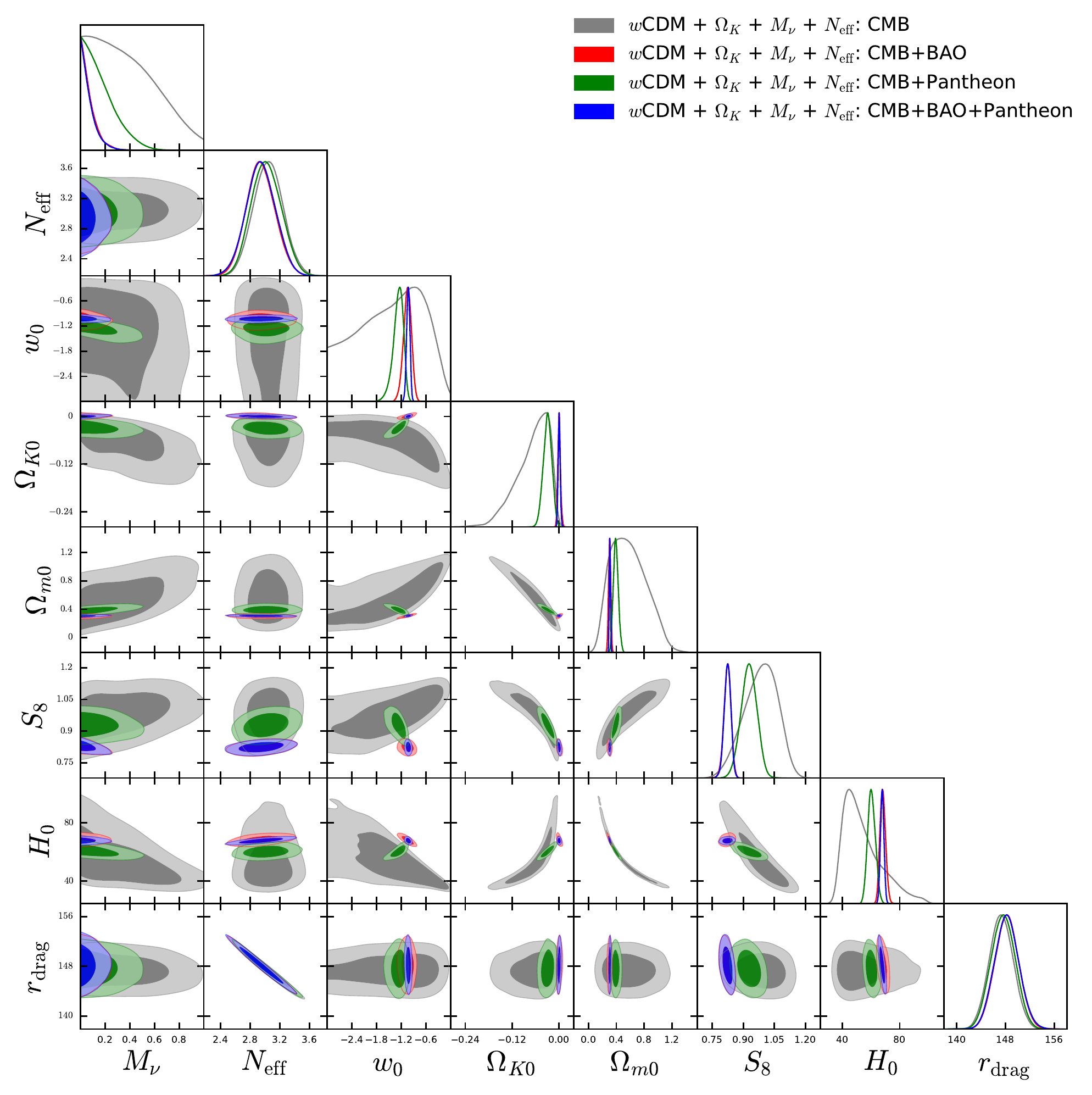}
    \caption{1-dimensional marginalized posterior distributions and the 2-dimensional joint contours for the most relevant parameters of cosmological scenario  $w$CDM $+$ $\Omega_K$ $+$ $M_{\nu}$ $+$ $N_{\rm eff}$  for various observational datasets. }
    \label{fig:wCDM-Omegak-Mnu-Neff}
\end{figure*}

In this last extension of $w$CDM $+$ $\Omega_K$, we simultaneously vary the total neutrino mass and the effective number of relativistic degrees of freedom. We refer to this model as $w$CDM $+$ $\Omega_K$ $+$ $M_{\nu}$ $+$ $N_{\rm eff}$ providing the results in \autoref{tab:wCDM-Mnu-Neff}, and showing the 1D and 2D posterior distributions of different cosmological parameters in \autoref{fig:wCDM-Omegak-Mnu-Neff}. 

In this extended scenario the Planck preference for a closed Universe is reduced below $2\sigma$ and the CMB data constrain $\Omega_{K0}=-0.067^{+0.068}_{-0.086}$ at 95\% CL. Conversely, a preference for a curved geometry is found analyzing CMB+Pantheon, as in this case we get $\Omega_{K0}=-0.029^{+0.020}_{-0.022}$ at 95\% CL. This is the same behavior observed studying the previous extensions. Indeed also the results for $H_0$ remain stable with respect to the case without neutrinos. In particular from the CMB measurements we get $H_0=54^{+29}_{-22}$ km/s/Mpc at 68\% CL, reducing the tension with the SH0ES collaboration value at less than $\sim 2\sigma$ due to the large error, and not to a better overlapping. Combining the CMB data with BAO or Pantheon we obtain $H_0=68.2^{+1.8}_{-2.2}$ km/s/Mpc and $H_0=60.3^{+2.5}_{-2.9}$ km/s/Mpc both at 68\% CL, respectively, producing basically the same tensions among the different datasets discussed in the previous model. As concerns the equation of state of the dark energy, analyzing CMB+Pantheon we still observe a preference at more then 95\% CL for phantom models, while all the other datasets are in agreement with the cosmological constant within one standard deviation. 
As concerns the constraints on the total neutrino mass, the upper bounds are not changed significantly and the most constraining bound obtained for CMB+BAO ($M_{\nu}<0.20$ eV at 95\% CL) is only slightly relaxed varying also $N_{\rm eff}$ in the sample. Similarly, no significant change is observed in the observational constraints for $N_{\rm eff}$. 

\begingroup                                                                                                                     
\squeezetable                                                                                                                   
\begin{center}                                                                                                                  
\begin{table*}[htb]                                                                                                                  
\resizebox{\textwidth}{!}{   
\begin{tabular}{ccccccccccccc}                                                                                                            
\hline\hline                                                                                                                    
Parameters & CMB & CMB+BAO & CMB+Pantheon &  CMB+BAO+Pantheon \\ \hline 

$\Omega_c h^2$ & $    0.1183_{-    0.0033-    0.0059}^{+    0.0030 +    0.0061}$ & $    0.1182_{-    0.0030-    0.0061}^{+    0.0031+    0.0060}$  & $    0.1180_{-    0.0033-    0.0058}^{+    0.0030+    0.0062}$  & $    0.1182_{-    0.0031-    0.0059}^{+    0.0031+    0.0062}$ \\ 

$\Omega_b h^2$ & $    0.02253_{-    0.00025-    0.00049}^{+    0.00025 +    0.00049}$ & $  0.02230_{-    0.00023-    0.00044}^{+    0.00024+    0.00046}$  & $    0.02252_{-    0.00025-    0.00050}^{+    0.00025+    0.00050}$ & $    0.02231_{-    0.00023-    0.00047}^{+    0.00023+    0.00045}$   \\ 

$100\theta_{MC}$ & $    1.04103_{-    0.00045-    0.00088}^{+    0.00045+    0.00091}$ & $    1.04113_{-    0.00046-    0.00085}^{+    0.00044+    0.00088}$ & $    1.04112_{-    0.00046-    0.00089}^{+    0.00046+    0.00091}$ & $    1.04113_{-    0.00045-    0.00085}^{+    0.00044+    0.00088}$   \\ 

$\tau$ & $    0.0476_{-    0.0077-    0.017}^{+    0.0084+    0.016}$ &  $    0.0541_{-    0.0076-    0.015}^{+    0.0078+    0.016}$  & $    0.0495_{-    0.0075-    0.017}^{+    0.0080+    0.016}$ & $    0.055_{-    0.0083-    0.016}^{+    0.0073+    0.017}$ \\ 

$n_s$ & $    0.9689_{-    0.0092-    0.0187}^{+    0.0092 +    0.0183}$ &  $    0.9618_{-    0.0095-    0.0168}^{+    0.0086+    0.0176}$  & $    0.9681_{-    0.00940989-    0.018}^{+    0.0093+    0.018}$  & $    0.9621_{-    0.0088-    0.0178}^{+    0.0089+    0.0176}$ \\ 

${\rm{ln}}(10^{10} A_s)$ & $    3.025_{-    0.019-    0.039}^{+    0.019+    0.036}$ & $    3.039_{-    0.019-    0.037}^{+    0.019+    0.037}$  &  $    3.030_{-    0.019-    0.039}^{+    0.019+    0.038}$ &  $    3.040_{-    0.018-    0.037}^{+    0.019+    0.038}$  \\ 

$w$ & $   -1.5_{-    0.6-    1.5}^{+    1.0+    1.1}$ &  $   -1.05_{-    0.09-    0.20}^{+    0.11+    0.19}$ & $   -1.28_{-    0.09-    0.26}^{+    0.15+    0.24}$  &  $   -1.027_{-    0.039-    0.079}^{+    0.040+    0.075}$  \\ 

$\Omega_{K0}$ & $   -0.067_{-    0.024-    0.086}^{+    0.055+    0.068}$ & $    0.0003_{-    0.0037-    0.0066}^{+    0.0030+    0.0071}$ & $   -0.029_{-    0.010-    0.022}^{+    0.012+    0.020}$  &  $    0.0005_{-    0.0027-    0.0052}^{+    0.0024+    0.0052}$  \\ 

$\Omega_{m0}$ & $    0.58_{-    0.34-    0.44}^{+    0.20+    0.48}$ & $    0.304_{-    0.015-    0.031}^{+    0.015+    0.030}$ & $    0.394_{-    0.039-    0.068}^{+    0.032+    0.072}$  &  $    0.307_{-    0.0085-    0.016}^{+    0.0080+    0.016}$ \\ 

$\sigma_8$ & $    0.76_{-    0.17-    0.22}^{+    0.08+    0.28}$ & $    0.818_{-    0.028-    0.054}^{+    0.027+    0.055}$   & $    0.811_{-    0.020-    0.058}^{+    0.031+    0.050}$  &  $    0.814_{-    0.016-    0.037}^{+    0.019+    0.035}$  \\ 

$H_0$ [Km/s/Mpc] & $   54_{-   17-   22}^{+    7+   29}$ & $   68.2_{-    2.2-    3.8}^{+    1.8+    3.9}$  & $   60.3_{-    2.9-    5.0}^{+    2.5+    5.4}$  & $   67.8_{-    1.2-    2.4}^{+    1.2+    2.4}$ \\ 

$M_\nu$ [eV] & $   <0.49\,<0.81 $ & $   <0.084\,<0.20 $  & $ <0.203\,<0.41   $  & $ <0.080\,<0.20  $  \\ 

$N_{\rm eff}$ & $   3.04_{-    0.20-    0.39}^{+    0.20+    0.40}$ & $    2.94_{-    0.20-    0.37}^{+    0.19+    0.39}$   & $    3.02_{-    0.22-    0.37}^{+    0.20+    0.39}$ & $    2.94_{-    0.19-    0.39}^{+    0.20+    0.39}$ \\

$S_8$ & $    0.984_{-    0.071-    0.16}^{+    0.096+    0.15}$ &  $    0.823_{-    0.016-    0.035}^{+    0.018+    0.032}$ & $    0.927_{-    0.038-    0.076}^{+    0.038+    0.072}$   & $    0.823_{-    0.015-    0.033}^{+    0.017+    0.030}$ \\ 

$r_{\rm{drag}}$ [Mpc] &  $  147.3_{-    2.0-    3.8}^{+    1.9+    3.9}$  &  $  148.2_{-    1.9-    3.7}^{+    1.9+    3.8}$  & $  147.6_{-    2.0-    3.8}^{+    2.0+    3.8}$   & $  148.2_{-    2.0-    3.8}^{+    2.0+    3.9}$ \\ 

\hline                                                  
%$\chi^2$ &  2762.922  & 2777.502  &  3799.036  &  3813.066  \\
\hline                                                             
\end{tabular}                                                   }                                                                
\caption{68\% and 95\% CL constraints on various free and derived parameters of the $w$CDM $+$ $\Omega_K$ $+$ $M_{\nu}$ $+$ $N_{\rm eff}$ scenario using several observational datasets.  }
\label{tab:wCDM-Mnu-Neff}                                                                                                   
\end{table*}                                                                                                                     
\end{center}                                                                                                                    
\endgroup 

\subsection{Non-flat $w_0w_a$CDM and its extensions}

Here we discuss the observational constraints on the non-flat $w_0w_a$CDM model and its various extensions through the inclusion of the neutrino sector. As written before, we have used exactly the same datasets and their combinations and for each parameter we report either its 68\% and 95\% CL constraints or the upper/lower limits at 68\% and 95\% CL.  

\subsubsection{$w_0w_a$CDM $+$ $\Omega_K$}

\begin{figure*}
    \centering
    \includegraphics[width=0.8\textwidth]{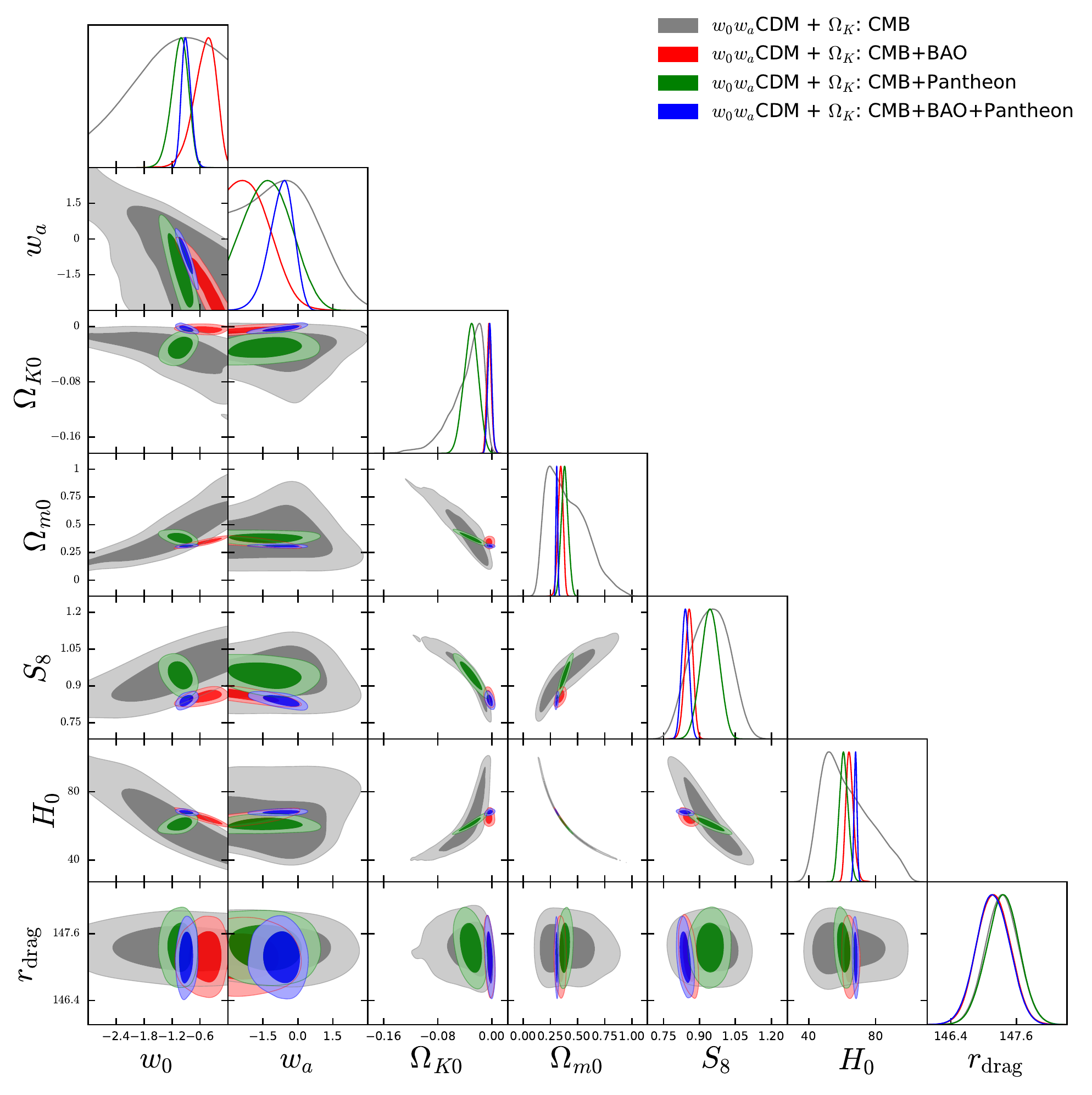}
    \caption{This figure corresponds to the 1-dimensional posterior distributions and the 2-dimensional joint contours for the most important parameters of cosmological scenario labeled by $w_0w_a$CDM $+$ $\Omega_K$ using various observational datasets. }
    \label{fig:w0waCDM-Omegak}
\end{figure*}

In \autoref{tab:CPL} we have summarized the observational constraints on the free and derived parameters of this model and in \autoref{fig:w0waCDM-Omegak} we show the 1D and 2D posterior distributions on a few interesting parameters. 
Let us describe the impacts of CMB alone and its combination with other non-CMB datasets. 

As displayed in the second column of \autoref{tab:CPL}, we see that CMB alone indicates the preference of a closed Universe at more than 68\% CL, but $\Omega_K$ is in agreement with zero within $2\sigma$ ($\Omega_K =  -0.038_{-    0.012}^{+    0.031}$ at 68\% CL). And as usual we have a lower mean value of $H_0$ for this case ($H_0 = 63_{-19}^{+    10}$ km/s/Mpc at 68\% CL) but with very large error bars. Due to such high error bars, the tension with SH0ES is reduced down to $\sim 1 \sigma$. The current value of the dark energy equation of state even though has its mean value in the phantom state ($w_0 =  -1.3_{-  0.4}^{+    1.2}$ at 68\% CL) within 68\% CL is in agreement with the cosmological constant.  Also, the dynamical character of the dark energy equation of state quantified through $w_a$ ($= -0.6_{-    1.8}^{+    1.2}$ at 68\% CL) is in agreement with zero, i.e. the dynamical DE is not favoured.

When BAO data are added to the CMB, the possibility of nonzero curvature of the Universe is present only at 68\% CL ($\Omega_K=-0.0042^{+0.0030}_{-0.0036}$ at 68\% CL for CMB+BAO) and the usual shift of $H_0$ towards lower mean values is present as well, with a reduction of its error bars with respect to the CMB only case ($H_0 = 64.6_{-    2.6}^{+    1.9}$ km/s/Mpc at 68\% CL), restoring the tension with local measurement at $\sim 3.9 \sigma$. 
Concerning the free parameters $w_0$ and $w_a$ we see that $w_0$ is strictly in the quintessence regime at more than 95\% CL while $w_a$ is non-zero at more than 95\% CL, that means we have an evidence of the dynamical quintessence dark energy for this case at more than $2\sigma$.

When Pantheon data are added to CMB, we recover the evidence of nonzero curvature at more than 95\% CL ($\Omega_K =  -0.031_{-0.021}^{+  0.020}$ at 95\% CL), while the tension of $H_0$ with SH0ES exacerbates. About the properties of the dark energy parameters, we see that the mean value of $w_0$ is perfectly in agreement with the cosmological constant $w_0 = -1$ ( $w_0 = -1.04_{-    0.16}^{+    0.20}$ at 68\% CL for CMB+Pantheon), but the dynamical nature of the DE is signaled through $w_a$ which is found to be nonzero within 68\% CL ($w_a = -1.2_{-    1.1}^{+    0.9}$ at 68\% CL for CMB+Pantheon) while it becomes consistent to zero at 95\% CL.  

For the combined analysis CMB+BAO+Pantheon we are able to break all of the degeneracies and have a very mild improvement of the Hubble constant agreement, i.e. $H_0 = 68.01_{-    0.85}^{+    0.83}$ km/s/Mpc at 68\% CL for this case. The curvature parameter is consistent with zero at slightly more than 68\% CL. Additionally, we recover the quintessence nature of $w_0$ at slightly more than 68\% CL, similar to what we observed for the CMB+BAO analysis, and the same we can say for $w_a$, that is only found to be nonzero at more than 68\% CL, showing a hint for a dynamical nature, while within the 95\% CL $w_a =0$ is back in agreement with the data.

\begingroup                                                                                                                     
\squeezetable                                                                                                                   
\begin{center}                                                                                                                  
\begin{table*}[htb]                                                                                                                  
\resizebox{\textwidth}{!}{   
\begin{tabular}{ccccccccccccc}                        
\hline\hline                                                                                                                    
Parameters & CMB & CMB+BAO & CMB+Pantheon &  CMB+BAO+Pantheon \\ \hline 

$\Omega_c h^2$ & $    0.1181_{-    0.0015-    0.0028}^{+    0.0015+    0.0029}$ & $    0.1194_{-    0.0014-    0.0028-    0.0037}^{+    0.0014+    0.0028}$ & $    0.1181_{-    0.0014-    0.0029}^{+    0.0014+    0.0029}$  & $    0.1195_{-    0.0014-    0.0028}^{+    0.0014+    0.0028}$\\ 

$\Omega_b h^2$ & $    0.02261_{-    0.00017-    0.00034}^{+    0.00017+    0.00033}$ &   $    0.02244_{-    0.00015-    0.00030}^{+    0.00015+    0.00031}$  & $    0.02260_{-    0.00017-    0.00034}^{+    0.00017+    0.00033}$ & $    0.02243_{-    0.00016-    0.00031}^{+    0.00016+    0.00032}$ \\ 

$100\theta_{MC}$ & $    1.04117_{-    0.00033-    0.00065}^{+    0.00033+    0.00063}$ &  $    1.04100_{-    0.00031-    0.00062}^{+    0.00031+    0.00062}$  & $    1.04118_{-    0.00032-    0.00065}^{+    0.00033+    0.00064}$ & $    1.04099_{-    0.00032-    0.00063}^{+    0.00032+    0.00064}$   \\ 

$\tau$ & $    0.0484_{-    0.0075-    0.017}^{+    0.0078+    0.016}$ &  $    0.0535_{-    0.0080-    0.015}^{+    0.0075+    0.016}$  &  $    0.0483_{-    0.0075-    0.016}^{+    0.0077+    0.016}$ & $    0.054_{-    0.0075-    0.015}^{+    0.0075+    0.016}$  \\ 

$n_s$ & $    0.9705_{-    0.0048-    0.0092}^{+    0.0047+    0.0095}$ &  $    0.9670_{-    0.0043-    0.0085}^{+    0.0044+    0.0088}$ & $    0.9707_{-    0.0046-    0.0094}^{+    0.0046+    0.0093}$ & $    0.9666_{-    0.0044-    0.0087}^{+    0.0044+    0.0090}$ \\ 

${\rm{ln}}(10^{10} A_s)$ & $    3.028_{-    0.016-    0.034}^{+    0.016 +    0.033}$ & $    3.041_{-    0.016-    0.031}^{+    0.016+    0.032}$   &  $    3.028_{-    0.016-    0.035}^{+    0.016+    0.032}$ & $    3.043_{-    0.015-    0.031}^{+    0.016+    0.032}$  \\ 

$w_0$ & $   -1.3_{-    0.4-    1.7}^{+    1.2+    1.3}$ & $   -0.51_{-    0.19-    0.48}^{+    0.29+    0.44}$ & $   -1.04_{-    0.16-    0.36}^{+    0.20+    0.34}$  & $   -0.89_{-    0.11-    0.20}^{+    0.09+    0.21}$  \\ 

$w_a$ & $   -0.6_{-    1.8-    2.4}^{+    1.2+    2.3}$ &  $   -1.8_{-    1.2-    1.2}^{+    0.3+    1.4}$ & $   -1.2_{-    1.1}^{+    0.9}\,<0.4$ &  $   -0.73_{-    0.46-    1.1}^{+    0.61+    1.0}$ \\ 

$\Omega_{K0}$ & $   -0.038_{-    0.012-    0.052}^{+    0.031+    0.039}$ &  $   -0.0042_{-    0.0036-    0.0066}^{+    0.0030+    0.0067}$   & $   -0.031_{-    0.011-    0.021}^{+    0.011+    0.020}$ & $   -0.0032_{-    0.0030-    0.0065}^{+    0.0030+    0.0062}$  \\ 

$\Omega_{m0}$ & $    0.41_{-    0.24-    0.30}^{+    0.12+    0.35}$ & $    0.342_{-    0.022-    0.047}^{+    0.025 +    0.047}$ & $    0.382_{-    0.035-    0.058}^{+    0.030+    0.063}$ & $    0.308_{-    0.0076-    0.015}^{+    0.0077+    0.016}$ \\ 

$\sigma_8$ & $    0.86_{-    0.18-    0.22}^{+    0.10+    0.26}$ &  $    0.802_{-    0.026-    0.048}^{+    0.023+    0.049}$  & $    0.837_{-    0.013-    0.029}^{+    0.015+    0.026}$ & $    0.829_{-    0.014-    0.028}^{+    0.014+    0.027}$  \\

$H_0$ [Km/s/Mpc] & $   63_{-   19-   24}^{+    10+   30}$ &  $   64.6_{-    2.6-    4.4}^{+    1.9+    4.6}$  & $   61.0_{-    2.8-    5.1}^{+    2.5+    5.1}$  & $   68.01_{-    0.85-    1.7}^{+    0.83+    1.7}$  \\ 

$S_8$ & $    0.944_{-    0.079-    0.15}^{+    0.088+    0.14}$ &  $    0.855_{-    0.016-    0.036}^{+    0.019+    0.032}$ & $    0.944_{-    0.038-    0.074}^{+    0.039+    0.072}$ &  $    0.841_{-    0.015-    0.030}^{+    0.015+    0.030}$ \\

$r_{\rm{drag}}$ [Mpc] & $  147.34_{-    0.30-    0.59}^{+    0.30+    0.59}$ & $  147.18_{-    0.30-    0.59}^{+    0.31+    0.59}$ & $  147.35_{-    0.29-    0.59}^{+    0.31+    0.59}$ &  $  147.17_{-    0.30-    0.59}^{+    0.30+    0.58}$ \\ 

\hline                                                  
%$\chi^2$ &  2762.360 & 2776.006  & 3798.260 & 3810.866 \\
\hline                                                        
\end{tabular}  }                                            
\caption{68\% and 95\% CL constraints on various free and derived parameters of the $w_0w_a$CDM $+$ $\Omega_{K}$ scenario using several observational datasets. }
\label{tab:CPL}                                                                                                   
\end{table*}                                                                                                                     
\end{center}                                                                                                                    
\endgroup                                                                                                

\subsubsection{$w_0w_a$CDM $+$ $\Omega_K$ $+$ $M_{\nu}$}

\begin{figure*}
    \centering
    \includegraphics[width=0.8\textwidth]{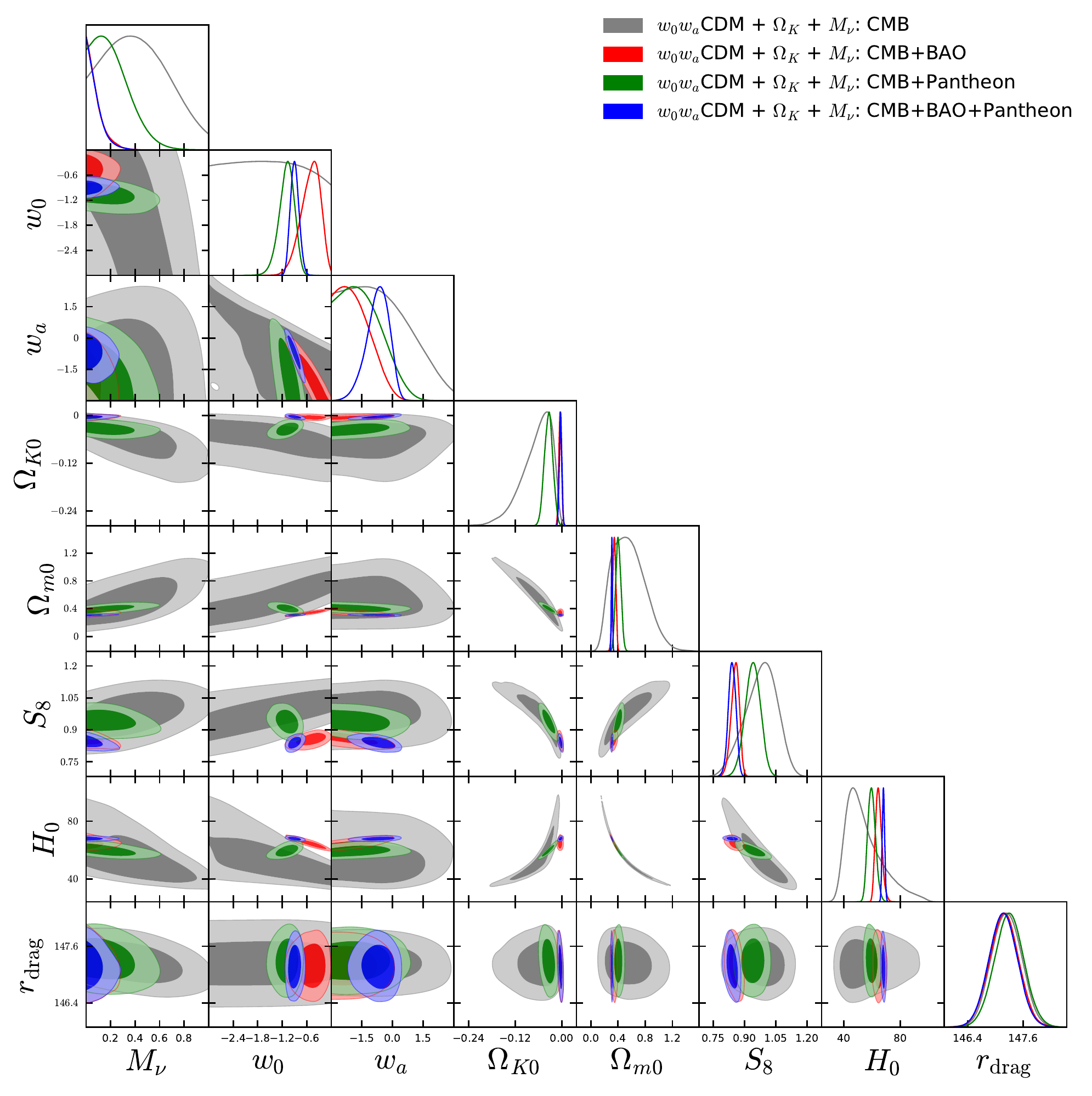}
    \caption{This figure corresponds to the 1-dimensional posterior distributions and the 2-dimensional joint contours for the most important parameters of cosmological scenario labeled by $w_0w_a$CDM $+$ $\Omega_K$ $+$ $M_{\nu}$ using various observational datasets.  }
    \label{fig:w0waCDM-Omegak-Mnu}
\end{figure*}

As usual, we start extending the parameter space of the model considering massive neutrinos. We refer to this model as $w_0w_a$CDM $+$ $\Omega_K$ $+$ $M_{\nu}$ and we show the cosmological constraints on the parameters  in \autoref{tab:CPL-Mnu} and in ~\autoref{fig:w0waCDM-Omegak-Mnu}. 

Considering only the CMB data the value found for the curvature parameter is $\Omega_{K0}=-0.065^{+0.066}_{-0.081}$ at 95\% CL and one can see that while flatness is disfavored by the 68\% CL limit it is consistent within the 95\% CL bounds, similarly to the baseline case discussed above. Interestingly, including BAO measurements, we still find a preference for a closed Universe at the level of 68\% CL even though flatness is still consistent with the 95\% CL limits. Conversely, when the Pantheon data are added to the CMB measurements, we recover the evidence of nonzero curvature at more than 95\% CL ($\Omega_{K0} =  -0.034_{-0.020}^{+0.021}$ at 95\% CL for CMB+Pantheon). As concerns the Hubble parameter, from the CMB data we obtain a lower mean value with respect to the baseline case without neutrinos and our 95\% CL now reads $H_0=55^{+29}_{-21}$ km/s/Mpc.  Due to the large error-bars this result is in agreement with SH0ES within $2\sigma$. On the other hand, combining CMB+BAO we find $H_0=64.6^{+1.9}_{-2.6}$ km/s/Mpc at 68\% CL and in this case, due to the smaller uncertainty, the tension is increased to $3.9\sigma$. Finally, when we combine CMB+Pantheon, the combined effects of curvature and neutrinos, lead to smaller values for the expansion rate, $H_0=59.7^{+2.4}_{-3.0}$ km/s/Mpc, at 68\% CL and the result is in strong tension with local estimation at $5\sigma$. The different datasets analyzed in this extended model provide discordant predictions also for the dark energy sector. In particular, exploiting only the Planck satellite measurements $w_0$ turns out to be unconstrained while only an uninformative upper bound $w_a<1.9$ at 95\% CL can be derived for the parameter who carries information about the dynamical nature of DE. Conversely, including also the BAO measurements we find both a strong preference for a quintessential dark energy ($w_0=-0.51^{+0.43}_{-0.47}$ at 95\%) with a dynamical behavior, as the upper bound $w_a<-0.4$ rules out the case $w_a=0$ at 95\% CL. On the other hand, CMB+Pantheon is again in agreement with a cosmological constant within 68\% CL, and the upper bound  $w_a<0.24$ at 95\% is in agreement with a non-dynamical equation of state for the dark energy. Finally, combining CMB+BAO+Phantom all together we find that quintessential dynamical models are preferred at more than $1\sigma$ but both phantom models and non-dynamical dark energy are allowed within the 95\% CL constraints. As concerns neutrinos, the limits on their total mass are not drastically changed by the modification in the dark energy sector and the 95\% CL bound obtain for CMB+BAO ($M_{\nu}<0.22$ eV at 95\% CL) is only slightly larger than the respective limit obtained within curved $\Lambda$CDM or $w$CDM cosmologies. This full dataset is completely in agreement with a flat universe.

\begingroup                                                                                                                     
\squeezetable                                                                                                                   
\begin{center}                                                                                                                  
\begin{table*}[htb]                                                                                                                   
\resizebox{\textwidth}{!}{   
\begin{tabular}{ccccccccccccc}                                                                                                            
\hline\hline                                                                                                                    
Parameters & CMB & CMB+BAO & CMB+Pantheon &  CMB+BAO+Pantheon \\ \hline %& CMB+R19  \\ \hline

$\Omega_c h^2$ & $    0.1183_{-    0.0015-    0.0029}^{+    0.0015+    0.0029}$ & $    0.1193_{-    0.0014-    0.0029}^{+    0.0015+    0.0029}$ & $    0.1183_{-    0.0015-    0.0030}^{+    0.0015+    0.0030}$ & $    0.1195_{-    0.0014-    0.0029}^{+    0.0014+    0.0028}$ \\ 

$\Omega_b h^2$ & $    0.02252_{-    0.00019-    0.00035}^{+    0.00018 +    0.00036}$ & $    0.02244_{-    0.00016-    0.00032}^{+    0.00016+    0.00032}$ & $    0.02256_{-    0.00018-    0.00034}^{+    0.00017+    0.00035}$ & $    0.02242_{-    0.00016-    0.00031}^{+    0.00016+    0.00032}$ \\ 

$100\theta_{MC}$ & $    1.04097_{-    0.00034-    0.00067}^{+    0.00034 +    0.00067}$ & $    1.04098_{-    0.00032-    0.00063}^{+    0.00031+    0.00061}$   & $    1.04106_{-    0.00034-    0.00067}^{+    0.00034+    0.00065}$ & $    1.04097_{-    0.00032-    0.00064}^{+    0.00032+    0.00063}$  \\ 

$\tau$ & $    0.0471_{-    0.0076-    0.017}^{+    0.0078+    0.017}$ & $    0.0536_{-    0.0075-    0.015}^{+    0.0076+    0.016}$   & $    0.0493_{-    0.0074-    0.016}^{+    0.0076+    0.015}$ & $    0.054_{-    0.0075-    0.015}^{+    0.0075+    0.016}$  \\ 

$n_s$ & $    0.9690_{-    0.0050-    0.0098}^{+    0.0050+    0.0095}$ & $    0.9669_{-    0.0045-    0.0092}^{+    0.0046+    0.0091}$   & $    0.9696_{-    0.0048-    0.0095}^{+    0.0047+    0.0094}$ & $    0.9665_{-    0.0045-    0.0089}^{+    0.0046+    0.0090}$ \\ 

${\rm{ln}}(10^{10} A_s)$ & $    3.024_{-    0.017-    0.036}^{+    0.017+    0.033}$ & $    3.041_{-    0.016-    0.032}^{+    0.016+    0.033}$  & $    3.030_{-    0.015-    0.034}^{+    0.017+    0.032}$  & $    3.042_{-    0.016-    0.032}^{+    0.016+    0.032}$  \\ 

$w_0$ & $  unconstrained $ & $   -0.51_{-    0.19-    0.47}^{+    0.28+    0.43}$  & $   -1.09_{-    0.16-    0.38}^{+    0.21+    0.36}$  & $   -0.90_{-    0.12-    0.20}^{+    0.10+    0.21}$ \\ 

$w_a$ & $ <0.00\,< 1.9$ & $ <-1.50\,<-0.4$  &  $  <-1.03\,<0.24 $ & $   -0.73_{-    0.45-    1.1}^{+    0.65+    1.0}$ \\ 

$\Omega_{K0}$ & $   -0.065_{-    0.023-    0.081}^{+    0.051+    0.066}$ & $   -0.0041_{-    0.0034-    0.0067}^{+    0.0034+    0.0071}$   & $   -0.034_{-    0.011-    0.021}^{+    0.011+    0.020}$ & $   -0.0030_{-    0.0032-    0.0062}^{+    0.0032+    0.0063}$  \\ 

$\Omega_{m0}$ & $    0.56_{-    0.30-    0.43}^{+    0.19+    0.46}$ & $    0.343_{-    0.022-    0.048}^{+    0.025+    0.044}$  & $    0.404_{-    0.040-    0.071}^{+    0.037+    0.076}$  & $    0.309_{-    0.0079-    0.016}^{+    0.0080+    0.016}$  \\ 

$\sigma_8$ & $    0.76_{-    0.16-    0.21}^{+    0.08+    0.27}$ & $    0.800_{-    0.028-    0.050}^{+    0.025+    0.054}$  & $    0.810_{-    0.023-    0.062}^{+    0.037+    0.055}$ & $    0.827_{-    0.017-    0.036}^{+    0.020+    0.035}$ \\ 

$H_0$ [Km/s/Mpc] & $   55_{-   16-   21}^{+    7+   29}$ & $   64.6_{-    2.6-    4.3}^{+    1.9+    4.6}$  & $   59.7_{-    3.0-    5.2}^{+    2.4+    5.5}$  & $   67.98_{-    0.85-    1.7}^{+    0.86+    1.7}$  \\ 

$M_\nu$ [eV] & $    0.42_{-    0.32}^{+    0.21}\,<0.84$ & $  <0.092\,<0.22 $ & $  <0.272\,<0.49 $  & $  <0.090\,<0.21$  \\ 

$S_8$ & $    0.977_{-    0.067-    0.16}^{+    0.090+    0.14}$ &  $    0.854_{-    0.017-    0.041}^{+    0.022+    0.038}$ & $    0.938_{-    0.036-    0.075}^{+    0.036+    0.070}$  & $    0.840_{-    0.018-    0.036}^{+    0.018+    0.036}$ \\ 

$r_{\rm{drag}}$ [Mpc] &  $  147.24_{-    0.32-    0.62}^{+    0.32+    0.62}$ &  $  147.19_{-    0.31-    0.61}^{+    0.31+    0.60}$   & $  147.29_{-    0.31-    0.62}^{+    0.31+    0.62}$ &  $  147.17_{-    0.30-    0.60}^{+    0.30+    0.61}$ \\ 

\hline                                                  
%$\chi^2$ & 2762.916  &  2773.492  & 3797.768 &  3811.460    \\
\hline

\end{tabular}                                                   }                                                                
\caption{68\% and 95\% CL constraints on various free and derived parameters of the $w_0w_a$CDM $+$ $\Omega_K$ $+$ $M_{\nu}$ scenario using several observational datasets. }
\label{tab:CPL-Mnu}                                                                                                   
\end{table*}                                                                                                                     
\end{center}                                                                                                                    
\endgroup                 

\subsubsection{$w_0w_a$CDM $+$ $\Omega_K$ $+$ $N_{\rm eff}$}

\begin{figure*}
    \centering
    \includegraphics[width=0.8\textwidth]{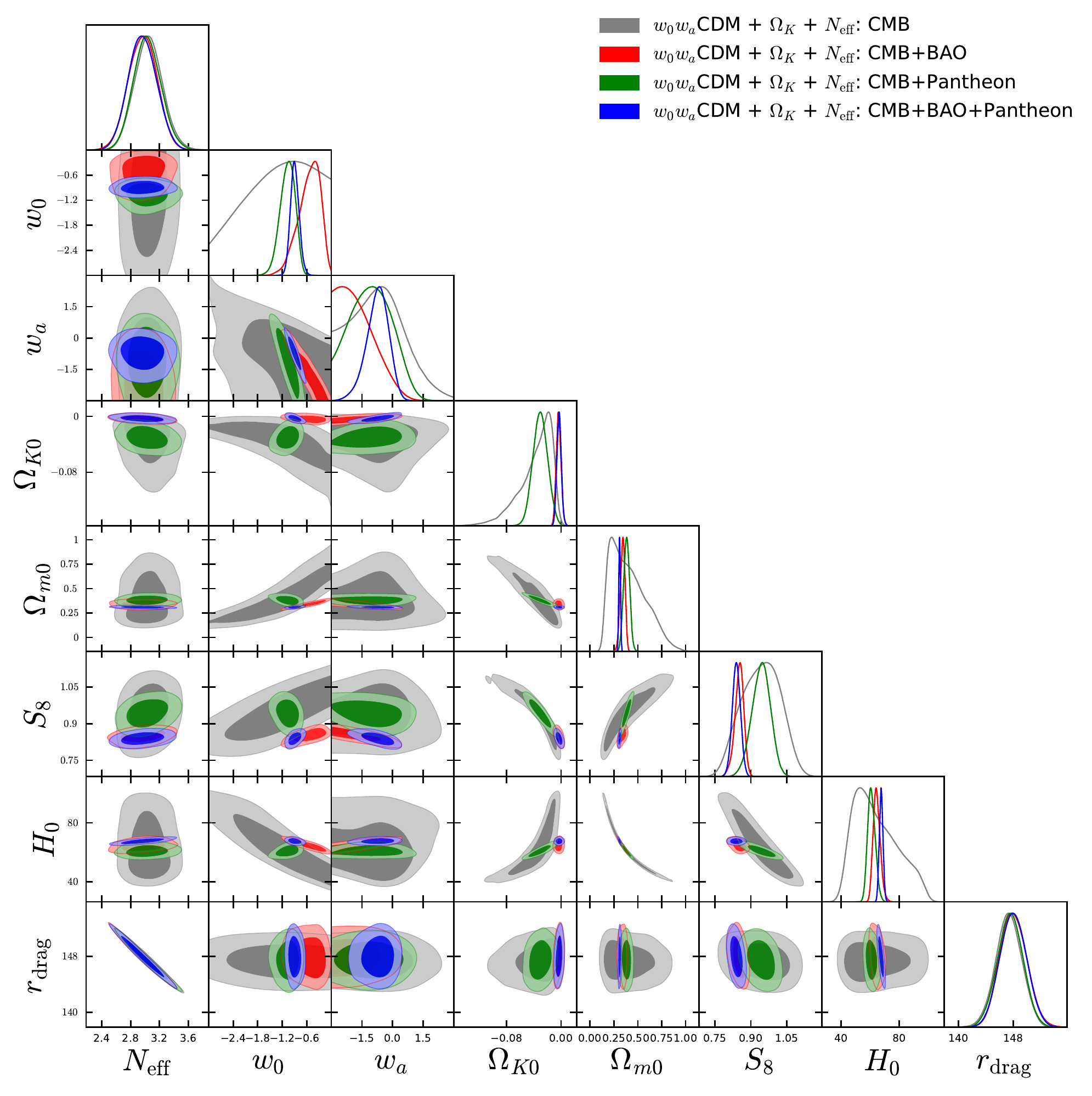}
    \caption{This figure corresponds to the 1-dimensional posterior distributions and the 2-dimensional joint contours for the most important parameters of cosmological scenario labeled by $w_0w_a$CDM $+$ $\Omega_K$ $+$ $N_{\rm eff}$ using various observational datasets.  }
    \label{fig:w0waCDM-Omegak-Neff}
\end{figure*}

We now allow the effective number of relativistic degrees of freedom to freely vary in curved cosmologies with dynamical models of dark energy. We refer to this scenario as $w_0w_a$CDM $+$ $\Omega_K$ $+$ $N_{\rm eff}$. The 68\% and 95\% CL constraints on the parameters of the model are given in \autoref{tab:CPL-Neff} and in \autoref{fig:w0waCDM-Omegak-Neff} we show the marginalized 1D and 2D posterior distributions of the parameters. 

Considering the CMB bounds, also in this case we find that flatness, while disfavored by the 68\% CL limit, is consistent within the 95\% CL bounds. Instead, for the Hubble constant we get $H_0=64^{+10}_{-20}$ km/s/Mpc at 68\% CL that, due to the large errors, is consistent with the SH0ES measurement within one standard deviation. Notice also that in this model the parameters involving modifications to the dark energy sector are basically unconstrained by the CMB measurements. 
Including the BAO data a curved background geometry is always slightly preferred at 68\% CL, but flatness is consistent within the  95\% CL limits, as well. The value of the Hubble constant  $H_0=64.6^{+2.1}_{-2.8}$ km/s/Mpc at 68\% CL remains stable with respect to the CMB result, but with a much smaller uncertainty. Consequently, the tension with SH0ES is increased at the level of $\sim 3.6\sigma$. Concerning the dark energy parameters, for this dataset  quintessential models are preferred and the constraints on $w_0$ reads at 95\% CL $w_0=-0.55^{+0.47}_{-0.54}$. Instead  the upper bound $w_a<-0.1$ excludes a non-dynamical evolution at 95\% CL. 
Combining CMB and Pantheon we find the usual evidence for a curved Universe ($\Omega_{K0}=-0.031\pm0.021$ at 95\% CL) resulting into a preference for smaller values of the expansion rate ($H_0=60.9^{+2.4}_{-2.9}$ km/s/Mpc at 68\% CL). This is in large tension with SH0ES at more than $4.6\sigma$. Furthermore in this case both phantom and quintessential models of dark energy are allowed, or in other words $w_0$ is in agreement with a cosmological constant, and the bounds on $w_a$ do not give substantial information about its dynamical evolution.
When instead  we combine CMB+BAO+Pantheon all together, we do not find evidence for curvature and for then expansion rate we obtain $H_0=67.6\pm 1.3$ km/s/Mpc at 68\% CL. This result is similar to the best fit value that can be obtained within the standard flat $\Lambda$CDM model. For the same dataset quintessential dynamical models of dark energy turns out to be preferred at 68\% CL even though both phantom models and non-dynamical evolution are allowed at 95\% CL and the bounds on the dark energy parameters reads $w_0=-0.89^{+0.22}_{-0.20}$ and $w_{a}=-0.75\pm1.1$ at 95\% CL, respectively. 
Finally, concerning the constraints on the effective number or relativistic degrees of freedom, it is worth pointing out that they remain almost unchanged with respect to the other curved models of dark energy analyzed in this work for all the different data combinations. 

\begingroup                                                                                                                     
\squeezetable                                                                                                                   
\begin{center}                                                                                                                  
\begin{table*}[htb]                                                                                                                   
\resizebox{\textwidth}{!}{   
\begin{tabular}{ccccccccccccc}                                                                                                            
\hline\hline                                                                                                                    
Parameters & CMB & CMB+BAO & CMB+Pantheon &  CMB+BAO+Pantheon \\ \hline 

$\Omega_c h^2$ & $    0.1181_{-    0.0030-    0.0056}^{+    0.0029+    0.0060}$ & $    0.1183_{-    0.0030-    0.0058}^{+    0.0030+    0.0058}$   & $    0.1179_{-    0.0029-    0.0057}^{+    0.0029+    0.0057}$ &  $    0.1184_{-    0.0030-    0.0058}^{+    0.0030+    0.0059}$  \\ 

$\Omega_b h^2$ & $    0.02261_{-    0.00025-    0.00049}^{+    0.00025+    0.00049}$ & $    0.02237_{-    0.00024-    0.00047}^{+    0.00024+    0.00047}$   & $    0.02257_{-    0.00025-    0.00049}^{+    0.00026+    0.00051}$ & $    0.02235_{-    0.00023-    0.00046}^{+    0.00023+    0.00046}$  \\ 

$100\theta_{MC}$ & $    1.04118_{-    0.00043-    0.00082}^{+    0.00044+    0.00087}$ & $    1.04114_{-    0.00044-    0.00086}^{+    0.00044+    0.00090}$   & $    1.04119_{-    0.00045-    0.00085}^{+    0.00045+    0.00086}$ & $    1.04111_{-    0.00046-    0.00083}^{+    0.00044+    0.00086}$ \\ 

$\tau$ & $    0.0483_{-    0.0077-    0.019}^{+    0.0085+    0.016}$ & $    0.0532_{-    0.0075-    0.015}^{+    0.0077+    0.016}$  & $    0.0487_{-    0.0078-    0.017}^{+    0.0085+    0.016}$ & $    0.0536_{-    0.0074-    0.015}^{+    0.0074+    0.015}$ \\ 

$n_s$ & $    0.9709_{-    0.0096-    0.0182}^{+    0.0094+    0.0183}$ & $    0.9639_{-    0.0088-    0.0186}^{+    0.0099+    0.0171}$  & $    0.9699_{-    0.0090-    0.018}^{+    0.0088+    0.018}$ & $    0.9632_{-    0.0090-    0.018}^{+    0.0091+    0.018}$   \\ 

${\rm{ln}}(10^{10} A_s)$ & $    3.027_{-    0.018-    0.043}^{+    0.020+    0.038}$ & $    3.037_{-    0.018-    0.036}^{+    0.018+    0.037}$ & $    3.027_{-    0.018-    0.037}^{+    0.018+    0.037}$ & $    3.038_{-    0.018-    0.035}^{+    0.017+    0.035}$  \\ 

$w_0$ & $ >-1.68,\,unconstr.$ & $   -0.55_{-    0.19-    0.54}^{+    0.34+    0.47}$   & $   -1.07_{-    0.16-    0.36}^{+    0.21+    0.34}$  & $   -0.89_{-    0.11-    0.20}^{+    0.10+    0.22}$  \\ 

$w_a$ & $   -0.7_{-    1.4}^{+    1.3}\,<1.4$ & $  <-1.36\,<-0.1$   & $   -1.0_{-    1.1}^{+    1.1} <0.6$ & $   -0.75_{-    0.47-    1.1}^{+    0.62+    1.1}$ \\ 

$\Omega_{K0}$ & $   -0.037_{-    0.012-    0.052}^{+    0.031+    0.038}$ &  $   -0.0036_{-    0.0038-    0.0066}^{+    0.0033+    0.0072}$   & $   -0.031_{-    0.011-    0.021}^{+    0.011+    0.021}$  & $   -0.0029_{-    0.0032-    0.0061}^{+    0.0031+    0.0065}$  \\ 

$\Omega_{m0}$ & $    0.40_{-    0.25-    0.29}^{+    0.11+    0.36}$ & $    0.340_{-    0.024-    0.049}^{+    0.027+    0.048}$  & $    0.382_{-    0.031-    0.060}^{+    0.030+    0.060}$ & $    0.3094_{-    0.0083-    0.016}^{+    0.0085+    0.017}$ \\ 

$\sigma_8$ & $    0.87_{-    0.19-    0.23}^{+    0.11+    0.27}$ & $    0.800_{-    0.027-    0.046}^{+    0.024+    0.050}$ & $    0.836_{-    0.015-    0.032}^{+    0.017+    0.030}$   & $    0.826_{-    0.016-    0.032}^{+    0.016+    0.031}$ \\ 

$H_0$ [Km/s/Mpc] & $   64_{-   20-   24}^{+   10+   31}$ &  $   64.6_{-    2.8-    4.8}^{+    2.1+    5.0}$  & $   60.9_{-    2.9-    5.0}^{+    2.4+    5.3}$ & $   67.6_{-    1.3-    2.4}^{+    1.3+    2.5}$ \\ 

$N_{\rm eff}$ & $    3.05_{-    0.20-    0.37}^{+    0.19+    0.40}$ & $    2.97_{-    0.20-    0.38}^{+    0.20+    0.37}$   & $    3.03_{-    0.19-    0.36}^{+    0.19+    0.39}$  & $    2.97_{-    0.21-    0.37}^{+    0.19+    0.39}$  \\ 

$S_8$ & $    0.942_{-    0.079-    0.15}^{+    0.091+    0.14}$ & $    0.851_{-    0.017-    0.040}^{+    0.021+    0.038}$  & $    0.943_{-    0.039-    0.080}^{+    0.038+    0.072}$   & $    0.839_{-    0.017-    0.034}^{+    0.017+    0.034}$ \\ 

$r_{\rm{drag}}$ [Mpc] & $  147.34_{-    1.85-    3.77}^{+    1.91+    3.69}$  &  $  147.98_{-    2.09-    3.59}^{+    1.92+    3.85}$  & $  147.5_{-    1.9-    3.7}^{+    1.9+    3.6}$ & $  148.0_{-    1.9-    3.7}^{+    1.9+    3.8}$\\ 

\hline                                                  
%$\chi^2$ &  2761.908  & 2775.292 & 3799.142 &  3812.190  \\
\hline

\end{tabular}                                                   }                                                                
\caption{68\% and 95\% CL constraints on various free and derived parameters of the $w_0w_a$CDM $+$ $\Omega_K$ $+$ $N_{\rm eff}$ scenario using several observational datasets. }
\label{tab:CPL-Neff}                                        
\end{table*}                                                
\end{center}                                                                                                                    
\endgroup    

\subsubsection{$w_0w_a$CDM $+$ $\Omega_K$ $+$ $M_{\nu}$ $+$ $N_{\rm eff}$}

\begin{figure*}
    \centering
    \includegraphics[width=0.8\textwidth]{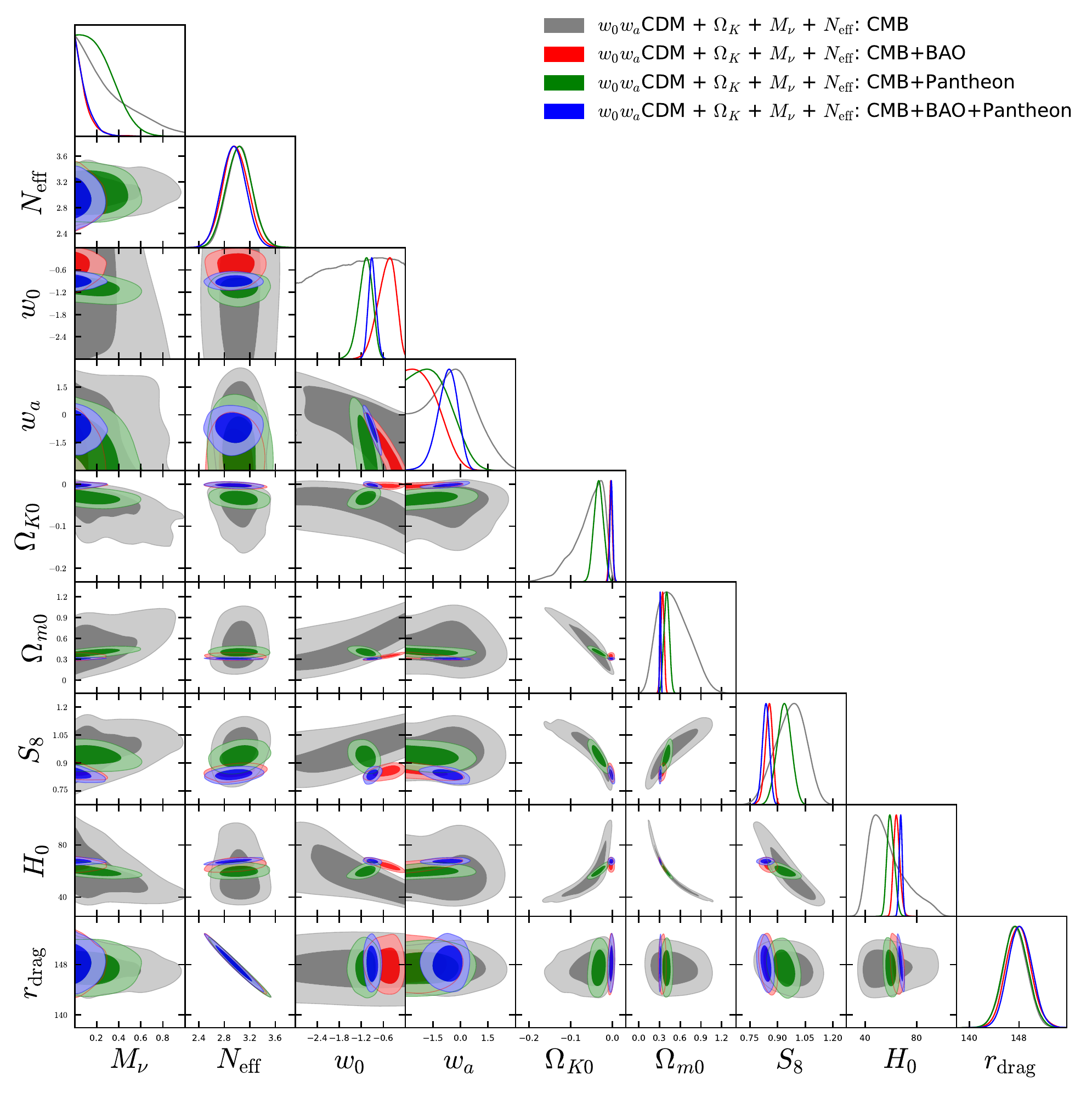}
    \caption{This figure corresponds to the 1-dimensional posterior distributions and the 2-dimensional joint contours for the most important parameters of cosmological scenario labeled by $w_0w_a$CDM $+$ $\Omega_K$ $+$ $M_{\nu}$ $+$ $N_{\rm eff}$ using various observational datasets. }
    \label{fig:w0waCDM-Omegak-Mnu-Neff}
\end{figure*}

In this last extension we simultaneously vary the total neutrino mass and the effective number of relativistic degrees of freedom. We refer to this scenario as $w_0w_a$CDM $+$ $\Omega_K$ $+$ $M_{\nu}$ $+$ $N_{\rm eff}$, providing the results in \autoref{tab:CPL-Mnu-Neff}, and the contour plots in \autoref{fig:w0waCDM-Omegak-Mnu-Neff}.

In the large parameter-space allowed in this model, the results for curvature does not change significantly. In particular, as in the previous extensions, the CMB data always prefer curved spaces at 68\% CL, while flatness is always allowed within the 95\% CL results 
($\Omega_{K0}=0.056^{+0.058}_{-0.076}$). As concerns the Hubble parameter, CMB data give $H_0=57^{+7}_{-19}$ km/s/Mpc at 68\%  which is similar to the bound obtained without including the effective number of relativistic degrees of freedom in the sample. Indeed, as we already pointed out, the preference for smaller values of $H_0$ is mostly driven by the combined effects of curvature and massive neutrinos and adding $N_{\rm eff}$ does not reduce the tension between the different datasets discussed so far. Concerning dark energy, both $w_0$ and $w_a$ turns out to be basically unconstrained for these dataset and we can derive only a 95\% CL upper bound $w_a<1.6$ which is quite uninformative about the possible dynamical nature of the late time expansion.
Including BAO in the picture, we basically recover the same results discussed without varying $N_{\rm eff}$, as well. In particular we find $\Omega_{K0}=-0.0037^{+0.0073}_{-0.0068}$ at 95\% CL and it should be noted that, while this bound is very shrink around zero, flatness is still slightly disfavored at 68\% CL. Similarly for $H_0$ we find $64.4^{+2.1}_{-2.7}$ at 68\% CL producing the same tensions discussed in the case without $N_{\rm eff}$.  On the other hand, for the constraints $w_0=-0.52^{+0.43}_{-0.48}$ and $w_a<-0.4$ (both at 95\% CL) we can observe a strong indication for quintessential models of dark energy with a dynamical evolution that are indeed preferred at more than 95\% CL. 
When CMB+Pantheon are combined together, we recover the usual evidence for a closed Universe at more than 95\% CL ($\Omega_{K0}=-0.033\pm21$ at 95\% CL) and we get a lower value $H_0=59.8^{+2.5}_{-3.2}$ km/s/Mpc at 68\% CL for the Hubble parameter, in strong tension with local measurements as usual. In this case both quintessential and phantom models of dark energy are both allowed within the 68\% CL ($w_{0}=-1.09^{+0.21}_{.-0.17}$ at 68\% CL), i.e. $w_0$ is consistent with a cosmological constant,  and only an upper bound $w_a<0.24$ can be derived on $w_a$ at 95\% CL. On the other hand, combining CMB+BAO+Pantheon all together the evidence for a closed Universe disappears, and the value of the expansion rate is shifted towards higher values $H0=67.6\pm1.2$ km/s/Mpc at 68\% CL, similar to the standard $\Lambda$CDM case. Furthermore in this case $w_0$ is consistent with a cosmological constant within the 68\% CL, and $w_a=0$ within the 95\% CL.
Finally, concerning the mass or relic neutrinos, exploiting the CMB temperature and polarization measurements only, we get $M_{\nu}<0.753$ eV at 95\% CL, while combining the CMB and BAO data we can improve this limit to $M_{\nu}<0.231$ eV at 95\% CL. This bound remains almost unchanged including also Pantheon data and the results are very similar to those discussed for the previous cosmological models. The same discussion applies also to the effective number of relativistic degrees of freedom and in particular we can appreciate that additional contribution always constrain $\Delta N_{\rm eff}\lesssim 0.4$, independently from the parametrization introduced in the dark energy sector.

\begingroup                                                                                                                     
\squeezetable                                                                                                                   
\begin{center}                                                                                                                  
\begin{table*}[htb]                                                                                                                   
\resizebox{\textwidth}{!}{   
\begin{tabular}{ccccccccccccc}                                                                                                            
\hline\hline                                                                                                                    
Parameters & CMB & CMB+BAO & CMB+Pantheon &  CMB+BAO+Pantheon \\ \hline 

$\Omega_c h^2$ & $    0.1181_{-    0.0030-    0.0057}^{+    0.0030+    0.0061}$ & $    0.1184_{-    0.0030-    0.0060}^{+    0.0030+    0.0059}$  & $    0.1182_{-    0.0030-    0.0060}^{+    0.0031+    0.0062}$ & $    0.1182_{-    0.0029-    0.0057}^{+    0.0030+    0.0059}$  \\ 

$\Omega_b h^2$ & $    0.02255_{-    0.00025-    0.00050}^{+    0.00025+    0.00049}$ & $    0.02236_{-    0.00024-    0.00048}^{+    0.00024+    0.00047}$   & $    0.02254_{-    0.00026-    0.00051}^{+    0.00025+    0.00050}$ & $    0.02232_{-    0.00023-    0.00046}^{+    0.00023+    0.00046}$  \\ 

$100\theta_{MC}$ & $    1.04106_{-    0.00045-    0.00086}^{+    0.00045 +    0.00087}$ & $    1.04111_{-    0.00047-    0.00082}^{+    0.00041+    0.00090}$  & $    1.04109_{-    0.00048-    0.00088}^{+    0.00044+    0.00092}$ & $    1.04113_{-    0.00044-    0.00082}^{+    0.00043+    0.00087}$  \\ 

$\tau$ & $    0.0477_{-    0.0069-    0.017}^{+    0.0085+    0.015}$ & $    0.0535_{-    0.0075-    0.015}^{+    0.0073+    0.016}$ & $    0.0491_{-    0.0076-    0.017}^{+    0.0077+    0.016}$  & $    0.0530_{-    0.0074-    0.016}^{+    0.0074+    0.016}$   \\ 

$n_s$ & $    0.9695_{-    0.0094-    0.018}^{+    0.0093+    0.019}$ & $    0.9640_{-    0.0092-    0.018}^{+    0.0092+    0.018}$  & $    0.9689_{-    0.0093-    0.018}^{+    0.0098+    0.019}$ & $    0.9623_{-    0.0087-    0.018}^{+    0.0088+    0.017}$ \\ 

${\rm{ln}}(10^{10} A_s)$ & $    3.025_{-    0.017-    0.039}^{+    0.020+    0.036}$ & $    3.038_{-    0.018-    0.036}^{+    0.018+    0.037}$  & $    3.029_{-    0.018-    0.038}^{+    0.018+    0.037}$  & $    3.037_{-    0.018-    0.037}^{+    0.018+    0.037}$ \\ 

$w_0$ & $ unconstrained $ & $   -0.52_{-    0.19-    0.48}^{+    0.29+    0.43}$  & $   -1.09_{-    0.17-    0.38}^{+    0.21+    0.36}$ & $   -0.90_{-    0.11-    0.20}^{+    0.10+    0.21}$ \\ 

$w_a$ & $   -0.5_{-    1.5}^{+    1.3}\,<1.6$ & $ <-1.49\,<-0.4$  & $  <-0.96\,<0.24$ & $   -0.72_{-    0.48-    1.1}^{+    0.63+    1.1}$ \\ 

$\Omega_{K0}$ & $   -0.056_{-    0.019-    0.076}^{+    0.047+    0.058}$ & $   -0.0037_{-    0.0035-    0.0068}^{+    0.0035+    0.0073}$  & $   -0.033_{-    0.011-    0.021}^{+    0.011+    0.021}$  & $   -0.0025_{-    0.0033-    0.0065}^{+    0.0033+    0.0069}$  \\

$\Omega_{m0}$ & $    0.51_{-    0.31-    0.39}^{+    0.16+    0.44}$ & $    0.342_{-    0.023-    0.047}^{+    0.026+    0.044}$   & $    0.403_{-    0.039-    0.071}^{+    0.038+    0.074}$ & $    0.3099_{-    0.0092-    0.016}^{+    0.0081+    0.017}$ \\ 

$\sigma_8$ & $    0.79_{-    0.18-    0.22}^{+    0.08+    0.28}$ & $    0.799_{-    0.027-    0.054}^{+    0.026+    0.054}$  & $    0.809_{-    0.025-    0.064}^{+    0.038+    0.058}$ & $    0.823_{-    0.018-    0.041}^{+    0.022+    0.039}$ \\ 

$H_0$ [Km/s/Mpc] & $   57_{-   19-   23}^{+    7+   31}$ & $   64.4_{-    2.7-    4.6}^{+    2.1+    4.8}$  & $   59.8_{-    3.2-    5.3}^{+    2.5+    5.7}$ & $   67.6_{-    1.2-    2.4}^{+    1.2+    2.4}$ \\ 

$M_\nu$ [eV] & $  <0.365\,<0.753  $ & $  <0.092\,<0.231$  & $  <0.279\,<0.49 $ & $  <0.098\,<0.22  $  \\ 

$N_{\rm eff}$ & $    3.04_{-    0.20-    0.37}^{+    0.19+    0.40}$ & $    2.97_{-    0.20-    0.39}^{+    0.20+    0.39}$ & $    3.03_{-    0.20-    0.39}^{+    0.20+    0.40}$ & $    2.95_{-    0.19-    0.36}^{+    0.19+    0.37}$ \\   

$S_8$ & $    0.970_{-    0.076-    0.17}^{+    0.095+    0.14}$ & $    0.852_{-    0.018-    0.042}^{+    0.023+    0.040}$  & $    0.936_{-    0.038-    0.076}^{+    0.040+    0.073}$ & $    0.836_{-    0.018-    0.040}^{+    0.021+    0.039}$ \\ 

$r_{\rm{drag}}$ [Mpc] & $  147.39_{-    1.87-    3.76}^{+    1.91+    3.74}$  & $  147.94_{-    2.10-    3.76}^{+    1.89+    4.04}$  & $  147.5_{-    1.9-    3.8}^{+    2.0+    4.0}$ &   $ 148.2_{-    1.9-    3.6}^{+    1.9+    3.7}$ \\    \\

\hline                                                  
%$\chi^2$ & 2762.066  &  2773.772  & 3800.122  &   3811.708  \\
\hline                                                           
\end{tabular}                                                      }                                                             
\caption{68\% and 95\% CL constraints on various free and derived parameters of the $w_0w_a$CDM $+$ $\Omega_K$ $+$ $M_{\nu}$ $+$ $N_{\rm eff}$ scenario using several observational datasets. }
\label{tab:CPL-Mnu-Neff}                                                                                                   
\end{table*}                                        
\end{center}                                                
\endgroup              

\subsection{Non-flat PEDE and its extensions}

Here we describe the final model of this article and its extensions. In a similar fashion we have considered exactly the same datasets and their combinations and concerning the observational constraints, we report either their 68\% and 95\% CL constraints or the upper/lower limits at 68\% and 95\% CL.

\subsubsection{PEDE $+$ $\Omega_K$}

\begin{figure*}
    \centering
    \includegraphics[width=0.8\textwidth]{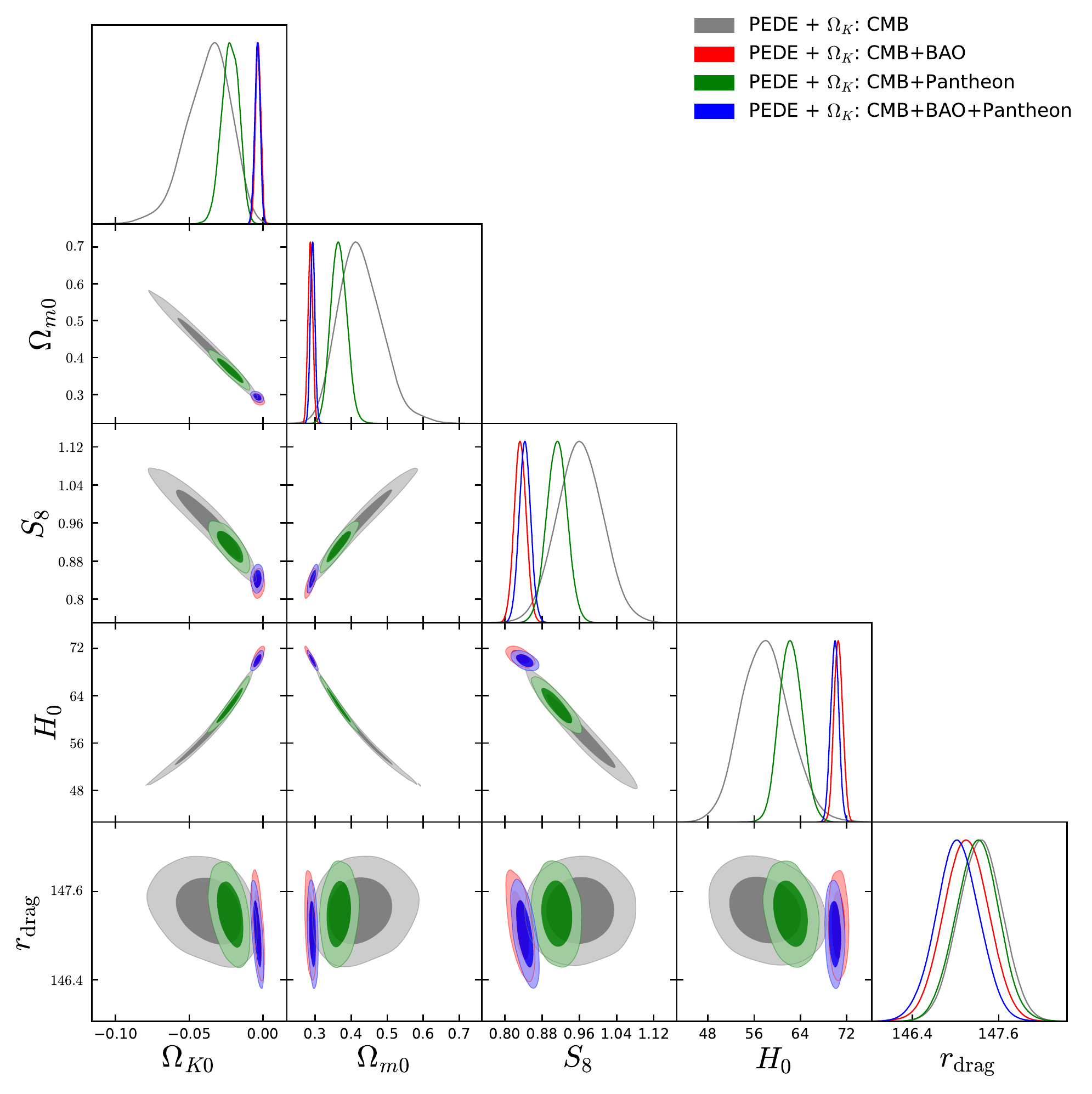}
    \caption{This figure corresponds to the 1-dimensional posterior distributions and the 2-dimensional joint contours for the most important parameters of cosmological scenario labeled by PEDE $+$ $\Omega_K$ using various observational datasets. }
    \label{fig:PEDE-Omegak}
\end{figure*}

In \autoref{tab:PEDE} we have summarized the observational constraints on this scenario, and in \autoref{fig:PEDE-Omegak} we show the 1-dimensional posterior distributions of some parameters and the 2-dimensional joint contours between several parameters.   

We have some interesting results in this case. We see that all the datasets indicate the preference of a nonzero spatial curvature. Precisely, we observe that $\Omega_K$ remains nonzero at more than 95\% CL for CMB, CMB+Pantheon and CMB+BAO+Pantheon while for CMB+BAO, $\Omega_K$ remains nonzero at more than 68\% CL. Additionally, for all the cases explored here, the mean values of $\Omega_K$
are negative which indicate the preference of a closed Universe. 

Concerning the estimated values of $H_0$, we see that CMB alone predicts a very low value ($H_0 = 58^{+3.8}_{-4.8}$ km/s/Mpc at 68\% CL) and hence the tension is about $\sim 3.5 \sigma$. However, the inclusion of BAO quite surprisingly increases the mean value of $H_0$ and decreases the error bars leading to $H_0 = 70.57^{+0.73}_{-0.81}$ km/s/Mpc at 68\% CL
and thus the tension with SH0ES is reduced down to  $~2\sigma$. 
In the contrary, the inclusion of Pantheon to CMB slightly increases the Hubble constant compared to CMB alone leading to $H_0 = 62.3 \pm 2$ km/s/Mpc at 68\% CL (CMB+Pantheon) and again the tension with SH0ES is found to be at $\sim 4.8 \sigma$. So, effectively for CMB and CMB+Pantheon we can see a clear tension. 
However, similar to the previous case with CMB+BAO, the combined analysis CMB+BAO+Pantheon also leads to a slightly higher value of $H_0$ ($~= 69.94^{+0.70}_{-0.72}$ km/s/Mpc at 68\% CL for CMB+BAO+Pantheon) compared to Planck alone (within $\Lambda$CDM) and the tension with SH0ES is reduced down to $\sim 2.5 \sigma$. 

\begingroup                                                     
\squeezetable                                                   
\begin{center}                                                  
\begin{table*}[htb]                                                  
\resizebox{\textwidth}{!}{   
\begin{tabular}{ccccccccccccc}                                                                                                            
\hline\hline                                                                                                                    
Parameters & CMB & CMB+BAO & CMB+Pantheon &  CMB+BAO+Pantheon \\ \hline 

$\Omega_c h^2$ & $    0.1181_{-    0.0015-    0.0028}^{+    0.0015+    0.0029}$ & $    0.1196_{-    0.0014-    0.0027}^{+    0.0014+    0.0028}$   & $    0.1184_{-    0.0014-    0.0028}^{+    0.0014+    0.0028}$  & $    0.1203_{-    0.0014-    0.0027}^{+    0.0014+    0.0028}$ \\ 

$\Omega_b h^2$ & $    0.02260_{-    0.00017-    0.00033}^{+    0.00017 +    0.00033}$ & $    0.02242_{-    0.00015-    0.00030}^{+    0.00015+    0.00030}$  & $    0.02255_{-    0.00016-    0.00032}^{+    0.00016+    0.000328}$  & $    0.02237_{-    0.00015-    0.00030}^{+    0.00015+    0.00030}$ \\ 

$100\theta_{MC}$ & $    1.041164_{-    0.00033-    0.00066}^{+    0.00033 +    0.00063}$ & $    1.04098_{-    0.00031-    0.00060}^{+    0.00031+    0.00062}$  & $    1.04112_{-    0.00032-    0.00064}^{+    0.00032+    0.00063}$  & $    1.04090_{-    0.00031-    0.00062}^{+    0.00031+    0.00063}$ \\ 

$\tau$ & $    0.0485_{-    0.0074-    0.016}^{+    0.0076+    0.017}$ & $    0.0537_{-    0.0086-    0.015}^{+    0.0075+    0.017}$  &$    0.0509_{-    0.0071-    0.0145}^{+    0.0071+    0.014}$ & $    0.0525_{-    0.0074-    0.015}^{+    0.0074+    0.015}$ \\ 

$n_s$ & $    0.9706_{-    0.0047-    0.0094}^{+    0.0048+    0.0090}$ & $    0.9662_{-    0.0044-    0.0084}^{+    0.0044+    0.0089}$  & $    0.9695_{-    0.0045-    0.0089}^{+    0.0045+    0.0089}$ & $    0.9648_{-    0.0044-    0.0090}^{+    0.0045+    0.0089}$  \\ 

${\rm{ln}}(10^{10} A_s)$ & $    3.028_{-    0.016-    0.036}^{+    0.017+    0.033}$ & $    3.042_{-    0.017-    0.032}^{+    0.016+    0.033}$  & $    3.034_{-    0.015-    0.030}^{+    0.015+    0.030}$  & $    3.041_{-    0.015-    0.031}^{+    0.015+    0.031}$  \\

$\Omega_{K0}$ & $   -0.037_{-    0.013-    0.031}^{+    0.018+    0.029}$ & $   -0.0033_{-    0.0018-    0.0035}^{+    0.0018+    0.0035}$   & $   -0.0222_{-    0.0055-    0.012}^{+    0.0066+    0.011}$  & $   -0.0038_{-    0.0017-    0.0037}^{+    0.0020+    0.0034}$  \\

$\Omega_{m0}$ & $    0.426_{-    0.069-    0.12}^{+    0.054+    0.12}$ &  $    0.2866_{-    0.0065-    0.013}^{+    0.0066+    0.013}$  & $    0.366_{-    0.024-    0.042}^{+    0.022+    0.044}$ & $    0.2930_{-    0.0062-    0.012}^{+    0.0062+    0.012}$   \\ 

$\sigma_8$ & $    0.810_{-    0.019-    0.038}^{+    0.018+    0.035}$ &  $    0.852_{-    0.0084-    0.017}^{+    0.0086+    0.017}$   & $    0.826_{-    0.011-    0.021}^{+    0.011+    0.021}$ & $    0.8530_{-    0.0084-    0.017}^{+    0.0086+    0.017}$ \\ 

$H_0$ [Km/s/Mpc] & $   58.0_{-    4.8-    7.8}^{+    3.8+    8.4}$ &  $   70.57_{-    0.81-    1.4}^{+    0.73+    1.5}$  & $   62.3_{-    2.0-    3.7}^{+    2.0+    4.0}$ & $   69.94_{-    0.72-    1.4}^{+    0.70+    1.4}$  \\ 

$S_8$ & $    0.961_{-    0.049-    0.096}^{+    0.049+    0.097}$ &  $    0.833_{-    0.013-    0.025}^{+    0.013+    0.025}$ & $    0.912_{-    0.022-    0.043}^{+    0.021+    0.043}$ & $    0.843_{-    0.012-    0.024}^{+    0.012+    0.025}$ \\ 

$r_{\rm{drag}}$ [Mpc] & $  147.34_{-    0.31-    0.59}^{+    0.30+    0.59}$  & $  147.14_{-    0.31-    0.60}^{+    0.31+    0.60}$ & $  147.30_{-    0.30-    0.58}^{+    0.29+    0.58}$ &  $  147.03_{-    0.29-    0.60}^{+    0.30+    0.59}$ \\ 

\hline                                                  
%$\chi^2$ & 2763.656 &  2778.252  & 3799.930  &  3824.144 \\
\hline                                                              
\end{tabular}                                                          }                                                         
\caption{68\% and 95\% CL constraints on various free and derived parameters of the PEDE $+$ $\Omega_K$ scenario using several observational datasets.}
\label{tab:PEDE}                                                                                                   
\end{table*}                                                                                                                     
\end{center}                                                                                                                    
\endgroup     

\subsubsection{PEDE $+$ $\Omega_K$ $+$ $M_{\nu}$}

\begin{figure*}
    \centering
    \includegraphics[width=0.8\textwidth]{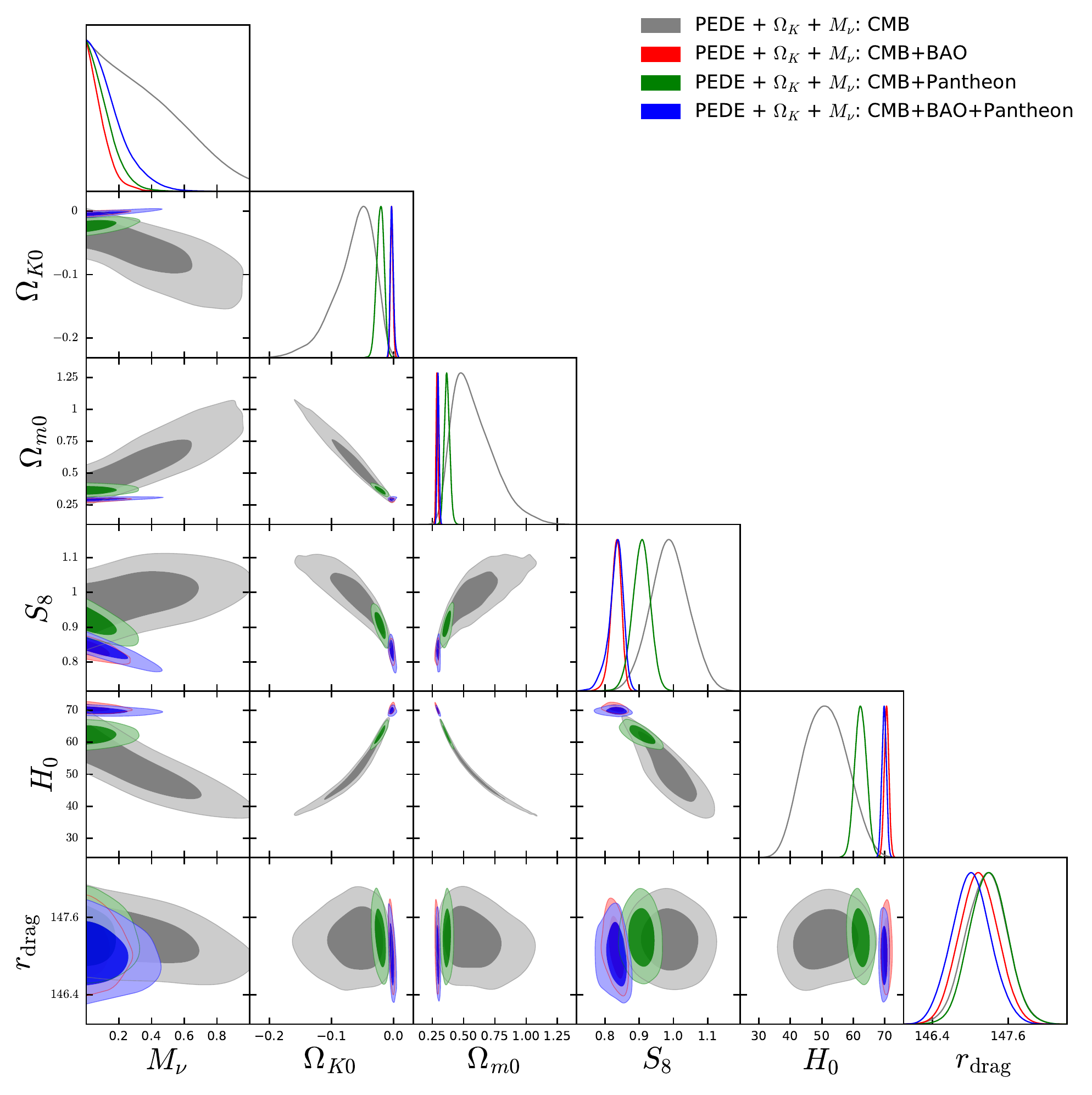}
    \caption{This figure corresponds to the 1-dimensional posterior distributions and the 2-dimensional joint contours for the most important parameters of cosmological scenario labeled by PEDE $+$ $\Omega_K$ $+$ $M_{\nu}$ using various observational datasets.  }
    \label{fig:PEDE-Omegak-Mnu}
\end{figure*}

As usual, we start expanding the parameter space introducing massive neutrinos in the cosmological model. We summarize the results obtained in this case in \autoref{tab:PEDE-Mnu} and in \autoref{fig:PEDE-Omegak-Mnu}.

Including neutrinos we find that the CMB and CMB+Pantheon data always prefer a closed Universe at more than 95\% CL, while this preference is reduced to 68\%  CL both for CMB+BAO and for CMB+BAO+Pantheon. As concerns the Hubble parameter, in this case, due to the degeneracy with massive neutrinos, the CMB data alone show  a preference for lower values giving $H_0=51.2^{+6.7}_{-7.6} $ km/s/Mpc at 68\% CL. Despite the large error-bars, this value is in tension with SH0ES at $3.2\sigma$. Interestingly, including BAO measurements, we find back the same preference for larger values discussed in the baseline case with the 68\% CL result reading $H_0=70^{+0.71}_{-0.72}$ km/s/Mpc. In this case the tension with the SH0ES result is reduced down to $2.5\sigma$. Conversely for CMB+Pantheon, we get $H_0=62.3\pm1.9$ km/s/Mpc at 68\% CL and the tension with SH0ES is back to be statistically significant. Finally combining CMB+BAO+Pantheon all together we obtain an intermediate value  $H_0=69.78^{+0.71}_{-0.73}$ km/s/Mpc at 68\% CL which is very similar to the result that one can derive within the standard $\Lambda$CDM model of cosmology. Notice that, as already discussed for all the previous models, also in this case the $H_0$ tension has an impact on the estimation of the matter content of the universe, resulting into a tension for $S_8$.  As concerns relics neutrinos, the constraints on their total mass remain almost unchanged with respect to the previous scenarios analyzed in this work, but it is worth noting that in this case we observe an improvement in the constraining power of CMB+Pantheon that give $M_{\nu}<0.26$eV at 95\% CL. When CMB and BAO  are combined together we can improve this upper limit to  $M_{\nu}<0.22$ eV at 95\% CL, while combining CMB+BAO+Pantheon all together we get  a less tight bound  $M_{\nu}<0.37$ eV at 95\% CL. 

\begingroup                                                                                                                     
\squeezetable                                                                                                                   
\begin{center}                                                                                                                  
\begin{table*}[htb]                                                                                                                   
\resizebox{\textwidth}{!}{   
\begin{tabular}{ccccccccccccc}                                                                                                            
\hline\hline                                                                                                                    
Parameters & CMB & CMB+BAO & CMB+Pantheon &  CMB+BAO+Pantheon \\ \hline

$\Omega_c h^2$ & $    0.1183_{-    0.0016-    0.0029}^{+    0.0015+    0.0030}$ & $    0.1197_{-    0.0014-    0.0028}^{+    0.0014+    0.0028}$  &$    0.1186_{-    0.0014-    0.0028}^{+    0.0014+    0.0028}$ & $    0.1203_{-    0.0014-    0.0028}^{+    0.0014+    0.0028}$ \\ 

$\Omega_b h^2$ & $    0.02254_{-    0.00018-    0.00035}^{+    0.00019+    0.00035}$ & $    0.02240_{-    0.00015-    0.00030}^{+    0.00016+    0.00029}$  &$    0.02253_{-    0.00016-    0.00034}^{+    0.00017+    0.00032}$ & $    0.02235_{-    0.00016-    0.00031}^{+    0.00016+    0.00031}$ \\ 

$100\theta_{MC}$ & $    1.04100_{-    0.00035-    0.00068}^{+    0.00035+    0.00068}$ & $    1.04096_{-    0.00031-    0.00064}^{+    0.00033+    0.00059}$  & $    1.04108_{-    0.00033-    0.00063}^{+    0.00032+    0.00066}$ & $    1.04088_{-    0.00033-    0.00066}^{+    0.00035+    0.00061}$ \\ 

$\tau$ & $    0.0478_{-    0.0075-    0.016}^{+    0.0083+    0.015}$ & $    0.0541_{-    0.0074-    0.015}^{+    0.0074+    0.016}$  & $    0.0512_{-    0.0075-    0.015}^{+    0.0076+    0.015}$ & $    0.0527_{-    0.0078-    0.015}^{+    0.0076+    0.016}$\\ 

$n_s$ & $    0.9692_{-    0.0049-    0.0096}^{+    0.0049+    0.0093}$ &  $    0.9659_{-    0.0045-    0.0090}^{+    0.0045+    0.0089}$  & $    0.9694_{-    0.0050-    0.0088}^{+    0.0045+    0.0093}$ & $    0.9644_{-    0.0046-    0.0087}^{+    0.0045+    0.0090}$ \\ 

${\rm{ln}}(10^{10} A_s)$ & $    3.026_{-    0.016-    0.034}^{+    0.016+    0.032}$ &  $    3.043_{-    0.015-    0.031}^{+    0.015+    0.031}$   & $    3.035_{-    0.016-    0.031}^{+    0.016+    0.031}$ & $    3.042_{-    0.016-    0.032}^{+    0.016+    0.033}$ \\ 

$\Omega_{K0}$ & $   -0.064_{-    0.020-    0.069}^{+    0.041+    0.057}$ &  $   -0.0029_{-    0.0023-    0.0042}^{+    0.0019+    0.0043}$    & $   -0.0213_{-    0.0057-    0.012}^{+    0.0064+    0.011}$ & $   -0.0026_{-    0.0030-    0.0054}^{+    0.0022+    0.0058}$ \\ 

$\Omega_{m0}$ & $    0.58_{-    0.21-    0.29}^{+    0.11+    0.35}$ &  $    0.2872_{-    0.0078-    0.014}^{+    0.0066+    0.015}$  & $    0.367_{-    0.024-    0.042}^{+    0.020+    0.044}$ & $    0.2963_{-    0.0084-    0.015}^{+    0.0074+    0.016}$ \\ 

$\sigma_8$ & $    0.726_{-    0.061-    0.14}^{+    0.094+    0.12}$ &  $    0.850_{-    0.010-    0.036}^{+    0.019+    0.030}$  &$    0.820_{-    0.014-    0.041}^{+    0.023+    0.036}$ & $    0.836_{-    0.016-    0.056}^{+    0.032+    0.045}$ \\ 

$H_0$ [Km/s/Mpc] & $   51.2_{-    7.6-   12}^{+    6.7+   13}$ &  $   70.57_{-    0.72-    1.5}^{+    0.71+    1.5}$  & $   62.3_{-    1.9-    3.6}^{+    1.9+    3.7}$ & $   69.78_{-    0.73-    1.4}^{+    0.71+    1.4}$ \\ 

$M_\nu$ [eV] & $  <0.459\,<0.81 $ & $   <0.095\,<0.22 $ & $   <0.123\,<0.26 $ & $ <0.167\,< 0.37 $ \\ 

$S_8$ & $    0.989_{-    0.055-    0.11}^{+    0.056+    0.11}$ &  $    0.832_{-    0.013-    0.033}^{+    0.017+    0.031}$ &  $    0.906_{-    0.025-    0.053}^{+    0.026+    0.050}$ & $    0.831_{-    0.016-    0.050}^{+    0.026+    0.042}$ \\ 

$r_{\rm{drag}}$ [Mpc] & $  147.25_{-    0.32-    0.63}^{+    0.32+    0.62}$  & $  147.13_{-    0.31-    0.58}^{+    0.31+    0.60}$  & $  147.29_{-    0.30-    0.57}^{+    0.29+    0.58}$ & $  147.01_{-    0.31-    0.59}^{+    0.30+    0.62}$ \\ 

\hline                                                  
%$\chi^2$ & 2763.232 & 2779.398  & 3799.268  &  3824.140 \\
\hline  

\end{tabular}                                                    }                                                               
\caption{68\% and 95\% CL constraints on various free and derived parameters of the PEDE $+$ $\Omega_K$ $+$ $M_{\nu}$ scenario using several observational datasets. }
\label{tab:PEDE-Mnu}                                                                                                   
\end{table*}                                                                                                                     
\end{center}                                                                                                                    
\endgroup              

\subsubsection{PEDE $+$ $\Omega_K$ $+$ $N_{\rm eff}$}

\begin{figure*}
    \centering
    \includegraphics[width=0.8\textwidth]{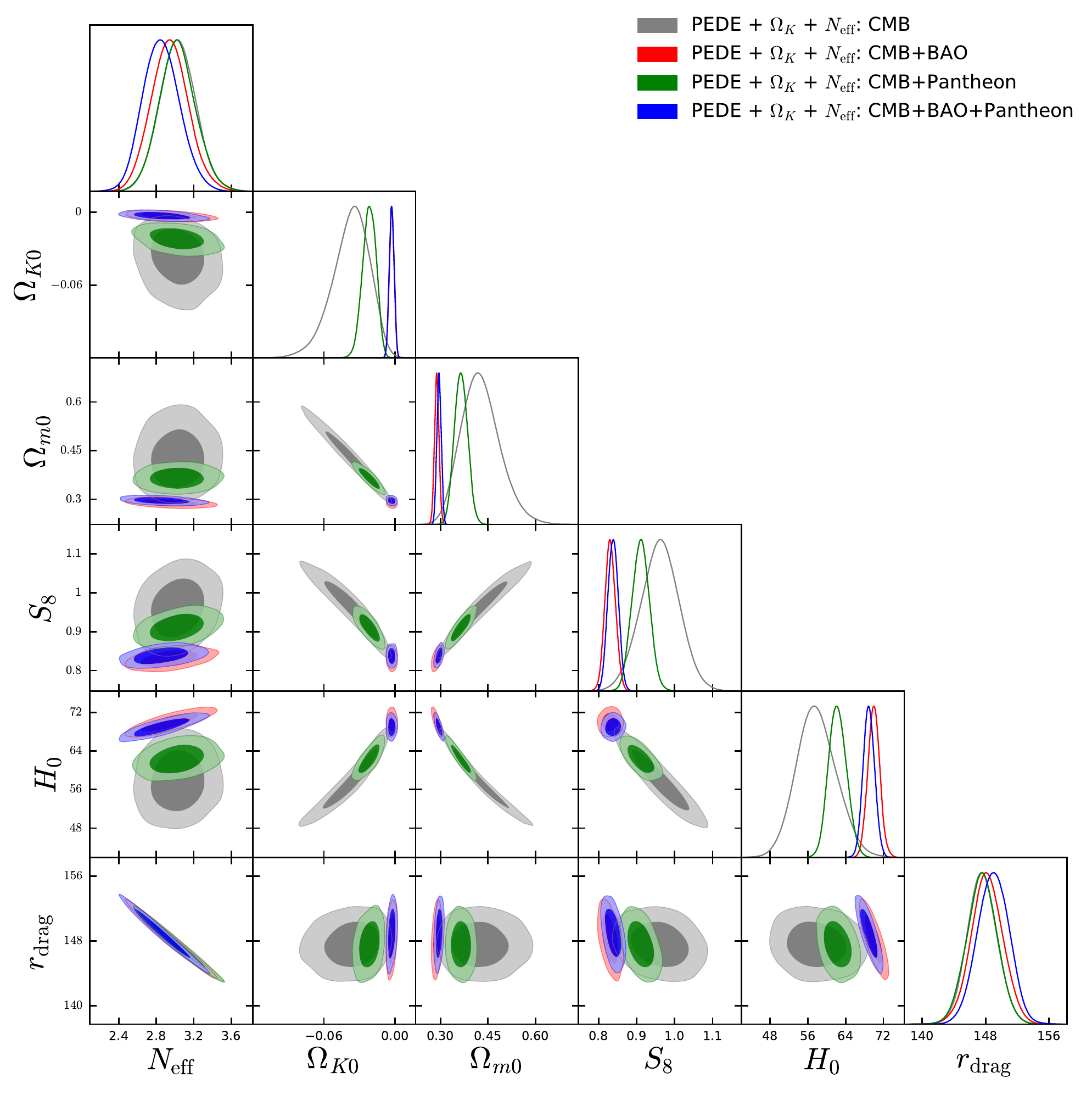}
    \caption{This figure corresponds to the 1-dimensional posterior distributions and the 2-dimensional joint contours for the most important parameters of cosmological scenario labeled by PEDE $+$ $\Omega_K$ $+$ $N_{\rm eff}$ using various observational datasets.  }
    \label{fig:PEDE-Omegak-Neff}
\end{figure*}

We now replace massive neutrinos with the effective number of relativistic degrees of freedom $N_{\rm eff}$ in the cosmological model. We summarize the results obtained within this scenario in \autoref{tab:PEDE-Neff} and in \autoref{fig:PEDE-Omegak-Neff}.

Analyzing the Planck satellite measurements of CMB temperature anisotropies and polarization only, in this case we observe an indication for a closed Universe at more than $95\%$ CL, with the curvature parameter reading $\Omega_{K0}=-0.038^{+0.030}_{-0.032}$ (at $95\%$ CL). Such an indication remains true for CMB+Pantheon while combining CMB+BAO and CMB+BAO+Pantheon it is reduced to the level of 68\% CL, with flatness always allowed within the 95\% CL results. Similarly to the baseline case,  from the CMB we get a 68\% CL value $H_0=57.9^{+4.0}_{-4.6}$ km/s/Mpc for the Hubble parameter which is about $3.7\sigma$ in tension with the independent local measurement provided by the SH0ES collaboration. Interestingly, also in this case, combining CMB+BAO the tension is reduced at $1.9\sigma$ due to the higher value $H_0=70.0\pm1.2$ km/s/Mpc at 68\% CL preferred for this dataset. Conversely CMB+Pantheon prefer again smaller values of the  Hubble parameter ($H_0=62.3^{+1.8}_{-2.1}$ km/s/Mpc at 68\% CL) in contrast with SH0ES at $5.1\sigma$. Finally combining CMB+BAO+Pantheon all together we find an intermediate value between the two cases which, being very similar to the value obtained within the standard cosmological model, is in tension with local measurements as well. As concerns the effective number of relativistic degrees of freedom, we see that also in this case the constraints are almost unchanged with respect to the results derived analyzing different parametrizations of the dark energy sector and additional contributions to dark radiation are always constrained to be $\Delta N_{\rm eff}\lesssim 0.4$ at 95\% CL. Anyway it is worth pointing out that for the data combination CMB+BAO+Pantheon, the reference value $N_{\rm eff}=3.044$ is disfavored at 68\% CL and a preference for smaller values is observed. This smaller value is the same that seems can solve the ``CMB tension'' between the different CMB datasets~\cite{DiValentino:2022rdg}.

\begingroup                                                                                                                     
\squeezetable                                                                                                                   
\begin{center}                                                                                                                  
\begin{table*}[htb]                                                                                                                   
\resizebox{\textwidth}{!}{   
\begin{tabular}{ccccccccccccc}                                                                                                            
\hline\hline                                                                                                                    
Parameters & CMB & CMB+BAO & CMB+Pantheon &  CMB+BAO+Pantheon \\ \hline 
$\Omega_c h^2$ & $    0.1178_{-    0.0029-    0.0057}^{+    0.0029+    0.0058}$ &  $    0.1182_{-    0.0031-    0.0062}^{+    0.0030+    0.0064}$  & $    0.1182_{-    0.0030-    0.0056}^{+    0.0030+    0.0060}$ & $    0.1176_{-    0.0032-    0.0058}^{+    0.0032+    0.0060}$ \\ 

$\Omega_b h^2$ & $    0.02259_{-    0.00024-    0.00048}^{+    0.00024 +    0.00050}$ & $    0.02233_{-    0.00023-    0.00044}^{+    0.00023+    0.00046}$  & $    0.02253_{-    0.00024-    0.00047}^{+    0.00023+    0.00046}$ & $    0.02219_{-    0.00025-    0.00043}^{+    0.00022+    0.00045}$ \\ 

$100\theta_{MC}$ & $    1.04120_{-    0.00046-    0.00084}^{+    0.00042 +    0.00088}$ & $    1.04115_{-    0.00045-    0.00087}^{+    0.00044+    0.00090}$  & $    1.04116_{-    0.00043-    0.00084}^{+    0.00043+    0.00085}$ & $    1.04122_{-    0.00048-    0.00084}^{+    0.00044+    0.00086}$ \\ 

$\tau$ & $    0.0487_{-    0.0076-    0.016}^{+    0.0078+    0.016}$ & $    0.0528_{-    0.0080-    0.015}^{+    0.0072+    0.016}$  & $    0.0513_{-    0.0073-    0.015}^{+    0.0072+    0.015}$ & $    0.0516_{-    0.0074-    0.015}^{+    0.0073+    0.014}$\\ 

$n_s$ & $    0.9702_{-    0.0090-    0.0177}^{+    0.0090+    0.0182}$ & $    0.9624_{-    0.0090-    0.0181}^{+    0.0088+    0.0177}$  & $    0.9686_{-    0.0088-    0.018}^{+    0.0090+    0.018}$ & $    0.9573_{-    0.0097-    0.017}^{+    0.0088+    0.018}$ \\ 

${\rm{ln}}(10^{10} A_s)$ & $    3.028_{-    0.018-    0.038}^{+    0.018+    0.036}$ &  $    3.036_{-    0.018-    0.036}^{+    0.018+    0.037}$  & $    3.034_{-    0.017-    0.036}^{+    0.017+    0.035}$ & $    3.032_{-    0.018-    0.036}^{+    0.018+    0.035}$ \\ 

$\Omega_{K0}$ & $   -0.038_{-    0.013-    0.032}^{+    0.018+    0.030}$ &  $   -0.0029_{-    0.0020-    0.0038}^{+    0.0020+    0.0038}$   & $   -0.0218_{-    0.0053-    0.012}^{+    0.0065+    0.011}$ & $   -0.0029_{-    0.0020-    0.0039}^{+    0.0020+    0.0039}$ \\ 

$\Omega_{m0}$ & $    0.428_{-    0.069-    0.12}^{+    0.053+    0.13}$ &  $    0.2883_{-    0.0069-    0.014}^{+    0.0069+    0.014}$  & $    0.365_{-    0.021-    0.040}^{+    0.021+    0.041}$ & $    0.2956_{-    0.0069-    0.013}^{+    0.0069+    0.013}$  \\ 

$\sigma_8$ & $    0.809_{-   0.019-    0.041}^{+    0.021+    0.037}$ &  $    0.847_{-    0.012-    0.024}^{+    0.012+    0.025}$  & $    0.826_{-    0.012-    0.023}^{+    0.012+    0.025}$ & $    0.845_{-    0.012-    0.024}^{+    0.013+    0.023}$\\ 

$H_0$ [Km/s/Mpc] & $   57.9_{-    4.6-    8.0}^{+    4.0+    8.7}$ &  $   70.0_{-    1.2-    2.6}^{+    1.2+    2.5}$   & $   62.3_{-    2.1-    3.5}^{+    1.8+    3.9}$  & $   68.9_{-    1.3-    2.3}^{+    1.2+    2.4}$ \\ 

$N_{\rm eff}$ & $    3.03_{-    0.19-    0.38}^{+    0.19+    0.39}$ &  $    2.94_{-    0.20-    0.40}^{+    0.19+    0.40}$  & $    3.02_{-    0.21-    0.38}^{+    0.18+    0.38}$ & $    2.85_{-    0.21-    0.38}^{+    0.19+    0.40}$ \\ 

$S_8$ & $    0.962_{-    0.051-    0.10}^{+    0.050+    0.10}$ &  $    0.831_{-    0.014-    0.026}^{+    0.014+    0.028}$ &  $    0.911_{-    0.023-    0.044}^{+    0.022+    0.044}$ & $    0.838_{-    0.013-    0.026}^{+    0.013+    0.026}$ \\ 

$r_{\rm{drag}}$ [Mpc] & $  147.56_{-    1.85-    3.66}^{+    1.86+    3.74}$  & $  148.17_{-    1.94-    4.02}^{+    1.97+    3.94}$  & $  147.5_{-    1.9-    3.7}^{+    1.9+    3.7}$ & $  149.0_{-    1.9-    3.9}^{+    2.0+    3.7}$ \\ 

\hline                                                  
%$\chi^2$ & 2762.932 & 2780.224  & 3798.782  & 3825.294   \\
\hline                                                           
\end{tabular}                                                   }                                                                
\caption{68\% and 95\% CL constraints on various free and derived parameters of the PEDE $+$ $\Omega_K$ $+$ $N_{\rm eff}$ scenario using several observational datasets. }
\label{tab:PEDE-Neff}                                                                                                   
\end{table*}                                                                                                                     
\end{center}                                                                                                                    
\endgroup             

\subsubsection{PEDE $+$ $\Omega_K$ $+$ $M_{\nu}$ $+$ $N_{\rm eff}$}

\begin{figure*}
    \centering
    \includegraphics[width=0.8\textwidth]{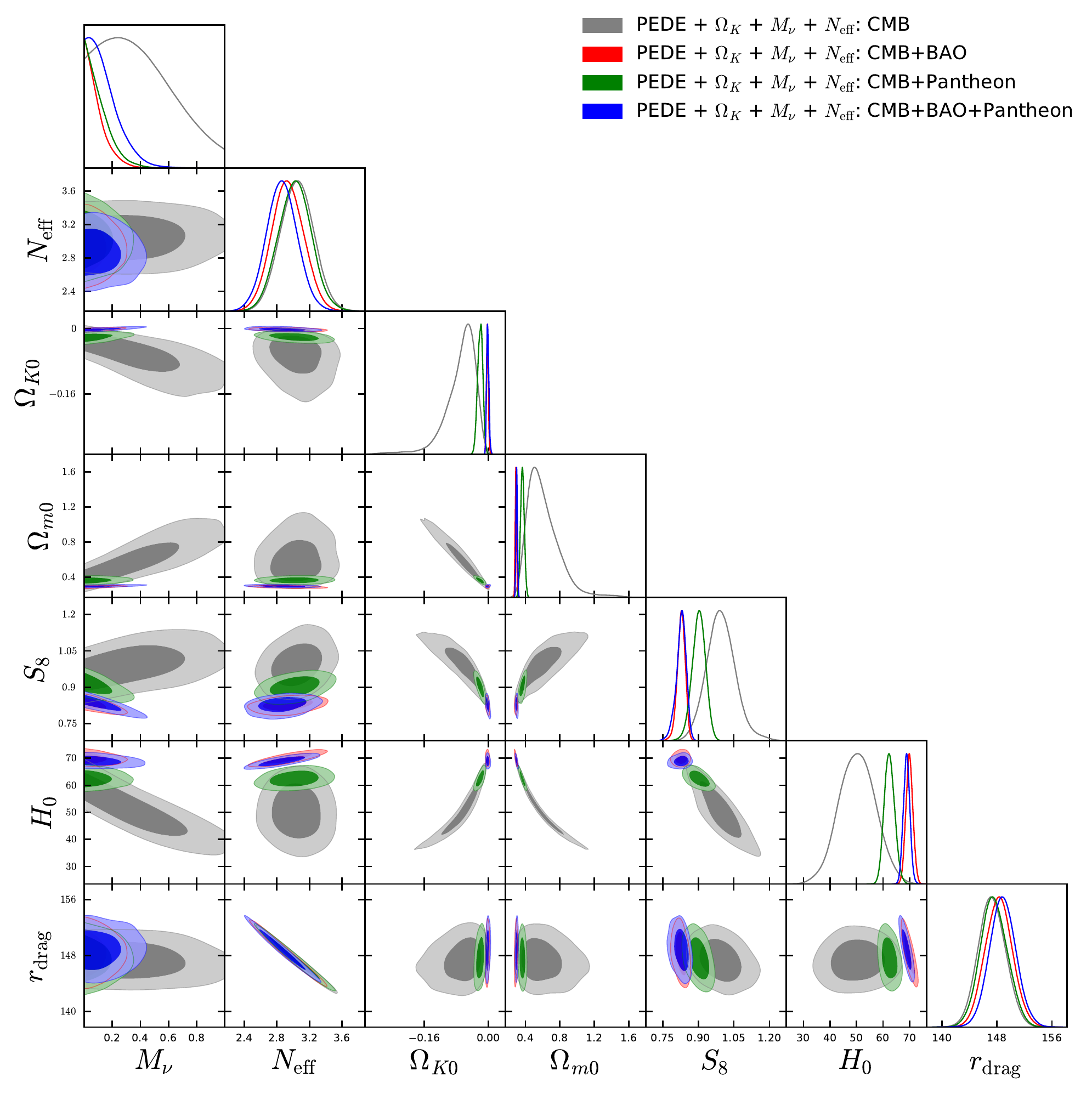}
    \caption{This figure corresponds to the 1-dimensional posterior distributions and the 2-dimensional joint contours for the most important parameters of cosmological scenario labeled by PEDE $+$ $\Omega_K$ $+$ $M_{\nu}$ $+$ $N_{\rm eff}$ using various observational datasets. }
    \label{fig:PEDE-Omegak-Mnu-Neff}
\end{figure*}

Finally, we simultaneously vary the total neutrino mass and the effective number of relativistic degrees of freedom, reporting the results in \autoref{tab:PEDE-Mnu-Neff} and in \autoref{fig:PEDE-Omegak-Mnu-Neff}. 

Exploiting only the Planck observations of the CMB radiation, in the largest parameter space sampled in this model, we obtain a preference for a curved cosmological spacetime at more than 95\% CL, obtaining a very small value of the expansion rate $H_0=50.3^{+6.8}_{-6.9}$ km/s/Mpc at 95\% CL. The preference for a such small value is always due to the combined effect of curvature and the degeneracy with the mass of relic neutrinos that produce also a large uncertainty. Combining the CMB and BAO data, the preference for a closed Universe is reduced at slightly more than 68\% CL and flatness becomes consistent at 95\% CL.  On the other hand the constraints on $H_0$ are drastically changed and, like for the previous extensions, we instead obtain preference for higher values ($H_0=69.9\pm1.3$ km/s/Mpc at 68\% CL) reducing the tension with the SH0ES measurement at $\sim 1.9\sigma$. Conversely, combining CMB and Pantheon we find back a strong preference for a curved spacetime (with the curvature parameter reading $\Omega_{K0}=-0.021^{+0.012}_{-0.013}$ at 95\% CL) and smaller values of the Hubble parameter ($H_0=62.4^{+1.8}_{-2.1}$ km/s/Mpc at 68\% CL) increasing again the tension between the different estimations of $H_0$ and $S_8$ basically in the same way, and within the same statistical significance reported for the previous models. Interestingly, in contrast with the other extensions of this baseline model, here combining CMB+BAO+Pantheon all together the preference for a closed Universe disappears and flatness is always consistent within the 68\% bound. On the other hand for the Hubble parameter we derive results similar to those obtained within the standard $\Lambda$CDM model, namely $H_0=68.9^{+1.1}_{-1.2}$ km/s/Mpc at 68\% CL. Notice that also in this extended scenarios large disagreements among the different datasets are observed and adding additional parameters to the baseline case does not lead to any significant reduction of such tensions.  Finally, concerning the neutrino sector, also in this case the CMB bound $M_{\nu}<0.84$ eV at 95\% can be improved up to $M_{\nu}<0.26$ eV and $M_{\nu}<0.24$ eV including the Pantheon and BAO measurements, respectively. Conversely, combining CMB+BAO+Pantheon the resulting bound is less constrained ($M_{\nu}<0.35$ eV). 
These results are almost the same derived without varying the effective number of relativistic degrees of freedom and, similarly, also the bound on $N_{\rm eff}$ are not affected by the inclusion of neutrinos. In particular, combining CMB+BAO+Pantheon we find the same slight preference for smaller values of $N_{\rm eff}$ with respect to the prediction of the standard model of elementary particles. Anyway everything remains consistent with the standard model within the 95\% CL as well as additional contributions are always constrained to be $\Delta N_{\rm eff}\lesssim0.4$ at 95\% CL for all the different dataset.

\begingroup                                                                                                                     
\squeezetable                                                                                                                   
\begin{center}                                                                                                                  
\begin{table*}[htb]                                                                                                                   
\resizebox{\textwidth}{!}{   
\begin{tabular}{ccccccccccccc}                                                                                                            
\hline\hline                                                                                                                    
Parameters & CMB & CMB+BAO & CMB+Pantheon &  CMB+BAO+Pantheon \\ \hline

$\Omega_c h^2$ & $    0.1182_{-    0.0030-    0.0059}^{+    0.0030+    0.0059}$ & $    0.1181_{-    0.0031-    0.0060}^{+    0.0031+    0.0063}$  &  $    0.1182_{-    0.0030-    0.0059}^{+    0.0030+    0.0059}$  &  $    0.1178_{-    0.0029-    0.0058}^{+    0.0029+    0.0059}$ \\ 

$\Omega_b h^2$ & $    0.02254_{-    0.00026-    0.00053}^{+    0.00026 +    0.00052}$ &  $    0.02232_{-    0.00023-    0.00044}^{+    0.00023+    0.00046}$  & $    0.02251_{-    0.00025-    0.000450}^{+    0.00025+    0.00051}$ &  $    0.02218_{-    0.00024-    0.00045}^{+    0.00024+    0.00045}$  \\ 

$100\theta_{MC}$ & $    1.04102_{-    0.00045-    0.00092}^{+    0.00046+    0.00089}$ &  $    1.04115_{-    0.00045-    0.00086}^{+    0.00045+    0.00088}$  & $    1.04112_{-    0.00047-    0.00082}^{+    0.00042+    0.00088}$ & $    1.04116_{-    0.00046-    0.00091}^{+    0.00045+    0.00091}$ \\ 

$\tau$ & $    0.0476_{-    0.0080-    0.018}^{+    0.0090+    0.017}$ &  $    0.0535_{-    0.0078-    0.016}^{+    0.0077+    0.017}$  & $    0.0510_{-    0.0075-    0.016}^{+    0.0074+    0.015}$ & $    0.0525_{-    0.0081-    0.015}^{+    0.0075+    0.017}$\\ 

$n_s$ & $    0.9693_{-    0.0096-    0.0196}^{+    0.0104+    0.0186}$ &  $    0.9618_{-    0.0086-    0.0177}^{+    0.0088+    0.0175}$   & $    0.9679_{-    0.0094-    0.018}^{+    0.0095+    0.019}$ & $    0.9568_{-    0.0086-    0.017}^{+    0.0087+    0.017}$ \\ 

${\rm{ln}}(10^{10} A_s)$ & $    3.025_{-    0.018-    0.042}^{+    0.021 +    0.037}$ & $    3.038_{-    0.019-    0.036}^{+    0.018+    0.039}$   & $    3.033_{-    0.018-    0.037}^{+    0.018+    0.036}$ &  $    3.035_{-    0.020-    0.038}^{+    0.018+    0.038}$ \\ 

$\Omega_{K0}$ & $   -0.069_{-    0.020-    0.078}^{+    0.045+    0.066}$ &  $   -0.0025_{-    0.0025-    0.0046}^{+    0.0021+    0.0047}$  & $   -0.0206_{-    0.0061-    0.013}^{+    0.0068+    0.012}$ & $   -0.0017_{-    0.0032-    0.0054}^{+    0.0024+    0.0058}$ \\ 

$\Omega_{m0}$ & $    0.61_{-    0.23-    0.32}^{+    0.11+    0.38}$ & $    0.2890_{-    0.0083-    0.014}^{+    0.0074+    0.016}$  & $    0.365_{-    0.022-    0.040}^{+    0.021+    0.043}$ & $    0.2983_{-    0.0085-    0.015}^{+    0.0078+    0.017}$ \\ 

$\sigma_8$ & $    0.716_{-    0.064-    0.14}^{+    0.089+    0.13}$ &  $    0.844_{-    0.015-    0.042}^{+    0.023+    0.038}$  & $    0.818_{-    0.016-    0.046}^{+    0.025+    0.041}$  & $    0.830_{-    0.018-    0.054}^{+    0.031+    0.047}$ \\ 

$H_0$ [Km/s/Mpc] & $   50.3_{-    6.9-   13}^{+    6.8+   14}$ & $   69.9_{-    1.3-    2.5}^{+    1.3+    2.6}$  & $   62.4_{-    2.1-    3.6}^{+    1.8+    3.9}$  & $   68.9_{-    1.2-    2.3}^{+    1.1+    2.4}$ \\ 

$M_\nu$ [eV] & $    0.39_{-    0.37}^{+    0.12}\,<0.84$ &  $  <0.103\,<0.24 $   & $  <0.128\,<0.28$ & $  <0.172\,<0.35$ \\ 

$N_{\rm eff}$ & $    3.05_{-    0.20-    0.40}^{+    0.20+    0.40}$ &  $    2.93_{-    0.19-    0.38}^{+    0.20+    0.39}$  & $    3.02_{-    0.20-    0.39}^{+    0.20+    0.40}$  & $    2.86_{-    0.20-    0.38}^{+    0.18+    0.38}$ \\ 

$S_8$ & $    0.996_{-    0.060-    0.13}^{+    0.057+    0.12}$ &  $    0.828_{-    0.015-    0.034}^{+    0.019+    0.034}$ & $    0.902_{-    0.027-    0.058}^{+    0.030+    0.053}$ & $    0.827_{-    0.017-    0.046}^{+    0.025+    0.041}$ \\ 

$r_{\rm{drag}}$ [Mpc] & $  147.28_{-    2.10-    3.80}^{+    1.95+    4.00}$   & $  148.27_{-    1.99-    3.78}^{+    1.95+    3.96}$ & $  147.6_{-    2.1-    3.8}^{+    1.9+    3.9}$  & $  148.8_{-    1.9-    3.8}^{+    1.9+    3.9}$ \\ 

\hline                                                  
%$\chi^2$ &  2763.42 &  2779.648  &  3799.746   &  3823.974   \\
\hline 

\end{tabular} 
}
\caption{68\% and 95\% CL constraints on various free and derived parameters of the PEDE $+$ $\Omega_K$ $+$ $M_{\nu}$ $+$ $N_{\rm eff}$ scenario using several observational datasets. }
\label{tab:PEDE-Mnu-Neff}                                                                                                   
\end{table*}                                                                                                                     
\end{center}                                                                                                                    
\endgroup          

\section{Summary and Conclusions}
\label{sec-sumamry}

Modern cosmology has witnessed a remarkable success over the last several years and the credit goes a large amount of observational data coming from different astronomical surveys. With the increasing sensitivity in the astronomical data, undoubtedly, we have been able to put stringent constraints on the cosmological parameters, and at the same time we are facing new challenges. The $H_0$ and $S_8$ tensions have emerged as two biggest challenges for the standard $\Lambda$CDM cosmology demanding its revision, and the Planck preference of a closed universe model (see Refs.~\cite{Aghanim:2018eyx,Handley:2019tkm,DiValentino:2019qzk,DiValentino:2020hov,Efstathiou:2020wem,Efstathiou:2019mdh}) further argues that the curvature of the universe can play a vital role in the estimations of the cosmological parameters. Even though the earlier investigations were in agreement with a spatially flat universe, however, there is no reason to stick to this point~\cite{Anselmi:2022uvj} and the effects of curvature on the cosmological models and the key cosmological parameters should be investigated.  
Thus, the main theme of this work is the curvature of the universe and its impacts on extended cosmologies with a special focus on the cosmological tensions. 

In order to investigate these issue,  we have considered 16 extended cosmological scenarios emerging from 4 well known cosmological models ($\Lambda$CDM, $w$CDM, $w_0w_a$CDM, PEDE, see \autoref{sec-2} for their descriptions) and performed a systematic study with the use of  CMB temperature and polarization spectra from Planck 2018 and its combination with other astronomical data, e.g. BAO and Pantheon SNIa catalogue. In particular, these  {\it sixteen} extended cosmological scenarios have been constrained making use of CMB, CMB+BAO, CMB+Pantheon and CMB+BAO+Pantheon. In the following we describe the constraints on the curvature and then present the effects of the a varying curvature on other key cosmological parameters.

\subsection{Summary}

\subsubsection{Curvature}

\begin{figure*}
	\centering
	\includegraphics[width=0.8\linewidth]{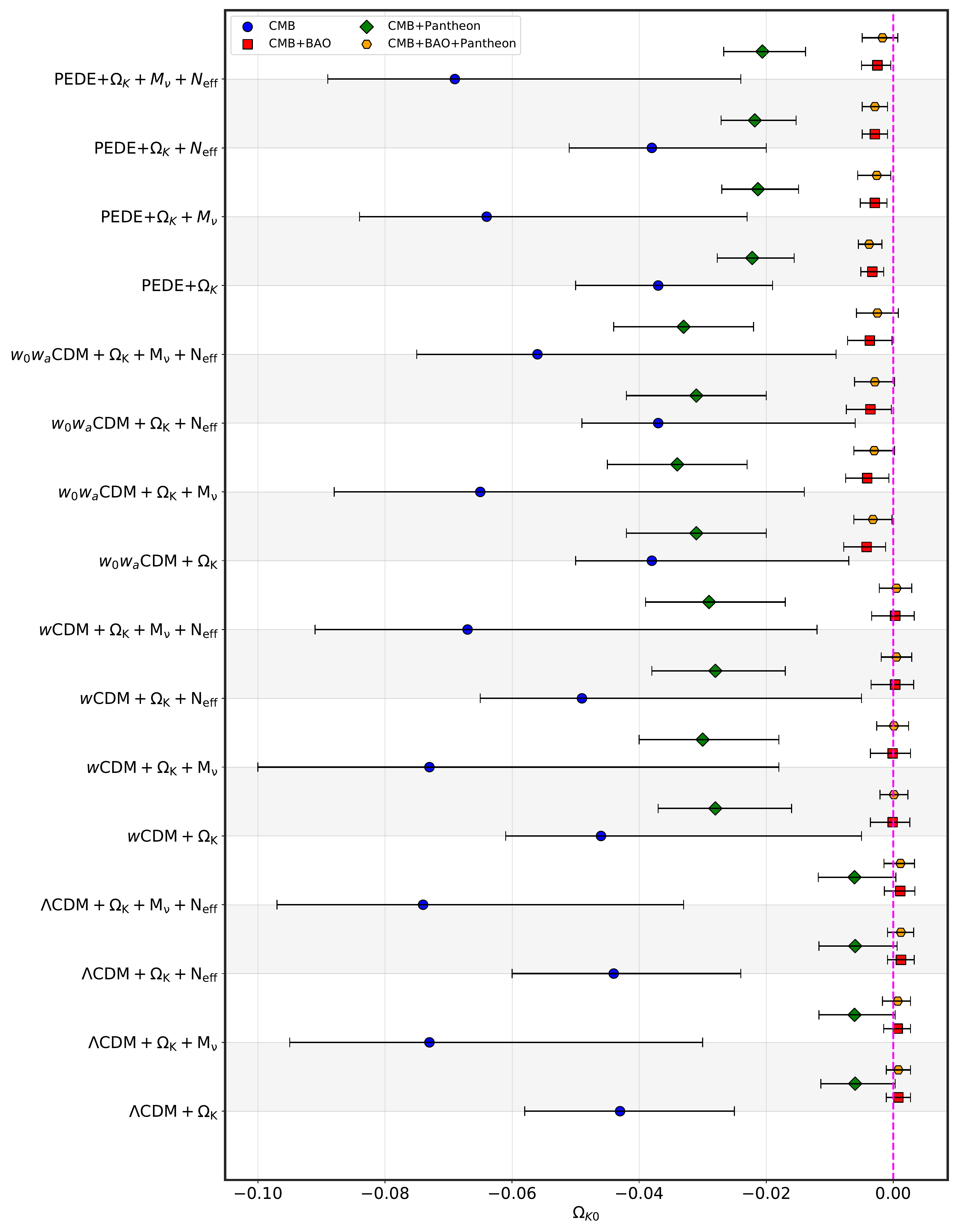}
	\caption{Whisker plot with 68\% CL constraints on the curvature parameter ($\Omega_{K0}$) obtained in various cosmological scenarios considered in this article has been displayed for several observational datasets, namely, CMB, CMB+BAO, CMB+Pantheon and CMB+BAO+Pantheon.  The magenta vertical line corresponds to the flat case ($\Omega_{K0} = 0$).  }
	\label{fig:omega_k}
\end{figure*}

The results for the curvature density parameter are summarized in \autoref{fig:omega_k} for all the different cosmological models and data-combinations. 

The Planck indication for a closed universe observed within the baseline $\Lambda$CDM case persists in all the non-flat extended cosmologies analyzed in this work, although at different statistical level because of the large error-bars obtained by extending the parameter-space and/or considering different parameterizations for the dark energy sector. Notice that this loss of constraining power is typically due to the strong geometrical degeneracy among different parameters but in particular among the DE equation of state and the curvature density parameter $\Omega_{K0}$. Such degeneracy can be broken by combining the CMB data with other complementary datasets, such as the large scale structure information from BAO or the distance moduli measurements from the Pantheon sample. In the first case, for the combination of CMB and BAO (i.e. CMB+BAO), the preference for a closed Universe disappears and, regardless from the dark energy model, the constraints on curvature shrink around $\Omega_{K0}=0$, always pointing towards spatial flatness (within one standard deviation or slightly more for a dynamical dark energy). However, it is worth recalling that BAOs are in general tension with Planck when curvature is allowed to vary and so this data-combination is not robust. On the other hand, combining CMB+Pantheon, the results on $\Omega_{K0}$ become sensitive to the assumptions in the dark energy sector. Indeed fixing the dark energy equation of state to $w= - 1$, this dataset still indicates a spatially flat Universe at one standard deviation, while in all the other dynamical and non-dynamical dark energy parameterizations it systematically shows a preference for a spatially closed geometry. 

\subsubsection{$H_0$}

\begin{figure*}
	\centering
	\includegraphics[width=0.8\linewidth]{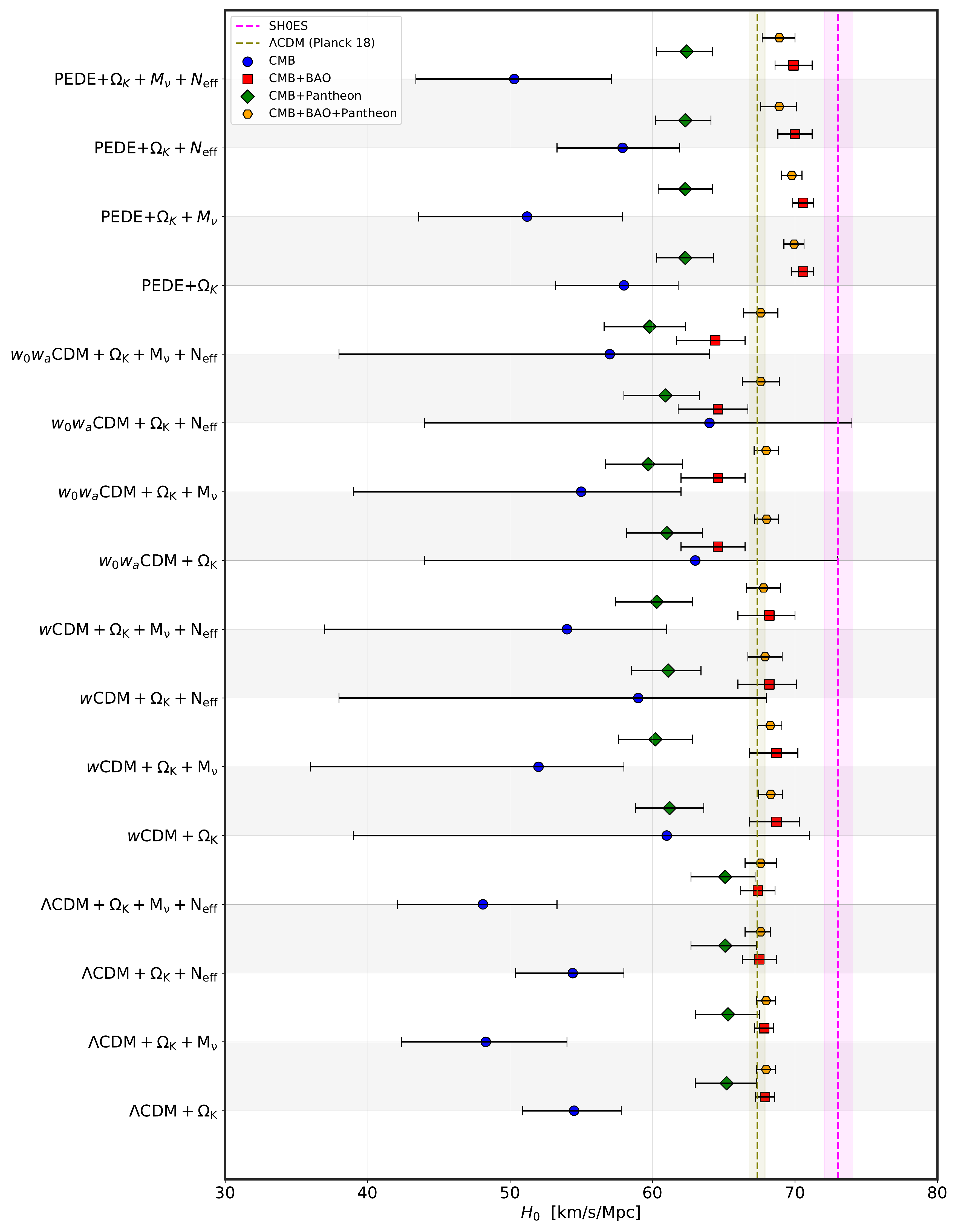}
	\caption{Whisker plot with 68\% CL constraints on the Hubble constant $H_0$ for a variety of cosmological scenarios in the presence of a non-zero curvature of the universe has been shown considering CMB, CMB+BAO, CMB+Pantheon and CMB+BAO+Pantheon. The yellow vertical dotted line corresponds to the $H_0$ value from Planck 2018 team~\cite{Aghanim:2018eyx} assuming a $\Lambda$CDM model and the  magenta vertical band corresponds to the $H_0$ value from SH0ES team ($H_0 = 73.04 \pm 0.99$ km/s/Mpc at $68\%$ CL)~\cite{Riess:2022mme}. }
	\label{fig:h0}
\end{figure*}

\autoref{fig:h0} summarizes the constraints on the Hubble constant $H_0$ for all cosmological models and for all data combinations showing clearly how different cosmological models are returning different values of  $H_0$ when the curvature is taken into account. Throughout the present article, we find that the Hubble constant has  a deep degeneracy with the curvature parameter for the proposed cosmological models.

In all extensions of $\Lambda$CDM, we have an evidence of a closed Universe at more than 95\% CL for CMB alone dataset. Since $H_0$ and $\Omega_{K0}$ are positively correlated in all these extended models (see \autoref{fig:LCDM-Omegak}, \autoref{fig:LCDM-Omegak-Mnu}, \autoref{fig:LCDM-Omegak-Neff} and \autoref{fig:LCDM-Omegak-Mnu-Neff}) and we notice that as long as $\Omega_{K0}$ starts deviating from its null value (i.e. $\Omega_{K0} = 0$) towards the negative region, $H_0$ takes lower values due to this correlation which as a result increases the tension with SH0ES (see \autoref{fig:h0}). This increased tension is solely due to the evidence of a closed Universe for CMB alone in such scenarios. For CMB+BAO and CMB+BAO+Pantheon, the estimated values of $H_0$ in all such scenarios are almost same with the Planck value (within flat $\Lambda$CDM), as $\Omega_{K0}$ is consistent to zero, with the exception of the PEDE parametrization for which $H_0$ is slightly higher alleviating the tension with SH0ES.  For CMB+Pantheon, the $H_0$ values in all four extended scenarios are relatively lower than the Planck value obtained within the flat $\Lambda$CDM scenario. Thus, within these four extended scenarios of $\Lambda$CDM, we do not observe any effective resolution of the $H_0$ tension.

Focusing on all the extended scenarios of the $w$CDM cosmology, we find that CMB alone again returns lower values of $H_0$ compared to Planck's estimation (within the flat $\Lambda$CDM model) due to an evidence of the closed Universe at more than 68\% CL (for $w$CDM + $\Omega_K$ + $M_{\nu}$, closed Universe is evident at more than 95\% CL). However, the error bars in $H_0$  are found to be very large and due to this large error bars the tension  with the SH0ES collaboration is reduced down to $1.2\sigma$ (for $w$CDM + $\Omega_K$), 
$< 2 \sigma$ ($w$CDM + $\Omega_K$ + $N_{\rm eff}$), $2.7 \sigma$ ($w$CDM + $\Omega_K$ + $M_{\nu}$ + $N_{\rm eff}$).  However, such alleviation of the $H_0$ tension is mainly driven by the volume effect of the parameter space. 
The constraints on $H_0$ from CMB+BAO are almost similar to the Planck values (within flat $\Lambda$CDM), however, the mean values of $H_0$ in all the cases and the error bars are slightly large than the Planck values and as a result we see that for the CMB+BAO analysis, the tension with the SH0ES for such extended scenarios are reduced down to $2.3\sigma$ (for $w$CDM + $\Omega_K$), $2.4 \sigma$ (for $w$CDM + $\Omega_K$ + $M_{\nu}$), $2.2 \sigma$ (for $w$CDM + $\Omega_K$ + $N_{\rm eff}$), $ 2.3 \sigma$ (for $w$CDM + $\Omega_K$ + $M_{\nu}$ + $N_{\rm eff}$).  For CMB+Pantheon, we obtain relatively lower values of $H_0$ than Planck (within flat $\Lambda$CDM) due to the preference of a closed Universe at more than 95\% CL. Finally, the constraints on $H_0$ from CMB+BAO+Pantheon are almost similar to the CMB+BAO analysis for all the extended cosmological scenarios in this category. 

For $w_0w_a$CDM + $\Omega_K$, and its other extensions with neutrinos, we see that CMB alone returns lower values of $H_0$ compared to Planck's estimation (within flat $\Lambda$CDM) but with very large error bars which as a result help to reduce the tension with SH0ES down to $1\sigma$, $2.5\sigma$, $< 1 \sigma$, $2.3 \sigma$, respectively, for $w_0w_a$CDM + $\Omega_K$, $w_0w_a$CDM + $\Omega_K$ + $M_{\nu}$, $w_0w_a$CDM + $\Omega_K$ + $N_{\rm eff}$ and $w_0w_a$CDM + $\Omega_K$ + $M_{\nu}$ + $N_{\rm eff}$. 
For CMB+BAO, in all extended scenarios we obtain a preference for a closed Universe at more than 68\% CL and as a consequence we get lower values of $H_0$ compared to Planck (within flat $\Lambda$CDM) as an effect of this closed geometry. 
For CMB+Pantheon, the estimated values of $H_0$ are lower than Planck's value and they are also lower than the CMB+BAO estimations in these extended scenarios. This is due to the fact that CMB+Pantheon supports a closed Universe at more than 95\% CL. For CMB+BAO+Pantheon, the $H_0$ values obtained in these extended scenarios match almost Planck values (within flat $\Lambda$CDM).  

In the extended PEDE scenarios, we see that for both CMB alone and CMB+Pantheon, 
$H_0$ assumes lower values similar to the previously explored scenarios and this is due to the fact that in both the cases, a closed Universe scenario is preferred at more than 95\% CL. 
For CMB+BAO analysis, we see that even though all the extended scenarios indicate a preference of a closed Universe at more than 68\% CL, however, unlike to the previous scenarios where closed Universe preference worsens the $H_0$ tension, here on the contrary, the tension with SH0ES is reduced down to many standard deviations. In particular, we find that the $H_0$ tension is reduced down to $~ 2\sigma $ (PEDE + $\Omega_K$), $~ 2\sigma $ (PEDE + $\Omega_K$ + $M_{\nu}$), $~ 1.9\sigma $ (PEDE + $\Omega_K$ + $N_{\rm eff}$), $~ 1.9\sigma $ (PEDE + $\Omega_K$+ $M_{\nu}$ + $N_{\rm eff}$).    
The results for CMB+BAO+Pantheon are almost similar to the CMB+BAO case with a hints of a closed Universe model at more than 68\% CL. 

 It is worth noticing here that models of new physics at low-redshift (or the late time solutions) cannot solve the $H_0$ tension once BAO and Pantheon data are considered (see Refs.~\cite{Knox:2019rjx,Keeley:2022ojz,Arendse:2019hev}), unless the energy density of DE becomes negative~\cite{DiValentino:2020naf,Chudaykin:2022rnl}.

\begin{figure*}
	\centering
	\includegraphics[width=0.8\linewidth]{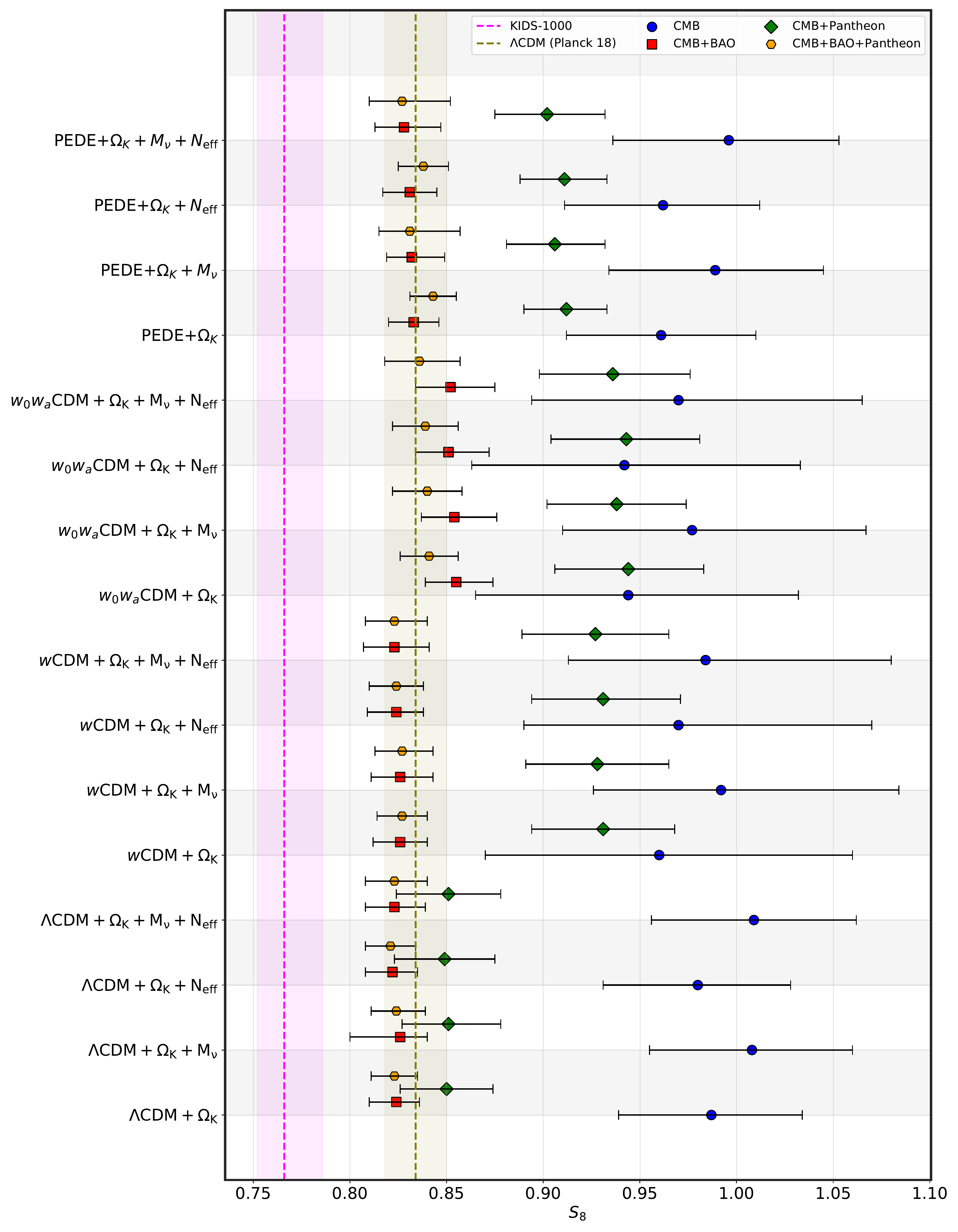}
	\caption{Whisker plot with 68\% CL constraints on the $S_8$ ($= \sigma_8 \sqrt{\Omega_{m0}/0.3}$) parameter obtained in various cosmological scenarios considered in this article has been displayed for several observational datasets, namely, CMB, CMB+BAO, CMB+Pantheon and CMB+BAO+Pantheon. The magenta vertical band corresponds to the estimated value by the Kilo-Degree Survey (KIDS-1000) leading to $S_8 = 0.766^{+0.020}_{-0.014}$~\cite{Heymans:2020gsg},
	while the green vertical band corresponds to the value provided by the Planck collaboration $S_8=0.834 \pm 0.016$~\cite{Aghanim:2018eyx}, all obtained by assuming a $\Lambda$CDM scenario. }  
	\label{fig:S8}
\end{figure*}

\subsubsection{$S_8$}

In \autoref{fig:S8} we have summarized the constraints on the $S_8$ parameter extracted out of all the extended cosmological models using  CMB, CMB+BAO, CMB+Pantheon and CMB+BAO+Pantheon together with the estimated values of $S_8$ from Kilo-Degree Survey (KIDS-1000) [$S_8 = 0.766^{+0.020}_{-0.014}$]~\cite{Heymans:2020gsg}, Dark Energy Survey (DES) Year 3 data (DES-Y3)
[$S_8 = 0.759^{+0.025}_{-0.023}$]~\cite{DES:2021vln} and Planck $S_8=0.834 \pm 0.016$~\cite{Aghanim:2018eyx}, all obtained by assuming a $\Lambda$CDM scenario as the underlying model.  

From \autoref{fig:S8} we observe two things.  For CMB and CMB+Pantheon, the estimated values of the $S_8$ parameter in all the extended cosmological scenarios are very far from both KIDS-1000~\cite{Heymans:2020gsg} and DES-Y3~\cite{DES:2021vln}, when they assume a vanilla $\Lambda$CDM model.\footnote{Both KiDS-100 and DES collaborations explored extended scenarios in Refs.~\cite{KiDS:2020ghu,DES:2022ygi}, but the combinations of parameters explored here are not considered, so we can't make a safe comparison.} Moreover, they are also far from Planck's estimation assuming a vanilla $\Lambda$CDM model, except in the extended scenarios assuming a cosmological constant, where for CMB+Pantheon case $S_8$ values are slightly close to it. For all the other cases, for these two datasets the tension in the $S_8$ parameter increases significantly. This is because of the indication for a closed Universe coming from these two datasets when it is not assumed a cosmological constant.   
On the other hand, when we look at the estimations of $S_8$ by CMB+BAO and CMB+BAO+Pantheon, we see that these estimations are almost similar to the Planck values assuming a vanilla $\Lambda$CDM model.    
This indicates that within such extended scenarios, the $S_8$ tension increases when there is an indication for a closed Universe.

\subsubsection{$M_{\nu}$}

\begin{figure*}
	\centering
	\includegraphics[width=0.8\linewidth]{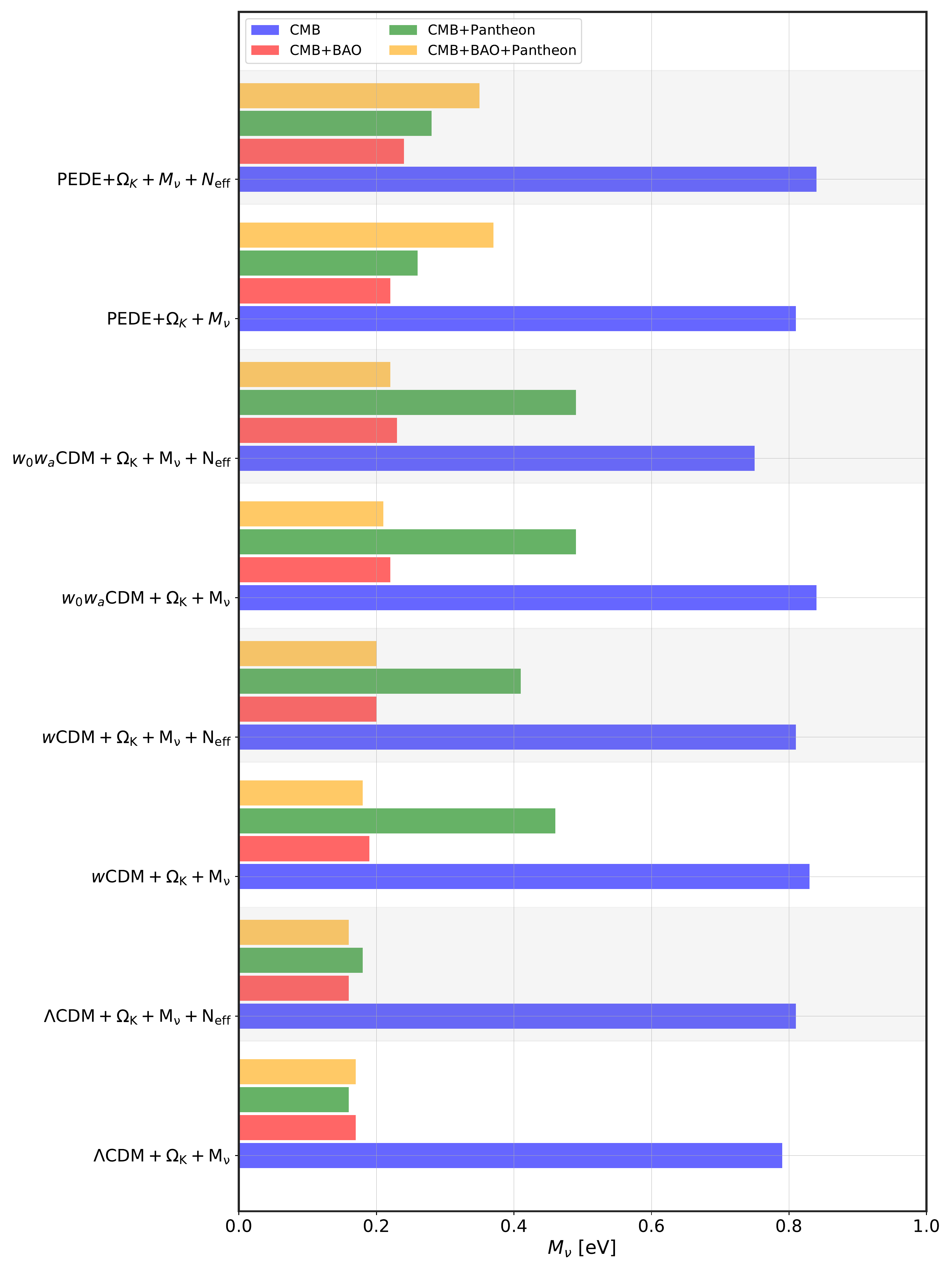}
	\caption{Whisker plot with 95\% CL upper bounds on $M_{\nu}$ [eV] obtained in various cosmological scenarios considered in this article has been displayed for several observational datasets, namely, CMB, CMB+BAO, CMB+Pantheon and CMB+BAO+Pantheon.  }
	\label{fig:Mnu}
\end{figure*}

From all the analyses, we find no evidence of a total neutrino mass. In particular, only in a few cases, namely, $\Lambda$CDM + $\Omega_K$ + $M_{\nu}$ + $N_{\rm eff}$, $w$CDM + $\Omega_K$ + $M_{\nu}$, $w_0w_a$CDM + $\Omega_K$ + $M_{\nu}$, PEDE + $\Omega_K$ + $M_{\nu}$ + $N_{\rm eff}$, for CMB alone, we find an indication of a total neutrino mass at 68\% CL but this evidence disappears when other external and complementary probes are added to CMB. However, in order to realize the bounds on $M_{\nu}$ in all these scenarios,  in \autoref{fig:Mnu}, we have presented a graphical presentation of the the upper limits of $M_{\nu}$ at 95\% CL for our extended cosmological models using CMB, CMB+BAO, CMB+Pantheon and CMB+BAO+Pantheon.

\subsubsection{$N_{\rm eff}$}

\begin{figure*}
	\centering
	\includegraphics[width=0.7\linewidth]{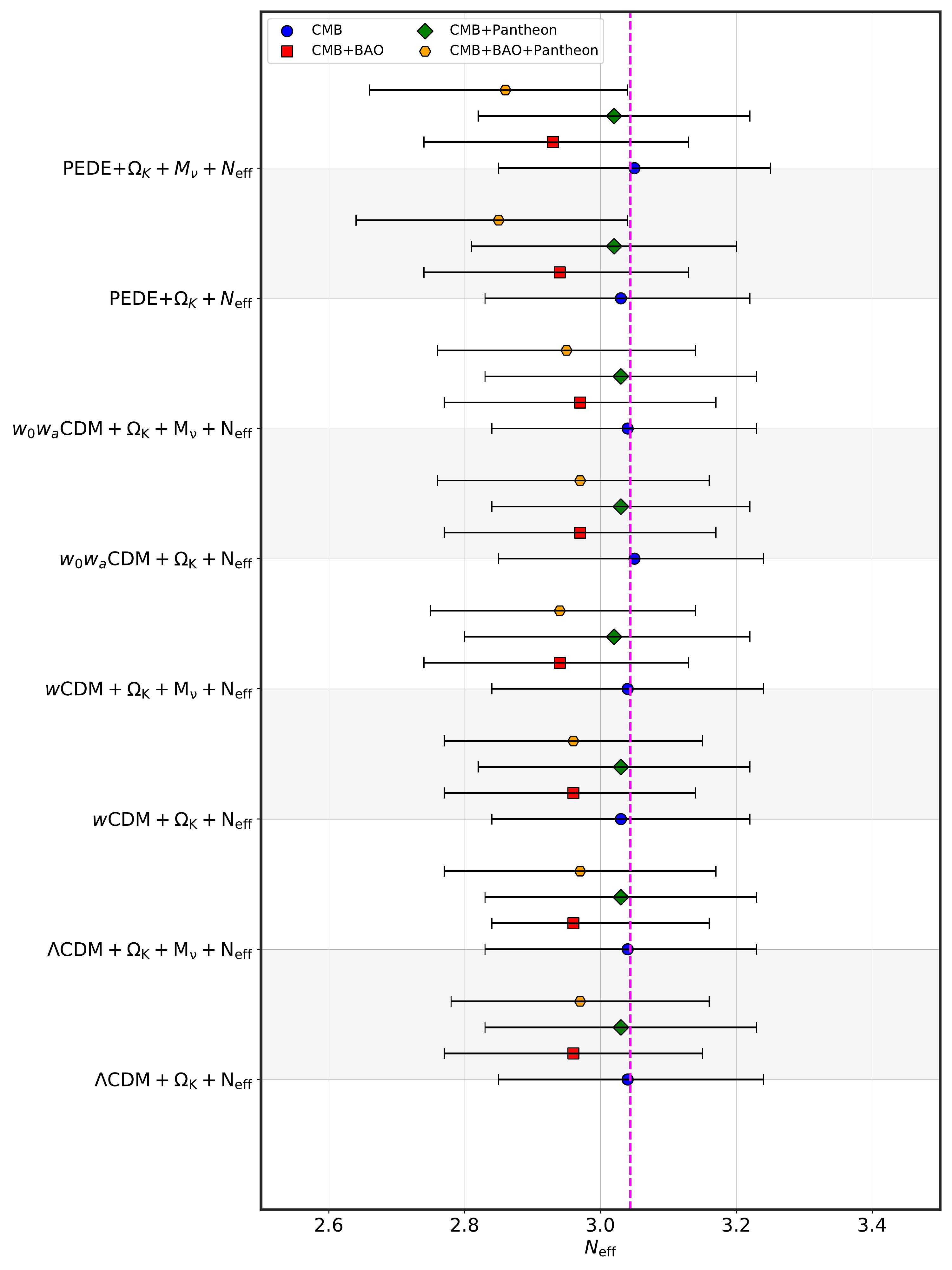}
	\caption{Whisker plot with 68\% CL constraints on $N_{\rm eff}$ obtained in various cosmological scenarios considered in this article has been displayed for several observational datasets, namely, CMB, CMB+BAO, CMB+Pantheon and CMB+BAO+Pantheon. The magenta vertical line corresponds to the standard value $N_{\rm eff}=3.044$.} 
	\label{fig:Neff}
\end{figure*}

We provide a summary picture  of $N_{\rm eff}$ in \autoref{fig:Neff} obtained in all our extended cosmological scenarios considered in this work for CMB, CMB+BAO, CMB+Pantheon, and CMB+BAO+Pantheon. We find that irrespective of all the cosmological models considered in this work, the estimated values of $N_{\rm eff}$ for CMB and CMB+Pantheon are almost same with that of the standard value $N_{\rm eff}=3.044$ and thus these estimations are consistent with this standard value. For the remaining two cases, namely, CMB+BAO and CMB+BAO+Pantheon, the values of  $N_{\rm eff}$ are slightly lower ($N_{\rm eff} < 3$) but within 68\% CL, they are also consistent with the standard value.

\subsubsection{$w_0$ and $w_a$}

\begin{figure*}
	\centering
	\includegraphics[width=\linewidth]{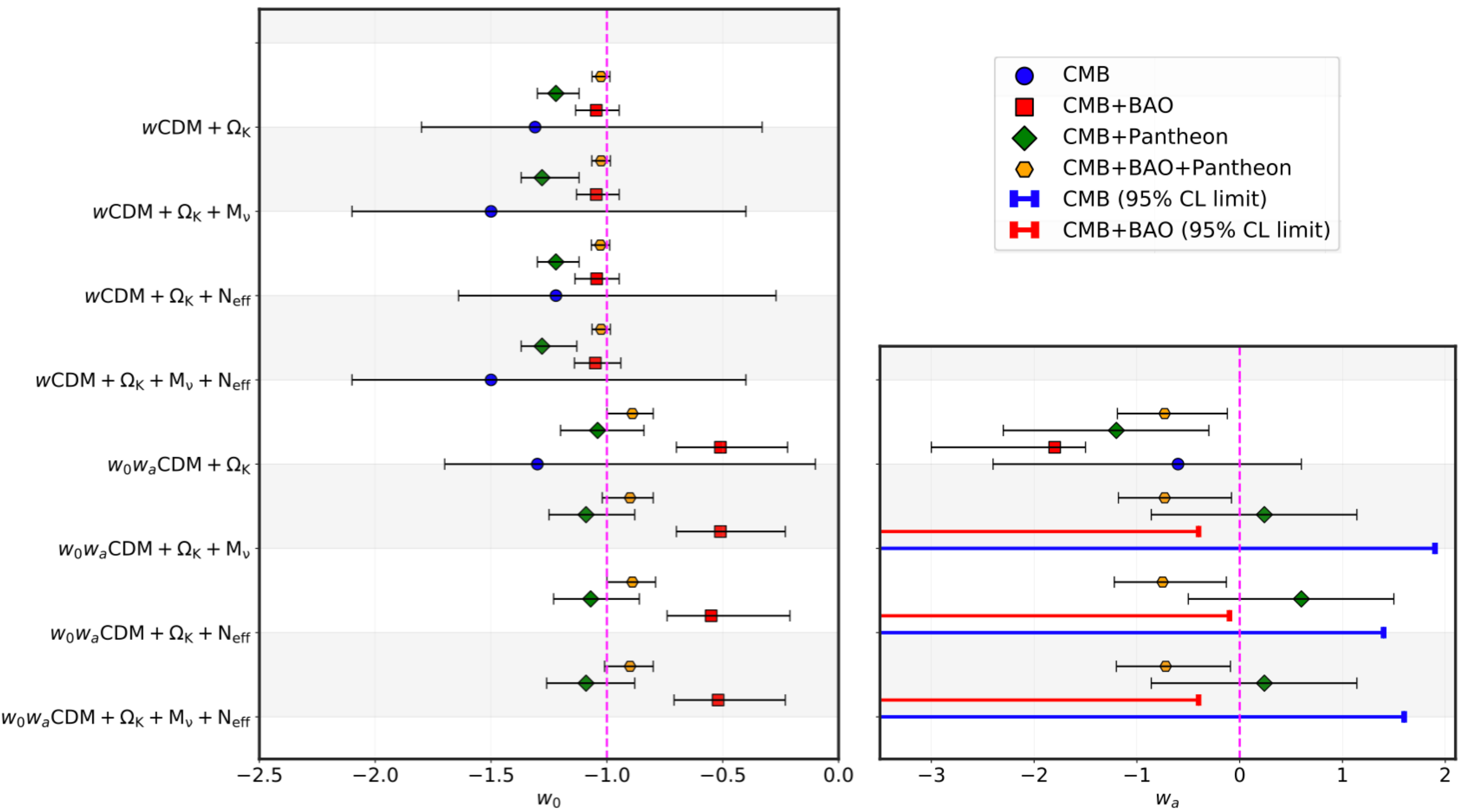}
	\caption{Whisker plot with 68\% CL constraints and 95\% CL limits on $w_0$ and $w_a$ obtained in various cosmological scenarios considered in this article has been displayed for several observational datasets, namely, CMB, CMB+BAO, CMB+Pantheon and CMB+BAO+Pantheon. The magenta vertical lines in the two panels represent to the standard cosmological constant value $w_0=-1$ and to the non dynamical $w_a=0$ limit, respectively. The missing blue lines (Planck only results) in the left panel are indicating that $w$ is unconstrained for those models.}
	\label{fig:w}
\end{figure*}

Here we summarize the main results on the dark energy equation of state parameter $w$ of the $w$CDM model and also on the $(w_0, w_a)$ parameters of the $w_0w_a$CDM model. 
In \autoref{fig:w}, we show the whisker plots with 68\% CL, and 95\% upper limits, on $w$ and $w_a$ for the $w$CDM and $w_0w_a$CDM models for all the observational datasets. 

One can clearly see that in the $w$CDM model, only the CMB+Pantheon dataset gives a strong indication of a phantom DE (i.e. $w< -1$) at more than 95\% CL, only when the equation of state of the DE is constant and not dynamical. For the remaining cases, $w =-1$ is allowed within 68\% CL.

However, for the $w_0w_a$CDM model in which the dark energy equation of state is dynamical, CMB+BAO gives an indication for a quintessence DE (i.e. $w_0>-1$), while all the other dataset combinations are in agreement with a cosmological constant\footnote{Note here that $w_0$ remains unconstrained in the scenarios $w_0w_a$CDM + $\Omega_K$ + $M_{\nu}$, $w_0w_a$CDM + $\Omega_K$ + $M_{\nu}$ + $N_{\rm eff}$ and for the scenario $w_0w_a$CDM + $\Omega_K$ + $N_{\rm eff}$, even though the 68\% CL upper bound on $w_0$ is available but it becomes unconstrained at 95\% CL. }. Regarding the dynamical nature of the DE quantified through the free parameter $w_a$, we find that CMB alone and CMB+Pantheon are very consistent with $w_a=0$ (except in the model $w_0w_a$CDM + $\Omega_K$ in which $w_a \neq 0$ at more than 68\% CL), however, when BAO data are included to CMB, an indication for a $w_a<0$ at more than $2\sigma$ is found. While this evidence is mildly diluted (but remains true at slightly more than 68\% CL) when the full combination CMB+BAO+Pantheon is explored.  In other words, for this $w_0w_a$CDM model, CMB+BAO is preferring a quintessence dynamical DE at more than 95\% CL.

\subsection{Concluding remarks and the road ahead}
\label{sec-final-remarks}

There is no doubt that without observational data, modern cosmology is incomplete.  From the detection of the cosmic microwave background radiation to the discovery of the late-time cosmic acceleration, we have witnessed how the observational data have been crucially important in the understanding of our Universe. 
Whenever we talk about cosmology, the six parameter $\Lambda$CDM cosmological model under the assumptions of a positive cosmological constant, a pressure-less DM and a flat Universe, naturally enters into the picture because of its super agreement with a series of astronomical data. However, the true nature of DE and DM within this simplest cosmological scenario has remained one of the longstanding debates in cosmology. In addition to that, the assumption of a flat Universe has created a further debate in the scientific community. A series of recent articles have argued that the present observational data  hint towards an evidence of a closed Universe~\cite{Aghanim:2018eyx,Handley:2019tkm,DiValentino:2019qzk,DiValentino:2020hov,Glanville:2022xes}. This evidence has sparked the entire scientific community because  any evidence of a closed Universe increases the tensions in the key cosmological parameters and hence the flat $\Lambda$CDM cosmology is further challenged.  
Generally, the curvature of the Universe is ``assumed'' to be zero when the cosmological models are analyzed. This is motivated because:  (i) the inflationary paradigm has been tremendously successful in demanding a spatially flat Universe, (ii) the observational data in the past preferred a small curvature of the Universe which is very close to zero, and further (iii) the inclusion of extra parameters, in terms of the curvature parameter for instance, may increase the degeneracies in the parameters.  However, over the last several years, the sensitivity of the astronomical data has improved a lot, and given such improvements,
it is very natural to allow the observational data to decide the nature of the curvature parameter. That means, the consideration of a flat Universe (i.e. $\Omega_{K} =0$) while analyzing the cosmological models is not a very realistic assumption. We further comment that forcing $\Omega_{K} =0$  could bias the results and the intrinsic nature of the cosmological models may remain hidden forever~\cite{Anselmi:2022uvj},  and at this stage where we have a large amount of astronomical data from various sources, having $\Omega_{K} =0$ is not expected.

Thus, in the present article
we  investigated the effects of curvature on the cosmological models and their associated (free and derived) parameters.  We considered some classical and well known cosmological models and their several extensions, including variations in the DE and neutrino sectors,
in presence of the curvature of our Universe. Our analyses clearly indicate that it is better not to fix $\Omega_{K} = 0$, to analyze the cosmological models. 
We trust that the upcoming observational data from various astronomical surveys (see for instance, Refs. \cite{LSSTScience:2009jmu,EUCLID:2011zbd,DESI:2013agm,SimonsObservatory:2018koc}) will be highly promising in this direction which will reveal the intrinsic nature of the dark sector as well as the curvature of the Universe.

%----------------------------------------------------

\begin{acknowledgments}
The authors thank the referee for some useful comments which improved the quality of discussion of the article. WY was supported by the National Natural Science Foundation of China under Grants No. 12175096 and No. 11705079, and Liaoning Revitalization Talents Program under Grant no. XLYC1907098.
WG and AM are supported by ``Theoretical Astroparticle Physics" (TAsP), iniziativa specifica INFN. SP acknowledges the financial support from the  Department of Science and Technology (DST), Govt. of India under the Scheme   ``Fund for Improvement of S\&T Infrastructure (FIST)'' (File No. SR/FST/MS-I/2019/41). EDV is supported by a Royal Society Dorothy Hodgkin Research Fellowship. 
\end{acknowledgments}

%--------------------------------------------------------
%\bibliographystyle{utphys} 
\bibliography{biblio}

\begin{thebibliography}{138}%
\makeatletter
\providecommand \@ifxundefined [1]{%
 \@ifx{#1\undefined}
}%
\providecommand \@ifnum [1]{%
 \ifnum #1\expandafter \@firstoftwo
 \else \expandafter \@secondoftwo
 \fi
}%
\providecommand \@ifx [1]{%
 \ifx #1\expandafter \@firstoftwo
 \else \expandafter \@secondoftwo
 \fi
}%
\providecommand \natexlab [1]{#1}%
\providecommand \enquote  [1]{``#1''}%
\providecommand \bibnamefont  [1]{#1}%
\providecommand \bibfnamefont [1]{#1}%
\providecommand \citenamefont [1]{#1}%
\providecommand \href@noop [0]{\@secondoftwo}%
\providecommand \href [0]{\begingroup \@sanitize@url \@href}%
\providecommand \@href[1]{\@@startlink{#1}\@@href}%
\providecommand \@@href[1]{\endgroup#1\@@endlink}%
\providecommand \@sanitize@url [0]{\catcode `\\12\catcode `\$12\catcode
  `\&12\catcode `\#12\catcode `\^12\catcode `\_12\catcode `\%12\relax}%
\providecommand \@@startlink[1]{}%
\providecommand \@@endlink[0]{}%
\providecommand \url  [0]{\begingroup\@sanitize@url \@url }%
\providecommand \@url [1]{\endgroup\@href {#1}{\urlprefix }}%
\providecommand \urlprefix  [0]{URL }%
\providecommand \Eprint [0]{\href }%
\providecommand \doibase [0]{http://dx.doi.org/}%
\providecommand \selectlanguage [0]{\@gobble}%
\providecommand \bibinfo  [0]{\@secondoftwo}%
\providecommand \bibfield  [0]{\@secondoftwo}%
\providecommand \translation [1]{[#1]}%
\providecommand \BibitemOpen [0]{}%
\providecommand \bibitemStop [0]{}%
\providecommand \bibitemNoStop [0]{.\EOS\space}%
\providecommand \EOS [0]{\spacefactor3000\relax}%
\providecommand \BibitemShut  [1]{\csname bibitem#1\endcsname}%
\let\auto@bib@innerbib\@empty
%</preamble>
\bibitem [{\citenamefont {Di~Valentino}\ \emph
  {et~al.}(2020{\natexlab{a}})\citenamefont {Di~Valentino} \emph
  {et~al.}}]{DiValentino:2020srs}%
  \BibitemOpen
  \bibfield  {author} {\bibinfo {author} {\bibfnamefont {E.}~\bibnamefont
  {Di~Valentino}} \emph {et~al.},\ }\href@noop {} {\  (\bibinfo {year}
  {2020}{\natexlab{a}})},\ \Eprint {http://arxiv.org/abs/2008.11286}
  {arXiv:2008.11286 [astro-ph.CO]} \BibitemShut {NoStop}%
\bibitem [{\citenamefont {Gaztanaga}\ \emph {et~al.}(2009)\citenamefont
  {Gaztanaga}, \citenamefont {Miquel},\ and\ \citenamefont
  {Sanchez}}]{Gaztanaga:2008de}%
  \BibitemOpen
  \bibfield  {author} {\bibinfo {author} {\bibfnamefont {E.}~\bibnamefont
  {Gaztanaga}}, \bibinfo {author} {\bibfnamefont {R.}~\bibnamefont {Miquel}}, \
  and\ \bibinfo {author} {\bibfnamefont {E.}~\bibnamefont {Sanchez}},\ }\href
  {\doibase 10.1103/PhysRevLett.103.091302} {\bibfield  {journal} {\bibinfo
  {journal} {Phys. Rev. Lett.}\ }\textbf {\bibinfo {volume} {103}},\ \bibinfo
  {pages} {091302} (\bibinfo {year} {2009})},\ \Eprint
  {http://arxiv.org/abs/0808.1921} {arXiv:0808.1921 [astro-ph]} \BibitemShut
  {NoStop}%
\bibitem [{\citenamefont {Mortonson}(2009)}]{Mortonson:2009nw}%
  \BibitemOpen
  \bibfield  {author} {\bibinfo {author} {\bibfnamefont {M.~J.}\ \bibnamefont
  {Mortonson}},\ }\href {\doibase 10.1103/PhysRevD.80.123504} {\bibfield
  {journal} {\bibinfo  {journal} {Phys. Rev. D}\ }\textbf {\bibinfo {volume}
  {80}},\ \bibinfo {pages} {123504} (\bibinfo {year} {2009})},\ \Eprint
  {http://arxiv.org/abs/0908.0346} {arXiv:0908.0346 [astro-ph.CO]} \BibitemShut
  {NoStop}%
\bibitem [{\citenamefont {Suyu}\ \emph {et~al.}(2014)\citenamefont {Suyu} \emph
  {et~al.}}]{Suyu:2013kha}%
  \BibitemOpen
  \bibfield  {author} {\bibinfo {author} {\bibfnamefont {S.~H.}\ \bibnamefont
  {Suyu}} \emph {et~al.},\ }\href {\doibase 10.1088/2041-8205/788/2/L35}
  {\bibfield  {journal} {\bibinfo  {journal} {Astrophys. J. Lett.}\ }\textbf
  {\bibinfo {volume} {788}},\ \bibinfo {pages} {L35} (\bibinfo {year}
  {2014})},\ \Eprint {http://arxiv.org/abs/1306.4732} {arXiv:1306.4732
  [astro-ph.CO]} \BibitemShut {NoStop}%
\bibitem [{\citenamefont {L'Huillier}\ and\ \citenamefont
  {Shafieloo}(2017)}]{LHuillier:2016mtc}%
  \BibitemOpen
  \bibfield  {author} {\bibinfo {author} {\bibfnamefont {B.}~\bibnamefont
  {L'Huillier}}\ and\ \bibinfo {author} {\bibfnamefont {A.}~\bibnamefont
  {Shafieloo}},\ }\href {\doibase 10.1088/1475-7516/2017/01/015} {\bibfield
  {journal} {\bibinfo  {journal} {JCAP}\ }\textbf {\bibinfo {volume} {01}},\
  \bibinfo {pages} {015} (\bibinfo {year} {2017})},\ \Eprint
  {http://arxiv.org/abs/1606.06832} {arXiv:1606.06832 [astro-ph.CO]}
  \BibitemShut {NoStop}%
\bibitem [{\citenamefont {Chudaykin}\ \emph {et~al.}(2021)\citenamefont
  {Chudaykin}, \citenamefont {Dolgikh},\ and\ \citenamefont
  {Ivanov}}]{Chudaykin:2020ghx}%
  \BibitemOpen
  \bibfield  {author} {\bibinfo {author} {\bibfnamefont {A.}~\bibnamefont
  {Chudaykin}}, \bibinfo {author} {\bibfnamefont {K.}~\bibnamefont {Dolgikh}},
  \ and\ \bibinfo {author} {\bibfnamefont {M.~M.}\ \bibnamefont {Ivanov}},\
  }\href {\doibase 10.1103/PhysRevD.103.023507} {\bibfield  {journal} {\bibinfo
   {journal} {Phys. Rev. D}\ }\textbf {\bibinfo {volume} {103}},\ \bibinfo
  {pages} {023507} (\bibinfo {year} {2021})},\ \Eprint
  {http://arxiv.org/abs/2009.10106} {arXiv:2009.10106 [astro-ph.CO]}
  \BibitemShut {NoStop}%
\bibitem [{\citenamefont {Acquaviva}\ \emph {et~al.}(2021)\citenamefont
  {Acquaviva}, \citenamefont {Akarsu}, \citenamefont {Katirci},\ and\
  \citenamefont {Vazquez}}]{Acquaviva:2021jov}%
  \BibitemOpen
  \bibfield  {author} {\bibinfo {author} {\bibfnamefont {G.}~\bibnamefont
  {Acquaviva}}, \bibinfo {author} {\bibfnamefont {O.}~\bibnamefont {Akarsu}},
  \bibinfo {author} {\bibfnamefont {N.}~\bibnamefont {Katirci}}, \ and\
  \bibinfo {author} {\bibfnamefont {J.~A.}\ \bibnamefont {Vazquez}},\ }\href
  {\doibase 10.1103/PhysRevD.104.023505} {\bibfield  {journal} {\bibinfo
  {journal} {Phys. Rev. D}\ }\textbf {\bibinfo {volume} {104}},\ \bibinfo
  {pages} {023505} (\bibinfo {year} {2021})},\ \Eprint
  {http://arxiv.org/abs/2104.02623} {arXiv:2104.02623 [astro-ph.CO]}
  \BibitemShut {NoStop}%
\bibitem [{\citenamefont {Aghanim}\ \emph
  {et~al.}(2020{\natexlab{a}})\citenamefont {Aghanim} \emph
  {et~al.}}]{Aghanim:2018eyx}%
  \BibitemOpen
  \bibfield  {author} {\bibinfo {author} {\bibfnamefont {N.}~\bibnamefont
  {Aghanim}} \emph {et~al.} (\bibinfo {collaboration} {Planck}),\ }\href
  {\doibase 10.1051/0004-6361/201833910} {\bibfield  {journal} {\bibinfo
  {journal} {Astron. Astrophys.}\ }\textbf {\bibinfo {volume} {641}},\ \bibinfo
  {pages} {A6} (\bibinfo {year} {2020}{\natexlab{a}})},\ \Eprint
  {http://arxiv.org/abs/1807.06209} {arXiv:1807.06209 [astro-ph.CO]}
  \BibitemShut {NoStop}%
\bibitem [{\citenamefont {Handley}(2019)}]{Handley:2019tkm}%
  \BibitemOpen
  \bibfield  {author} {\bibinfo {author} {\bibfnamefont {W.}~\bibnamefont
  {Handley}},\ }\href@noop {} {\bibfield  {journal} {\bibinfo  {journal}
  {arXiv:1908.09139}\ } (\bibinfo {year} {2019})},\ \Eprint
  {http://arxiv.org/abs/1908.09139} {arXiv:1908.09139 [astro-ph.CO]}
  \BibitemShut {NoStop}%
\bibitem [{\citenamefont {Di~Valentino}\ \emph {et~al.}(2019)\citenamefont
  {Di~Valentino}, \citenamefont {Melchiorri},\ and\ \citenamefont
  {Silk}}]{DiValentino:2019qzk}%
  \BibitemOpen
  \bibfield  {author} {\bibinfo {author} {\bibfnamefont {E.}~\bibnamefont
  {Di~Valentino}}, \bibinfo {author} {\bibfnamefont {A.}~\bibnamefont
  {Melchiorri}}, \ and\ \bibinfo {author} {\bibfnamefont {J.}~\bibnamefont
  {Silk}},\ }\href {\doibase 10.1038/s41550-019-0906-9} {\bibfield  {journal}
  {\bibinfo  {journal} {Nature Astron.}\ }\textbf {\bibinfo {volume} {4}},\
  \bibinfo {pages} {196} (\bibinfo {year} {2019})},\ \Eprint
  {http://arxiv.org/abs/1911.02087} {arXiv:1911.02087 [astro-ph.CO]}
  \BibitemShut {NoStop}%
\bibitem [{\citenamefont {Di~Valentino}\ \emph
  {et~al.}(2020{\natexlab{b}})\citenamefont {Di~Valentino}, \citenamefont
  {Melchiorri},\ and\ \citenamefont {Silk}}]{DiValentino:2020hov}%
  \BibitemOpen
  \bibfield  {author} {\bibinfo {author} {\bibfnamefont {E.}~\bibnamefont
  {Di~Valentino}}, \bibinfo {author} {\bibfnamefont {A.}~\bibnamefont
  {Melchiorri}}, \ and\ \bibinfo {author} {\bibfnamefont {J.}~\bibnamefont
  {Silk}},\ }\href@noop {} {\bibfield  {journal} {\bibinfo  {journal}
  {arXiv:2003.04935}\ } (\bibinfo {year} {2020}{\natexlab{b}})},\ \Eprint
  {http://arxiv.org/abs/2003.04935} {arXiv:2003.04935 [astro-ph.CO]}
  \BibitemShut {NoStop}%
\bibitem [{\citenamefont {Semenaite}\ \emph {et~al.}(2022)\citenamefont
  {Semenaite}, \citenamefont {S\'anchez}, \citenamefont {Pezzotta},
  \citenamefont {Hou}, \citenamefont {Eggemeier}, \citenamefont {Crocce},
  \citenamefont {Zhao}, \citenamefont {Brownstein}, \citenamefont {Rossi},\
  and\ \citenamefont {Schneider}}]{Semenaite:2022unt}%
  \BibitemOpen
  \bibfield  {author} {\bibinfo {author} {\bibfnamefont {A.}~\bibnamefont
  {Semenaite}}, \bibinfo {author} {\bibfnamefont {A.~G.}\ \bibnamefont
  {S\'anchez}}, \bibinfo {author} {\bibfnamefont {A.}~\bibnamefont {Pezzotta}},
  \bibinfo {author} {\bibfnamefont {J.}~\bibnamefont {Hou}}, \bibinfo {author}
  {\bibfnamefont {A.}~\bibnamefont {Eggemeier}}, \bibinfo {author}
  {\bibfnamefont {M.}~\bibnamefont {Crocce}}, \bibinfo {author} {\bibfnamefont
  {C.}~\bibnamefont {Zhao}}, \bibinfo {author} {\bibfnamefont {J.~R.}\
  \bibnamefont {Brownstein}}, \bibinfo {author} {\bibfnamefont
  {G.}~\bibnamefont {Rossi}}, \ and\ \bibinfo {author} {\bibfnamefont {D.~P.}\
  \bibnamefont {Schneider}},\ }\href@noop {} {\  (\bibinfo {year} {2022})},\
  \Eprint {http://arxiv.org/abs/2210.07304} {arXiv:2210.07304 [astro-ph.CO]}
  \BibitemShut {NoStop}%
\bibitem [{\citenamefont {Calabrese}\ \emph {et~al.}(2008)\citenamefont
  {Calabrese}, \citenamefont {Slosar}, \citenamefont {Melchiorri},
  \citenamefont {Smoot},\ and\ \citenamefont {Zahn}}]{Calabrese:2008rt}%
  \BibitemOpen
  \bibfield  {author} {\bibinfo {author} {\bibfnamefont {E.}~\bibnamefont
  {Calabrese}}, \bibinfo {author} {\bibfnamefont {A.}~\bibnamefont {Slosar}},
  \bibinfo {author} {\bibfnamefont {A.}~\bibnamefont {Melchiorri}}, \bibinfo
  {author} {\bibfnamefont {G.~F.}\ \bibnamefont {Smoot}}, \ and\ \bibinfo
  {author} {\bibfnamefont {O.}~\bibnamefont {Zahn}},\ }\href {\doibase
  10.1103/PhysRevD.77.123531} {\bibfield  {journal} {\bibinfo  {journal} {Phys.
  Rev.}\ }\textbf {\bibinfo {volume} {D77}},\ \bibinfo {pages} {123531}
  (\bibinfo {year} {2008})},\ \Eprint {http://arxiv.org/abs/0803.2309}
  {arXiv:0803.2309 [astro-ph]} \BibitemShut {NoStop}%
%%CITATION = ARXIV:0803.2309;%%
\bibitem [{\citenamefont {Di~Valentino}\ \emph
  {et~al.}(2020{\natexlab{c}})\citenamefont {Di~Valentino}, \citenamefont
  {Melchiorri},\ and\ \citenamefont {Silk}}]{DiValentino:2019dzu}%
  \BibitemOpen
  \bibfield  {author} {\bibinfo {author} {\bibfnamefont {E.}~\bibnamefont
  {Di~Valentino}}, \bibinfo {author} {\bibfnamefont {A.}~\bibnamefont
  {Melchiorri}}, \ and\ \bibinfo {author} {\bibfnamefont {J.}~\bibnamefont
  {Silk}},\ }\href {\doibase 10.1088/1475-7516/2020/01/013} {\bibfield
  {journal} {\bibinfo  {journal} {JCAP}\ }\textbf {\bibinfo {volume} {01}},\
  \bibinfo {pages} {013} (\bibinfo {year} {2020}{\natexlab{c}})},\ \Eprint
  {http://arxiv.org/abs/1908.01391} {arXiv:1908.01391 [astro-ph.CO]}
  \BibitemShut {NoStop}%
\bibitem [{\citenamefont {Efstathiou}\ and\ \citenamefont
  {Gratton}(2020)}]{Efstathiou:2020wem}%
  \BibitemOpen
  \bibfield  {author} {\bibinfo {author} {\bibfnamefont {G.}~\bibnamefont
  {Efstathiou}}\ and\ \bibinfo {author} {\bibfnamefont {S.}~\bibnamefont
  {Gratton}},\ }\href {\doibase 10.1093/mnrasl/slaa093} {\  (\bibinfo {year}
  {2020}),\ 10.1093/mnrasl/slaa093},\ \Eprint {http://arxiv.org/abs/2002.06892}
  {arXiv:2002.06892 [astro-ph.CO]} \BibitemShut {NoStop}%
\bibitem [{\citenamefont {Efstathiou}\ and\ \citenamefont
  {Gratton}(2019)}]{Efstathiou:2019mdh}%
  \BibitemOpen
  \bibfield  {author} {\bibinfo {author} {\bibfnamefont {G.}~\bibnamefont
  {Efstathiou}}\ and\ \bibinfo {author} {\bibfnamefont {S.}~\bibnamefont
  {Gratton}},\ }\href@noop {} {\  (\bibinfo {year} {2019})},\ \Eprint
  {http://arxiv.org/abs/1910.00483} {arXiv:1910.00483 [astro-ph.CO]}
  \BibitemShut {NoStop}%
\bibitem [{\citenamefont {Rosenberg}\ \emph {et~al.}(2022)\citenamefont
  {Rosenberg}, \citenamefont {Gratton},\ and\ \citenamefont
  {Efstathiou}}]{Rosenberg:2022sdy}%
  \BibitemOpen
  \bibfield  {author} {\bibinfo {author} {\bibfnamefont {E.}~\bibnamefont
  {Rosenberg}}, \bibinfo {author} {\bibfnamefont {S.}~\bibnamefont {Gratton}},
  \ and\ \bibinfo {author} {\bibfnamefont {G.}~\bibnamefont {Efstathiou}},\
  }\href@noop {} {\  (\bibinfo {year} {2022})},\ \Eprint
  {http://arxiv.org/abs/2205.10869} {arXiv:2205.10869 [astro-ph.CO]}
  \BibitemShut {NoStop}%
\bibitem [{\citenamefont {Glanville}\ \emph {et~al.}(2022)\citenamefont
  {Glanville}, \citenamefont {Howlett},\ and\ \citenamefont
  {Davis}}]{Glanville:2022xes}%
  \BibitemOpen
  \bibfield  {author} {\bibinfo {author} {\bibfnamefont {A.}~\bibnamefont
  {Glanville}}, \bibinfo {author} {\bibfnamefont {C.}~\bibnamefont {Howlett}},
  \ and\ \bibinfo {author} {\bibfnamefont {T.~M.}\ \bibnamefont {Davis}},\
  }\href@noop {} {\  (\bibinfo {year} {2022})},\ \Eprint
  {http://arxiv.org/abs/2205.05892} {arXiv:2205.05892 [astro-ph.CO]}
  \BibitemShut {NoStop}%
\bibitem [{\citenamefont {Anselmi}\ \emph {et~al.}(2022)\citenamefont
  {Anselmi}, \citenamefont {Carney}, \citenamefont {Giblin}, \citenamefont
  {Kumar}, \citenamefont {Mertens}, \citenamefont {ODwyer}, \citenamefont
  {Starkman},\ and\ \citenamefont {Tian}}]{Anselmi:2022uvj}%
  \BibitemOpen
  \bibfield  {author} {\bibinfo {author} {\bibfnamefont {S.}~\bibnamefont
  {Anselmi}}, \bibinfo {author} {\bibfnamefont {M.~F.}\ \bibnamefont {Carney}},
  \bibinfo {author} {\bibfnamefont {J.~T.}\ \bibnamefont {Giblin}}, \bibinfo
  {author} {\bibfnamefont {S.}~\bibnamefont {Kumar}}, \bibinfo {author}
  {\bibfnamefont {J.~B.}\ \bibnamefont {Mertens}}, \bibinfo {author}
  {\bibfnamefont {M.}~\bibnamefont {ODwyer}}, \bibinfo {author} {\bibfnamefont
  {G.~D.}\ \bibnamefont {Starkman}}, \ and\ \bibinfo {author} {\bibfnamefont
  {C.}~\bibnamefont {Tian}},\ }\href@noop {} {\  (\bibinfo {year} {2022})},\
  \Eprint {http://arxiv.org/abs/2207.06547} {arXiv:2207.06547 [astro-ph.CO]}
  \BibitemShut {NoStop}%
\bibitem [{\citenamefont {Dossett}\ and\ \citenamefont
  {Ishak}(2012)}]{Dossett:2012kd}%
  \BibitemOpen
  \bibfield  {author} {\bibinfo {author} {\bibfnamefont {J.}~\bibnamefont
  {Dossett}}\ and\ \bibinfo {author} {\bibfnamefont {M.}~\bibnamefont
  {Ishak}},\ }\href {\doibase 10.1103/PhysRevD.86.103008} {\bibfield  {journal}
  {\bibinfo  {journal} {Phys. Rev. D}\ }\textbf {\bibinfo {volume} {86}},\
  \bibinfo {pages} {103008} (\bibinfo {year} {2012})},\ \Eprint
  {http://arxiv.org/abs/1205.2422} {arXiv:1205.2422 [astro-ph.CO]} \BibitemShut
  {NoStop}%
\bibitem [{\citenamefont {Abdalla}\ \emph {et~al.}(2022)\citenamefont {Abdalla}
  \emph {et~al.}}]{Abdalla:2022yfr}%
  \BibitemOpen
  \bibfield  {author} {\bibinfo {author} {\bibfnamefont {E.}~\bibnamefont
  {Abdalla}} \emph {et~al.},\ }\href {\doibase 10.1016/j.jheap.2022.04.002}
  {\bibfield  {journal} {\bibinfo  {journal} {JHEAp}\ }\textbf {\bibinfo
  {volume} {34}},\ \bibinfo {pages} {49} (\bibinfo {year} {2022})},\ \Eprint
  {http://arxiv.org/abs/2203.06142} {arXiv:2203.06142 [astro-ph.CO]}
  \BibitemShut {NoStop}%
\bibitem [{\citenamefont {Cao}\ and\ \citenamefont
  {Ratra}(2022)}]{Cao:2022ugh}%
  \BibitemOpen
  \bibfield  {author} {\bibinfo {author} {\bibfnamefont {S.}~\bibnamefont
  {Cao}}\ and\ \bibinfo {author} {\bibfnamefont {B.}~\bibnamefont {Ratra}},\
  }\href {\doibase 10.1093/mnras/stac1184} {\bibfield  {journal} {\bibinfo
  {journal} {Mon. Not. Roy. Astron. Soc.}\ }\textbf {\bibinfo {volume} {513}},\
  \bibinfo {pages} {5686} (\bibinfo {year} {2022})},\ \Eprint
  {http://arxiv.org/abs/2203.10825} {arXiv:2203.10825 [astro-ph.CO]}
  \BibitemShut {NoStop}%
\bibitem [{\citenamefont {Qi}\ \emph {et~al.}(2022)\citenamefont {Qi},
  \citenamefont {Cui}, \citenamefont {Hu}, \citenamefont {Zhang}, \citenamefont
  {Cui},\ and\ \citenamefont {Zhang}}]{Qi:2022sxm}%
  \BibitemOpen
  \bibfield  {author} {\bibinfo {author} {\bibfnamefont {J.-Z.}\ \bibnamefont
  {Qi}}, \bibinfo {author} {\bibfnamefont {Y.}~\bibnamefont {Cui}}, \bibinfo
  {author} {\bibfnamefont {W.-H.}\ \bibnamefont {Hu}}, \bibinfo {author}
  {\bibfnamefont {J.-F.}\ \bibnamefont {Zhang}}, \bibinfo {author}
  {\bibfnamefont {J.-L.}\ \bibnamefont {Cui}}, \ and\ \bibinfo {author}
  {\bibfnamefont {X.}~\bibnamefont {Zhang}},\ }\href {\doibase
  10.1103/PhysRevD.106.023520} {\bibfield  {journal} {\bibinfo  {journal}
  {Phys. Rev. D}\ }\textbf {\bibinfo {volume} {106}},\ \bibinfo {pages}
  {023520} (\bibinfo {year} {2022})},\ \Eprint
  {http://arxiv.org/abs/2202.01396} {arXiv:2202.01396 [astro-ph.CO]}
  \BibitemShut {NoStop}%
\bibitem [{\citenamefont {Cao}\ \emph {et~al.}(2022{\natexlab{a}})\citenamefont
  {Cao}, \citenamefont {Khadka},\ and\ \citenamefont {Ratra}}]{Cao:2021irf}%
  \BibitemOpen
  \bibfield  {author} {\bibinfo {author} {\bibfnamefont {S.}~\bibnamefont
  {Cao}}, \bibinfo {author} {\bibfnamefont {N.}~\bibnamefont {Khadka}}, \ and\
  \bibinfo {author} {\bibfnamefont {B.}~\bibnamefont {Ratra}},\ }\href
  {\doibase 10.1093/mnras/stab3559} {\bibfield  {journal} {\bibinfo  {journal}
  {Mon. Not. Roy. Astron. Soc.}\ }\textbf {\bibinfo {volume} {510}},\ \bibinfo
  {pages} {2928} (\bibinfo {year} {2022}{\natexlab{a}})},\ \Eprint
  {http://arxiv.org/abs/2110.14840} {arXiv:2110.14840 [astro-ph.CO]}
  \BibitemShut {NoStop}%
\bibitem [{\citenamefont {Cao}\ \emph {et~al.}(2022{\natexlab{b}})\citenamefont
  {Cao}, \citenamefont {Ryan},\ and\ \citenamefont {Ratra}}]{Cao:2021cix}%
  \BibitemOpen
  \bibfield  {author} {\bibinfo {author} {\bibfnamefont {S.}~\bibnamefont
  {Cao}}, \bibinfo {author} {\bibfnamefont {J.}~\bibnamefont {Ryan}}, \ and\
  \bibinfo {author} {\bibfnamefont {B.}~\bibnamefont {Ratra}},\ }\href
  {\doibase 10.1093/mnras/stab3304} {\bibfield  {journal} {\bibinfo  {journal}
  {Mon. Not. Roy. Astron. Soc.}\ }\textbf {\bibinfo {volume} {509}},\ \bibinfo
  {pages} {4745} (\bibinfo {year} {2022}{\natexlab{b}})},\ \Eprint
  {http://arxiv.org/abs/2109.01987} {arXiv:2109.01987 [astro-ph.CO]}
  \BibitemShut {NoStop}%
\bibitem [{\citenamefont {Cao}\ \emph {et~al.}(2022{\natexlab{c}})\citenamefont
  {Cao}, \citenamefont {Zheng}, \citenamefont {Qi}, \citenamefont {Zhang},\
  and\ \citenamefont {Zhu}}]{Cao:2021zpf}%
  \BibitemOpen
  \bibfield  {author} {\bibinfo {author} {\bibfnamefont {M.-D.}\ \bibnamefont
  {Cao}}, \bibinfo {author} {\bibfnamefont {J.}~\bibnamefont {Zheng}}, \bibinfo
  {author} {\bibfnamefont {J.-Z.}\ \bibnamefont {Qi}}, \bibinfo {author}
  {\bibfnamefont {X.}~\bibnamefont {Zhang}}, \ and\ \bibinfo {author}
  {\bibfnamefont {Z.-H.}\ \bibnamefont {Zhu}},\ }\href {\doibase
  10.3847/1538-4357/ac7ce4} {\bibfield  {journal} {\bibinfo  {journal}
  {Astrophys. J.}\ }\textbf {\bibinfo {volume} {934}},\ \bibinfo {pages} {108}
  (\bibinfo {year} {2022}{\natexlab{c}})},\ \Eprint
  {http://arxiv.org/abs/2112.14564} {arXiv:2112.14564 [astro-ph.CO]}
  \BibitemShut {NoStop}%
\bibitem [{\citenamefont {Cao}\ \emph {et~al.}(2021)\citenamefont {Cao},
  \citenamefont {Ryan},\ and\ \citenamefont {Ratra}}]{Cao:2021ldv}%
  \BibitemOpen
  \bibfield  {author} {\bibinfo {author} {\bibfnamefont {S.}~\bibnamefont
  {Cao}}, \bibinfo {author} {\bibfnamefont {J.}~\bibnamefont {Ryan}}, \ and\
  \bibinfo {author} {\bibfnamefont {B.}~\bibnamefont {Ratra}},\ }\href
  {\doibase 10.1093/mnras/stab942} {\bibfield  {journal} {\bibinfo  {journal}
  {Mon. Not. Roy. Astron. Soc.}\ }\textbf {\bibinfo {volume} {504}},\ \bibinfo
  {pages} {300} (\bibinfo {year} {2021})},\ \Eprint
  {http://arxiv.org/abs/2101.08817} {arXiv:2101.08817 [astro-ph.CO]}
  \BibitemShut {NoStop}%
\bibitem [{\citenamefont {Qi}\ \emph {et~al.}(2021)\citenamefont {Qi},
  \citenamefont {Zhao}, \citenamefont {Cao}, \citenamefont {Biesiada},\ and\
  \citenamefont {Liu}}]{Qi:2020rmm}%
  \BibitemOpen
  \bibfield  {author} {\bibinfo {author} {\bibfnamefont {J.-Z.}\ \bibnamefont
  {Qi}}, \bibinfo {author} {\bibfnamefont {J.-W.}\ \bibnamefont {Zhao}},
  \bibinfo {author} {\bibfnamefont {S.}~\bibnamefont {Cao}}, \bibinfo {author}
  {\bibfnamefont {M.}~\bibnamefont {Biesiada}}, \ and\ \bibinfo {author}
  {\bibfnamefont {Y.}~\bibnamefont {Liu}},\ }\href {\doibase
  10.1093/mnras/stab638} {\bibfield  {journal} {\bibinfo  {journal} {Mon. Not.
  Roy. Astron. Soc.}\ }\textbf {\bibinfo {volume} {503}},\ \bibinfo {pages}
  {2179} (\bibinfo {year} {2021})},\ \Eprint {http://arxiv.org/abs/2011.00713}
  {arXiv:2011.00713 [astro-ph.CO]} \BibitemShut {NoStop}%
\bibitem [{\citenamefont {Cao}\ \emph {et~al.}(2020)\citenamefont {Cao},
  \citenamefont {Ryan},\ and\ \citenamefont {Ratra}}]{Cao:2020jgu}%
  \BibitemOpen
  \bibfield  {author} {\bibinfo {author} {\bibfnamefont {S.}~\bibnamefont
  {Cao}}, \bibinfo {author} {\bibfnamefont {J.}~\bibnamefont {Ryan}}, \ and\
  \bibinfo {author} {\bibfnamefont {B.}~\bibnamefont {Ratra}},\ }\href
  {\doibase 10.1093/mnras/staa2190} {\bibfield  {journal} {\bibinfo  {journal}
  {Mon. Not. Roy. Astron. Soc.}\ }\textbf {\bibinfo {volume} {497}},\ \bibinfo
  {pages} {3191} (\bibinfo {year} {2020})},\ \Eprint
  {http://arxiv.org/abs/2005.12617} {arXiv:2005.12617 [astro-ph.CO]}
  \BibitemShut {NoStop}%
\bibitem [{\citenamefont {Wang}\ \emph {et~al.}(2020)\citenamefont {Wang},
  \citenamefont {Qi}, \citenamefont {Zhang},\ and\ \citenamefont
  {Zhang}}]{Wang:2019yob}%
  \BibitemOpen
  \bibfield  {author} {\bibinfo {author} {\bibfnamefont {B.}~\bibnamefont
  {Wang}}, \bibinfo {author} {\bibfnamefont {J.-Z.}\ \bibnamefont {Qi}},
  \bibinfo {author} {\bibfnamefont {J.-F.}\ \bibnamefont {Zhang}}, \ and\
  \bibinfo {author} {\bibfnamefont {X.}~\bibnamefont {Zhang}},\ }\href
  {\doibase 10.3847/1538-4357/ab9b22} {\bibfield  {journal} {\bibinfo
  {journal} {Astrophys. J.}\ }\textbf {\bibinfo {volume} {898}},\ \bibinfo
  {pages} {100} (\bibinfo {year} {2020})},\ \Eprint
  {http://arxiv.org/abs/1910.12173} {arXiv:1910.12173 [astro-ph.CO]}
  \BibitemShut {NoStop}%
\bibitem [{\citenamefont {Riess}\ \emph
  {et~al.}(2022{\natexlab{a}})\citenamefont {Riess} \emph
  {et~al.}}]{Riess:2021jrx}%
  \BibitemOpen
  \bibfield  {author} {\bibinfo {author} {\bibfnamefont {A.~G.}\ \bibnamefont
  {Riess}} \emph {et~al.},\ }\href {\doibase 10.3847/2041-8213/ac5c5b}
  {\bibfield  {journal} {\bibinfo  {journal} {Astrophys. J. Lett.}\ }\textbf
  {\bibinfo {volume} {934}},\ \bibinfo {pages} {L7} (\bibinfo {year}
  {2022}{\natexlab{a}})},\ \Eprint {http://arxiv.org/abs/2112.04510}
  {arXiv:2112.04510 [astro-ph.CO]} \BibitemShut {NoStop}%
\bibitem [{\citenamefont {Riess}\ \emph
  {et~al.}(2022{\natexlab{b}})\citenamefont {Riess}, \citenamefont {Breuval},
  \citenamefont {Yuan}, \citenamefont {Casertano}, \citenamefont
  {\textasciitilde{}Macri}, \citenamefont {Scolnic}, \citenamefont
  {Cantat-Gaudin}, \citenamefont {Anderson},\ and\ \citenamefont
  {Reyes}}]{Riess:2022mme}%
  \BibitemOpen
  \bibfield  {author} {\bibinfo {author} {\bibfnamefont {A.~G.}\ \bibnamefont
  {Riess}}, \bibinfo {author} {\bibfnamefont {L.}~\bibnamefont {Breuval}},
  \bibinfo {author} {\bibfnamefont {W.}~\bibnamefont {Yuan}}, \bibinfo {author}
  {\bibfnamefont {S.}~\bibnamefont {Casertano}}, \bibinfo {author}
  {\bibfnamefont {L.~M.}\ \bibnamefont {\textasciitilde{}Macri}}, \bibinfo
  {author} {\bibfnamefont {D.}~\bibnamefont {Scolnic}}, \bibinfo {author}
  {\bibfnamefont {T.}~\bibnamefont {Cantat-Gaudin}}, \bibinfo {author}
  {\bibfnamefont {R.~I.}\ \bibnamefont {Anderson}}, \ and\ \bibinfo {author}
  {\bibfnamefont {M.~C.}\ \bibnamefont {Reyes}},\ }\href@noop {} {\  (\bibinfo
  {year} {2022}{\natexlab{b}})},\ \Eprint {http://arxiv.org/abs/2208.01045}
  {arXiv:2208.01045 [astro-ph.CO]} \BibitemShut {NoStop}%
\bibitem [{\citenamefont {Di~Valentino}\ \emph
  {et~al.}(2020{\natexlab{d}})\citenamefont {Di~Valentino} \emph
  {et~al.}}]{DiValentino:2020zio}%
  \BibitemOpen
  \bibfield  {author} {\bibinfo {author} {\bibfnamefont {E.}~\bibnamefont
  {Di~Valentino}} \emph {et~al.},\ }\href@noop {} {\bibfield  {journal}
  {\bibinfo  {journal} {arXiv:2008.11284}\ } (\bibinfo {year}
  {2020}{\natexlab{d}})},\ \Eprint {http://arxiv.org/abs/2008.11284}
  {arXiv:2008.11284 [astro-ph.CO]} \BibitemShut {NoStop}%
\bibitem [{\citenamefont {Di~Valentino}\ \emph
  {et~al.}(2021{\natexlab{a}})\citenamefont {Di~Valentino}, \citenamefont
  {Mena}, \citenamefont {Pan}, \citenamefont {Visinelli}, \citenamefont {Yang},
  \citenamefont {Melchiorri}, \citenamefont {Mota}, \citenamefont {Riess},\
  and\ \citenamefont {Silk}}]{DiValentino:2021izs}%
  \BibitemOpen
  \bibfield  {author} {\bibinfo {author} {\bibfnamefont {E.}~\bibnamefont
  {Di~Valentino}}, \bibinfo {author} {\bibfnamefont {O.}~\bibnamefont {Mena}},
  \bibinfo {author} {\bibfnamefont {S.}~\bibnamefont {Pan}}, \bibinfo {author}
  {\bibfnamefont {L.}~\bibnamefont {Visinelli}}, \bibinfo {author}
  {\bibfnamefont {W.}~\bibnamefont {Yang}}, \bibinfo {author} {\bibfnamefont
  {A.}~\bibnamefont {Melchiorri}}, \bibinfo {author} {\bibfnamefont {D.~F.}\
  \bibnamefont {Mota}}, \bibinfo {author} {\bibfnamefont {A.~G.}\ \bibnamefont
  {Riess}}, \ and\ \bibinfo {author} {\bibfnamefont {J.}~\bibnamefont {Silk}},\
  }\href {\doibase 10.1088/1361-6382/ac086d} {\bibfield  {journal} {\bibinfo
  {journal} {Class. Quant. Grav.}\ }\textbf {\bibinfo {volume} {38}},\ \bibinfo
  {pages} {153001} (\bibinfo {year} {2021}{\natexlab{a}})},\ \Eprint
  {http://arxiv.org/abs/2103.01183} {arXiv:2103.01183 [astro-ph.CO]}
  \BibitemShut {NoStop}%
\bibitem [{\citenamefont {Di~Valentino}\ \emph
  {et~al.}(2017{\natexlab{a}})\citenamefont {Di~Valentino}, \citenamefont
  {Melchiorri},\ and\ \citenamefont {Mena}}]{DiValentino:2017iww}%
  \BibitemOpen
  \bibfield  {author} {\bibinfo {author} {\bibfnamefont {E.}~\bibnamefont
  {Di~Valentino}}, \bibinfo {author} {\bibfnamefont {A.}~\bibnamefont
  {Melchiorri}}, \ and\ \bibinfo {author} {\bibfnamefont {O.}~\bibnamefont
  {Mena}},\ }\href {\doibase 10.1103/PhysRevD.96.043503} {\bibfield  {journal}
  {\bibinfo  {journal} {Phys. Rev. D}\ }\textbf {\bibinfo {volume} {96}},\
  \bibinfo {pages} {043503} (\bibinfo {year} {2017}{\natexlab{a}})},\ \Eprint
  {http://arxiv.org/abs/1704.08342} {arXiv:1704.08342 [astro-ph.CO]}
  \BibitemShut {NoStop}%
\bibitem [{\citenamefont {Kumar}\ and\ \citenamefont
  {Nunes}(2017)}]{Kumar:2017dnp}%
  \BibitemOpen
  \bibfield  {author} {\bibinfo {author} {\bibfnamefont {S.}~\bibnamefont
  {Kumar}}\ and\ \bibinfo {author} {\bibfnamefont {R.~C.}\ \bibnamefont
  {Nunes}},\ }\href {\doibase 10.1103/PhysRevD.96.103511} {\bibfield  {journal}
  {\bibinfo  {journal} {Phys. Rev.}\ }\textbf {\bibinfo {volume} {D96}},\
  \bibinfo {pages} {103511} (\bibinfo {year} {2017})},\ \Eprint
  {http://arxiv.org/abs/1702.02143} {arXiv:1702.02143 [astro-ph.CO]}
  \BibitemShut {NoStop}%
%%CITATION = ARXIV:1702.02143;%%
\bibitem [{\citenamefont {Verde}\ \emph {et~al.}(2019)\citenamefont {Verde},
  \citenamefont {Treu},\ and\ \citenamefont {Riess}}]{Verde:2019ivm}%
  \BibitemOpen
  \bibfield  {author} {\bibinfo {author} {\bibfnamefont {L.}~\bibnamefont
  {Verde}}, \bibinfo {author} {\bibfnamefont {T.}~\bibnamefont {Treu}}, \ and\
  \bibinfo {author} {\bibfnamefont {A.~G.}\ \bibnamefont {Riess}},\ }in\ \href
  {\doibase 10.1038/s41550-019-0902-0} {\emph {\bibinfo {booktitle} {{Nature
  Astronomy 2019}}}}\ (\bibinfo {year} {2019})\ \Eprint
  {http://arxiv.org/abs/1907.10625} {arXiv:1907.10625 [astro-ph.CO]}
  \BibitemShut {NoStop}%
%%CITATION = ARXIV:1907.10625;%%
\bibitem [{\citenamefont {Knox}\ and\ \citenamefont
  {Millea}(2020)}]{Knox:2019rjx}%
  \BibitemOpen
  \bibfield  {author} {\bibinfo {author} {\bibfnamefont {L.}~\bibnamefont
  {Knox}}\ and\ \bibinfo {author} {\bibfnamefont {M.}~\bibnamefont {Millea}},\
  }\href {\doibase 10.1103/PhysRevD.101.043533} {\bibfield  {journal} {\bibinfo
   {journal} {Phys. Rev. D}\ }\textbf {\bibinfo {volume} {101}},\ \bibinfo
  {pages} {043533} (\bibinfo {year} {2020})},\ \Eprint
  {http://arxiv.org/abs/1908.03663} {arXiv:1908.03663 [astro-ph.CO]}
  \BibitemShut {NoStop}%
\bibitem [{\citenamefont {Jedamzik}\ \emph {et~al.}(2021)\citenamefont
  {Jedamzik}, \citenamefont {Pogosian},\ and\ \citenamefont
  {Zhao}}]{Jedamzik:2020zmd}%
  \BibitemOpen
  \bibfield  {author} {\bibinfo {author} {\bibfnamefont {K.}~\bibnamefont
  {Jedamzik}}, \bibinfo {author} {\bibfnamefont {L.}~\bibnamefont {Pogosian}},
  \ and\ \bibinfo {author} {\bibfnamefont {G.-B.}\ \bibnamefont {Zhao}},\
  }\href {\doibase 10.1038/s42005-021-00628-x} {\bibfield  {journal} {\bibinfo
  {journal} {Commun. in Phys.}\ }\textbf {\bibinfo {volume} {4}},\ \bibinfo
  {pages} {123} (\bibinfo {year} {2021})},\ \Eprint
  {http://arxiv.org/abs/2010.04158} {arXiv:2010.04158 [astro-ph.CO]}
  \BibitemShut {NoStop}%
\bibitem [{\citenamefont {Di~Valentino}\ \emph {et~al.}(2018)\citenamefont
  {Di~Valentino}, \citenamefont {B{\o}ehm}, \citenamefont {Hivon},\ and\
  \citenamefont {Bouchet}}]{DiValentino:2017oaw}%
  \BibitemOpen
  \bibfield  {author} {\bibinfo {author} {\bibfnamefont {E.}~\bibnamefont
  {Di~Valentino}}, \bibinfo {author} {\bibfnamefont {C.}~\bibnamefont
  {B{\o}ehm}}, \bibinfo {author} {\bibfnamefont {E.}~\bibnamefont {Hivon}}, \
  and\ \bibinfo {author} {\bibfnamefont {F.~c.~R.}\ \bibnamefont {Bouchet}},\
  }\href {\doibase 10.1103/PhysRevD.97.043513} {\bibfield  {journal} {\bibinfo
  {journal} {Phys. Rev. D}\ }\textbf {\bibinfo {volume} {97}},\ \bibinfo
  {pages} {043513} (\bibinfo {year} {2018})},\ \Eprint
  {http://arxiv.org/abs/1710.02559} {arXiv:1710.02559 [astro-ph.CO]}
  \BibitemShut {NoStop}%
\bibitem [{\citenamefont {Yang}\ \emph
  {et~al.}(2018{\natexlab{a}})\citenamefont {Yang}, \citenamefont {Pan},
  \citenamefont {Di~Valentino}, \citenamefont {Nunes}, \citenamefont
  {Vagnozzi},\ and\ \citenamefont {Mota}}]{Yang:2018euj}%
  \BibitemOpen
  \bibfield  {author} {\bibinfo {author} {\bibfnamefont {W.}~\bibnamefont
  {Yang}}, \bibinfo {author} {\bibfnamefont {S.}~\bibnamefont {Pan}}, \bibinfo
  {author} {\bibfnamefont {E.}~\bibnamefont {Di~Valentino}}, \bibinfo {author}
  {\bibfnamefont {R.~C.}\ \bibnamefont {Nunes}}, \bibinfo {author}
  {\bibfnamefont {S.}~\bibnamefont {Vagnozzi}}, \ and\ \bibinfo {author}
  {\bibfnamefont {D.~F.}\ \bibnamefont {Mota}},\ }\href {\doibase
  10.1088/1475-7516/2018/09/019} {\bibfield  {journal} {\bibinfo  {journal}
  {JCAP}\ }\textbf {\bibinfo {volume} {09}},\ \bibinfo {pages} {019} (\bibinfo
  {year} {2018}{\natexlab{a}})},\ \Eprint {http://arxiv.org/abs/1805.08252}
  {arXiv:1805.08252 [astro-ph.CO]} \BibitemShut {NoStop}%
\bibitem [{\citenamefont {Yang}\ \emph
  {et~al.}(2018{\natexlab{b}})\citenamefont {Yang}, \citenamefont {Mukherjee},
  \citenamefont {Di~Valentino},\ and\ \citenamefont {Pan}}]{Yang:2018uae}%
  \BibitemOpen
  \bibfield  {author} {\bibinfo {author} {\bibfnamefont {W.}~\bibnamefont
  {Yang}}, \bibinfo {author} {\bibfnamefont {A.}~\bibnamefont {Mukherjee}},
  \bibinfo {author} {\bibfnamefont {E.}~\bibnamefont {Di~Valentino}}, \ and\
  \bibinfo {author} {\bibfnamefont {S.}~\bibnamefont {Pan}},\ }\href {\doibase
  10.1103/PhysRevD.98.123527} {\bibfield  {journal} {\bibinfo  {journal} {Phys.
  Rev.}\ }\textbf {\bibinfo {volume} {D98}},\ \bibinfo {pages} {123527}
  (\bibinfo {year} {2018}{\natexlab{b}})},\ \Eprint
  {http://arxiv.org/abs/1809.06883} {arXiv:1809.06883 [astro-ph.CO]}
  \BibitemShut {NoStop}%
%%CITATION = ARXIV:1809.06883;%%
\bibitem [{\citenamefont {Pan}\ \emph {et~al.}(2019{\natexlab{a}})\citenamefont
  {Pan}, \citenamefont {Yang}, \citenamefont {Singha},\ and\ \citenamefont
  {Saridakis}}]{Pan:2019jqh}%
  \BibitemOpen
  \bibfield  {author} {\bibinfo {author} {\bibfnamefont {S.}~\bibnamefont
  {Pan}}, \bibinfo {author} {\bibfnamefont {W.}~\bibnamefont {Yang}}, \bibinfo
  {author} {\bibfnamefont {C.}~\bibnamefont {Singha}}, \ and\ \bibinfo {author}
  {\bibfnamefont {E.~N.}\ \bibnamefont {Saridakis}},\ }\href {\doibase
  10.1103/PhysRevD.100.083539} {\bibfield  {journal} {\bibinfo  {journal}
  {Phys. Rev.}\ }\textbf {\bibinfo {volume} {D100}},\ \bibinfo {pages} {083539}
  (\bibinfo {year} {2019}{\natexlab{a}})},\ \Eprint
  {http://arxiv.org/abs/1903.10969} {arXiv:1903.10969 [astro-ph.CO]}
  \BibitemShut {NoStop}%
%%CITATION = ARXIV:1903.10969;%%
\bibitem [{\citenamefont {Yang}\ \emph
  {et~al.}(2020{\natexlab{a}})\citenamefont {Yang}, \citenamefont {Pan},
  \citenamefont {Nunes},\ and\ \citenamefont {Mota}}]{Yang:2019uog}%
  \BibitemOpen
  \bibfield  {author} {\bibinfo {author} {\bibfnamefont {W.}~\bibnamefont
  {Yang}}, \bibinfo {author} {\bibfnamefont {S.}~\bibnamefont {Pan}}, \bibinfo
  {author} {\bibfnamefont {R.~C.}\ \bibnamefont {Nunes}}, \ and\ \bibinfo
  {author} {\bibfnamefont {D.~F.}\ \bibnamefont {Mota}},\ }\href {\doibase
  10.1088/1475-7516/2020/04/008} {\bibfield  {journal} {\bibinfo  {journal}
  {JCAP}\ }\textbf {\bibinfo {volume} {04}},\ \bibinfo {pages} {008} (\bibinfo
  {year} {2020}{\natexlab{a}})},\ \Eprint {http://arxiv.org/abs/1910.08821}
  {arXiv:1910.08821 [astro-ph.CO]} \BibitemShut {NoStop}%
\bibitem [{\citenamefont {Pan}\ \emph {et~al.}(2019{\natexlab{b}})\citenamefont
  {Pan}, \citenamefont {Yang}, \citenamefont {Di~Valentino}, \citenamefont
  {Saridakis},\ and\ \citenamefont {Chakraborty}}]{Pan:2019gop}%
  \BibitemOpen
  \bibfield  {author} {\bibinfo {author} {\bibfnamefont {S.}~\bibnamefont
  {Pan}}, \bibinfo {author} {\bibfnamefont {W.}~\bibnamefont {Yang}}, \bibinfo
  {author} {\bibfnamefont {E.}~\bibnamefont {Di~Valentino}}, \bibinfo {author}
  {\bibfnamefont {E.~N.}\ \bibnamefont {Saridakis}}, \ and\ \bibinfo {author}
  {\bibfnamefont {S.}~\bibnamefont {Chakraborty}},\ }\href {\doibase
  10.1103/PhysRevD.100.103520} {\bibfield  {journal} {\bibinfo  {journal}
  {Phys. Rev. D}\ }\textbf {\bibinfo {volume} {100}},\ \bibinfo {pages}
  {103520} (\bibinfo {year} {2019}{\natexlab{b}})},\ \Eprint
  {http://arxiv.org/abs/1907.07540} {arXiv:1907.07540 [astro-ph.CO]}
  \BibitemShut {NoStop}%
\bibitem [{\citenamefont {Poulin}\ \emph {et~al.}(2019)\citenamefont {Poulin},
  \citenamefont {Smith}, \citenamefont {Karwal},\ and\ \citenamefont
  {Kamionkowski}}]{Poulin:2018cxd}%
  \BibitemOpen
  \bibfield  {author} {\bibinfo {author} {\bibfnamefont {V.}~\bibnamefont
  {Poulin}}, \bibinfo {author} {\bibfnamefont {T.~L.}\ \bibnamefont {Smith}},
  \bibinfo {author} {\bibfnamefont {T.}~\bibnamefont {Karwal}}, \ and\ \bibinfo
  {author} {\bibfnamefont {M.}~\bibnamefont {Kamionkowski}},\ }\href {\doibase
  10.1103/PhysRevLett.122.221301} {\bibfield  {journal} {\bibinfo  {journal}
  {Phys. Rev. Lett.}\ }\textbf {\bibinfo {volume} {122}},\ \bibinfo {pages}
  {221301} (\bibinfo {year} {2019})},\ \Eprint
  {http://arxiv.org/abs/1811.04083} {arXiv:1811.04083 [astro-ph.CO]}
  \BibitemShut {NoStop}%
\bibitem [{\citenamefont {Yang}\ \emph
  {et~al.}(2019{\natexlab{a}})\citenamefont {Yang}, \citenamefont {Pan},
  \citenamefont {Di~Valentino}, \citenamefont {Saridakis},\ and\ \citenamefont
  {Chakraborty}}]{Yang:2018qmz}%
  \BibitemOpen
  \bibfield  {author} {\bibinfo {author} {\bibfnamefont {W.}~\bibnamefont
  {Yang}}, \bibinfo {author} {\bibfnamefont {S.}~\bibnamefont {Pan}}, \bibinfo
  {author} {\bibfnamefont {E.}~\bibnamefont {Di~Valentino}}, \bibinfo {author}
  {\bibfnamefont {E.~N.}\ \bibnamefont {Saridakis}}, \ and\ \bibinfo {author}
  {\bibfnamefont {S.}~\bibnamefont {Chakraborty}},\ }\href {\doibase
  10.1103/PhysRevD.99.043543} {\bibfield  {journal} {\bibinfo  {journal} {Phys.
  Rev.}\ }\textbf {\bibinfo {volume} {D99}},\ \bibinfo {pages} {043543}
  (\bibinfo {year} {2019}{\natexlab{a}})},\ \Eprint
  {http://arxiv.org/abs/1810.05141} {arXiv:1810.05141 [astro-ph.CO]}
  \BibitemShut {NoStop}%
%%CITATION = ARXIV:1810.05141;%%
\bibitem [{\citenamefont {Pan}\ \emph {et~al.}(2020{\natexlab{a}})\citenamefont
  {Pan}, \citenamefont {Yang},\ and\ \citenamefont
  {Paliathanasis}}]{Pan:2020bur}%
  \BibitemOpen
  \bibfield  {author} {\bibinfo {author} {\bibfnamefont {S.}~\bibnamefont
  {Pan}}, \bibinfo {author} {\bibfnamefont {W.}~\bibnamefont {Yang}}, \ and\
  \bibinfo {author} {\bibfnamefont {A.}~\bibnamefont {Paliathanasis}},\ }\href
  {\doibase 10.1093/mnras/staa213} {\bibfield  {journal} {\bibinfo  {journal}
  {Mon. Not. Roy. Astron. Soc.}\ }\textbf {\bibinfo {volume} {493}},\ \bibinfo
  {pages} {3114} (\bibinfo {year} {2020}{\natexlab{a}})},\ \Eprint
  {http://arxiv.org/abs/2002.03408} {arXiv:2002.03408 [astro-ph.CO]}
  \BibitemShut {NoStop}%
\bibitem [{\citenamefont {Di~Valentino}\ \emph
  {et~al.}(2020{\natexlab{e}})\citenamefont {Di~Valentino}, \citenamefont
  {Melchiorri}, \citenamefont {Mena},\ and\ \citenamefont
  {Vagnozzi}}]{DiValentino:2019ffd}%
  \BibitemOpen
  \bibfield  {author} {\bibinfo {author} {\bibfnamefont {E.}~\bibnamefont
  {Di~Valentino}}, \bibinfo {author} {\bibfnamefont {A.}~\bibnamefont
  {Melchiorri}}, \bibinfo {author} {\bibfnamefont {O.}~\bibnamefont {Mena}}, \
  and\ \bibinfo {author} {\bibfnamefont {S.}~\bibnamefont {Vagnozzi}},\ }\href
  {\doibase 10.1016/j.dark.2020.100666} {\bibfield  {journal} {\bibinfo
  {journal} {Phys. Dark Univ.}\ }\textbf {\bibinfo {volume} {30}},\ \bibinfo
  {pages} {100666} (\bibinfo {year} {2020}{\natexlab{e}})},\ \Eprint
  {http://arxiv.org/abs/1908.04281} {arXiv:1908.04281 [astro-ph.CO]}
  \BibitemShut {NoStop}%
\bibitem [{\citenamefont {Di~Valentino}\ \emph
  {et~al.}(2020{\natexlab{f}})\citenamefont {Di~Valentino}, \citenamefont
  {Melchiorri}, \citenamefont {Mena},\ and\ \citenamefont
  {Vagnozzi}}]{DiValentino:2019jae}%
  \BibitemOpen
  \bibfield  {author} {\bibinfo {author} {\bibfnamefont {E.}~\bibnamefont
  {Di~Valentino}}, \bibinfo {author} {\bibfnamefont {A.}~\bibnamefont
  {Melchiorri}}, \bibinfo {author} {\bibfnamefont {O.}~\bibnamefont {Mena}}, \
  and\ \bibinfo {author} {\bibfnamefont {S.}~\bibnamefont {Vagnozzi}},\ }\href
  {\doibase 10.1103/PhysRevD.101.063502} {\bibfield  {journal} {\bibinfo
  {journal} {Phys. Rev. D}\ }\textbf {\bibinfo {volume} {101}},\ \bibinfo
  {pages} {063502} (\bibinfo {year} {2020}{\natexlab{f}})},\ \Eprint
  {http://arxiv.org/abs/1910.09853} {arXiv:1910.09853 [astro-ph.CO]}
  \BibitemShut {NoStop}%
\bibitem [{\citenamefont {Yao}\ and\ \citenamefont {Meng}(2020)}]{Yao:2020pji}%
  \BibitemOpen
  \bibfield  {author} {\bibinfo {author} {\bibfnamefont {Y.}~\bibnamefont
  {Yao}}\ and\ \bibinfo {author} {\bibfnamefont {X.}~\bibnamefont {Meng}},\
  }\href@noop {} {\  (\bibinfo {year} {2020})},\ \Eprint
  {http://arxiv.org/abs/2011.09160} {arXiv:2011.09160 [astro-ph.CO]}
  \BibitemShut {NoStop}%
\bibitem [{\citenamefont {Lucca}\ and\ \citenamefont
  {Hooper}(2020)}]{Lucca:2020zjb}%
  \BibitemOpen
  \bibfield  {author} {\bibinfo {author} {\bibfnamefont {M.}~\bibnamefont
  {Lucca}}\ and\ \bibinfo {author} {\bibfnamefont {D.~C.}\ \bibnamefont
  {Hooper}},\ }\href {\doibase 10.1103/PhysRevD.102.123502} {\bibfield
  {journal} {\bibinfo  {journal} {Phys. Rev. D}\ }\textbf {\bibinfo {volume}
  {102}},\ \bibinfo {pages} {123502} (\bibinfo {year} {2020})},\ \Eprint
  {http://arxiv.org/abs/2002.06127} {arXiv:2002.06127 [astro-ph.CO]}
  \BibitemShut {NoStop}%
\bibitem [{\citenamefont {Blinov}\ \emph {et~al.}(2020)\citenamefont {Blinov},
  \citenamefont {Keith},\ and\ \citenamefont {Hooper}}]{Blinov:2020uvz}%
  \BibitemOpen
  \bibfield  {author} {\bibinfo {author} {\bibfnamefont {N.}~\bibnamefont
  {Blinov}}, \bibinfo {author} {\bibfnamefont {C.}~\bibnamefont {Keith}}, \
  and\ \bibinfo {author} {\bibfnamefont {D.}~\bibnamefont {Hooper}},\ }\href
  {\doibase 10.1088/1475-7516/2020/06/005} {\bibfield  {journal} {\bibinfo
  {journal} {JCAP}\ }\textbf {\bibinfo {volume} {06}},\ \bibinfo {pages} {005}
  (\bibinfo {year} {2020})},\ \Eprint {http://arxiv.org/abs/2004.06114}
  {arXiv:2004.06114 [astro-ph.CO]} \BibitemShut {NoStop}%
\bibitem [{\citenamefont {Anchordoqui}\ \emph {et~al.}(2021)\citenamefont
  {Anchordoqui}, \citenamefont {Di~Valentino}, \citenamefont {Pan},\ and\
  \citenamefont {Yang}}]{Anchordoqui:2021gji}%
  \BibitemOpen
  \bibfield  {author} {\bibinfo {author} {\bibfnamefont {L.~A.}\ \bibnamefont
  {Anchordoqui}}, \bibinfo {author} {\bibfnamefont {E.}~\bibnamefont
  {Di~Valentino}}, \bibinfo {author} {\bibfnamefont {S.}~\bibnamefont {Pan}}, \
  and\ \bibinfo {author} {\bibfnamefont {W.}~\bibnamefont {Yang}},\ }\href
  {\doibase 10.1016/j.jheap.2021.08.001} {\bibfield  {journal} {\bibinfo
  {journal} {JHEAp}\ }\textbf {\bibinfo {volume} {32}},\ \bibinfo {pages} {28}
  (\bibinfo {year} {2021})},\ \Eprint {http://arxiv.org/abs/2107.13932}
  {arXiv:2107.13932 [astro-ph.CO]} \BibitemShut {NoStop}%
\bibitem [{\citenamefont {Karwal}\ \emph {et~al.}(2022)\citenamefont {Karwal},
  \citenamefont {Raveri}, \citenamefont {Jain}, \citenamefont {Khoury},\ and\
  \citenamefont {Trodden}}]{Karwal:2021vpk}%
  \BibitemOpen
  \bibfield  {author} {\bibinfo {author} {\bibfnamefont {T.}~\bibnamefont
  {Karwal}}, \bibinfo {author} {\bibfnamefont {M.}~\bibnamefont {Raveri}},
  \bibinfo {author} {\bibfnamefont {B.}~\bibnamefont {Jain}}, \bibinfo {author}
  {\bibfnamefont {J.}~\bibnamefont {Khoury}}, \ and\ \bibinfo {author}
  {\bibfnamefont {M.}~\bibnamefont {Trodden}},\ }\href {\doibase
  10.1103/PhysRevD.105.063535} {\bibfield  {journal} {\bibinfo  {journal}
  {Phys. Rev. D}\ }\textbf {\bibinfo {volume} {105}},\ \bibinfo {pages}
  {063535} (\bibinfo {year} {2022})},\ \Eprint
  {http://arxiv.org/abs/2106.13290} {arXiv:2106.13290 [astro-ph.CO]}
  \BibitemShut {NoStop}%
\bibitem [{\citenamefont {Freese}\ and\ \citenamefont
  {Winkler}(2021)}]{Freese:2021rjq}%
  \BibitemOpen
  \bibfield  {author} {\bibinfo {author} {\bibfnamefont {K.}~\bibnamefont
  {Freese}}\ and\ \bibinfo {author} {\bibfnamefont {M.~W.}\ \bibnamefont
  {Winkler}},\ }\href {\doibase 10.1103/PhysRevD.104.083533} {\bibfield
  {journal} {\bibinfo  {journal} {Phys. Rev. D}\ }\textbf {\bibinfo {volume}
  {104}},\ \bibinfo {pages} {083533} (\bibinfo {year} {2021})},\ \Eprint
  {http://arxiv.org/abs/2102.13655} {arXiv:2102.13655 [astro-ph.CO]}
  \BibitemShut {NoStop}%
\bibitem [{\citenamefont {Perivolaropoulos}\ and\ \citenamefont
  {Skara}(2022)}]{Perivolaropoulos:2021jda}%
  \BibitemOpen
  \bibfield  {author} {\bibinfo {author} {\bibfnamefont {L.}~\bibnamefont
  {Perivolaropoulos}}\ and\ \bibinfo {author} {\bibfnamefont {F.}~\bibnamefont
  {Skara}},\ }\href {\doibase 10.1016/j.newar.2022.101659} {\bibfield
  {journal} {\bibinfo  {journal} {New Astron. Rev.}\ }\textbf {\bibinfo
  {volume} {95}},\ \bibinfo {pages} {101659} (\bibinfo {year} {2022})},\
  \Eprint {http://arxiv.org/abs/2105.05208} {arXiv:2105.05208 [astro-ph.CO]}
  \BibitemShut {NoStop}%
\bibitem [{\citenamefont {Sch\"oneberg}\ \emph {et~al.}(2022)\citenamefont
  {Sch\"oneberg}, \citenamefont {Franco~Abell\'an}, \citenamefont
  {P\'erez~S\'anchez}, \citenamefont {Witte}, \citenamefont {Poulin},\ and\
  \citenamefont {Lesgourgues}}]{Schoneberg:2021qvd}%
  \BibitemOpen
  \bibfield  {author} {\bibinfo {author} {\bibfnamefont {N.}~\bibnamefont
  {Sch\"oneberg}}, \bibinfo {author} {\bibfnamefont {G.}~\bibnamefont
  {Franco~Abell\'an}}, \bibinfo {author} {\bibfnamefont {A.}~\bibnamefont
  {P\'erez~S\'anchez}}, \bibinfo {author} {\bibfnamefont {S.~J.}\ \bibnamefont
  {Witte}}, \bibinfo {author} {\bibfnamefont {V.}~\bibnamefont {Poulin}}, \
  and\ \bibinfo {author} {\bibfnamefont {J.}~\bibnamefont {Lesgourgues}},\
  }\href {\doibase 10.1016/j.physrep.2022.07.001} {\bibfield  {journal}
  {\bibinfo  {journal} {Phys. Rept.}\ }\textbf {\bibinfo {volume} {984}},\
  \bibinfo {pages} {1} (\bibinfo {year} {2022})},\ \Eprint
  {http://arxiv.org/abs/2107.10291} {arXiv:2107.10291 [astro-ph.CO]}
  \BibitemShut {NoStop}%
\bibitem [{\citenamefont {Reeves}\ \emph {et~al.}(2022)\citenamefont {Reeves},
  \citenamefont {Herold}, \citenamefont {Vagnozzi}, \citenamefont {Sherwin},\
  and\ \citenamefont {Ferreira}}]{Reeves:2022aoi}%
  \BibitemOpen
  \bibfield  {author} {\bibinfo {author} {\bibfnamefont {A.}~\bibnamefont
  {Reeves}}, \bibinfo {author} {\bibfnamefont {L.}~\bibnamefont {Herold}},
  \bibinfo {author} {\bibfnamefont {S.}~\bibnamefont {Vagnozzi}}, \bibinfo
  {author} {\bibfnamefont {B.~D.}\ \bibnamefont {Sherwin}}, \ and\ \bibinfo
  {author} {\bibfnamefont {E.~G.~M.}\ \bibnamefont {Ferreira}},\ }\href@noop {}
  {\  (\bibinfo {year} {2022})},\ \Eprint {http://arxiv.org/abs/2207.01501}
  {arXiv:2207.01501 [astro-ph.CO]} \BibitemShut {NoStop}%
\bibitem [{\citenamefont {Colg\'ain}\ \emph
  {et~al.}(2022{\natexlab{a}})\citenamefont {Colg\'ain}, \citenamefont
  {Sheikh-Jabbari}, \citenamefont {Solomon}, \citenamefont {Bargiacchi},
  \citenamefont {Capozziello}, \citenamefont {Dainotti},\ and\ \citenamefont
  {Stojkovic}}]{Colgain:2022nlb}%
  \BibitemOpen
  \bibfield  {author} {\bibinfo {author} {\bibfnamefont {E.~O.}\ \bibnamefont
  {Colg\'ain}}, \bibinfo {author} {\bibfnamefont {M.~M.}\ \bibnamefont
  {Sheikh-Jabbari}}, \bibinfo {author} {\bibfnamefont {R.}~\bibnamefont
  {Solomon}}, \bibinfo {author} {\bibfnamefont {G.}~\bibnamefont {Bargiacchi}},
  \bibinfo {author} {\bibfnamefont {S.}~\bibnamefont {Capozziello}}, \bibinfo
  {author} {\bibfnamefont {M.~G.}\ \bibnamefont {Dainotti}}, \ and\ \bibinfo
  {author} {\bibfnamefont {D.}~\bibnamefont {Stojkovic}},\ }\href {\doibase
  10.1103/PhysRevD.106.L041301} {\bibfield  {journal} {\bibinfo  {journal}
  {Phys. Rev. D}\ }\textbf {\bibinfo {volume} {106}},\ \bibinfo {pages}
  {L041301} (\bibinfo {year} {2022}{\natexlab{a}})},\ \Eprint
  {http://arxiv.org/abs/2203.10558} {arXiv:2203.10558 [astro-ph.CO]}
  \BibitemShut {NoStop}%
\bibitem [{\citenamefont {Naidoo}\ \emph {et~al.}(2022)\citenamefont {Naidoo},
  \citenamefont {Jaber}, \citenamefont {Hellwing},\ and\ \citenamefont
  {Bilicki}}]{Naidoo:2022rda}%
  \BibitemOpen
  \bibfield  {author} {\bibinfo {author} {\bibfnamefont {K.}~\bibnamefont
  {Naidoo}}, \bibinfo {author} {\bibfnamefont {M.}~\bibnamefont {Jaber}},
  \bibinfo {author} {\bibfnamefont {W.~A.}\ \bibnamefont {Hellwing}}, \ and\
  \bibinfo {author} {\bibfnamefont {M.}~\bibnamefont {Bilicki}},\ }\href@noop
  {} {\  (\bibinfo {year} {2022})},\ \Eprint {http://arxiv.org/abs/2209.08102}
  {arXiv:2209.08102 [astro-ph.CO]} \BibitemShut {NoStop}%
\bibitem [{\citenamefont {Cruz}\ \emph {et~al.}(2022)\citenamefont {Cruz},
  \citenamefont {Niedermann},\ and\ \citenamefont {Sloth}}]{Cruz:2022oqk}%
  \BibitemOpen
  \bibfield  {author} {\bibinfo {author} {\bibfnamefont {J.~S.}\ \bibnamefont
  {Cruz}}, \bibinfo {author} {\bibfnamefont {F.}~\bibnamefont {Niedermann}}, \
  and\ \bibinfo {author} {\bibfnamefont {M.~S.}\ \bibnamefont {Sloth}},\
  }\href@noop {} {\  (\bibinfo {year} {2022})},\ \Eprint
  {http://arxiv.org/abs/2209.02708} {arXiv:2209.02708 [astro-ph.CO]}
  \BibitemShut {NoStop}%
\bibitem [{\citenamefont {Escudero}\ \emph {et~al.}(2022)\citenamefont
  {Escudero}, \citenamefont {Kuo}, \citenamefont {Keeley},\ and\ \citenamefont
  {Abazajian}}]{Escudero:2022rbq}%
  \BibitemOpen
  \bibfield  {author} {\bibinfo {author} {\bibfnamefont {H.~G.}\ \bibnamefont
  {Escudero}}, \bibinfo {author} {\bibfnamefont {J.-L.}\ \bibnamefont {Kuo}},
  \bibinfo {author} {\bibfnamefont {R.~E.}\ \bibnamefont {Keeley}}, \ and\
  \bibinfo {author} {\bibfnamefont {K.~N.}\ \bibnamefont {Abazajian}},\
  }\href@noop {} {\  (\bibinfo {year} {2022})},\ \Eprint
  {http://arxiv.org/abs/2208.14435} {arXiv:2208.14435 [astro-ph.CO]}
  \BibitemShut {NoStop}%
\bibitem [{\citenamefont {G\'omez-Valent}\ \emph {et~al.}(2022)\citenamefont
  {G\'omez-Valent}, \citenamefont {Zheng}, \citenamefont {Amendola},
  \citenamefont {Wetterich},\ and\ \citenamefont
  {Pettorino}}]{Gomez-Valent:2022bku}%
  \BibitemOpen
  \bibfield  {author} {\bibinfo {author} {\bibfnamefont {A.}~\bibnamefont
  {G\'omez-Valent}}, \bibinfo {author} {\bibfnamefont {Z.}~\bibnamefont
  {Zheng}}, \bibinfo {author} {\bibfnamefont {L.}~\bibnamefont {Amendola}},
  \bibinfo {author} {\bibfnamefont {C.}~\bibnamefont {Wetterich}}, \ and\
  \bibinfo {author} {\bibfnamefont {V.}~\bibnamefont {Pettorino}},\ }\href@noop
  {} {\  (\bibinfo {year} {2022})},\ \Eprint {http://arxiv.org/abs/2207.14487}
  {arXiv:2207.14487 [astro-ph.CO]} \BibitemShut {NoStop}%
\bibitem [{\citenamefont {Colg\'ain}\ \emph
  {et~al.}(2022{\natexlab{b}})\citenamefont {Colg\'ain}, \citenamefont
  {Sheikh-Jabbari}, \citenamefont {Solomon}, \citenamefont {Dainotti},\ and\
  \citenamefont {Stojkovic}}]{Colgain:2022rxy}%
  \BibitemOpen
  \bibfield  {author} {\bibinfo {author} {\bibfnamefont {E.~O.}\ \bibnamefont
  {Colg\'ain}}, \bibinfo {author} {\bibfnamefont {M.~M.}\ \bibnamefont
  {Sheikh-Jabbari}}, \bibinfo {author} {\bibfnamefont {R.}~\bibnamefont
  {Solomon}}, \bibinfo {author} {\bibfnamefont {M.~G.}\ \bibnamefont
  {Dainotti}}, \ and\ \bibinfo {author} {\bibfnamefont {D.}~\bibnamefont
  {Stojkovic}},\ }\href@noop {} {\  (\bibinfo {year} {2022}{\natexlab{b}})},\
  \Eprint {http://arxiv.org/abs/2206.11447} {arXiv:2206.11447 [astro-ph.CO]}
  \BibitemShut {NoStop}%
\bibitem [{\citenamefont {Yang}\ \emph
  {et~al.}(2019{\natexlab{b}})\citenamefont {Yang}, \citenamefont {Pan},
  \citenamefont {Paliathanasis}, \citenamefont {Ghosh},\ and\ \citenamefont
  {Wu}}]{Yang:2019jwn}%
  \BibitemOpen
  \bibfield  {author} {\bibinfo {author} {\bibfnamefont {W.}~\bibnamefont
  {Yang}}, \bibinfo {author} {\bibfnamefont {S.}~\bibnamefont {Pan}}, \bibinfo
  {author} {\bibfnamefont {A.}~\bibnamefont {Paliathanasis}}, \bibinfo {author}
  {\bibfnamefont {S.}~\bibnamefont {Ghosh}}, \ and\ \bibinfo {author}
  {\bibfnamefont {Y.}~\bibnamefont {Wu}},\ }\href {\doibase
  10.1093/mnras/stz2753} {\bibfield  {journal} {\bibinfo  {journal} {Mon. Not.
  Roy. Astron. Soc.}\ }\textbf {\bibinfo {volume} {490}},\ \bibinfo {pages}
  {2071} (\bibinfo {year} {2019}{\natexlab{b}})},\ \Eprint
  {http://arxiv.org/abs/1904.10436} {arXiv:1904.10436 [gr-qc]} \BibitemShut
  {NoStop}%
%%CITATION = ARXIV:1904.10436;%%
\bibitem [{\citenamefont {Yang}\ \emph
  {et~al.}(2020{\natexlab{b}})\citenamefont {Yang}, \citenamefont
  {Di~Valentino}, \citenamefont {Pan}, \citenamefont {Basilakos},\ and\
  \citenamefont {Paliathanasis}}]{Yang:2020zuk}%
  \BibitemOpen
  \bibfield  {author} {\bibinfo {author} {\bibfnamefont {W.}~\bibnamefont
  {Yang}}, \bibinfo {author} {\bibfnamefont {E.}~\bibnamefont {Di~Valentino}},
  \bibinfo {author} {\bibfnamefont {S.}~\bibnamefont {Pan}}, \bibinfo {author}
  {\bibfnamefont {S.}~\bibnamefont {Basilakos}}, \ and\ \bibinfo {author}
  {\bibfnamefont {A.}~\bibnamefont {Paliathanasis}},\ }\href {\doibase
  10.1103/PhysRevD.102.063503} {\bibfield  {journal} {\bibinfo  {journal}
  {Phys. Rev. D}\ }\textbf {\bibinfo {volume} {102}},\ \bibinfo {pages}
  {063503} (\bibinfo {year} {2020}{\natexlab{b}})},\ \Eprint
  {http://arxiv.org/abs/2001.04307} {arXiv:2001.04307 [astro-ph.CO]}
  \BibitemShut {NoStop}%
\bibitem [{\citenamefont {Di~Valentino}\ \emph
  {et~al.}(2022{\natexlab{a}})\citenamefont {Di~Valentino}, \citenamefont
  {Gariazzo}, \citenamefont {Giunti}, \citenamefont {Mena}, \citenamefont
  {Pan},\ and\ \citenamefont {Yang}}]{DiValentino:2021rjj}%
  \BibitemOpen
  \bibfield  {author} {\bibinfo {author} {\bibfnamefont {E.}~\bibnamefont
  {Di~Valentino}}, \bibinfo {author} {\bibfnamefont {S.}~\bibnamefont
  {Gariazzo}}, \bibinfo {author} {\bibfnamefont {C.}~\bibnamefont {Giunti}},
  \bibinfo {author} {\bibfnamefont {O.}~\bibnamefont {Mena}}, \bibinfo {author}
  {\bibfnamefont {S.}~\bibnamefont {Pan}}, \ and\ \bibinfo {author}
  {\bibfnamefont {W.}~\bibnamefont {Yang}},\ }\href {\doibase
  10.1103/PhysRevD.105.103511} {\bibfield  {journal} {\bibinfo  {journal}
  {Phys. Rev. D}\ }\textbf {\bibinfo {volume} {105}},\ \bibinfo {pages}
  {103511} (\bibinfo {year} {2022}{\natexlab{a}})},\ \Eprint
  {http://arxiv.org/abs/2110.03990} {arXiv:2110.03990 [astro-ph.CO]}
  \BibitemShut {NoStop}%
\bibitem [{\citenamefont {Yang}\ \emph {et~al.}(2022)\citenamefont {Yang},
  \citenamefont {Pan}, \citenamefont {Mena},\ and\ \citenamefont
  {Di~Valentino}}]{Yang:2022csz}%
  \BibitemOpen
  \bibfield  {author} {\bibinfo {author} {\bibfnamefont {W.}~\bibnamefont
  {Yang}}, \bibinfo {author} {\bibfnamefont {S.}~\bibnamefont {Pan}}, \bibinfo
  {author} {\bibfnamefont {O.}~\bibnamefont {Mena}}, \ and\ \bibinfo {author}
  {\bibfnamefont {E.}~\bibnamefont {Di~Valentino}},\ }\href@noop {} {\
  (\bibinfo {year} {2022})},\ \Eprint {http://arxiv.org/abs/2209.14816}
  {arXiv:2209.14816 [astro-ph.CO]} \BibitemShut {NoStop}%
\bibitem [{\citenamefont {Di~Valentino}\ \emph
  {et~al.}(2021{\natexlab{b}})\citenamefont {Di~Valentino} \emph
  {et~al.}}]{DiValentino:2020vvd}%
  \BibitemOpen
  \bibfield  {author} {\bibinfo {author} {\bibfnamefont {E.}~\bibnamefont
  {Di~Valentino}} \emph {et~al.},\ }\href {\doibase
  10.1016/j.astropartphys.2021.102604} {\bibfield  {journal} {\bibinfo
  {journal} {Astropart. Phys.}\ }\textbf {\bibinfo {volume} {131}},\ \bibinfo
  {pages} {102604} (\bibinfo {year} {2021}{\natexlab{b}})},\ \Eprint
  {http://arxiv.org/abs/2008.11285} {arXiv:2008.11285 [astro-ph.CO]}
  \BibitemShut {NoStop}%
\bibitem [{\citenamefont {Heymans}\ \emph {et~al.}(2021)\citenamefont {Heymans}
  \emph {et~al.}}]{Heymans:2020gsg}%
  \BibitemOpen
  \bibfield  {author} {\bibinfo {author} {\bibfnamefont {C.}~\bibnamefont
  {Heymans}} \emph {et~al.},\ }\href {\doibase 10.1051/0004-6361/202039063}
  {\bibfield  {journal} {\bibinfo  {journal} {Astron. Astrophys.}\ }\textbf
  {\bibinfo {volume} {646}},\ \bibinfo {pages} {A140} (\bibinfo {year}
  {2021})},\ \Eprint {http://arxiv.org/abs/2007.15632} {arXiv:2007.15632
  [astro-ph.CO]} \BibitemShut {NoStop}%
\bibitem [{\citenamefont {Tr\"oster}\ \emph {et~al.}(2021)\citenamefont
  {Tr\"oster} \emph {et~al.}}]{KiDS:2020ghu}%
  \BibitemOpen
  \bibfield  {author} {\bibinfo {author} {\bibfnamefont {T.}~\bibnamefont
  {Tr\"oster}} \emph {et~al.} (\bibinfo {collaboration} {KiDS}),\ }\href
  {\doibase 10.1051/0004-6361/202039805} {\bibfield  {journal} {\bibinfo
  {journal} {Astron. Astrophys.}\ }\textbf {\bibinfo {volume} {649}},\ \bibinfo
  {pages} {A88} (\bibinfo {year} {2021})},\ \Eprint
  {http://arxiv.org/abs/2010.16416} {arXiv:2010.16416 [astro-ph.CO]}
  \BibitemShut {NoStop}%
\bibitem [{\citenamefont {Secco}\ \emph {et~al.}(2022)\citenamefont {Secco}
  \emph {et~al.}}]{DES:2021vln}%
  \BibitemOpen
  \bibfield  {author} {\bibinfo {author} {\bibfnamefont {L.~F.}\ \bibnamefont
  {Secco}} \emph {et~al.} (\bibinfo {collaboration} {DES}),\ }\href {\doibase
  10.1103/PhysRevD.105.023515} {\bibfield  {journal} {\bibinfo  {journal}
  {Phys. Rev. D}\ }\textbf {\bibinfo {volume} {105}},\ \bibinfo {pages}
  {023515} (\bibinfo {year} {2022})},\ \Eprint
  {http://arxiv.org/abs/2105.13544} {arXiv:2105.13544 [astro-ph.CO]}
  \BibitemShut {NoStop}%
\bibitem [{\citenamefont {Abbott}\ \emph {et~al.}(2022)\citenamefont {Abbott}
  \emph {et~al.}}]{DES:2022ygi}%
  \BibitemOpen
  \bibfield  {author} {\bibinfo {author} {\bibfnamefont {T.~M.~C.}\
  \bibnamefont {Abbott}} \emph {et~al.} (\bibinfo {collaboration} {DES}),\
  }\href@noop {} {\  (\bibinfo {year} {2022})},\ \Eprint
  {http://arxiv.org/abs/2207.05766} {arXiv:2207.05766 [astro-ph.CO]}
  \BibitemShut {NoStop}%
\bibitem [{\citenamefont {Franco~Abell\'an}\ \emph {et~al.}(2022)\citenamefont
  {Franco~Abell\'an}, \citenamefont {Murgia}, \citenamefont {Poulin},\ and\
  \citenamefont {Lavalle}}]{FrancoAbellan:2020xnr}%
  \BibitemOpen
  \bibfield  {author} {\bibinfo {author} {\bibfnamefont {G.}~\bibnamefont
  {Franco~Abell\'an}}, \bibinfo {author} {\bibfnamefont {R.}~\bibnamefont
  {Murgia}}, \bibinfo {author} {\bibfnamefont {V.}~\bibnamefont {Poulin}}, \
  and\ \bibinfo {author} {\bibfnamefont {J.}~\bibnamefont {Lavalle}},\ }\href
  {\doibase 10.1103/PhysRevD.105.063525} {\bibfield  {journal} {\bibinfo
  {journal} {Phys. Rev. D}\ }\textbf {\bibinfo {volume} {105}},\ \bibinfo
  {pages} {063525} (\bibinfo {year} {2022})},\ \Eprint
  {http://arxiv.org/abs/2008.09615} {arXiv:2008.09615 [astro-ph.CO]}
  \BibitemShut {NoStop}%
\bibitem [{\citenamefont {Franco~Abell\'an}\ \emph {et~al.}(2021)\citenamefont
  {Franco~Abell\'an}, \citenamefont {Murgia},\ and\ \citenamefont
  {Poulin}}]{FrancoAbellan:2021sxk}%
  \BibitemOpen
  \bibfield  {author} {\bibinfo {author} {\bibfnamefont {G.}~\bibnamefont
  {Franco~Abell\'an}}, \bibinfo {author} {\bibfnamefont {R.}~\bibnamefont
  {Murgia}}, \ and\ \bibinfo {author} {\bibfnamefont {V.}~\bibnamefont
  {Poulin}},\ }\href {\doibase 10.1103/PhysRevD.104.123533} {\bibfield
  {journal} {\bibinfo  {journal} {Phys. Rev. D}\ }\textbf {\bibinfo {volume}
  {104}},\ \bibinfo {pages} {123533} (\bibinfo {year} {2021})},\ \Eprint
  {http://arxiv.org/abs/2102.12498} {arXiv:2102.12498 [astro-ph.CO]}
  \BibitemShut {NoStop}%
\bibitem [{\citenamefont {Lucca}(2021)}]{Lucca:2021dxo}%
  \BibitemOpen
  \bibfield  {author} {\bibinfo {author} {\bibfnamefont {M.}~\bibnamefont
  {Lucca}},\ }\href {\doibase 10.1016/j.dark.2021.100899} {\bibfield  {journal}
  {\bibinfo  {journal} {Phys. Dark Univ.}\ }\textbf {\bibinfo {volume} {34}},\
  \bibinfo {pages} {100899} (\bibinfo {year} {2021})},\ \Eprint
  {http://arxiv.org/abs/2105.09249} {arXiv:2105.09249 [astro-ph.CO]}
  \BibitemShut {NoStop}%
\bibitem [{\citenamefont {de~Araujo}\ \emph {et~al.}(2021)\citenamefont
  {de~Araujo}, \citenamefont {De~Felice}, \citenamefont {Kumar},\ and\
  \citenamefont {Nunes}}]{deAraujo:2021cnd}%
  \BibitemOpen
  \bibfield  {author} {\bibinfo {author} {\bibfnamefont {J.~C.~N.}\
  \bibnamefont {de~Araujo}}, \bibinfo {author} {\bibfnamefont {A.}~\bibnamefont
  {De~Felice}}, \bibinfo {author} {\bibfnamefont {S.}~\bibnamefont {Kumar}}, \
  and\ \bibinfo {author} {\bibfnamefont {R.~C.}\ \bibnamefont {Nunes}},\ }\href
  {\doibase 10.1103/PhysRevD.104.104057} {\bibfield  {journal} {\bibinfo
  {journal} {Phys. Rev. D}\ }\textbf {\bibinfo {volume} {104}},\ \bibinfo
  {pages} {104057} (\bibinfo {year} {2021})},\ \Eprint
  {http://arxiv.org/abs/2106.09595} {arXiv:2106.09595 [astro-ph.CO]}
  \BibitemShut {NoStop}%
\bibitem [{\citenamefont {Clark}\ \emph {et~al.}(2021)\citenamefont {Clark},
  \citenamefont {Vattis}, \citenamefont {Fan},\ and\ \citenamefont
  {Koushiappas}}]{Clark:2021hlo}%
  \BibitemOpen
  \bibfield  {author} {\bibinfo {author} {\bibfnamefont {S.~J.}\ \bibnamefont
  {Clark}}, \bibinfo {author} {\bibfnamefont {K.}~\bibnamefont {Vattis}},
  \bibinfo {author} {\bibfnamefont {J.}~\bibnamefont {Fan}}, \ and\ \bibinfo
  {author} {\bibfnamefont {S.~M.}\ \bibnamefont {Koushiappas}},\ }\href@noop {}
  {\  (\bibinfo {year} {2021})},\ \Eprint {http://arxiv.org/abs/2110.09562}
  {arXiv:2110.09562 [astro-ph.CO]} \BibitemShut {NoStop}%
\bibitem [{\citenamefont {Beltr\'an~Jim\'enez}\ \emph
  {et~al.}(2021)\citenamefont {Beltr\'an~Jim\'enez}, \citenamefont {Bettoni},
  \citenamefont {Figueruelo}, \citenamefont {Teppa~Pannia},\ and\ \citenamefont
  {Tsujikawa}}]{BeltranJimenez:2021wbq}%
  \BibitemOpen
  \bibfield  {author} {\bibinfo {author} {\bibfnamefont {J.}~\bibnamefont
  {Beltr\'an~Jim\'enez}}, \bibinfo {author} {\bibfnamefont {D.}~\bibnamefont
  {Bettoni}}, \bibinfo {author} {\bibfnamefont {D.}~\bibnamefont {Figueruelo}},
  \bibinfo {author} {\bibfnamefont {F.~A.}\ \bibnamefont {Teppa~Pannia}}, \
  and\ \bibinfo {author} {\bibfnamefont {S.}~\bibnamefont {Tsujikawa}},\ }\href
  {\doibase 10.1103/PhysRevD.104.103503} {\bibfield  {journal} {\bibinfo
  {journal} {Phys. Rev. D}\ }\textbf {\bibinfo {volume} {104}},\ \bibinfo
  {pages} {103503} (\bibinfo {year} {2021})},\ \Eprint
  {http://arxiv.org/abs/2106.11222} {arXiv:2106.11222 [astro-ph.CO]}
  \BibitemShut {NoStop}%
\bibitem [{\citenamefont {Heimersheim}\ \emph {et~al.}(2020)\citenamefont
  {Heimersheim}, \citenamefont {Sch\"oneberg}, \citenamefont {Hooper},\ and\
  \citenamefont {Lesgourgues}}]{Heimersheim:2020aoc}%
  \BibitemOpen
  \bibfield  {author} {\bibinfo {author} {\bibfnamefont {S.}~\bibnamefont
  {Heimersheim}}, \bibinfo {author} {\bibfnamefont {N.}~\bibnamefont
  {Sch\"oneberg}}, \bibinfo {author} {\bibfnamefont {D.~C.}\ \bibnamefont
  {Hooper}}, \ and\ \bibinfo {author} {\bibfnamefont {J.}~\bibnamefont
  {Lesgourgues}},\ }\href {\doibase 10.1088/1475-7516/2020/12/016} {\bibfield
  {journal} {\bibinfo  {journal} {JCAP}\ }\textbf {\bibinfo {volume} {12}},\
  \bibinfo {pages} {016} (\bibinfo {year} {2020})},\ \Eprint
  {http://arxiv.org/abs/2008.08486} {arXiv:2008.08486 [astro-ph.CO]}
  \BibitemShut {NoStop}%
\bibitem [{\citenamefont {Poulin}\ \emph {et~al.}(2022)\citenamefont {Poulin},
  \citenamefont {Bernal}, \citenamefont {Kovetz},\ and\ \citenamefont
  {Kamionkowski}}]{Poulin:2022sgp}%
  \BibitemOpen
  \bibfield  {author} {\bibinfo {author} {\bibfnamefont {V.}~\bibnamefont
  {Poulin}}, \bibinfo {author} {\bibfnamefont {J.~L.}\ \bibnamefont {Bernal}},
  \bibinfo {author} {\bibfnamefont {E.}~\bibnamefont {Kovetz}}, \ and\ \bibinfo
  {author} {\bibfnamefont {M.}~\bibnamefont {Kamionkowski}},\ }\href@noop {} {\
   (\bibinfo {year} {2022})},\ \Eprint {http://arxiv.org/abs/2209.06217}
  {arXiv:2209.06217 [astro-ph.CO]} \BibitemShut {NoStop}%
\bibitem [{\citenamefont {Amon}\ and\ \citenamefont
  {Efstathiou}(2022)}]{Amon:2022azi}%
  \BibitemOpen
  \bibfield  {author} {\bibinfo {author} {\bibfnamefont {A.}~\bibnamefont
  {Amon}}\ and\ \bibinfo {author} {\bibfnamefont {G.}~\bibnamefont
  {Efstathiou}},\ }\href {\doibase 10.1093/mnras/stac2429} {\  (\bibinfo {year}
  {2022}),\ 10.1093/mnras/stac2429},\ \Eprint {http://arxiv.org/abs/2206.11794}
  {arXiv:2206.11794 [astro-ph.CO]} \BibitemShut {NoStop}%
\bibitem [{\citenamefont {Chevallier}\ and\ \citenamefont
  {Polarski}(2001)}]{Chevallier:2000qy}%
  \BibitemOpen
  \bibfield  {author} {\bibinfo {author} {\bibfnamefont {M.}~\bibnamefont
  {Chevallier}}\ and\ \bibinfo {author} {\bibfnamefont {D.}~\bibnamefont
  {Polarski}},\ }\href {\doibase 10.1142/S0218271801000822} {\bibfield
  {journal} {\bibinfo  {journal} {Int. J. Mod. Phys.}\ }\textbf {\bibinfo
  {volume} {D10}},\ \bibinfo {pages} {213} (\bibinfo {year} {2001})},\ \Eprint
  {http://arxiv.org/abs/gr-qc/0009008} {arXiv:gr-qc/0009008 [gr-qc]}
  \BibitemShut {NoStop}%
%%CITATION = GR-QC/0009008;%%
\bibitem [{\citenamefont {Linder}(2003)}]{Linder:2002et}%
  \BibitemOpen
  \bibfield  {author} {\bibinfo {author} {\bibfnamefont {E.~V.}\ \bibnamefont
  {Linder}},\ }\href {\doibase 10.1103/PhysRevLett.90.091301} {\bibfield
  {journal} {\bibinfo  {journal} {Phys. Rev. Lett.}\ }\textbf {\bibinfo
  {volume} {90}},\ \bibinfo {pages} {091301} (\bibinfo {year} {2003})},\
  \Eprint {http://arxiv.org/abs/astro-ph/0208512} {arXiv:astro-ph/0208512
  [astro-ph]} \BibitemShut {NoStop}%
%%CITATION = ASTRO-PH/0208512;%%
\bibitem [{\citenamefont {Li}\ and\ \citenamefont
  {Shafieloo}(2019)}]{Li:2019yem}%
  \BibitemOpen
  \bibfield  {author} {\bibinfo {author} {\bibfnamefont {X.}~\bibnamefont
  {Li}}\ and\ \bibinfo {author} {\bibfnamefont {A.}~\bibnamefont {Shafieloo}},\
  }\href {\doibase 10.3847/2041-8213/ab3e09} {\bibfield  {journal} {\bibinfo
  {journal} {Astrophys. J. Lett.}\ }\textbf {\bibinfo {volume} {883}},\
  \bibinfo {pages} {L3} (\bibinfo {year} {2019})},\ \Eprint
  {http://arxiv.org/abs/1906.08275} {arXiv:1906.08275 [astro-ph.CO]}
  \BibitemShut {NoStop}%
\bibitem [{\citenamefont {Pan}\ \emph {et~al.}(2020{\natexlab{b}})\citenamefont
  {Pan}, \citenamefont {Yang}, \citenamefont {Di~Valentino}, \citenamefont
  {Shafieloo},\ and\ \citenamefont {Chakraborty}}]{Pan:2019hac}%
  \BibitemOpen
  \bibfield  {author} {\bibinfo {author} {\bibfnamefont {S.}~\bibnamefont
  {Pan}}, \bibinfo {author} {\bibfnamefont {W.}~\bibnamefont {Yang}}, \bibinfo
  {author} {\bibfnamefont {E.}~\bibnamefont {Di~Valentino}}, \bibinfo {author}
  {\bibfnamefont {A.}~\bibnamefont {Shafieloo}}, \ and\ \bibinfo {author}
  {\bibfnamefont {S.}~\bibnamefont {Chakraborty}},\ }\href {\doibase
  10.1088/1475-7516/2020/06/062} {\bibfield  {journal} {\bibinfo  {journal}
  {JCAP}\ }\textbf {\bibinfo {volume} {06}},\ \bibinfo {pages} {062} (\bibinfo
  {year} {2020}{\natexlab{b}})},\ \Eprint {http://arxiv.org/abs/1907.12551}
  {arXiv:1907.12551 [astro-ph.CO]} \BibitemShut {NoStop}%
\bibitem [{\citenamefont {Perenon}\ \emph {et~al.}(2022)\citenamefont
  {Perenon}, \citenamefont {Martinelli}, \citenamefont {Maartens},
  \citenamefont {Camera},\ and\ \citenamefont {Clarkson}}]{Perenon:2022fgw}%
  \BibitemOpen
  \bibfield  {author} {\bibinfo {author} {\bibfnamefont {L.}~\bibnamefont
  {Perenon}}, \bibinfo {author} {\bibfnamefont {M.}~\bibnamefont {Martinelli}},
  \bibinfo {author} {\bibfnamefont {R.}~\bibnamefont {Maartens}}, \bibinfo
  {author} {\bibfnamefont {S.}~\bibnamefont {Camera}}, \ and\ \bibinfo {author}
  {\bibfnamefont {C.}~\bibnamefont {Clarkson}},\ }\href {\doibase
  10.1016/j.dark.2022.101119} {\bibfield  {journal} {\bibinfo  {journal} {Phys.
  Dark Univ.}\ }\textbf {\bibinfo {volume} {37}},\ \bibinfo {pages} {101119}
  (\bibinfo {year} {2022})},\ \Eprint {http://arxiv.org/abs/2206.12375}
  {arXiv:2206.12375 [astro-ph.CO]} \BibitemShut {NoStop}%
\bibitem [{\citenamefont {Yang}\ \emph
  {et~al.}(2021{\natexlab{a}})\citenamefont {Yang}, \citenamefont
  {Di~Valentino}, \citenamefont {Pan}, \citenamefont {Shafieloo},\ and\
  \citenamefont {Li}}]{Yang:2021eud}%
  \BibitemOpen
  \bibfield  {author} {\bibinfo {author} {\bibfnamefont {W.}~\bibnamefont
  {Yang}}, \bibinfo {author} {\bibfnamefont {E.}~\bibnamefont {Di~Valentino}},
  \bibinfo {author} {\bibfnamefont {S.}~\bibnamefont {Pan}}, \bibinfo {author}
  {\bibfnamefont {A.}~\bibnamefont {Shafieloo}}, \ and\ \bibinfo {author}
  {\bibfnamefont {X.}~\bibnamefont {Li}},\ }\href {\doibase
  10.1103/PhysRevD.104.063521} {\bibfield  {journal} {\bibinfo  {journal}
  {Phys. Rev. D}\ }\textbf {\bibinfo {volume} {104}},\ \bibinfo {pages}
  {063521} (\bibinfo {year} {2021}{\natexlab{a}})},\ \Eprint
  {http://arxiv.org/abs/2103.03815} {arXiv:2103.03815 [astro-ph.CO]}
  \BibitemShut {NoStop}%
\bibitem [{\citenamefont {Yang}\ \emph
  {et~al.}(2021{\natexlab{b}})\citenamefont {Yang}, \citenamefont
  {Di~Valentino}, \citenamefont {Pan}, \citenamefont {Wu},\ and\ \citenamefont
  {Lu}}]{Yang:2021flj}%
  \BibitemOpen
  \bibfield  {author} {\bibinfo {author} {\bibfnamefont {W.}~\bibnamefont
  {Yang}}, \bibinfo {author} {\bibfnamefont {E.}~\bibnamefont {Di~Valentino}},
  \bibinfo {author} {\bibfnamefont {S.}~\bibnamefont {Pan}}, \bibinfo {author}
  {\bibfnamefont {Y.}~\bibnamefont {Wu}}, \ and\ \bibinfo {author}
  {\bibfnamefont {J.}~\bibnamefont {Lu}},\ }\href {\doibase
  10.1093/mnras/staa3914} {\bibfield  {journal} {\bibinfo  {journal} {Mon. Not.
  Roy. Astron. Soc.}\ }\textbf {\bibinfo {volume} {501}},\ \bibinfo {pages}
  {5845} (\bibinfo {year} {2021}{\natexlab{b}})},\ \Eprint
  {http://arxiv.org/abs/2101.02168} {arXiv:2101.02168 [astro-ph.CO]}
  \BibitemShut {NoStop}%
\bibitem [{\citenamefont {Menci}\ \emph {et~al.}(2020)\citenamefont {Menci}
  \emph {et~al.}}]{Menci:2020ybl}%
  \BibitemOpen
  \bibfield  {author} {\bibinfo {author} {\bibfnamefont {N.}~\bibnamefont
  {Menci}} \emph {et~al.},\ }\href {\doibase 10.3847/1538-4357/aba9d2}
  {\bibfield  {journal} {\bibinfo  {journal} {Astrophys. J.}\ }\textbf
  {\bibinfo {volume} {900}},\ \bibinfo {pages} {108} (\bibinfo {year}
  {2020})},\ \Eprint {http://arxiv.org/abs/2007.12453} {arXiv:2007.12453
  [astro-ph.CO]} \BibitemShut {NoStop}%
\bibitem [{\citenamefont {Zhao}\ \emph {et~al.}(2020)\citenamefont {Zhao},
  \citenamefont {Li}, \citenamefont {Qi}, \citenamefont {Gao}, \citenamefont
  {Zhang},\ and\ \citenamefont {Zhang}}]{Zhao:2020ole}%
  \BibitemOpen
  \bibfield  {author} {\bibinfo {author} {\bibfnamefont {Z.-W.}\ \bibnamefont
  {Zhao}}, \bibinfo {author} {\bibfnamefont {Z.-X.}\ \bibnamefont {Li}},
  \bibinfo {author} {\bibfnamefont {J.-Z.}\ \bibnamefont {Qi}}, \bibinfo
  {author} {\bibfnamefont {H.}~\bibnamefont {Gao}}, \bibinfo {author}
  {\bibfnamefont {J.-F.}\ \bibnamefont {Zhang}}, \ and\ \bibinfo {author}
  {\bibfnamefont {X.}~\bibnamefont {Zhang}},\ }\href {\doibase
  10.3847/1538-4357/abb8ce} {\bibfield  {journal} {\bibinfo  {journal}
  {Astrophys. J.}\ }\textbf {\bibinfo {volume} {903}},\ \bibinfo {pages} {83}
  (\bibinfo {year} {2020})},\ \Eprint {http://arxiv.org/abs/2006.01450}
  {arXiv:2006.01450 [astro-ph.CO]} \BibitemShut {NoStop}%
\bibitem [{\citenamefont {Zimdahl}\ \emph {et~al.}(2020)\citenamefont
  {Zimdahl}, \citenamefont {Fabris}, \citenamefont {Velten},\ and\
  \citenamefont {Herrera}}]{Zimdahl:2019pqg}%
  \BibitemOpen
  \bibfield  {author} {\bibinfo {author} {\bibfnamefont {W.}~\bibnamefont
  {Zimdahl}}, \bibinfo {author} {\bibfnamefont {J.}~\bibnamefont {Fabris}},
  \bibinfo {author} {\bibfnamefont {H.}~\bibnamefont {Velten}}, \ and\ \bibinfo
  {author} {\bibfnamefont {R.}~\bibnamefont {Herrera}},\ }\href {\doibase
  10.1016/j.dark.2020.100681} {\bibfield  {journal} {\bibinfo  {journal} {Phys.
  Dark Univ.}\ }\textbf {\bibinfo {volume} {30}},\ \bibinfo {pages} {100681}
  (\bibinfo {year} {2020})},\ \Eprint {http://arxiv.org/abs/1911.12084}
  {arXiv:1911.12084 [astro-ph.CO]} \BibitemShut {NoStop}%
\bibitem [{\citenamefont {Du}\ \emph {et~al.}(2019)\citenamefont {Du},
  \citenamefont {Yang}, \citenamefont {Xu}, \citenamefont {Pan},\ and\
  \citenamefont {Mota}}]{Du:2018tia}%
  \BibitemOpen
  \bibfield  {author} {\bibinfo {author} {\bibfnamefont {M.}~\bibnamefont
  {Du}}, \bibinfo {author} {\bibfnamefont {W.}~\bibnamefont {Yang}}, \bibinfo
  {author} {\bibfnamefont {L.}~\bibnamefont {Xu}}, \bibinfo {author}
  {\bibfnamefont {S.}~\bibnamefont {Pan}}, \ and\ \bibinfo {author}
  {\bibfnamefont {D.~F.}\ \bibnamefont {Mota}},\ }\href {\doibase
  10.1103/PhysRevD.100.043535} {\bibfield  {journal} {\bibinfo  {journal}
  {Phys. Rev. D}\ }\textbf {\bibinfo {volume} {100}},\ \bibinfo {pages}
  {043535} (\bibinfo {year} {2019})},\ \Eprint
  {http://arxiv.org/abs/1812.01440} {arXiv:1812.01440 [astro-ph.CO]}
  \BibitemShut {NoStop}%
\bibitem [{\citenamefont {Vagnozzi}\ \emph {et~al.}(2018)\citenamefont
  {Vagnozzi}, \citenamefont {Dhawan}, \citenamefont {Gerbino}, \citenamefont
  {Freese}, \citenamefont {Goobar},\ and\ \citenamefont
  {Mena}}]{Vagnozzi:2018jhn}%
  \BibitemOpen
  \bibfield  {author} {\bibinfo {author} {\bibfnamefont {S.}~\bibnamefont
  {Vagnozzi}}, \bibinfo {author} {\bibfnamefont {S.}~\bibnamefont {Dhawan}},
  \bibinfo {author} {\bibfnamefont {M.}~\bibnamefont {Gerbino}}, \bibinfo
  {author} {\bibfnamefont {K.}~\bibnamefont {Freese}}, \bibinfo {author}
  {\bibfnamefont {A.}~\bibnamefont {Goobar}}, \ and\ \bibinfo {author}
  {\bibfnamefont {O.}~\bibnamefont {Mena}},\ }\href {\doibase
  10.1103/PhysRevD.98.083501} {\bibfield  {journal} {\bibinfo  {journal} {Phys.
  Rev. D}\ }\textbf {\bibinfo {volume} {98}},\ \bibinfo {pages} {083501}
  (\bibinfo {year} {2018})},\ \Eprint {http://arxiv.org/abs/1801.08553}
  {arXiv:1801.08553 [astro-ph.CO]} \BibitemShut {NoStop}%
\bibitem [{\citenamefont {Li}\ \emph {et~al.}(2018)\citenamefont {Li},
  \citenamefont {Sabiu}, \citenamefont {Park}, \citenamefont {Wang},
  \citenamefont {Zhao}, \citenamefont {Park}, \citenamefont {Shafieloo},
  \citenamefont {Kim},\ and\ \citenamefont {Hong}}]{Li:2018nlh}%
  \BibitemOpen
  \bibfield  {author} {\bibinfo {author} {\bibfnamefont {X.-D.}\ \bibnamefont
  {Li}}, \bibinfo {author} {\bibfnamefont {C.~G.}\ \bibnamefont {Sabiu}},
  \bibinfo {author} {\bibfnamefont {C.}~\bibnamefont {Park}}, \bibinfo {author}
  {\bibfnamefont {Y.}~\bibnamefont {Wang}}, \bibinfo {author} {\bibfnamefont
  {G.-b.}\ \bibnamefont {Zhao}}, \bibinfo {author} {\bibfnamefont
  {H.}~\bibnamefont {Park}}, \bibinfo {author} {\bibfnamefont {A.}~\bibnamefont
  {Shafieloo}}, \bibinfo {author} {\bibfnamefont {J.}~\bibnamefont {Kim}}, \
  and\ \bibinfo {author} {\bibfnamefont {S.~E.}\ \bibnamefont {Hong}},\ }\href
  {\doibase 10.3847/1538-4357/aab42e} {\bibfield  {journal} {\bibinfo
  {journal} {Astrophys. J.}\ }\textbf {\bibinfo {volume} {856}},\ \bibinfo
  {pages} {88} (\bibinfo {year} {2018})},\ \Eprint
  {http://arxiv.org/abs/1803.01851} {arXiv:1803.01851 [astro-ph.CO]}
  \BibitemShut {NoStop}%
\bibitem [{\citenamefont {Di~Valentino}\ \emph
  {et~al.}(2017{\natexlab{b}})\citenamefont {Di~Valentino}, \citenamefont
  {Melchiorri}, \citenamefont {Linder},\ and\ \citenamefont
  {Silk}}]{DiValentino:2017zyq}%
  \BibitemOpen
  \bibfield  {author} {\bibinfo {author} {\bibfnamefont {E.}~\bibnamefont
  {Di~Valentino}}, \bibinfo {author} {\bibfnamefont {A.}~\bibnamefont
  {Melchiorri}}, \bibinfo {author} {\bibfnamefont {E.~V.}\ \bibnamefont
  {Linder}}, \ and\ \bibinfo {author} {\bibfnamefont {J.}~\bibnamefont
  {Silk}},\ }\href {\doibase 10.1103/PhysRevD.96.023523} {\bibfield  {journal}
  {\bibinfo  {journal} {Phys. Rev. D}\ }\textbf {\bibinfo {volume} {96}},\
  \bibinfo {pages} {023523} (\bibinfo {year} {2017}{\natexlab{b}})},\ \Eprint
  {http://arxiv.org/abs/1704.00762} {arXiv:1704.00762 [astro-ph.CO]}
  \BibitemShut {NoStop}%
\bibitem [{\citenamefont {Pan}\ \emph {et~al.}(2018)\citenamefont {Pan},
  \citenamefont {Saridakis},\ and\ \citenamefont {Yang}}]{Pan:2017zoh}%
  \BibitemOpen
  \bibfield  {author} {\bibinfo {author} {\bibfnamefont {S.}~\bibnamefont
  {Pan}}, \bibinfo {author} {\bibfnamefont {E.~N.}\ \bibnamefont {Saridakis}},
  \ and\ \bibinfo {author} {\bibfnamefont {W.}~\bibnamefont {Yang}},\ }\href
  {\doibase 10.1103/PhysRevD.98.063510} {\bibfield  {journal} {\bibinfo
  {journal} {Phys. Rev. D}\ }\textbf {\bibinfo {volume} {98}},\ \bibinfo
  {pages} {063510} (\bibinfo {year} {2018})},\ \Eprint
  {http://arxiv.org/abs/1712.05746} {arXiv:1712.05746 [astro-ph.CO]}
  \BibitemShut {NoStop}%
\bibitem [{\citenamefont {Yang}\ \emph
  {et~al.}(2018{\natexlab{c}})\citenamefont {Yang}, \citenamefont {Pan},\ and\
  \citenamefont {Paliathanasis}}]{Yang:2017alx}%
  \BibitemOpen
  \bibfield  {author} {\bibinfo {author} {\bibfnamefont {W.}~\bibnamefont
  {Yang}}, \bibinfo {author} {\bibfnamefont {S.}~\bibnamefont {Pan}}, \ and\
  \bibinfo {author} {\bibfnamefont {A.}~\bibnamefont {Paliathanasis}},\ }\href
  {\doibase 10.1093/mnras/sty019} {\bibfield  {journal} {\bibinfo  {journal}
  {Mon. Not. Roy. Astron. Soc.}\ }\textbf {\bibinfo {volume} {475}},\ \bibinfo
  {pages} {2605} (\bibinfo {year} {2018}{\natexlab{c}})},\ \Eprint
  {http://arxiv.org/abs/1708.01717} {arXiv:1708.01717 [gr-qc]} \BibitemShut
  {NoStop}%
\bibitem [{\citenamefont {Yang}\ \emph {et~al.}(2017)\citenamefont {Yang},
  \citenamefont {Nunes}, \citenamefont {Pan},\ and\ \citenamefont
  {Mota}}]{Yang:2017amu}%
  \BibitemOpen
  \bibfield  {author} {\bibinfo {author} {\bibfnamefont {W.}~\bibnamefont
  {Yang}}, \bibinfo {author} {\bibfnamefont {R.~C.}\ \bibnamefont {Nunes}},
  \bibinfo {author} {\bibfnamefont {S.}~\bibnamefont {Pan}}, \ and\ \bibinfo
  {author} {\bibfnamefont {D.~F.}\ \bibnamefont {Mota}},\ }\href {\doibase
  10.1103/PhysRevD.95.103522} {\bibfield  {journal} {\bibinfo  {journal} {Phys.
  Rev. D}\ }\textbf {\bibinfo {volume} {95}},\ \bibinfo {pages} {103522}
  (\bibinfo {year} {2017})},\ \Eprint {http://arxiv.org/abs/1703.02556}
  {arXiv:1703.02556 [astro-ph.CO]} \BibitemShut {NoStop}%
\bibitem [{\citenamefont {Zhao}\ \emph {et~al.}(2017)\citenamefont {Zhao} \emph
  {et~al.}}]{Zhao:2017cud}%
  \BibitemOpen
  \bibfield  {author} {\bibinfo {author} {\bibfnamefont {G.-B.}\ \bibnamefont
  {Zhao}} \emph {et~al.},\ }\href {\doibase 10.1038/s41550-017-0216-z}
  {\bibfield  {journal} {\bibinfo  {journal} {Nature Astron.}\ }\textbf
  {\bibinfo {volume} {1}},\ \bibinfo {pages} {627} (\bibinfo {year} {2017})},\
  \Eprint {http://arxiv.org/abs/1701.08165} {arXiv:1701.08165 [astro-ph.CO]}
  \BibitemShut {NoStop}%
\bibitem [{\citenamefont {Ma}\ and\ \citenamefont {Zhang}(2011)}]{Ma:2011nc}%
  \BibitemOpen
  \bibfield  {author} {\bibinfo {author} {\bibfnamefont {J.-Z.}\ \bibnamefont
  {Ma}}\ and\ \bibinfo {author} {\bibfnamefont {X.}~\bibnamefont {Zhang}},\
  }\href {\doibase 10.1016/j.physletb.2011.04.013} {\bibfield  {journal}
  {\bibinfo  {journal} {Phys. Lett. B}\ }\textbf {\bibinfo {volume} {699}},\
  \bibinfo {pages} {233} (\bibinfo {year} {2011})},\ \Eprint
  {http://arxiv.org/abs/1102.2671} {arXiv:1102.2671 [astro-ph.CO]} \BibitemShut
  {NoStop}%
\bibitem [{\citenamefont {Di~Valentino}\ \emph
  {et~al.}(2022{\natexlab{b}})\citenamefont {Di~Valentino}, \citenamefont
  {Giar\`e}, \citenamefont {Melchiorri},\ and\ \citenamefont
  {Silk}}]{DiValentino:2022oon}%
  \BibitemOpen
  \bibfield  {author} {\bibinfo {author} {\bibfnamefont {E.}~\bibnamefont
  {Di~Valentino}}, \bibinfo {author} {\bibfnamefont {W.}~\bibnamefont
  {Giar\`e}}, \bibinfo {author} {\bibfnamefont {A.}~\bibnamefont {Melchiorri}},
  \ and\ \bibinfo {author} {\bibfnamefont {J.}~\bibnamefont {Silk}},\
  }\href@noop {} {\  (\bibinfo {year} {2022}{\natexlab{b}})},\ \Eprint
  {http://arxiv.org/abs/2209.12872} {arXiv:2209.12872 [astro-ph.CO]}
  \BibitemShut {NoStop}%
\bibitem [{\citenamefont {Aghanim}\ \emph
  {et~al.}(2020{\natexlab{b}})\citenamefont {Aghanim} \emph
  {et~al.}}]{Aghanim:2019ame}%
  \BibitemOpen
  \bibfield  {author} {\bibinfo {author} {\bibfnamefont {N.}~\bibnamefont
  {Aghanim}} \emph {et~al.} (\bibinfo {collaboration} {Planck}),\ }\href
  {\doibase 10.1051/0004-6361/201936386} {\bibfield  {journal} {\bibinfo
  {journal} {Astron. Astrophys.}\ }\textbf {\bibinfo {volume} {641}},\ \bibinfo
  {pages} {A5} (\bibinfo {year} {2020}{\natexlab{b}})},\ \Eprint
  {http://arxiv.org/abs/1907.12875} {arXiv:1907.12875 [astro-ph.CO]}
  \BibitemShut {NoStop}%
\bibitem [{\citenamefont {Beutler}\ \emph {et~al.}(2011)\citenamefont
  {Beutler}, \citenamefont {Blake}, \citenamefont {Colless}, \citenamefont
  {Jones}, \citenamefont {Staveley-Smith}, \citenamefont {Campbell},
  \citenamefont {Parker}, \citenamefont {Saunders},\ and\ \citenamefont
  {Watson}}]{Beutler:2011hx}%
  \BibitemOpen
  \bibfield  {author} {\bibinfo {author} {\bibfnamefont {F.}~\bibnamefont
  {Beutler}}, \bibinfo {author} {\bibfnamefont {C.}~\bibnamefont {Blake}},
  \bibinfo {author} {\bibfnamefont {M.}~\bibnamefont {Colless}}, \bibinfo
  {author} {\bibfnamefont {D.}~\bibnamefont {Jones}}, \bibinfo {author}
  {\bibfnamefont {L.}~\bibnamefont {Staveley-Smith}}, \bibinfo {author}
  {\bibfnamefont {L.}~\bibnamefont {Campbell}}, \bibinfo {author}
  {\bibfnamefont {Q.}~\bibnamefont {Parker}}, \bibinfo {author} {\bibfnamefont
  {W.}~\bibnamefont {Saunders}}, \ and\ \bibinfo {author} {\bibfnamefont
  {F.}~\bibnamefont {Watson}},\ }\href {\doibase
  10.1111/j.1365-2966.2011.19250.x} {\bibfield  {journal} {\bibinfo  {journal}
  {Mon. Not. Roy. Astron. Soc.}\ }\textbf {\bibinfo {volume} {416}},\ \bibinfo
  {pages} {3017} (\bibinfo {year} {2011})},\ \Eprint
  {http://arxiv.org/abs/1106.3366} {arXiv:1106.3366 [astro-ph.CO]} \BibitemShut
  {NoStop}%
\bibitem [{\citenamefont {Ross}\ \emph {et~al.}(2015)\citenamefont {Ross},
  \citenamefont {Samushia}, \citenamefont {Howlett}, \citenamefont {Percival},
  \citenamefont {Burden},\ and\ \citenamefont {Manera}}]{Ross:2014qpa}%
  \BibitemOpen
  \bibfield  {author} {\bibinfo {author} {\bibfnamefont {A.~J.}\ \bibnamefont
  {Ross}}, \bibinfo {author} {\bibfnamefont {L.}~\bibnamefont {Samushia}},
  \bibinfo {author} {\bibfnamefont {C.}~\bibnamefont {Howlett}}, \bibinfo
  {author} {\bibfnamefont {W.~J.}\ \bibnamefont {Percival}}, \bibinfo {author}
  {\bibfnamefont {A.}~\bibnamefont {Burden}}, \ and\ \bibinfo {author}
  {\bibfnamefont {M.}~\bibnamefont {Manera}},\ }\href {\doibase
  10.1093/mnras/stv154} {\bibfield  {journal} {\bibinfo  {journal} {Mon. Not.
  Roy. Astron. Soc.}\ }\textbf {\bibinfo {volume} {449}},\ \bibinfo {pages}
  {835} (\bibinfo {year} {2015})},\ \Eprint {http://arxiv.org/abs/1409.3242}
  {arXiv:1409.3242 [astro-ph.CO]} \BibitemShut {NoStop}%
\bibitem [{\citenamefont {Alam}\ \emph {et~al.}(2017)\citenamefont {Alam} \emph
  {et~al.}}]{Alam:2016hwk}%
  \BibitemOpen
  \bibfield  {author} {\bibinfo {author} {\bibfnamefont {S.}~\bibnamefont
  {Alam}} \emph {et~al.} (\bibinfo {collaboration} {BOSS}),\ }\href {\doibase
  10.1093/mnras/stx721} {\bibfield  {journal} {\bibinfo  {journal} {Mon. Not.
  Roy. Astron. Soc.}\ }\textbf {\bibinfo {volume} {470}},\ \bibinfo {pages}
  {2617} (\bibinfo {year} {2017})},\ \Eprint {http://arxiv.org/abs/1607.03155}
  {arXiv:1607.03155 [astro-ph.CO]} \BibitemShut {NoStop}%
\bibitem [{\citenamefont {Scolnic}\ \emph {et~al.}(2018)\citenamefont {Scolnic}
  \emph {et~al.}}]{Scolnic:2017caz}%
  \BibitemOpen
  \bibfield  {author} {\bibinfo {author} {\bibfnamefont {D.}~\bibnamefont
  {Scolnic}} \emph {et~al.},\ }\href {\doibase 10.3847/1538-4357/aab9bb}
  {\bibfield  {journal} {\bibinfo  {journal} {Astrophys. J.}\ }\textbf
  {\bibinfo {volume} {859}},\ \bibinfo {pages} {101} (\bibinfo {year}
  {2018})},\ \Eprint {http://arxiv.org/abs/1710.00845} {arXiv:1710.00845
  [astro-ph.CO]} \BibitemShut {NoStop}%
\bibitem [{\citenamefont {Lewis}\ and\ \citenamefont
  {Bridle}(2002)}]{Lewis:2002ah}%
  \BibitemOpen
  \bibfield  {author} {\bibinfo {author} {\bibfnamefont {A.}~\bibnamefont
  {Lewis}}\ and\ \bibinfo {author} {\bibfnamefont {S.}~\bibnamefont {Bridle}},\
  }\href {\doibase 10.1103/PhysRevD.66.103511} {\bibfield  {journal} {\bibinfo
  {journal} {Phys. Rev. D}\ }\textbf {\bibinfo {volume} {66}},\ \bibinfo
  {pages} {103511} (\bibinfo {year} {2002})},\ \Eprint
  {http://arxiv.org/abs/astro-ph/0205436} {arXiv:astro-ph/0205436} \BibitemShut
  {NoStop}%
\bibitem [{\citenamefont {Lewis}\ \emph {et~al.}(2000)\citenamefont {Lewis},
  \citenamefont {Challinor},\ and\ \citenamefont {Lasenby}}]{Lewis:1999bs}%
  \BibitemOpen
  \bibfield  {author} {\bibinfo {author} {\bibfnamefont {A.}~\bibnamefont
  {Lewis}}, \bibinfo {author} {\bibfnamefont {A.}~\bibnamefont {Challinor}}, \
  and\ \bibinfo {author} {\bibfnamefont {A.}~\bibnamefont {Lasenby}},\ }\href
  {\doibase 10.1086/309179} {\bibfield  {journal} {\bibinfo  {journal}
  {Astrophys. J.}\ }\textbf {\bibinfo {volume} {538}},\ \bibinfo {pages} {473}
  (\bibinfo {year} {2000})},\ \Eprint {http://arxiv.org/abs/astro-ph/9911177}
  {arXiv:astro-ph/9911177} \BibitemShut {NoStop}%
\bibitem [{\citenamefont {Gelman}\ and\ \citenamefont
  {Rubin}(1992)}]{Gelman:1992zz}%
  \BibitemOpen
  \bibfield  {author} {\bibinfo {author} {\bibfnamefont {A.}~\bibnamefont
  {Gelman}}\ and\ \bibinfo {author} {\bibfnamefont {D.~B.}\ \bibnamefont
  {Rubin}},\ }\href {\doibase 10.1214/ss/1177011136} {\bibfield  {journal}
  {\bibinfo  {journal} {Statist. Sci.}\ }\textbf {\bibinfo {volume} {7}},\
  \bibinfo {pages} {457} (\bibinfo {year} {1992})}\BibitemShut {NoStop}%
\bibitem [{\citenamefont {Vagnozzi}\ \emph {et~al.}(2021)\citenamefont
  {Vagnozzi}, \citenamefont {Di~Valentino}, \citenamefont {Gariazzo},
  \citenamefont {Melchiorri}, \citenamefont {Mena},\ and\ \citenamefont
  {Silk}}]{Vagnozzi:2020rcz}%
  \BibitemOpen
  \bibfield  {author} {\bibinfo {author} {\bibfnamefont {S.}~\bibnamefont
  {Vagnozzi}}, \bibinfo {author} {\bibfnamefont {E.}~\bibnamefont
  {Di~Valentino}}, \bibinfo {author} {\bibfnamefont {S.}~\bibnamefont
  {Gariazzo}}, \bibinfo {author} {\bibfnamefont {A.}~\bibnamefont
  {Melchiorri}}, \bibinfo {author} {\bibfnamefont {O.}~\bibnamefont {Mena}}, \
  and\ \bibinfo {author} {\bibfnamefont {J.}~\bibnamefont {Silk}},\ }\href
  {\doibase 10.1016/j.dark.2021.100851} {\bibfield  {journal} {\bibinfo
  {journal} {Phys. Dark Univ.}\ }\textbf {\bibinfo {volume} {33}},\ \bibinfo
  {pages} {100851} (\bibinfo {year} {2021})},\ \Eprint
  {http://arxiv.org/abs/2010.02230} {arXiv:2010.02230 [astro-ph.CO]}
  \BibitemShut {NoStop}%
\bibitem [{\citenamefont {Mangano}\ \emph {et~al.}(2005)\citenamefont
  {Mangano}, \citenamefont {Miele}, \citenamefont {Pastor}, \citenamefont
  {Pinto}, \citenamefont {Pisanti},\ and\ \citenamefont
  {Serpico}}]{Mangano:2005cc}%
  \BibitemOpen
  \bibfield  {author} {\bibinfo {author} {\bibfnamefont {G.}~\bibnamefont
  {Mangano}}, \bibinfo {author} {\bibfnamefont {G.}~\bibnamefont {Miele}},
  \bibinfo {author} {\bibfnamefont {S.}~\bibnamefont {Pastor}}, \bibinfo
  {author} {\bibfnamefont {T.}~\bibnamefont {Pinto}}, \bibinfo {author}
  {\bibfnamefont {O.}~\bibnamefont {Pisanti}}, \ and\ \bibinfo {author}
  {\bibfnamefont {P.~D.}\ \bibnamefont {Serpico}},\ }\href {\doibase
  10.1016/j.nuclphysb.2005.09.041} {\bibfield  {journal} {\bibinfo  {journal}
  {Nucl. Phys. B}\ }\textbf {\bibinfo {volume} {729}},\ \bibinfo {pages} {221}
  (\bibinfo {year} {2005})},\ \Eprint {http://arxiv.org/abs/hep-ph/0506164}
  {arXiv:hep-ph/0506164} \BibitemShut {NoStop}%
\bibitem [{\citenamefont {de~Salas}\ and\ \citenamefont
  {Pastor}(2016)}]{deSalas:2016ztq}%
  \BibitemOpen
  \bibfield  {author} {\bibinfo {author} {\bibfnamefont {P.~F.}\ \bibnamefont
  {de~Salas}}\ and\ \bibinfo {author} {\bibfnamefont {S.}~\bibnamefont
  {Pastor}},\ }\href {\doibase 10.1088/1475-7516/2016/07/051} {\bibfield
  {journal} {\bibinfo  {journal} {JCAP}\ }\textbf {\bibinfo {volume} {07}},\
  \bibinfo {pages} {051} (\bibinfo {year} {2016})},\ \Eprint
  {http://arxiv.org/abs/1606.06986} {arXiv:1606.06986 [hep-ph]} \BibitemShut
  {NoStop}%
\bibitem [{\citenamefont {Akita}\ and\ \citenamefont
  {Yamaguchi}(2020)}]{Akita:2020szl}%
  \BibitemOpen
  \bibfield  {author} {\bibinfo {author} {\bibfnamefont {K.}~\bibnamefont
  {Akita}}\ and\ \bibinfo {author} {\bibfnamefont {M.}~\bibnamefont
  {Yamaguchi}},\ }\href {\doibase 10.1088/1475-7516/2020/08/012} {\bibfield
  {journal} {\bibinfo  {journal} {JCAP}\ }\textbf {\bibinfo {volume} {08}},\
  \bibinfo {pages} {012} (\bibinfo {year} {2020})},\ \Eprint
  {http://arxiv.org/abs/2005.07047} {arXiv:2005.07047 [hep-ph]} \BibitemShut
  {NoStop}%
\bibitem [{\citenamefont {Froustey}\ \emph {et~al.}(2020)\citenamefont
  {Froustey}, \citenamefont {Pitrou},\ and\ \citenamefont
  {Volpe}}]{Froustey:2020mcq}%
  \BibitemOpen
  \bibfield  {author} {\bibinfo {author} {\bibfnamefont {J.}~\bibnamefont
  {Froustey}}, \bibinfo {author} {\bibfnamefont {C.}~\bibnamefont {Pitrou}}, \
  and\ \bibinfo {author} {\bibfnamefont {M.~C.}\ \bibnamefont {Volpe}},\ }\href
  {\doibase 10.1088/1475-7516/2020/12/015} {\bibfield  {journal} {\bibinfo
  {journal} {JCAP}\ }\textbf {\bibinfo {volume} {12}},\ \bibinfo {pages} {015}
  (\bibinfo {year} {2020})},\ \Eprint {http://arxiv.org/abs/2008.01074}
  {arXiv:2008.01074 [hep-ph]} \BibitemShut {NoStop}%
\bibitem [{\citenamefont {Bennett}\ \emph {et~al.}(2020)\citenamefont
  {Bennett}, \citenamefont {Buldgen}, \citenamefont {de~Salas}, \citenamefont
  {Drewes}, \citenamefont {Gariazzo}, \citenamefont {Pastor},\ and\
  \citenamefont {Wong}}]{Bennett:2020zkv}%
  \BibitemOpen
  \bibfield  {author} {\bibinfo {author} {\bibfnamefont {J.~J.}\ \bibnamefont
  {Bennett}}, \bibinfo {author} {\bibfnamefont {G.}~\bibnamefont {Buldgen}},
  \bibinfo {author} {\bibfnamefont {P.~F.}\ \bibnamefont {de~Salas}}, \bibinfo
  {author} {\bibfnamefont {M.}~\bibnamefont {Drewes}}, \bibinfo {author}
  {\bibfnamefont {S.}~\bibnamefont {Gariazzo}}, \bibinfo {author}
  {\bibfnamefont {S.}~\bibnamefont {Pastor}}, \ and\ \bibinfo {author}
  {\bibfnamefont {Y.~Y.~Y.}\ \bibnamefont {Wong}},\ }\href@noop {} {\enquote
  {\bibinfo {title} {{Towards a precision calculation of $N_{\rm eff}$ in the
  Standard Model II: Neutrino decoupling in the presence of flavour
  oscillations and finite-temperature QED}},}\ } (\bibinfo {year}
  {2020})\BibitemShut {NoStop}%
\bibitem [{\citenamefont {Archidiacono}\ \emph {et~al.}(2011)\citenamefont
  {Archidiacono}, \citenamefont {Calabrese},\ and\ \citenamefont
  {Melchiorri}}]{Archidiacono:2011gq}%
  \BibitemOpen
  \bibfield  {author} {\bibinfo {author} {\bibfnamefont {M.}~\bibnamefont
  {Archidiacono}}, \bibinfo {author} {\bibfnamefont {E.}~\bibnamefont
  {Calabrese}}, \ and\ \bibinfo {author} {\bibfnamefont {A.}~\bibnamefont
  {Melchiorri}},\ }\href {\doibase 10.1103/PhysRevD.84.123008} {\bibfield
  {journal} {\bibinfo  {journal} {Phys. Rev. D}\ }\textbf {\bibinfo {volume}
  {84}},\ \bibinfo {pages} {123008} (\bibinfo {year} {2011})},\ \Eprint
  {http://arxiv.org/abs/1109.2767} {arXiv:1109.2767 [astro-ph.CO]} \BibitemShut
  {NoStop}%
\bibitem [{\citenamefont {Di~Valentino}\ \emph {et~al.}(2012)\citenamefont
  {Di~Valentino}, \citenamefont {Lattanzi}, \citenamefont {Mangano},
  \citenamefont {Melchiorri},\ and\ \citenamefont
  {Serpico}}]{DiValentino:2011sv}%
  \BibitemOpen
  \bibfield  {author} {\bibinfo {author} {\bibfnamefont {E.}~\bibnamefont
  {Di~Valentino}}, \bibinfo {author} {\bibfnamefont {M.}~\bibnamefont
  {Lattanzi}}, \bibinfo {author} {\bibfnamefont {G.}~\bibnamefont {Mangano}},
  \bibinfo {author} {\bibfnamefont {A.}~\bibnamefont {Melchiorri}}, \ and\
  \bibinfo {author} {\bibfnamefont {P.}~\bibnamefont {Serpico}},\ }\href
  {\doibase 10.1103/PhysRevD.85.043511} {\bibfield  {journal} {\bibinfo
  {journal} {Phys. Rev. D}\ }\textbf {\bibinfo {volume} {85}},\ \bibinfo
  {pages} {043511} (\bibinfo {year} {2012})},\ \Eprint
  {http://arxiv.org/abs/1111.3810} {arXiv:1111.3810 [astro-ph.CO]} \BibitemShut
  {NoStop}%
\bibitem [{\citenamefont {Di~Valentino}\ \emph {et~al.}(2013)\citenamefont
  {Di~Valentino}, \citenamefont {Melchiorri},\ and\ \citenamefont
  {Mena}}]{DiValentino:2013qma}%
  \BibitemOpen
  \bibfield  {author} {\bibinfo {author} {\bibfnamefont {E.}~\bibnamefont
  {Di~Valentino}}, \bibinfo {author} {\bibfnamefont {A.}~\bibnamefont
  {Melchiorri}}, \ and\ \bibinfo {author} {\bibfnamefont {O.}~\bibnamefont
  {Mena}},\ }\href {\doibase 10.1088/1475-7516/2013/11/018} {\bibfield
  {journal} {\bibinfo  {journal} {JCAP}\ }\textbf {\bibinfo {volume} {11}},\
  \bibinfo {pages} {018} (\bibinfo {year} {2013})},\ \Eprint
  {http://arxiv.org/abs/1304.5981} {arXiv:1304.5981 [astro-ph.CO]} \BibitemShut
  {NoStop}%
\bibitem [{\citenamefont {Di~Valentino}\ \emph {et~al.}(2016)\citenamefont
  {Di~Valentino}, \citenamefont {Giusarma}, \citenamefont {Lattanzi},
  \citenamefont {Mena}, \citenamefont {Melchiorri},\ and\ \citenamefont
  {Silk}}]{DiValentino:2015wba}%
  \BibitemOpen
  \bibfield  {author} {\bibinfo {author} {\bibfnamefont {E.}~\bibnamefont
  {Di~Valentino}}, \bibinfo {author} {\bibfnamefont {E.}~\bibnamefont
  {Giusarma}}, \bibinfo {author} {\bibfnamefont {M.}~\bibnamefont {Lattanzi}},
  \bibinfo {author} {\bibfnamefont {O.}~\bibnamefont {Mena}}, \bibinfo {author}
  {\bibfnamefont {A.}~\bibnamefont {Melchiorri}}, \ and\ \bibinfo {author}
  {\bibfnamefont {J.}~\bibnamefont {Silk}},\ }\href {\doibase
  10.1016/j.physletb.2015.11.025} {\bibfield  {journal} {\bibinfo  {journal}
  {Phys. Lett. B}\ }\textbf {\bibinfo {volume} {752}},\ \bibinfo {pages} {182}
  (\bibinfo {year} {2016})},\ \Eprint {http://arxiv.org/abs/1507.08665}
  {arXiv:1507.08665 [astro-ph.CO]} \BibitemShut {NoStop}%
\bibitem [{\citenamefont {Giar\`e}\ \emph
  {et~al.}(2021{\natexlab{a}})\citenamefont {Giar\`e}, \citenamefont
  {Di~Valentino}, \citenamefont {Melchiorri},\ and\ \citenamefont
  {Mena}}]{Giare:2020vzo}%
  \BibitemOpen
  \bibfield  {author} {\bibinfo {author} {\bibfnamefont {W.}~\bibnamefont
  {Giar\`e}}, \bibinfo {author} {\bibfnamefont {E.}~\bibnamefont
  {Di~Valentino}}, \bibinfo {author} {\bibfnamefont {A.}~\bibnamefont
  {Melchiorri}}, \ and\ \bibinfo {author} {\bibfnamefont {O.}~\bibnamefont
  {Mena}},\ }\href {\doibase 10.1093/mnras/stab1442} {\bibfield  {journal}
  {\bibinfo  {journal} {Mon. Not. Roy. Astron. Soc.}\ }\textbf {\bibinfo
  {volume} {505}},\ \bibinfo {pages} {2703} (\bibinfo {year}
  {2021}{\natexlab{a}})},\ \Eprint {http://arxiv.org/abs/2011.14704}
  {arXiv:2011.14704 [astro-ph.CO]} \BibitemShut {NoStop}%
\bibitem [{\citenamefont {Giar\`e}\ \emph
  {et~al.}(2021{\natexlab{b}})\citenamefont {Giar\`e}, \citenamefont {Renzi},
  \citenamefont {Melchiorri}, \citenamefont {Mena},\ and\ \citenamefont
  {Di~Valentino}}]{Giare:2021cqr}%
  \BibitemOpen
  \bibfield  {author} {\bibinfo {author} {\bibfnamefont {W.}~\bibnamefont
  {Giar\`e}}, \bibinfo {author} {\bibfnamefont {F.}~\bibnamefont {Renzi}},
  \bibinfo {author} {\bibfnamefont {A.}~\bibnamefont {Melchiorri}}, \bibinfo
  {author} {\bibfnamefont {O.}~\bibnamefont {Mena}}, \ and\ \bibinfo {author}
  {\bibfnamefont {E.}~\bibnamefont {Di~Valentino}},\ }\href {\doibase
  10.1093/mnras/stac126} {\bibfield  {journal} {\bibinfo  {journal} {Mon. Not.
  Roy. Astron. Soc.}\ }\textbf {\bibinfo {volume} {511}},\ \bibinfo {pages}
  {1373} (\bibinfo {year} {2021}{\natexlab{b}})},\ \Eprint
  {http://arxiv.org/abs/2110.00340} {arXiv:2110.00340 [astro-ph.CO]}
  \BibitemShut {NoStop}%
\bibitem [{\citenamefont {D'Eramo}\ \emph {et~al.}(2022)\citenamefont
  {D'Eramo}, \citenamefont {Di~Valentino}, \citenamefont {Giar\`e},
  \citenamefont {Hajkarim}, \citenamefont {Melchiorri}, \citenamefont {Mena},
  \citenamefont {Renzi},\ and\ \citenamefont {Yun}}]{DEramo:2022nvb}%
  \BibitemOpen
  \bibfield  {author} {\bibinfo {author} {\bibfnamefont {F.}~\bibnamefont
  {D'Eramo}}, \bibinfo {author} {\bibfnamefont {E.}~\bibnamefont
  {Di~Valentino}}, \bibinfo {author} {\bibfnamefont {W.}~\bibnamefont
  {Giar\`e}}, \bibinfo {author} {\bibfnamefont {F.}~\bibnamefont {Hajkarim}},
  \bibinfo {author} {\bibfnamefont {A.}~\bibnamefont {Melchiorri}}, \bibinfo
  {author} {\bibfnamefont {O.}~\bibnamefont {Mena}}, \bibinfo {author}
  {\bibfnamefont {F.}~\bibnamefont {Renzi}}, \ and\ \bibinfo {author}
  {\bibfnamefont {S.}~\bibnamefont {Yun}},\ }\href {\doibase
  10.1088/1475-7516/2022/09/022} {\bibfield  {journal} {\bibinfo  {journal}
  {JCAP}\ }\textbf {\bibinfo {volume} {09}},\ \bibinfo {pages} {022} (\bibinfo
  {year} {2022})},\ \Eprint {http://arxiv.org/abs/2205.07849} {arXiv:2205.07849
  [astro-ph.CO]} \BibitemShut {NoStop}%
\bibitem [{\citenamefont {Baumann}\ \emph {et~al.}(2016)\citenamefont
  {Baumann}, \citenamefont {Green},\ and\ \citenamefont
  {Wallisch}}]{Baumann:2016wac}%
  \BibitemOpen
  \bibfield  {author} {\bibinfo {author} {\bibfnamefont {D.}~\bibnamefont
  {Baumann}}, \bibinfo {author} {\bibfnamefont {D.}~\bibnamefont {Green}}, \
  and\ \bibinfo {author} {\bibfnamefont {B.}~\bibnamefont {Wallisch}},\ }\href
  {\doibase 10.1103/PhysRevLett.117.171301} {\bibfield  {journal} {\bibinfo
  {journal} {Phys. Rev. Lett.}\ }\textbf {\bibinfo {volume} {117}},\ \bibinfo
  {pages} {171301} (\bibinfo {year} {2016})},\ \Eprint
  {http://arxiv.org/abs/1604.08614} {arXiv:1604.08614 [astro-ph.CO]}
  \BibitemShut {NoStop}%
\bibitem [{\citenamefont {Gariazzo}\ \emph {et~al.}(2016)\citenamefont
  {Gariazzo}, \citenamefont {Giunti}, \citenamefont {Laveder}, \citenamefont
  {Li},\ and\ \citenamefont {Zavanin}}]{Gariazzo:2015rra}%
  \BibitemOpen
  \bibfield  {author} {\bibinfo {author} {\bibfnamefont {S.}~\bibnamefont
  {Gariazzo}}, \bibinfo {author} {\bibfnamefont {C.}~\bibnamefont {Giunti}},
  \bibinfo {author} {\bibfnamefont {M.}~\bibnamefont {Laveder}}, \bibinfo
  {author} {\bibfnamefont {Y.~F.}\ \bibnamefont {Li}}, \ and\ \bibinfo {author}
  {\bibfnamefont {E.~M.}\ \bibnamefont {Zavanin}},\ }\href {\doibase
  10.1088/0954-3899/43/3/033001} {\bibfield  {journal} {\bibinfo  {journal} {J.
  Phys. G}\ }\textbf {\bibinfo {volume} {43}},\ \bibinfo {pages} {033001}
  (\bibinfo {year} {2016})},\ \Eprint {http://arxiv.org/abs/1507.08204}
  {arXiv:1507.08204 [hep-ph]} \BibitemShut {NoStop}%
\bibitem [{\citenamefont {Archidiacono}\ and\ \citenamefont
  {Gariazzo}(2022)}]{Archidiacono:2022ich}%
  \BibitemOpen
  \bibfield  {author} {\bibinfo {author} {\bibfnamefont {M.}~\bibnamefont
  {Archidiacono}}\ and\ \bibinfo {author} {\bibfnamefont {S.}~\bibnamefont
  {Gariazzo}},\ }\href@noop {} {\enquote {\bibinfo {title} {{Two sides of the
  same coin: sterile neutrinos and dark radiation. Status and perspectives}},}\
  } (\bibinfo {year} {2022}),\ \Eprint {http://arxiv.org/abs/2201.10319}
  {arXiv:2201.10319 [hep-ph]} \BibitemShut {NoStop}%
\bibitem [{\citenamefont {An}\ \emph {et~al.}(2022)\citenamefont {An},
  \citenamefont {Gluscevic}, \citenamefont {Calabrese},\ and\ \citenamefont
  {Hill}}]{An:2022sva}%
  \BibitemOpen
  \bibfield  {author} {\bibinfo {author} {\bibfnamefont {R.}~\bibnamefont
  {An}}, \bibinfo {author} {\bibfnamefont {V.}~\bibnamefont {Gluscevic}},
  \bibinfo {author} {\bibfnamefont {E.}~\bibnamefont {Calabrese}}, \ and\
  \bibinfo {author} {\bibfnamefont {J.~C.}\ \bibnamefont {Hill}},\ }\href@noop
  {} {\enquote {\bibinfo {title} {{What does cosmology tell us about the mass
  of thermal-relic dark matter?}}}\ } (\bibinfo {year} {2022}),\ \Eprint
  {http://arxiv.org/abs/2202.03515} {arXiv:2202.03515 [astro-ph.CO]}
  \BibitemShut {NoStop}%
\bibitem [{\citenamefont {Abitbol}\ \emph {et~al.}(2019)\citenamefont {Abitbol}
  \emph {et~al.}}]{SimonsObservatory:2019qwx}%
  \BibitemOpen
  \bibfield  {author} {\bibinfo {author} {\bibfnamefont {M.~H.}\ \bibnamefont
  {Abitbol}} \emph {et~al.} (\bibinfo {collaboration} {Simons Observatory}),\
  }\href@noop {} {\bibfield  {journal} {\bibinfo  {journal} {Bull. Am. Astron.
  Soc.}\ }\textbf {\bibinfo {volume} {51}},\ \bibinfo {pages} {147} (\bibinfo
  {year} {2019})},\ \Eprint {http://arxiv.org/abs/1907.08284} {arXiv:1907.08284
  [astro-ph.IM]} \BibitemShut {NoStop}%
\bibitem [{\citenamefont {Di~Valentino}\ \emph
  {et~al.}(2022{\natexlab{c}})\citenamefont {Di~Valentino}, \citenamefont
  {Giar\`e}, \citenamefont {Melchiorri},\ and\ \citenamefont
  {Silk}}]{DiValentino:2022rdg}%
  \BibitemOpen
  \bibfield  {author} {\bibinfo {author} {\bibfnamefont {E.}~\bibnamefont
  {Di~Valentino}}, \bibinfo {author} {\bibfnamefont {W.}~\bibnamefont
  {Giar\`e}}, \bibinfo {author} {\bibfnamefont {A.}~\bibnamefont {Melchiorri}},
  \ and\ \bibinfo {author} {\bibfnamefont {J.}~\bibnamefont {Silk}},\
  }\href@noop {} {\  (\bibinfo {year} {2022}{\natexlab{c}})},\ \Eprint
  {http://arxiv.org/abs/2209.14054} {arXiv:2209.14054 [astro-ph.CO]}
  \BibitemShut {NoStop}%
\bibitem [{\citenamefont {Keeley}\ and\ \citenamefont
  {Shafieloo}(2022)}]{Keeley:2022ojz}%
  \BibitemOpen
  \bibfield  {author} {\bibinfo {author} {\bibfnamefont {R.~E.}\ \bibnamefont
  {Keeley}}\ and\ \bibinfo {author} {\bibfnamefont {A.}~\bibnamefont
  {Shafieloo}},\ }\href@noop {} {\  (\bibinfo {year} {2022})},\ \Eprint
  {http://arxiv.org/abs/2206.08440} {arXiv:2206.08440 [astro-ph.CO]}
  \BibitemShut {NoStop}%
\bibitem [{\citenamefont {Arendse}\ \emph {et~al.}(2020)\citenamefont {Arendse}
  \emph {et~al.}}]{Arendse:2019hev}%
  \BibitemOpen
  \bibfield  {author} {\bibinfo {author} {\bibfnamefont {N.}~\bibnamefont
  {Arendse}} \emph {et~al.},\ }\href {\doibase 10.1051/0004-6361/201936720}
  {\bibfield  {journal} {\bibinfo  {journal} {Astron. Astrophys.}\ }\textbf
  {\bibinfo {volume} {639}},\ \bibinfo {pages} {A57} (\bibinfo {year}
  {2020})},\ \Eprint {http://arxiv.org/abs/1909.07986} {arXiv:1909.07986
  [astro-ph.CO]} \BibitemShut {NoStop}%
\bibitem [{\citenamefont {Di~Valentino}\ \emph
  {et~al.}(2021{\natexlab{c}})\citenamefont {Di~Valentino}, \citenamefont
  {Mukherjee},\ and\ \citenamefont {Sen}}]{DiValentino:2020naf}%
  \BibitemOpen
  \bibfield  {author} {\bibinfo {author} {\bibfnamefont {E.}~\bibnamefont
  {Di~Valentino}}, \bibinfo {author} {\bibfnamefont {A.}~\bibnamefont
  {Mukherjee}}, \ and\ \bibinfo {author} {\bibfnamefont {A.~A.}\ \bibnamefont
  {Sen}},\ }\href {\doibase 10.3390/e23040404} {\bibfield  {journal} {\bibinfo
  {journal} {Entropy}\ }\textbf {\bibinfo {volume} {23}},\ \bibinfo {pages}
  {404} (\bibinfo {year} {2021}{\natexlab{c}})},\ \Eprint
  {http://arxiv.org/abs/2005.12587} {arXiv:2005.12587 [astro-ph.CO]}
  \BibitemShut {NoStop}%
\bibitem [{\citenamefont {Chudaykin}\ \emph {et~al.}(2022)\citenamefont
  {Chudaykin}, \citenamefont {Gorbunov},\ and\ \citenamefont
  {Nedelko}}]{Chudaykin:2022rnl}%
  \BibitemOpen
  \bibfield  {author} {\bibinfo {author} {\bibfnamefont {A.}~\bibnamefont
  {Chudaykin}}, \bibinfo {author} {\bibfnamefont {D.}~\bibnamefont {Gorbunov}},
  \ and\ \bibinfo {author} {\bibfnamefont {N.}~\bibnamefont {Nedelko}},\
  }\href@noop {} {\  (\bibinfo {year} {2022})},\ \Eprint
  {http://arxiv.org/abs/2203.03666} {arXiv:2203.03666 [astro-ph.CO]}
  \BibitemShut {NoStop}%
\bibitem [{\citenamefont {Abell}\ \emph {et~al.}(2009)\citenamefont {Abell}
  \emph {et~al.}}]{LSSTScience:2009jmu}%
  \BibitemOpen
  \bibfield  {author} {\bibinfo {author} {\bibfnamefont {P.~A.}\ \bibnamefont
  {Abell}} \emph {et~al.} (\bibinfo {collaboration} {LSST Science, LSST
  Project}),\ }\href@noop {} {\  (\bibinfo {year} {2009})},\ \Eprint
  {http://arxiv.org/abs/0912.0201} {arXiv:0912.0201 [astro-ph.IM]} \BibitemShut
  {NoStop}%
\bibitem [{\citenamefont {Laureijs}\ \emph {et~al.}(2011)\citenamefont
  {Laureijs} \emph {et~al.}}]{EUCLID:2011zbd}%
  \BibitemOpen
  \bibfield  {author} {\bibinfo {author} {\bibfnamefont {R.}~\bibnamefont
  {Laureijs}} \emph {et~al.} (\bibinfo {collaboration} {EUCLID}),\ }\href@noop
  {} {\  (\bibinfo {year} {2011})},\ \Eprint {http://arxiv.org/abs/1110.3193}
  {arXiv:1110.3193 [astro-ph.CO]} \BibitemShut {NoStop}%
\bibitem [{\citenamefont {Levi}\ \emph {et~al.}(2013)\citenamefont {Levi} \emph
  {et~al.}}]{DESI:2013agm}%
  \BibitemOpen
  \bibfield  {author} {\bibinfo {author} {\bibfnamefont {M.}~\bibnamefont
  {Levi}} \emph {et~al.} (\bibinfo {collaboration} {DESI}),\ }\href@noop {} {\
  (\bibinfo {year} {2013})},\ \Eprint {http://arxiv.org/abs/1308.0847}
  {arXiv:1308.0847 [astro-ph.CO]} \BibitemShut {NoStop}%
\bibitem [{\citenamefont {Ade}\ \emph {et~al.}(2019)\citenamefont {Ade} \emph
  {et~al.}}]{SimonsObservatory:2018koc}%
  \BibitemOpen
  \bibfield  {author} {\bibinfo {author} {\bibfnamefont {P.}~\bibnamefont
  {Ade}} \emph {et~al.} (\bibinfo {collaboration} {Simons Observatory}),\
  }\href {\doibase 10.1088/1475-7516/2019/02/056} {\bibfield  {journal}
  {\bibinfo  {journal} {JCAP}\ }\textbf {\bibinfo {volume} {02}},\ \bibinfo
  {pages} {056} (\bibinfo {year} {2019})},\ \Eprint
  {http://arxiv.org/abs/1808.07445} {arXiv:1808.07445 [astro-ph.CO]}
  \BibitemShut {NoStop}%
\end{thebibliography}%

\end{document}